\documentclass[fleqn,usenatbib]{mnras}

\usepackage{newtxtext,newtxmath}

\usepackage[T1]{fontenc}
\usepackage{ae,aecompl}

\usepackage{graphicx}	
\usepackage{amsmath}	
\usepackage{amssymb}	
\usepackage{booktabs}
\usepackage{subfig}
\usepackage{enumitem}
\usepackage{lscape}
\usepackage{adjustbox}
\usepackage{rotating}
\usepackage{multicol}
\usepackage[dvipsnames]{xcolor}
\usepackage[none]{hyphenat}
\usepackage{pgfplotstable,filecontents}
\usepackage{ulem}
\pgfplotsset{compat=1.9}

\title[Hardness of ionizing radiation fields]{Hardness of ionizing radiation fields in MaNGA star-forming galaxies}

\author[Kumari et al.]{
Nimisha Kumari$^{1}$\thanks{E-mail: kumari@stsci.edu(NK)},
Ricardo Amor\'in$^{2,3}$\thanks{E-mail: ricardo.amorin@userena.cl(RA)},
Enrique P{\'e}rez-Montero$^{4}$,
Jose V\'ilchez$^{4}$
\and
Roberto Maiolino$^{5,6}$\\
$^{1}$AURA for the European Space Agency (ESA), Space Telescope Science Institute, 3700 San Martin Drive, Baltimore, MD 21218, USA\\
$^{2}$Departamento de Astronom\'ia, Universidad de La Serena, Av. Juan Cisternas 1200 Norte, La Serena, Chile\\
$^{3}$Instituto de Investigaci\'on Multidisciplinar en Ciencia y Tecnolog\'ia, Universidad de La Serena, Ra\'ul Bitr\'an 1305, La Serena, Chile\\
$^{4}$Instituto de Astrof\'isica de Andaluc\'ia, CSIC, Apartado de correos 3004, E-18080 Granada, Spain\\
$^{5}$Kavli Institute for Cosmology, University of Cambridge, Cambridge CB3 0HA, UK\\
$^{6}$Cavendish Laboratory, University of Cambridge, Cambridge CB3 0HE, UK\\}

\date{Accepted XXX. Received YYY; in original form ZZZ}

\pubyear{2018}

\begin{document}

\label{firstpage}
\pagerange{\pageref{firstpage}--\pageref{lastpage}}
\maketitle

\begin{abstract}

We investigate radiation hardness within a representative sample of 67 nearby  (0.02~$\lesssim~$z$~\lesssim$~0.06) star-forming (SF) galaxies using the integral field spectroscopic data from the MaNGA survey. The softness parameter $\eta$ = $\frac{O^{+}/O^{2+}}{S^{+}/S^{2+}}$ is sensitive to the spectral energy distribution of the ionizing radiation. We study $\eta$ via the observable quantity $\eta\prime$ ($=\frac{[O\textsc{ii}]/[O\textsc{iii}]}{[S\textsc{ii}][S\textsc{iii}]}$)  We analyse the relation between radiation hardness (traced by $\eta$ and $\eta\prime$) and diagnostics sensitive to gas-phase metallicity, electron temperature, density, ionization parameter, effective temperature and age of ionizing populations. It is evident that low metallicity is accompanied by low log $\eta\prime$, i.e. hard radiation field. No direct relation is found between radiation hardness and other nebular parameters though such relations can not be ruled out. We provide empirical relations between log $\rm\eta$ and strong emission line ratios N$_2$, O$_3$N$_2$ and Ar$_3$O$_3$ which will allow future studies of radiation hardness in  SF galaxies where weak auroral lines are undetected. We compare the variation of [O \textsc{iii}]/[O \textsc{ii}] and [S~\textsc{iii}]/[S~\textsc{ii}] for MaNGA data with SF galaxies and H \textsc{ii} regions within spiral galaxies from literature, and find that the similarity and differences between different data set is mainly due to the metallicity. We find that predictions from photoionizaion models considering young and evolved stellar populations as ionizing sources in good agreement with the MaNGA data.  This comparison also suggests that hard radiation fields from hot and old low-mass stars within or around SF  regions might significantly contribute to the observed $\eta$ values. 

\end{abstract}




\section{Introduction}
\label{sect:intro}

\indent An in-depth analysis of star-forming galaxies requires the characterisation of the interstellar medium (ISM) in galaxies, which constitutes matter and radiation field. While the matter is composed of gas and dust, the radiation is produced by both stars and the interstellar matter. The hardness and intensity of radiation field are the fundamental parameters which affect the overall shape of the spectrum of a region consisting of stars and ionized gas. The hardness of ionizing radiation field has been studied via different definitions in different works \citep[e.g.][]{Vilchez1988, Morisset2016, Nakajima2018, Perez-Montero2020}. More fundamentally, radiation hardness is the shape of the spectral energy distribution (SED) or the slope of the extreme ultraviolet spectrum \citep[see e.g.][]{Kewley2015, Nakajima2018} and hence can be probed by the effective temperature (T$\rm_{eff}$) of stars producing ionizing photons \citep[e.g.][]{Vilchez1988, Steidel2014}. The ratio of photons capable of ionizing neutral hydrogen (H$^0$) and helium (He$^0$), i.e. Q$_{0/1}$ is also used to probe radiation hardness \citep{Morisset2016}. In photoionization models, it is assumed that radiation hardness (parameterized by T$\rm_{eff}$ or stellar population distribution and age), ionization parameter $\mathcal{U}$\footnote{$\mathcal{U} = \frac{Q}{4\pi r^2 n_e}$ for a spherical H \textsc{ii} region, where Q is the rate at which stars produce Lyman continuum photons, r is the distance from the central star or stellar clusters and n$_e$ is the volume density of neutral or ionized hydrogen.} and chemical abundance of ionized gas are independent quantities which affect the output spectrum of a photoionized nebulae. \citet{Steidel2014} states that the more generalized form of ionization parameter $\Gamma$\footnote{$\Gamma = \frac{n_{\gamma}}{n_e} \approx \frac{n_{\gamma}}{n_H}$ where n$_H$ is the number density of hydrogen atoms and n$_\gamma$ is the equivalent density hydrogen ionizing photons.}, depends on radiation hardness, because the number of hydrogen ionizing photons can be changed by changing the shape or intensity of ionizing radiation \citep[][]{Kewley2006}. Hence, the hardness of radiation field may be related to various parameters such as initial mass function, age of the stellar population, equivalent effective temperature, ionization parameter and metallicity, and can be probed via emission lines emanating from the ionized gas component of the ISM.

\indent Collisionally-excited emission lines (CELs) in optical wavelength range (e.g. [O \textsc{ii}] and [O \textsc{iii}]) are widely used to study the properties of the ISM. However, only a few works have also explored the use of near-infrared (NIR) CELs such as [S \textsc{iii}] $\lambda\lambda$9069, 9532 \citep[see e.g.][]{Vilchez1988, Vilchez1996, Diaz2000, Stasinska2006, Perez-Montero2005, Perez-Montero2009b, Fernandez2018, Mingozzi2020, Perez-Montero2020} mainly due to the following two reasons. First, only a few spectrographs used in galaxy surveys include useful wavelengths above 9000\AA. Secondly, the NIR wavelength range is strongly affected by telluric absorption and sky lines thus complicating the analysis of these sulphur emission lines. Nonetheless, these lines have enormous potential for determining the characteristic properties of ionized gas from which they emanate. For example, the emission line ratios involving  [S \textsc{iii}] $\lambda$9069, such as S$_{23}$ = [S \textsc{iii}] $\lambda\lambda$9069, 9532 + [S~\textsc{ii}]~$\lambda\lambda$~6717,~6731)/H$\beta$ and S$_3$O$_3$ = [S \textsc{iii}] $\lambda\lambda$9069, 9532/[O \textsc{iii}] $\lambda\lambda$4959, 5007 have been proposed as metallicity diagnostics \citep[see e.g.][]{Diaz2000, Perez-Montero2005, Stasinska2006}. Similarly, the emission line ratios involving sulphur lines, [S~\textsc{ii}] ~$\lambda\lambda$6717, 6731/H$\alpha$ versus  [S \textsc{iii}]~$\lambda\lambda$9069, 9532/H$\alpha$ are also proposed to identify the ionization mechanisms in the Low Ionization Nuclear Emission Regions \citep[LINERs,][]{Diaz1985}. The photoionization models have shown that the line ratio [S~\textsc{ii}]~$\lambda\lambda$~6717,~6731/{[}S \textsc{iii}{]}~ $\lambda\lambda$9069,~9532 is a good indicator of ionization parameter \citep{Mathis1985, Diaz1991, Morisset2016, Kewley2019}, which is critical in understanding the state of plasma in an H \textsc{ii} region. \cite{Mathis1982, Mathis1985} further pointed out the importance of sulphur lines such as [S~\textsc{iii}] $\lambda$9069, 9532 to determine the relative temperatures of hot stars within nebulae by comparing observations with nebular models on the S$^{+}/S^{++}$ versus O$^+$/O diagram. \citet{Vilchez1988} modified the procedure from \cite{Mathis1982} and introduced the so-called softness parameter

\begin{equation}
\eta = \frac{O^+/O^{2+}}{S^+/S^{2+}}.
\label{eq:eta}
\end{equation}

$\rm\eta$ is sensitive to the spectral energy distribution of the ionizing radiation because of the large difference in the ionization potentials of O$^{+}$ (35.1 eV) and S$^{+}$ (23.2 eV) \citep{Bresolin1999, Perez-Montero2009b}. However, $\rm\eta$ is not a directly observable quantity 
and can be studied via the observable line ratio $\rm\eta\prime$ which, in optical, is defined as

\begin{equation}
	\eta\prime = \frac{[O \textsc{ii}] \lambda 3727/[O \textsc{iii}] \lambda\lambda4959, 5007}{[S \textsc{ii}] \lambda\lambda6717, 6731/[S \textsc{iii}] \lambda\lambda 9069, 9532}.
	\label{eq:eta prime}
\end{equation}

The mid-infrared fine structure lines of Ne, Ar and S are also used to determine $\rm\eta$ \citep[see e.g.,][]{Martin-Hernandez2002, Morisset2004, Perez-Montero2009b}. The softness parameter $\rm\eta$ is related $\rm\eta\prime$ and electron temperature (T$_e$) as proposed by \citet{Vilchez1988}. The following revised relation is obtained using \textsc{pyneb} and its default values for atomic data: 

\begin{equation}
	\centering
	\rm log ~\eta = log ~\eta \prime + \frac{0.16}{t} + 0.22,
	\label{eq:eta tes}
\end{equation} 

\noindent where t = T$_e$([O \textsc{iii}])/10$^4$.

\indent The $\eta \prime$ parameter has been used to study the ionization structure and the relative hardness of the ionizing sources in the H \textsc{ii} regions within the Milky Way and  Magellanic Clouds \citep{Martin-Hernandez2002, Morisset2004}, in the star-forming galaxies \citep{Hagele2006, Kehrig2006, Hagele2008, Perez-Montero2020} and the radial variation of the hardness of the ionizing radiation of H~\textsc{ii} regions in the discs of spiral galaxies \citep{Perez-Montero2009b, Perez-Montero2019}. However, it is important to consider other variables while interpreting $\eta \prime$ as radiation hardness. For example, log $\rm\eta\prime$ is inversely proportional to equivalent effective temperature \citep[T$\rm_{eff}$][]{Kennicutt2000}, and log $\rm\eta$ can be expressed as a linear function of 1/T$\rm_{eff}$ \citep{Vilchez1988} for a blackbody spectrum. Similarly, log $\rm\eta$ and log $\rm\eta\prime$ may also be related to nebular parameters such as metallicity \citep{Morisset2004} and ionization parameter \citep{Perez-Montero2009b, Fernandez-Martin2017}.   
\indent Previous studies of $\rm\eta$ using long-slit or fibre spectra have been limited to a small number of local (z$\sim$0) galaxies \citep{Hagele2006, Hagele2008} due to the limitations of spectrographs to reach $\lambda >$ 9000~\AA. The data obtained from spectrographs like that of Sloan Digital Sky Survey (SDSS) typically cover 3800--9200\AA, which cover [S \textsc{iii}] $\lambda$9069 up to only a redshift of 0.01 and does not cover [O \textsc{ii}]$\lambda 3727$ required for studying $\eta\prime$. In addition, previous long-slit spectra lack spatially-resolved information. In comparison to global galaxy-scale analysis, a spatially-resolved investigation provides insight into the local environment within nebulae. Integral Field Spectroscopy  (IFS) is the best available technique to carry out such a study as it allows us to map various properties encoded in the emission lines emanating from the ionized gas component of the ISM within galaxies, thus facilitating studies of correlations between different  hardness of radiation fields and nebular and stellar properties at local scales. 
 \citet{Zinchenko2019} utilized the IFS data from the Calar Alto Legacy Integral Field Area \citep[CALIFA,][]{Sanchez2012} survey and performed an indirect study of radiation field hardness at local scales by analysing the relation between equivalent effective temperature (from [O \textsc{ii}] and [O \textsc{iii}] lines), ionization parameter and oxygen abundance. Since the [S \textsc{iii}] lines lie beyond the wavelength range of CALIFA, their lack thereof prevented this study to break the degeneracy between radiation hardness and the ionization parameter (see Equation \ref{eq:eta prime}). Moreover, the results of \citet{Zinchenko2019} are derived and hence applicable to a restrictive sample of H~\textsc{ii} regions within spiral galaxies. The metallicity estimates in their work are based on strong line methods rather than the robust direct T$_e$ method because the CALIFA survey is not sensitive enough to detect the weak auroral lines within high-metallicity environments.  
 
 

 \indent This work is the first spatially-resolved study of $\rm\eta$ and $\rm\eta\prime$ on a large sample of 67 star-forming galaxies, aimed at understanding the relation between various nebular parameters and the hardness of radiation field 
 at local scales. In this work, we  use data set from the MaNGA survey to address and overcome the issues faced by previous surveys and instruments. MaNGA is best-suited for the current analysis as its wavelength range is wide enough to cover the sulphur emission lines [S~\textsc{iii}]~$\lambda\lambda$9069,~9532 crucial to this study.  
 The MaNGA survey also allows us to include relatively low-metallicity star-forming galaxies which increases our odds to detect and map the auroral line [O \textsc{iii}] $\lambda$4363 enabling us to map T$_e$ and study its relation with radiation hardness at spatially-resolved scales. 
 Such a detailed study on radiation hardness within local star-forming galaxies exhibiting a wide range of ionization conditions, is imperative to understand various factors which regulate radiation hardness.



 \indent The paper is organised as follows. Section \ref{sect:data} describes the data set and the criteria for selecting sample galaxies from the MaNGA survey. In Section \ref{sect:results}, we focus on the relation between [O\textsc{iii}]/[O\textsc{ii}] and [S\textsc{iii}]/[S\textsc{ii}] and its dependence on several measurables related to gas-phase metallicity, age of stellar population and ionization parameter. 
 In Section \ref{sect:discussion}, we compare our results with the previous observations and existing photoionization models. We also discuss the relation between radiation hardness and helium lines in the handful of galaxies where He\textsc{ii} 4686 are detected. Section \ref{sect:summary} summarises our main results. Throughout this study, we use the following shorthand notation for the strong line ratios for a compact presentation:
 \setlength{\belowdisplayskip}{-1.5pt} \setlength{\belowdisplayshortskip}{-1.5pt}
 \setlength{\abovedisplayskip}{-1.5pt} \setlength{\abovedisplayshortskip}{-1.5pt}

 \begin{equation}
 	\rm N_2 = log([N \textsc{ii}]/H\alpha) = log ([N \textsc{ii}] \lambda 6584/H\alpha)
 	\label{eq:n2}
 \end{equation}

 \begin{multline}
 	\rm O_3N_2 = log(([O \textsc{iii}]/H\beta)/[N \textsc{ii}]/H\alpha)\\
 	\rm=log(([O \textsc{iii}] \lambda 5007/H\beta)/[N \textsc{ii}] \lambda 6584/H\alpha)
 	\label{eq:o3n2}
 \end{multline}

 \begin{multline}
 	\rm O_3O_2 = log([O \textsc{iii}]/[O \textsc{ii}])\\
 	\rm= log([O \textsc{iii}]\lambda\lambda 4959,5007/[O \textsc{ii}] \lambda 3727)
 	\label{eq:o3o2}
 \end{multline}

 \begin{multline}
 	\rm S_3S_2 = log([S \textsc{iii}]/[S \textsc{ii}]) \\
 	\rm = log([S \textsc{iii}] \lambda\lambda 9069,9532/[S \textsc{ii}] \lambda\lambda 6717, 6731)
 	\label{eq:s3s2}
 \end{multline}

 \begin{multline}
 \rm S_3O_3 = log([S \textsc{iii}]/[O \textsc{iii}]) \\
 \rm = log([S \textsc{iii}] \lambda\lambda 9069,9532/log([O \textsc{iii}]\lambda\lambda 4959,5007)
 \label{eq:s3o3}
 \end{multline}

 \begin{multline}
\rm Ar_3O_3 = log([Ar \textsc{iii}]/[O \textsc{iii}]) \\
\rm = log([Ar \textsc{iii}] \lambda\lambda 7135/log([O \textsc{iii}]\lambda\lambda 4959,5007)
\label{eq:ar3o3}
\end{multline}

 \begin{multline}
 \rm S_{23} = log(([S \textsc{iii}] + [S \textsc{ii}])/H\beta)  \\
 \rm = log(([S \textsc{iii}] \lambda\lambda 9069,9532 + [S \textsc{ii}] \lambda\lambda 6717, 6731)/H\beta )
 \label{eq:s32}
 \end{multline}

\begin{multline}
 \rm R_{23} = log(([O \textsc{iii}] + [O \textsc{ii}])/H\beta)  \\
 \rm = log(([O \textsc{iii}] \lambda\lambda 4959,5007 + [O \textsc{ii}] \lambda 3727)/H\beta )
 \label{eq:r32}
 \end{multline}

\indent

\indent In this paper, we adopt a standard cosmology assuming the parameters, H$_{0}$ = 67.3 $\pm$ 1.2 km s$^{-1}$ Mpc$^{-1}$ and $\Omega_{m}$ = 0.315$\pm$0.017, presented by \citet{Ade2014} and are consistent with \citet{Ade2016}.

\begin{table*}
	\centering
	\caption{General properties of sample MaNGA galaxies.}
	\label{tab:properties}
	\begin{tabular}{lcccccc}
			\toprule
		Plate-IFU & RA & Dec & z & log M$_{\star}$ & log SFR (H$\alpha$) & kpc/arcsec\\
		&(J2000)&(J2000)&& (M$_{\odot}$)& (M$_{\odot}$ yr$^{-1}$) & \\
		\midrule
		7495-6102$^{\star}$ & 204.51292 & 26.338177 & 0.0268 & 8.80 & -0.05 & 0.56 \\
		7975-1901 & 323.65747 & 11.421048 & 0.0227 & 8.95 & -0.09 & 0.47 \\
		7990-3703 & 262.09933 & 57.545418 & 0.0291 & 9.68 & 0.81 & 0.60 \\
		7992-6102 & 253.88911 & 63.242126 & 0.0228 & 9.36 & 0.28 & 0.48 \\
		8078-3703 & 42.387463 & -0.78446174 & 0.0239 & 9.28 & -0.35 & 0.50 \\
		8081-3704 & 49.821426 & -0.9696393 & 0.0540 & 9.85 & 0.97 & 1.09 \\
		8131-9101 & 112.57339 & 39.94194 & 0.0503 & 9.87 & 0.87 & 1.01 \\
		8132-3702 & 110.55611 & 42.18362 & 0.0446 & 9.63 & 0.21 & 0.91 \\
		8133-3704 & 112.51493 & 43.379227 & 0.0269 & 8.53 & -0.39 & 0.56 \\
		8135-3704 & 114.89731 & 37.751534 & 0.0307 & 9.36 & 0.28 & 0.64 \\
		8149-12701 & 120.26999 & 26.80237 & 0.0423 & 9.50 & 0.15 & 0.86 \\
		8156-3701 & 55.593178 & -0.5828109 & 0.0527 & 9.86 & 0.84 & 1.06 \\
		8241-6101$^{\dagger}$ & 127.04849 & 17.374716 & 0.0218 & 8.70 & -0.17 & 0.45 \\
		8243-9101 & 128.1783 & 52.416805 & 0.0435 & 9.54 & 0.50 & 0.89 \\
		8250-3703 & 139.73997 & 43.500603 & 0.0402 & 9.38 & 0.24 & 0.82 \\
		8250-6101 & 138.75304 & 42.024357 & 0.0281 & 10.05 & 0.98 & 0.58 \\
		8252-9102 & 145.54156 & 48.01285 & 0.0565 & 9.97 & 1.30 & 1.13 \\
		8257-3704$^{\star}$  & 165.55362 & 45.30387 & 0.0207 & 8.72 & -0.45 & 0.43 \\
		8258-3704 & 167.02502 & 43.894623 & 0.0585 & 9.87 & 1.00 & 1.17 \\
		8259-9101 & 178.3442 & 44.92035 & 0.0197 & 8.74 & -0.58 & 0.41 \\
		8261-12703 & 184.35774 & 46.566887 & 0.0240 & 9.31 & 0.18 & 0.50 \\
		8262-3701 & 183.57898 & 43.535275 & 0.0245 & 9.45 & -0.00 & 0.51 \\
		8313-1901$^{\dagger\star}$ & 240.28712 & 41.880753 & 0.0249 & 9.00 & 0.33 & 0.52 \\
		8320-12703 & 206.63098 & 23.122137 & 0.0305 & 9.45 & 0.14 & 0.63 \\
		8320-9101 & 206.31384 & 23.316532 & 0.0299 & 9.86 & 0.65 & 0.62 \\
		8325-12702 & 209.89511 & 47.14765 & 0.0424 & 9.33 & 0.38 & 0.86 \\
		8325-3701$^{\star}$ & 209.42442 & 46.714428 & 0.0407 & 9.14 & 0.23 & 0.83 \\
		8338-1901 & 172.16444 & 23.670927 & 0.0217 & 8.84 & -0.70 & 0.45 \\
		8439-9102 & 143.75383 & 48.97659 & 0.0253 & 9.08 & -0.06 & 0.53 \\
		8440-1902 & 136.27797 & 41.230907 & 0.0254 & 9.34 & -0.03 & 0.53 \\
		8440-6102 & 135.72 & 40.346703 & 0.0418 & 9.17 & -0.05 & 0.85 \\
		8448-1901 & 166.32912 & 21.62011 & 0.0215 & 9.08 & -0.42 & 0.45 \\
		8449-3703 & 169.29921 & 23.585657 & 0.0425 & 9.19 & 0.38 & 0.87 \\
		8456-12702 & 149.96826 & 45.283123 & 0.0237 & 9.29 & -0.04 & 0.49 \\
		8458-3702$^{\dagger}$ & 147.56242 & 45.958275 & 0.0252 & 9.52 & 0.72 & 0.52 \\
		8462-3703 & 146.42278 & 37.45126 & 0.0226 & 8.95 & -0.49 & 0.47 \\
		8465-3701 & 195.32036 & 48.060432 & 0.0301 & 9.95 & 0.81 & 0.62 \\
		8465-6102 & 197.5497 & 48.623394 & 0.0290 & 9.24 & 0.08 & 0.60 \\
		8485-3702 & 233.72545 & 47.761852 & 0.0232 & 9.38 & 0.14 & 0.48 \\
		8548-1902 & 243.33995 & 48.155037 & 0.0203 & 9.06 & -0.33 & 0.42 \\
		8548-3702 & 243.32684 & 48.391827 & 0.0206 & 8.74 & -0.17 & 0.43 \\
		8548-9102 & 244.91478 & 47.873234 & 0.0214 & 8.69 & -0.19 & 0.45 \\
		8549-6104$^{\dagger}$ & 244.4016 & 46.081997 & 0.0198 & 9.30 & -0.21 & 0.42 \\
		8551-1902$^{\star}$ & 234.59167 & 45.801952 & 0.0217 & 8.79 & -0.32 & 0.45 \\
		8553-3704$^{\dagger}$ & 234.97032 & 56.368366 & 0.0462 & 9.59 & 0.62 & 0.94 \\
		8566-3704 & 115.22481 & 40.06964 & 0.0405 & 9.50 & 0.78 & 0.83 \\
		8568-3703 & 155.69289 & 37.673573 & 0.0229 & 9.25 & -0.21 & 0.48 \\
		8601-3703 & 250.36378 & 40.20992 & 0.0329 & 8.92 & -0.58 & 0.68 \\
		8604-9102 & 246.45528 & 40.345215 & 0.0292 & 9.89 & 0.93 & 0.60 \\
		8613-12703$^{\dagger}$ & 256.81775 & 34.822598 & 0.0369 & 9.72 & 0.54 & 0.76 \\
		8615-1901$^{\star}$ & 321.0722 & 1.0283599 & 0.0204 & 8.76 & -0.03 & 0.43 \\
		8626-12704$^{\dagger\star}$ & 263.7552 & 57.052376 & 0.0478 & 8.63 & 1.06 & 0.97 \\
		8711-3704 & 119.62351 & 52.416985 & 0.0408 & 9.59 & 0.26 & 0.83 \\
		8712-6103 & 120.22973 & 53.670692 & 0.0409 & 9.59 & 0.28 & 0.84 \\
		8713-1902 & 118.86885 & 39.4202 & 0.0199 & 8.82 & -0.53 & 0.42 \\
		8715-12704 & 121.17238 & 50.718517 & 0.0231 & 8.77 & -0.24 & 0.48 \\
		8717-3703 & 118.31802 & 35.57258 & 0.0460 & 9.77 & 0.46 & 0.93 \\
		8718-3703$^{\dagger}$ & 122.22992 & 45.68758 & 0.0407 & 9.43 & 0.03 & 0.83 \\
		8719-12702 & 120.19929 & 46.69053 & 0.0196 & 9.52 & 0.16 & 0.41 \\
		8721-6101 & 135.16385 & 53.98325 & 0.0385 & 9.23 & -0.05 & 0.79 \\
		8725-3701 & 125.27413 & 45.848335 & 0.0378 & 9.38 & 0.15 & 0.77 \\
		8942-3703$^{\star}$ & 124.38507 & 28.357828 & 0.0203 & 8.79 & -0.32 & 0.42 \\
		8942-6103 & 124.33717 & 28.318333 & 0.0203 & 9.07 & -0.29 & 0.43 \\
		8945-3702 & 173.3619 & 47.286724 & 0.0461 & 9.52 & 0.66 & 0.93 \\
		8947-12702 & 171.71817 & 50.542404 & 0.0236 & 9.47 & -0.57 & 0.49 \\
		8987-1901 & 136.24005 & 27.727627 & 0.0217 & 8.83 & -0.79 & 0.45 \\
		8987-3701 & 136.2499 & 28.34773 & 0.0489 & 9.77 & 0.77 & 0.99 \\
		\bottomrule
	\end{tabular}
\\
Notes:
$^{\dagger}$: MaNGA galaxies with He \textsc{ii} $\lambda$4686 detection. $^{\star}$: MaNGA galaxies with [O \textsc{iii}] $\lambda$4363 detection. Stellar Mass is taken from the MaNGA Firefly VAC  \citep{Goddard2017, Parikh2018},  while rest of the quantities are taken from the Pipe3D VAC.
\end{table*}

\begin{figure}
	\centering
	\includegraphics[width=0.4\textwidth]{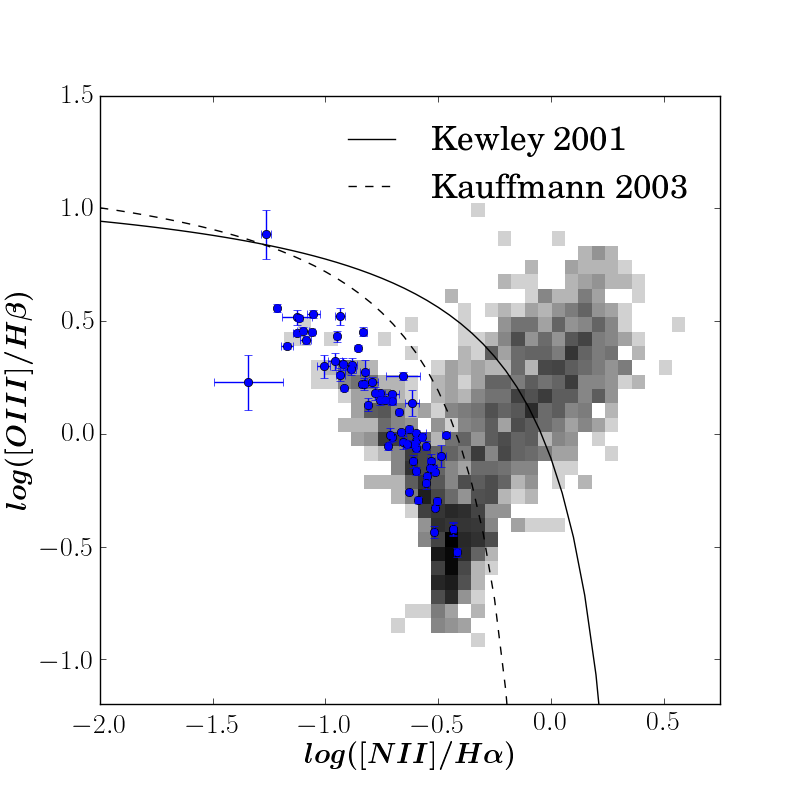}
	\caption{The emission line ratio diagnostic diagram showing [O~\textsc{iii}]/H$\beta$ versus [N~\textsc{ii}]/H$\alpha$ estimated from the central spectra of 67 sample galaxies (blue points). The dashed black curve and the solid black curve represent the empirical Kauffmann line \citep{Kauffmann2003} and the theoretical maximum starburst Kewley line \citep{Kewley2001}, respectively, used for separating the left star-forming sequence from the right mixing sequence comprised of AGN and LINERs. The grey density diagram in the background represents the galaxies in DR14 with finite values of both line ratios, i.e. a total of 2572 galaxies.}
	\label{fig:selection}
\end{figure}

\section{Galaxy Sample and Data}
\label{sect:data}

\subsection{Galaxy Sample}
\label{sect:sample}
We analyse the IFS data of a sample of 67 star-forming galaxies observed as a part of MaNGA survey \citep{Bundy2015}. Observations were taken with the Baryonic Oscillation Spectroscopic Survey (BOSS) spectrographs \citep{Smee2013} on the SDSS 2.5 m telescope \citep{Gunn2006} at the Apache Point Observatory (APO). The MaNGA data cover a wavelength range of 3600 \AA--10300 \AA, have a spectral resolution R $\sim$ 2000 corresponding to the instrumental full width half maximum (FWHM) $\sim$ 70 km s$^{-1}$ at the H$\alpha$ emission line. The reduced data cubes have the spatial-sampling of 0.5\arcsec ~and the effective spatial resolution of $\sim$ 2.5\arcsec~ FWHM. See \citet{Yan2016} for the Survey design, \citet{Law2015} for observing strategy and \citet{Law2016} for data reduction pipeline.

\indent The sample of 67 star-forming (SF) galaxies is a subset of a larger sample of $\sim$1400 galaxies in MaNGA data set from Data Release 14 (DR14)\footnote{\label{fnote:manga}\url{https://www.sdss.org/dr14/manga/}} which include SF galaxies, Active Galactic Nuclei (AGN) and LINERs with reliable [S \textsc{iii}] line detections in the redshift range of 0.02--0.06 (Amorin et al in prep.). Since these analyses depend on the use of [S \textsc{iii}] lines, the redshift cuts are imposed in the parent sample so that at least one of the two [S \textsc{iii}] $\lambda\lambda$9069,9531 lines lie outside the wavelength range of 9300--9700\AA ~which is most affected by a strong telluric absorption band. To address this issue, we include only those galaxies for which the emission line ratio [S \textsc{iii}]$\lambda$9069/9532 have a maximum deviation of 50\% of the theoretical line ratio \citep[=2.5, ][]{FroeseFischer2006} for all spaxels with signal-to-noise (S/N) ratio $>$ 3.

\indent For the current sample, we define a galaxy as SF if the emission line ratios of the central 2.5\arcsec~fibre region of the galaxy falls in the SF region of the classical emission line diagnostic \citep[BPT, ][]{Baldwin1981} diagram of [O \textsc{iii}]/H$\beta$ versus [N \textsc{ii}]/H$\alpha$. 
Figure \ref{fig:selection} shows that the emission line ratios estimated from the central spectrum (2.5\arcsec/diameter) of galaxy sample follows the SF sequence lying below the empirical demarcation line of  \citet{Kauffmann2003}. One galaxy, MaNGA-8626-12704 (also catalogued as SHOC~579), is an interesting exception which will be useful to probe more extreme environments. This is a well-known  extreme emission-line galaxy \citep[see e.g.][]{Kniazev2004, Fernandez2018} which falls slightly above (but still consistent within errors) the demarcation lines in the BPT diagrams due to its unusually high ionization properties. Note that similar H {\sc ii} galaxies with high excitation and low metallicity, such as the Green Peas \citep{Cardamone2009, Amorin2010} which are known to be local analogues of high-redshift galaxies, often lie in the upper-left part of the BPT diagnostics, quite offset with respect to the SF sequence and sometimes exceeding the demarcations set by photoionization models  \citep{Perez-Montero2009a, Feltre2016, Xiao2018}. 

Table \ref{tab:properties} lists general properties of all sample galaxies along with their plate identifications as mentioned in the DR14 catalogue \citep{Sanchez2016a, Sanchez2016b, Sanchez2018}. The sample spans the stellar mass range of 8.53 $\lesssim$ log(M$_\star$/M$_{\odot}$) $\lesssim$ 10.05 , SFR range of $-$0.79 $\lesssim$ log(SFR/M$_{\odot}$yr$^{-1}$) $\lesssim$ 1.30. The equivalent width (EW) of H$\alpha$ of the central spectrum of the sample lie in the range of $\sim$20--1000\AA, further ensuring that the galaxies are star-forming. Our selection criteria is intended to assemble a representative sample of SDSS-like SF galaxies along the [N\textsc{ii}]-BPT diagram, including galaxies with reliable spaxel data in all the relevant emission lines. Note that our sample is not complete by any means. \citet{Mingozzi2020} presents a larger sample of Manga SFGs with [S \textsc{iii}] measurements. 
Figures \ref{fig:cutout1}-\ref{fig:cutout3} show the SDSS cutouts of all 67 galaxies in the sample on which hexagonal field-of-view (FOV) of MaNGA is overlaid. 

\begin{figure*}
	\centering
	\includegraphics[width=0.45\textwidth]{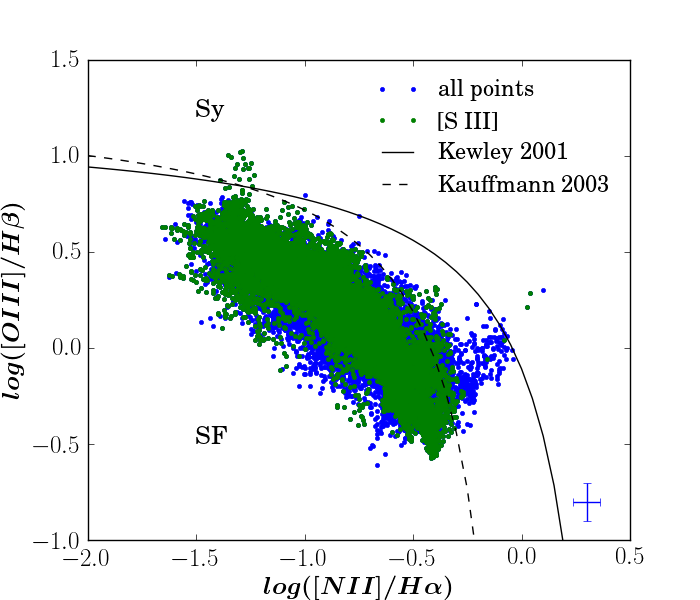}
	\includegraphics[width=0.45\textwidth]{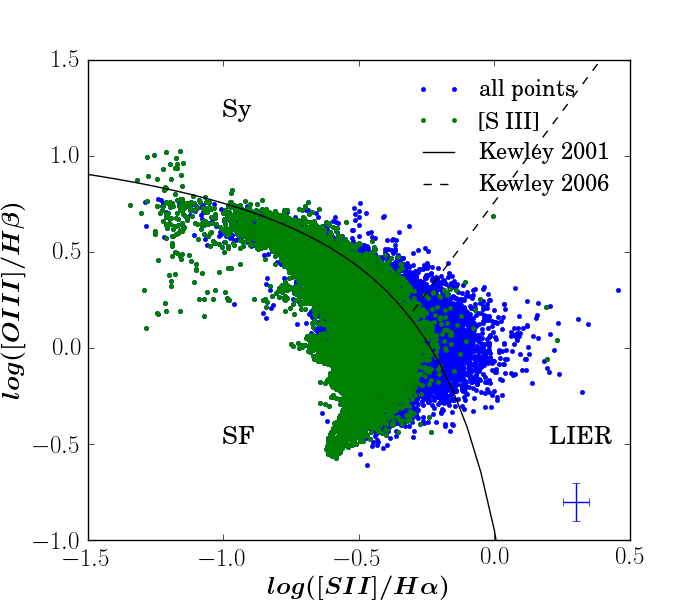}
	\caption{Left-hand panel: [O \textsc{iii}]/H$\beta$ versus [N \textsc{ii}]/H$\alpha$. Right-hand panel: [O \textsc{iii}]/H$\beta$ versus [S \textsc{ii}]/H$\alpha$. On both panels, spatially-resolved (spaxel-by-spaxel basis) emission line diagnostic diagrams  for all sample galaxies. Blue data points correspond to spatially-resolved (spaxel-by-spaxel basis) emission line ratios for all spaxels where all emission lines used in the line ratios on the two axes are detected with S/N $>$ 3, where as green data points represent a subset of blue data points which show the detection of [S \textsc{iii}] $\lambda$9532 emission line with S/N $>$ 3. The typical uncertainties on emission line ratios for all data points (i.e. blue data set) are shown in the bottom right corner of each panel. Solid black curve and dashed black curve correspond to the maximum starburst line from \citet{Kewley2001} and \citet{Kauffmann2003}, respectively, which provide a classification based on excitation mechanisms. The dashed straight line on right-hand panel separates Seyferts from LI(N)ERs derived by \citet{Kewley2006}. We find that a few green data points lie beyond the Kauffmann line and Kewley line in the left-hand and right-hand panels, respectively. However, the typical uncertainties are large enough for these points to put them in star-forming region of the BPT diagrams.}
	\label{fig:BPT all}
\end{figure*}

\subsection{Data}
\label{VAC}

\begin{figure*}
	\centering
	\includegraphics[width=0.45\textwidth, trim={0 1.5cm 0 1.5cm},clip]{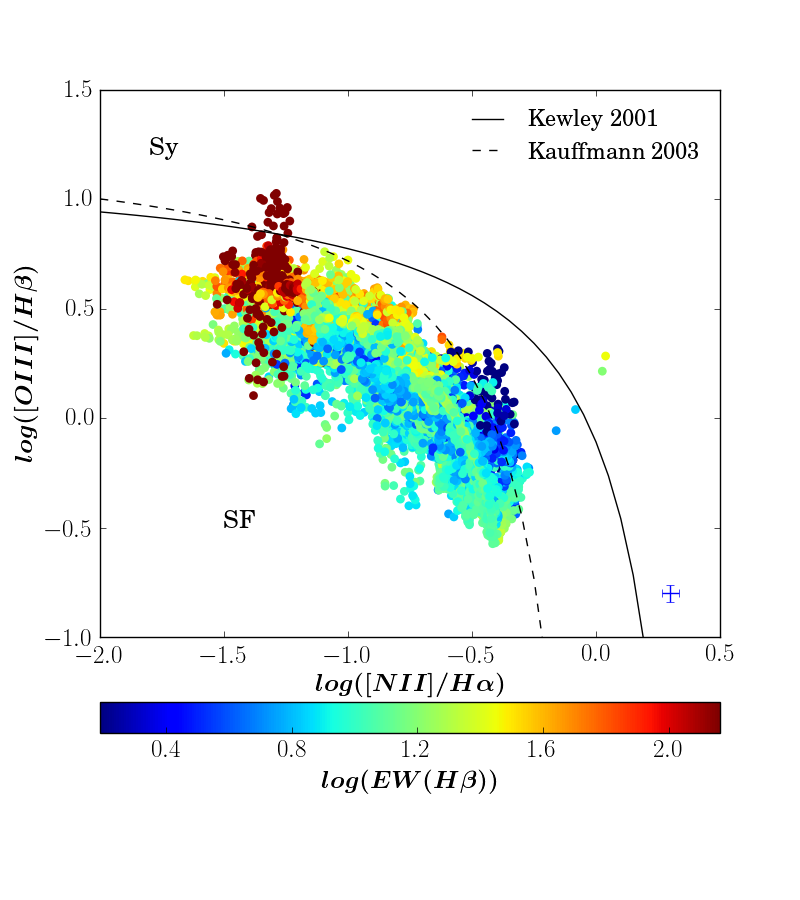}
	\includegraphics[width=0.45\textwidth, trim={0 1.5cm 0 1.5cm},clip]{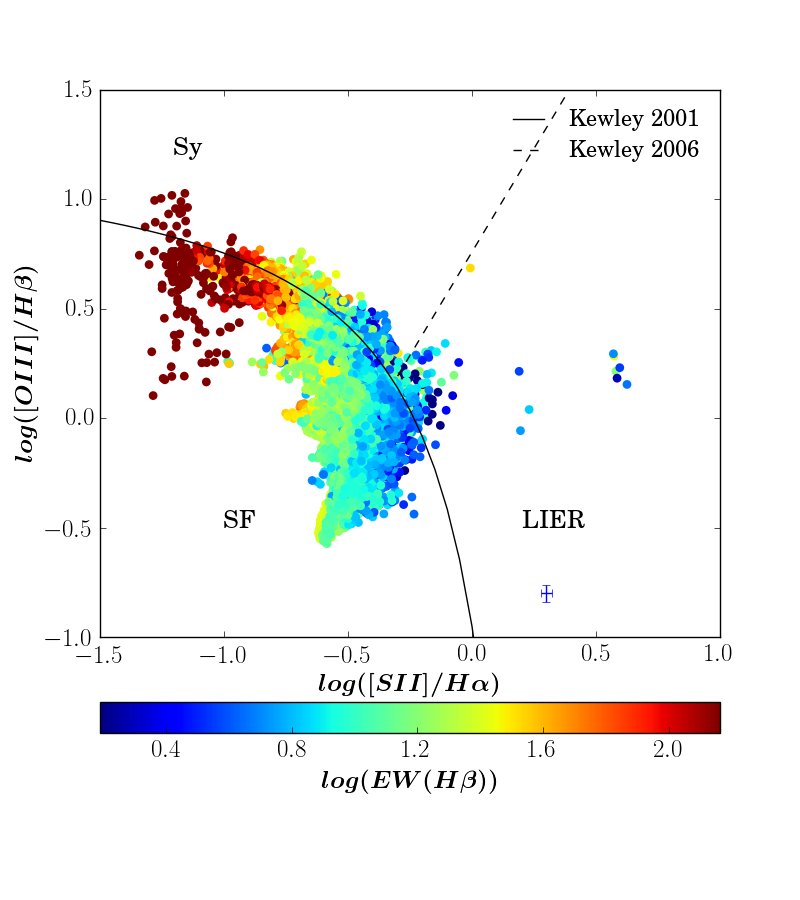}
	\includegraphics[width=0.45\textwidth, trim={0 1.5cm 0 1.5cm},clip]{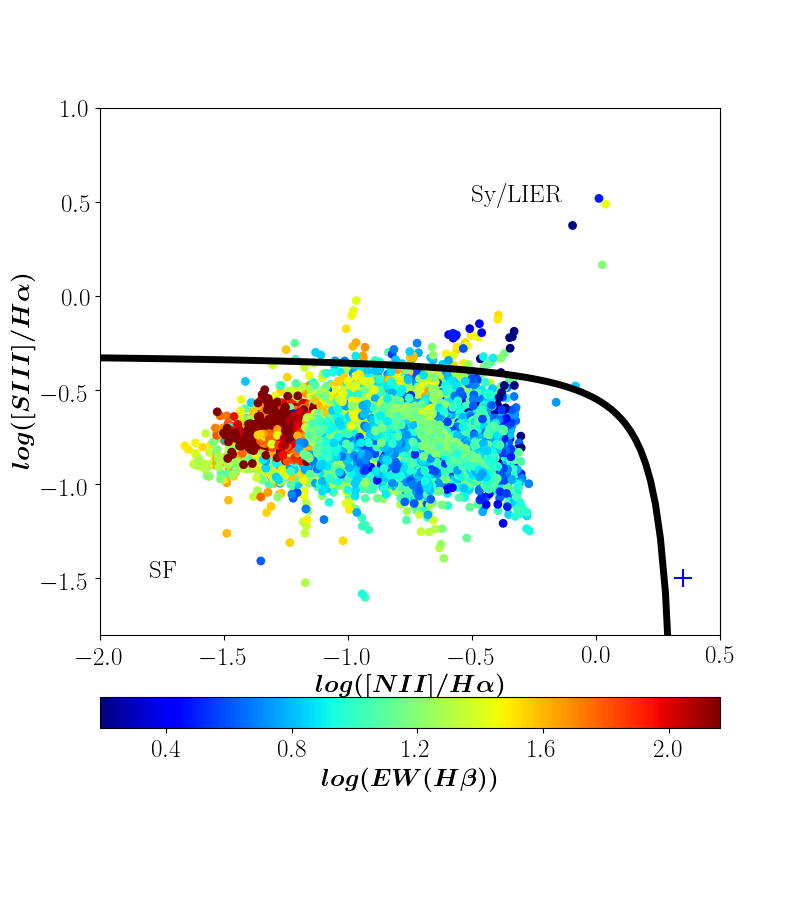}
	\includegraphics[width=0.45\textwidth, trim={0 1.5cm 0 1.5cm},clip]{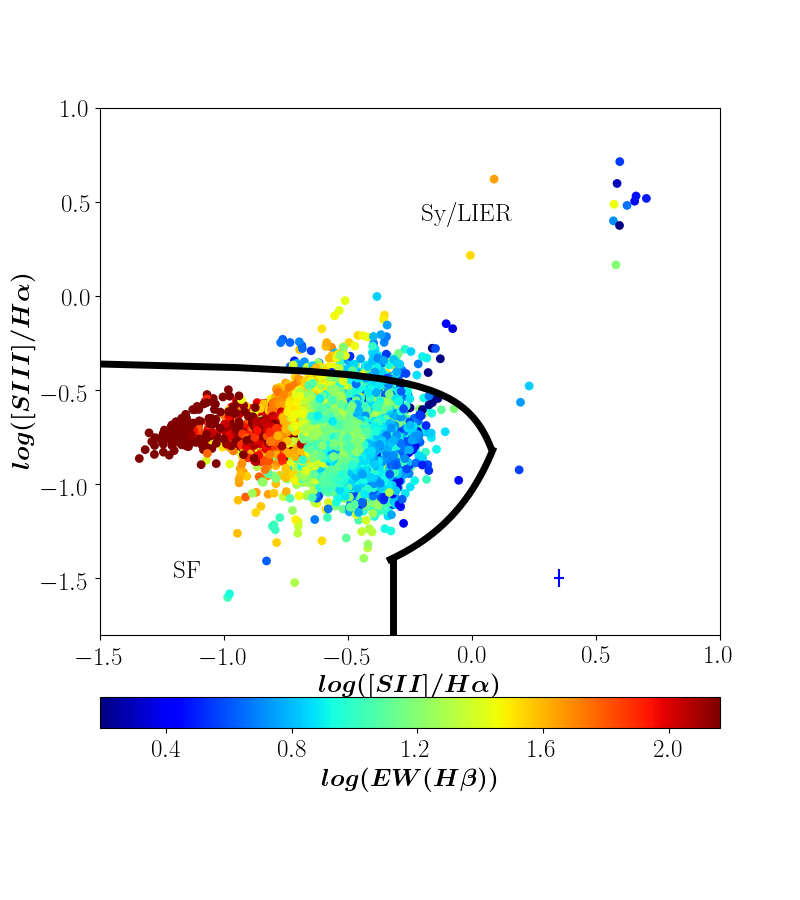}
	\caption{Upper panel: [O \textsc{iii}]/H$\beta$ versus [N \textsc{ii}]/H$\alpha$ (left-hand panel) and [O \textsc{iii}]/H$\beta$ versus [S \textsc{ii}]/H$\alpha$ (right-hand panel). The solid black curve on both panels correspond to the maximum starburst line given in \citet{Kewley2001}, while the dashed curve on the left-hand panel corresponds to the demarcation line between star-forming and Seyfert galaxies derived by \citet{Kauffmann2003}. The dashed line on the right-hand panel separates Seyferts from LI(N)ERs as derived in \citet{Kewley2006}. Lower panel: [S \textsc{iii}]/H$\alpha$ versus [N \textsc{ii}]/H$\alpha$ (left-hand panel) and [S \textsc{iii}]/H$\alpha$ versus [S \textsc{ii}]/H$\alpha$ (right-hand panel). The black solid curve on the lower panels represent the maximum starburst line derived empirically in Amorin et al. (in prep), and presented as equations \ref{eq:s3s2-bpt}--\ref{eq:s3n2-3}. On all panels, data points represent the spatially-resolved emission line ratios in all 67 galaxies but only for those spaxels which have [S \textsc{iii}] $\lambda$9532 detected with S/N $>$ 3, and are colour-coded with respect to log EW(H$\beta$). The typical uncertainties on x- and y-axes are shown in the bottom-right corner of each panel. Note that these data points correspond to green data points in classical emission line diagnostic diagrams presented in Figure \ref{fig:BPT all}.}
	\label{fig:SIII-BPT all}
\end{figure*}

\indent For this work, we use data products included in the MaNGA DR14 Pipe3D value added catalog (VAC)\footnote{\label{fnote:manga-vac}\url{https://www.sdss.org/dr14/manga/manga-data/manga-pipe3d-value-added-catalog/}} \citep{Sanchez2016a, Sanchez2016b, Sanchez2018}.  
\textsc{Pipe3D} is a spectroscopic analysis pipeline based on a package called \textsc{fit3d}\footnote{\url{http://www.astroscu.unam.mx/~sfsanchez/FIT3D}}, and is developed to analyse the properties of stellar populations and ionized gas via emission lines in the spatially-resolved optical spectra. In short, this pipeline performs continuum fitting on binned spaxels of data cubes using the single stellar population templates from the MIUSCAT library \citep{Vazdekis2012} which is an extension of MILES library \citep{Sanchez-Blazquez2006, Vazdekis2010, Falcon-Barroso2011}. 
For analysing strong emission lines, single Gaussian profiles are fit to estimate properties such as flux intensity, velocity, velocity dispersion and equivalent width. However, the same properties of weak emission lines is based on a direct estimation procedure based on a prior estimate of gas kinematics and Monte Carlo realisations. In addition, stellar indices such as D$\rm{_N}$(4000) are also estimated to characterise properties of stellar populations.

\indent We downloaded data products of sample galaxies from the links provided on the MaNGA website\textsuperscript{\ref{fnote:manga}}. The data products include data cubes, equivalent width maps and flux maps of emission lines of interest along with their uncertainty maps. We perform a S/N cut of 3 on all flux maps for the subsequent analysis. The emission line flux maps available from \textsc{Pipe3D} are corrected for Galactic foreground extinction but not for the internal reddening. We use the extinction curve of Large Magellanic Cloud \citep[LMC,][]{Fitzpatrick1986} and theoretical value of Balmer decrement (H$\alpha$/H$\beta$ = 2.86 assuming electron temperature T$_e$ = 10$^4$ K and electron density N$_e$ = 100 cm$^{-3}$, see e.g. \citealt{Osterbrock2006}), to first estimate the colour excess (i.e. E(B$-$V)) for each galaxy, and then combine with the observed flux maps of all emission lines of interest to estimate the extinction-corrected flux maps. A few studies recommend using the closer Paschen lines for estimating reddening correction for [S{\sc iii}] lines \citep[e.g.][]{Perez-Montero2019}, however, we do not adopt that methodology because Paschen line maps are not available for this sample in the DR14 MaNGA data set.  We use the intrinsic flux maps for further analysis except for line ratios with emission lines close in wavelengths. Uncertainties on these intrinsic maps are estimated by propagating errors on the observed flux maps, and then on all quantities of interest, for example, the emission line ratios.

\indent We use theoretical line ratios of [O\textsc{iii}]$\lambda$5007/[O\textsc{iii}]$\lambda$4959 \citep[=3, see e.g.][]{Osterbrock2006} and [S\textsc{iii}]$\lambda$9532/[S\textsc{iii}]$\lambda$9069 \citep[=2.5, ][]{FroeseFischer2006} to estimate intrinsic fluxes of emission lines [S~\textsc{iii}]~$\lambda$9069 and [O~\textsc{iii}]~$\lambda$4959 from  [S~\textsc{iii}]~$\lambda$9532 and [O~\textsc{iii}]~$\lambda$5007, respectively. 
For this work, we prefer to use [S \textsc{iii}] $\lambda$9532 instead of [S \textsc{iii}] $\lambda$9069 because it has better S/N ratio and is generally found outside the wavelength regions most affected by residuals from the telluric absorption correction at the redshift of this sample. It is worth noting that the continuum fitting  performed for the Pipe3D MaNGA VAC is found to be highly reliable out to $\sim$9470\AA. While [S \textsc{iii}] $\lambda$9069 is mostly within that limit, the [S \textsc{iii}] $\lambda$9531 is in a wavelength range where the continuum subtraction relies on an extrapolation of the MIUSCAT models beyond 9470\AA. Thus, the [S \textsc{iii}] $\lambda$9531 may in principle be subject of larger uncertainty. To minimise potential biases, first we have checked that our sample have excellent continuum fitting, i.e. quality flags $QCflag=0$ in the Pipe3D catalogs, and second, our sample selection was constrained to galaxies with spaxel data where the [S \textsc{iii}]$\lambda$9069/9532 ratio is consistent within 50\%  with the theoretical ratio. Our tests show that for galaxies with spaxel data with S/N([S \textsc{iii}] $\lambda$9531)$\gtrsim$\,10 such consistency is actually better than 15\%. Thus, we estimate the relative uncertainty due to continuum subtraction issues to be a factor of 1.5 at most. Overall, this translates into a maximum expected uncertainty of about 0.2 dex for line ratios involving [S~\textsc{iii}].




\subsection{Classical and Novel Emission line diagnostic diagrams}
\label{sect:BPTs}
\indent We use classical and novel emission line diagnostic diagrams to verify that the spaxels used in this analysis are predominantly star-forming. Figure \ref{fig:BPT all} shows spatially-resolved emission line diagnostic diagrams, [O~\textsc{iii}]/H$\beta$ versus [N~\textsc{ii}]/H$\alpha$ (left-hand panel) and   [O~\textsc{iii}]/H$\beta$ versus [S~\textsc{ii}]/H$\alpha$ (right-hand panel). 
The blue data points correspond to all spaxels considered within the galaxy sample, where the involved emission lines (i.e. H$\beta$, [O \textsc{iii}] $\lambda 5007$, [N \textsc{ii}] $\lambda$6584, H$\alpha$ and [S \textsc{ii}] $\lambda\lambda$6717, 6731) have a S/N $>$ 3. Superimposed green data points are a subset of blue ones corresponding to those spaxels where the  [S~\textsc{iii}] $\lambda$9532 emission line is also detected with S/N $>$ 3. On both panels, black solid curves denote the theoretical maximum starburst line from \citet{Kewley2001}. On the left-hand panel, we also show a black dashed curve which corresponds to the demarcation line from \citet{Kauffmann2003} derived empirically using $\sim$10$^5$ SDSS galaxies. 
The right-hand panel also shows a dashed straight line separating the spaxels with line ratios exhibited by Seyferts (Sy) and low ionization (nuclear) emission regions \citep[LI(N)ERs,][]{Belfiore2016a}. On both panels, we find that the spatially-resolved blue data points not only lie in the star-forming sequence but also spill into the region beyond the maximum star-burst lines. On the contrary, green data points (corresponding to spaxels with [S \textsc{iii}] $\lambda$9532 detection) appear to lie on the star-forming sequence better than blue data points. 
In the following, we solely concentrate on these green data points.


\indent In Figure \ref{fig:SIII-BPT all}, the upper panel presents the relations between  [O~\textsc{iii}]/H$\beta$ versus [N~\textsc{ii}]/H$\alpha$ (left-hand panel) and  [O \textsc{iii}]/H$\beta$ versus [S~\textsc{ii}]/H$\alpha$ (right-hand panel), and the lower panel presents  the relation between [S~\textsc{iii}]/H$\alpha$ versus [N~\textsc{ii}]/H$\alpha$ (left-hand panel) and [S~\textsc{iii}]/H$\alpha$ versus [S~\textsc{ii}]/H$\alpha$ (right-hand panel). On all panels, data are colour-coded with respect to log EW(H$\beta$) which is an age indicator for the young stellar populations \citep{Leitherer1999}. We find that EW(H$\beta$) varies with [S~\textsc{ii}]/H$\alpha$ and [N~\textsc{ii}]/H$\alpha$, with high EW regions showing higher excitation and lower [N~\textsc{ii}]/H$\alpha$ and [S~\textsc{ii}]/H$\alpha$, while there is no such trend with respect to [S~\textsc{iii}]/H$\alpha$ suggesting that [S~\textsc{iii}]/H$\alpha$ is not correlated with age. 

\indent We introduce here novel forms of classical BPT diagrams replacing [O \textsc{iii}]/H$\beta$ with [S\textsc{iii}]/H$\alpha$, that will serve to classify star-forming regions from LI(N)ER/Sy-like regions for future studies which lack blue end of optical spectrum. The black curves in Figure \ref{fig:SIII-BPT all} (lower panel) are the maximum starburst curves derived empirically from a larger sample of MaNGA galaxies  
(Amorin et al, in prep). The region lying below these curves are dominated by stellar photoionization while the region beyond the curves are mostly ionized by shocks and non-thermal sources. 
From Figure \ref{fig:SIII-BPT all}, we find that a majority of spaxels identified as SF on classical BPT diagrams (upper panel) show similar ionization on the novel [S~\textsc{iii}]-BPT diagrams in (lower panel). The equation for the maximum starburst curve for the novel [S \textsc{iii}]-BPT diagnostic diagrams are mentioned as below:

\indent For [S \textsc{iii}]/H$\alpha$ versus [N \textsc{ii}]/H$\alpha$ (Figure \ref{fig:SIII-BPT all}, lower-left panel), SF regions satisfy the following relation:
\vspace{0.25cm}
\begin{equation} 
    [S \textsc{iii}]/H\alpha < \frac{0.09}{[N \textsc{ii}]/H\alpha -0.35} - 0.29
    \label{eq:s3s2-bpt}
\end{equation}
\vspace{0.25cm}

\indent For [S \textsc{iii}]/H$\alpha$ versus [S \textsc{ii}]/H$\alpha$ (Figure \ref{fig:SIII-BPT all}, lower-right panel), SF regions satisfy the following relations:
\vspace{0.25cm}
\begin{equation}
      [S \textsc{iii}]/H\alpha < \frac{0.07}{( [S \textsc{ii}]/H\alpha -0.22)} - 0.32, \\
    \rm{for} ~[S \textsc{iii}]/H\alpha \geq -0.8\\
    \label{eq:s3n2-1}
\end{equation}
 \vspace{0.25cm}
 \begin{multline}
     [S \textsc{iii}]/H\alpha < \frac{-0.34}{( [S \textsc{ii}]/H\alpha -0.41)} - 1.86, \\
    \rm{for} -1.4 \leq [S \textsc{iii}]/H\alpha < -0.8
    \label{eq:s3n2-2}
 \end{multline}
 
 \begin{equation}
     [S \textsc{ii}]/H\alpha < -0.32, 
     \rm{for} -2.0 \leq [S \textsc{iii}]/H\alpha < -1.4
     \label{eq:s3n2-3}
  \end{equation}

\section{Results: Relations between radiation hardness and fundamental nebular parameters}
\label{sect:results}


\indent Understanding the hardness of the ionizing radiation field is important as, along with nebular geometry and gas density, it determines the ionization structure of an H \textsc{ii} region. As explained in Section \ref{sect:intro}, radiation hardness may correlate with various properties of ionizing stars and ionized nebula such as equivalent effective temperature, stellar age, electron temperature, density and chemical abundances within a nebula, while nebular structure can be characterized by ionization parameter. 
In an H \textsc{ii} region, the ratio of the number density of ions of same element in successive ionization state (e.g. S$^{2+}$/S$^+$, O$^{2+}$/O$^+$) depends on the hardness of the ionizing spectrum and the effective ionization parameter. The emission line ratios such as  [O~\textsc{iii}]/[O~\textsc{ii}] and [S~\textsc{iii}]/[S~\textsc{ii}] are known to be ionization parameter diagnostics \citep[see e.g.][]{Kewley2002, Morisset2016}. Thus, by studying the variation of [O \textsc{iii}]/[O \textsc{ii}] versus [S~\textsc{iii}]/[S~\textsc{ii}] (referred to as O$_3$O$_2$-S$_3$S$_2$ plane hereafter), we can in principle remove the effects of ionization parameter and study the hardness of radiation field to a first order approximation \citep{Vilchez1988}. In this section, we analyse the  O$_3$O$_2$-S$_3$S$_2$ plane with respect to several 
other properties which might be related to the hardness of radiation field.  We have performed this study on spaxel-by-spaxel basis for individual galaxies as well as for all spaxels combined from all 67 galaxies. We note that, despite some spaxels lie slightly above the SF empirical demarcation lines in the [S \textsc{ii}]-BPT diagnostic, the overall results presented in this section remain unchanged if we limit ourselves to those spaxels whose line ratios lie below the maximum theoretical starburst line \citep{Kewley2001}  on [S \textsc{ii}]-BPT. 
In the following we present the results from the combined sample for all spaxels, though same analysis for five individual galaxies are also presented in Appendix \ref{sect:individual} where [O \textsc{iii}] $\lambda$4363 detections are spatially-extended, enough for a comparison with other line ratios, in particular log $\eta\prime$. We also note that the spaxels presented in these diagrams are correlated over a few pixels. However, the presence of trends will not be affected by the correlated pixels, though the absence of trends should be interpreted with caution.  

\begin{figure*}
	\centering
	\includegraphics[width=0.33\textwidth, trim={0 2.5cm 0 0},clip]{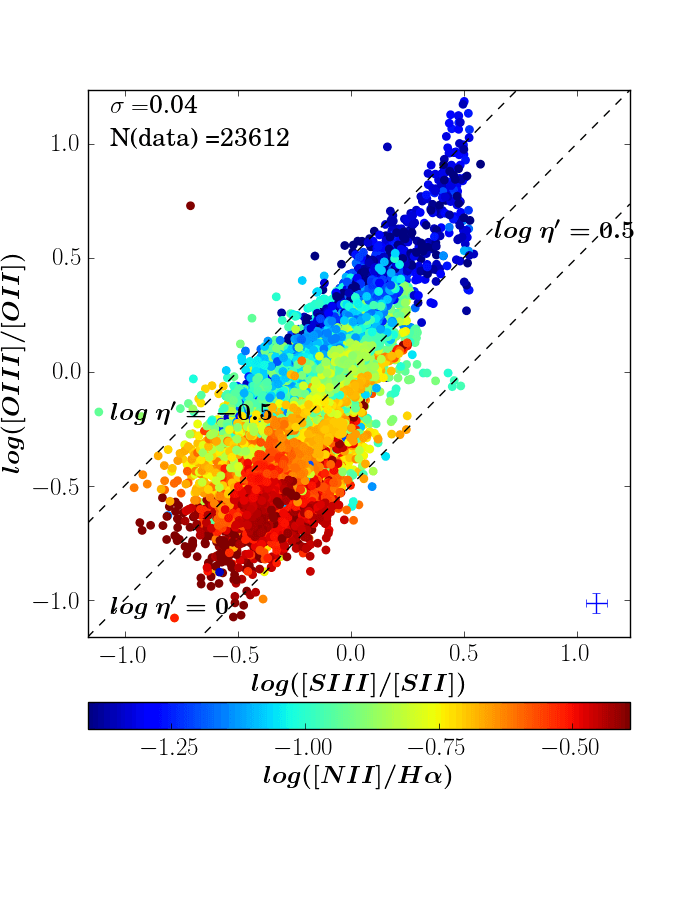}
	\includegraphics[width=0.33\textwidth, trim={0 2.5cm 0 0},clip]{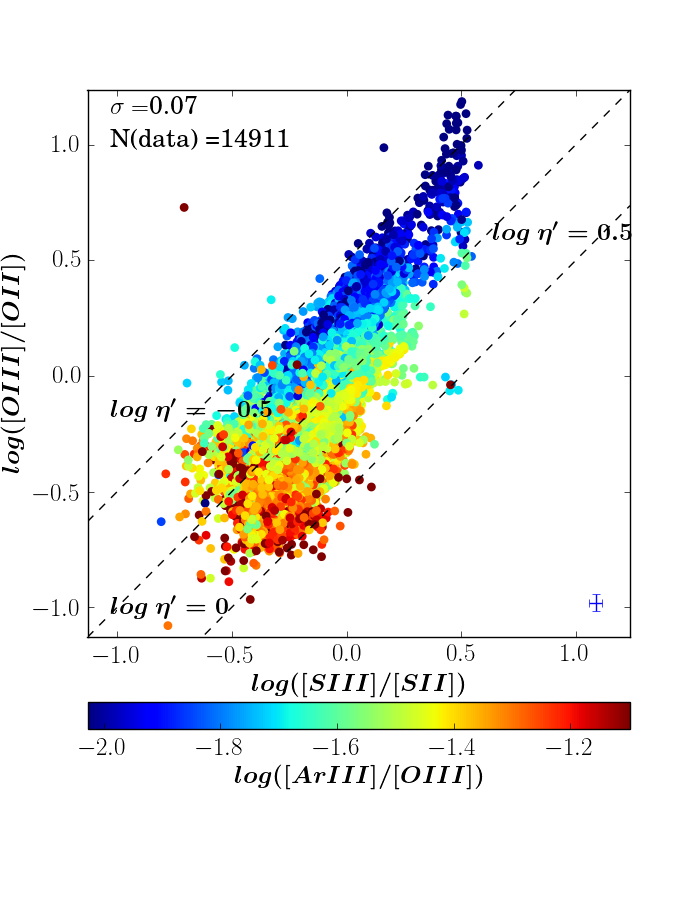}
	\includegraphics[width=0.33\textwidth, trim={0 2.5cm 0 0},clip]{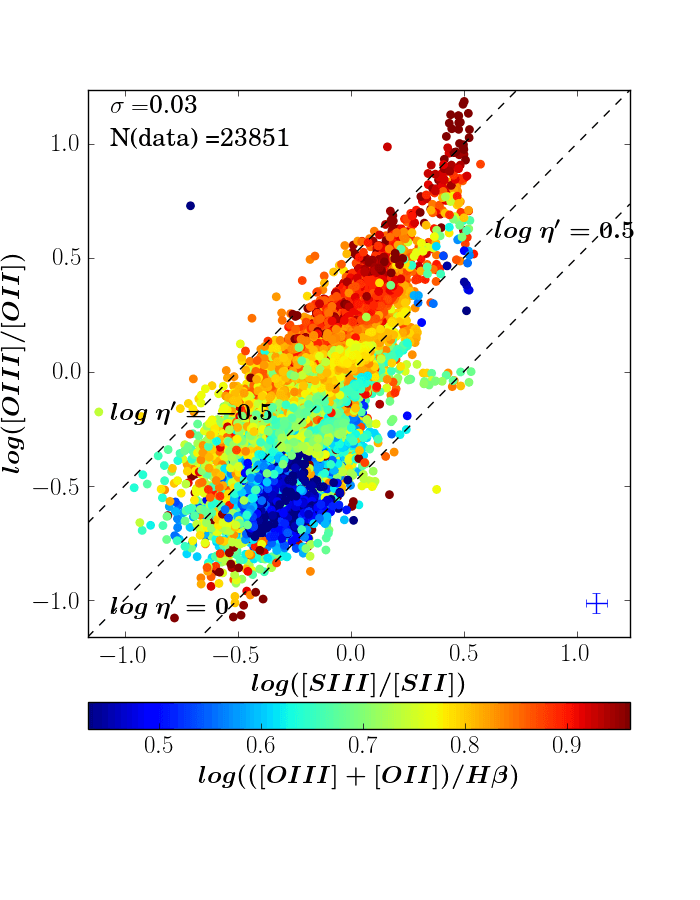}
	\includegraphics[width=0.33\textwidth, trim={0 2.5cm 0 0},clip]{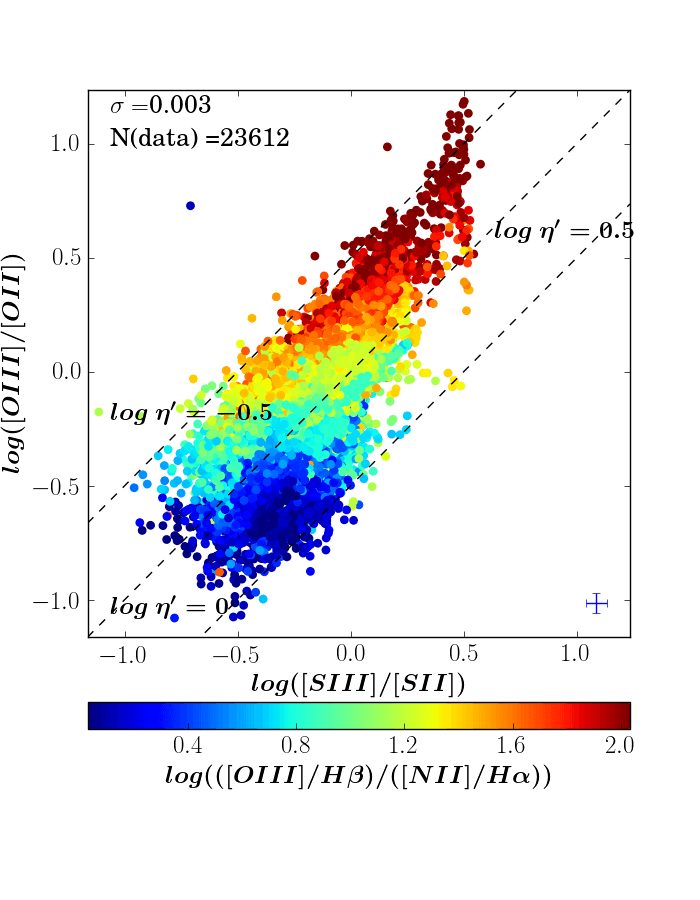}
	\includegraphics[width=0.33\textwidth, trim={0 2.5cm 0 0},clip]{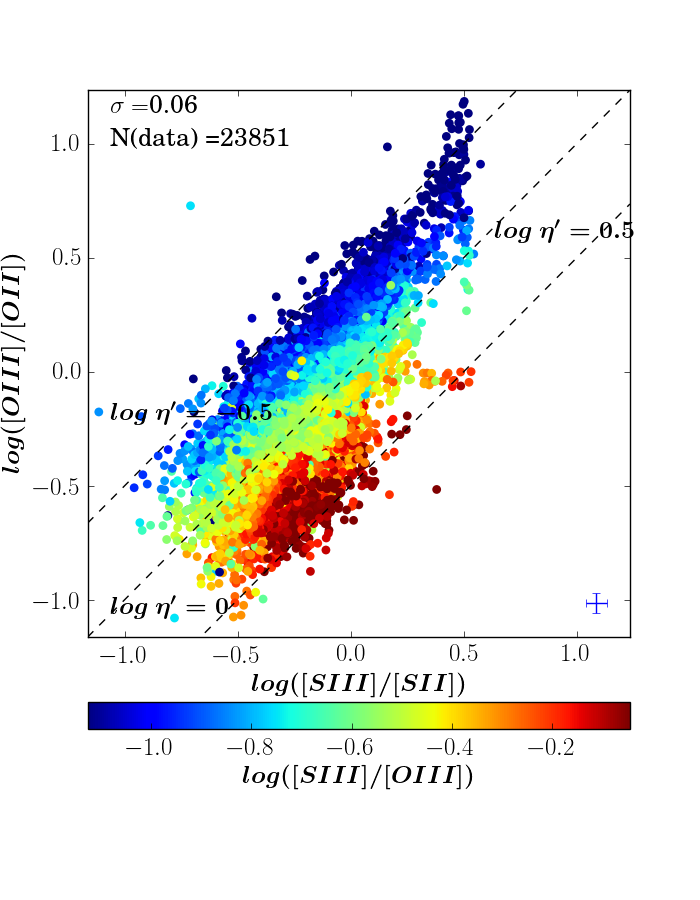}
	\includegraphics[width=0.33\textwidth, trim={0 2.5cm 0 0},clip]{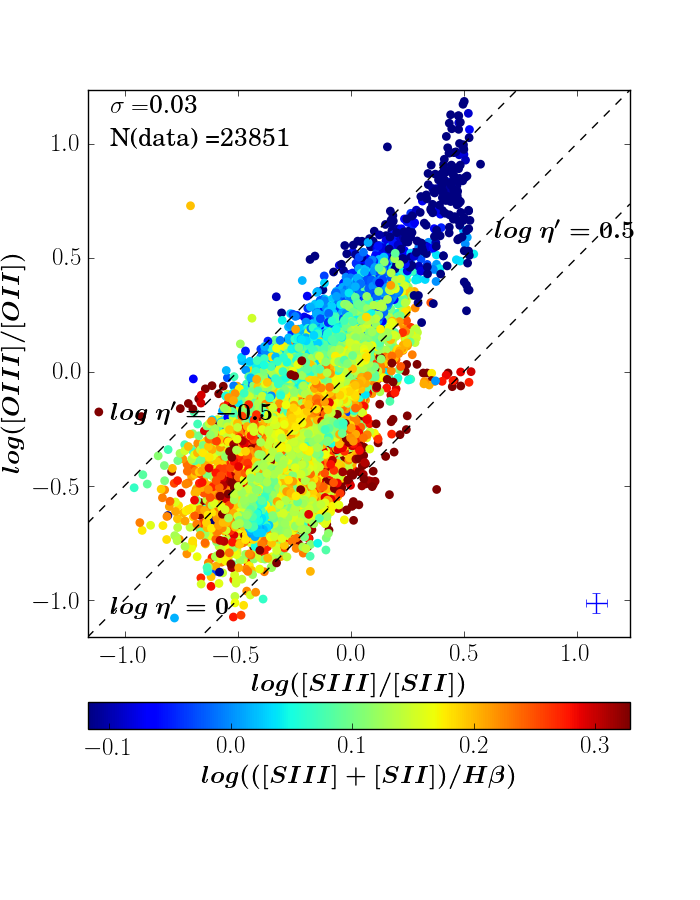}
	\caption{Spatially-resolved data of all 67 galaxies plotted on O$_3$O$_2$-S$_3$S$_2$ plane, where data points are colour-coded with respect to abundance-sensitive emission line diagnostics, N$_2$ (upper-left-hand panel), O$_3$N$_2$ (lower-left-hand panel),  Ar$_3$O$_3$ (upper-middle panel), S$_3$O$_3$ (lower-middle panel), R$_{23}$ (upper-right-hand panel) and S$_{23}$ (lower-right-hand panel). On each panel, the diagonal dashed lines represent the constant values of log $\eta \prime$ = $-$0.5, 0, 0.5. The typical uncertainties on O$_3$O$_2$ and S$_3$S$_2$ are shown in the lower-right corner of each panel. In the upper-left corner of each panel, $\sigma$ denotes the typical uncertainties on the variable (i.e. N$_2$, O$_3$N$_2$, Ar$_3$O$_3$, S$_3$O$_3$, S$_{32}$ and R$_{32}$) with respect to which data points are colour-coded, and `N(data)' represents the total number of plotted data points. Note that N(data) in the first two panels are determined by the number of spaxels with enough S/N ($>$ 3) for [S \textsc{iii}]$\lambda$9532 emission line. However, number of spaxels with enough S/N ($>$ 3) for [Ar \textsc{iii}]$\lambda$7135 determine N(data) in the last panels and is significantly lower than the first two panels as [Ar \textsc{iii}]$\lambda$7135 is more difficult to detect than [S \textsc{iii}]$\lambda$9532, since the ionization potential of [S \textsc{iii}] (34.83 eV) is lower than that of [Ar \textsc{iii}] (40.74 eV).}
	\label{fig:hardening metallicity}
\end{figure*}

\begin{figure}
	\centering
	\includegraphics[width=0.33\textwidth, trim={0 2.5cm 0 0},clip]{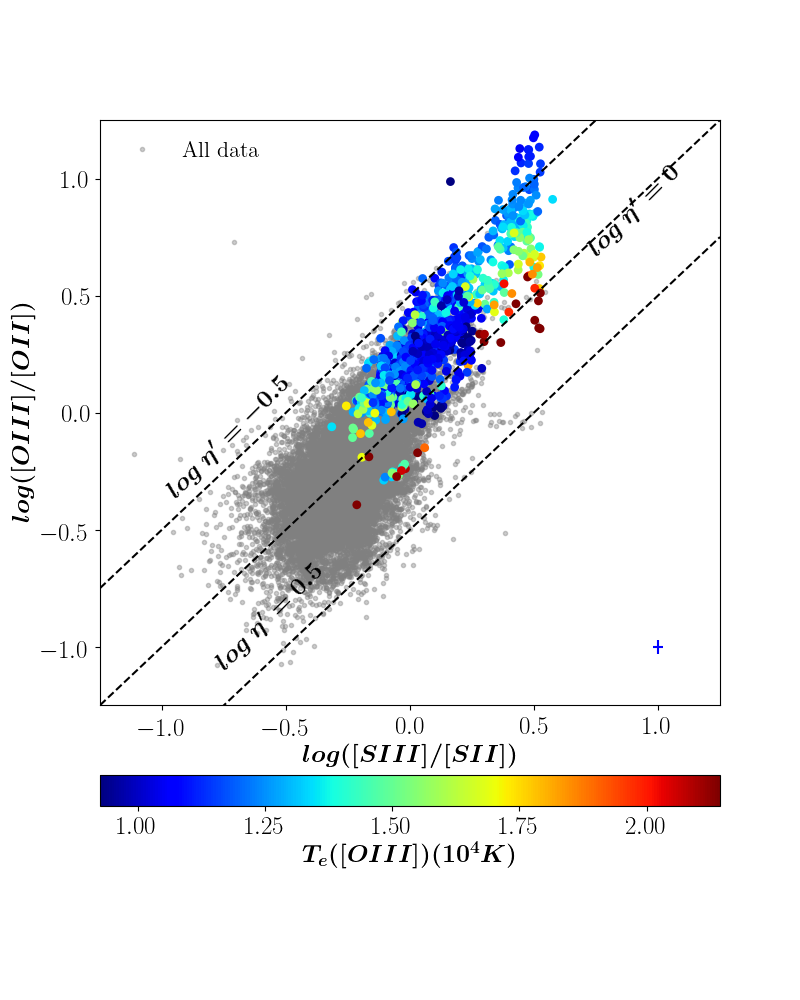}
	\caption{Spatially-resolved data of all 67 galaxies plotted on O$_3$O$_2$-S$_3$S$_2$ plane, where coloured data points have T$_e$([O \textsc{iii}]) measurements while grey data points indicate the entire data set. The diagonal dashed lines represent the constant values of log $\eta \prime$ = $-$0.5, 0, 0.5. The typical uncertainties on both axes (i.e. O$_3$O$_2$ and S$_3$S$_2$) are shown in the lower-right corner  while  the typical uncertainty on T$_e$([O \textsc{iii}]) is given as $\sigma$ in the upper-left corner of each panel.}
	\label{fig:hardening Te}
\end{figure}

\begin{figure*}
	\centering
	\includegraphics[width=0.33\textwidth, trim={0 2.5cm 0 0},clip]{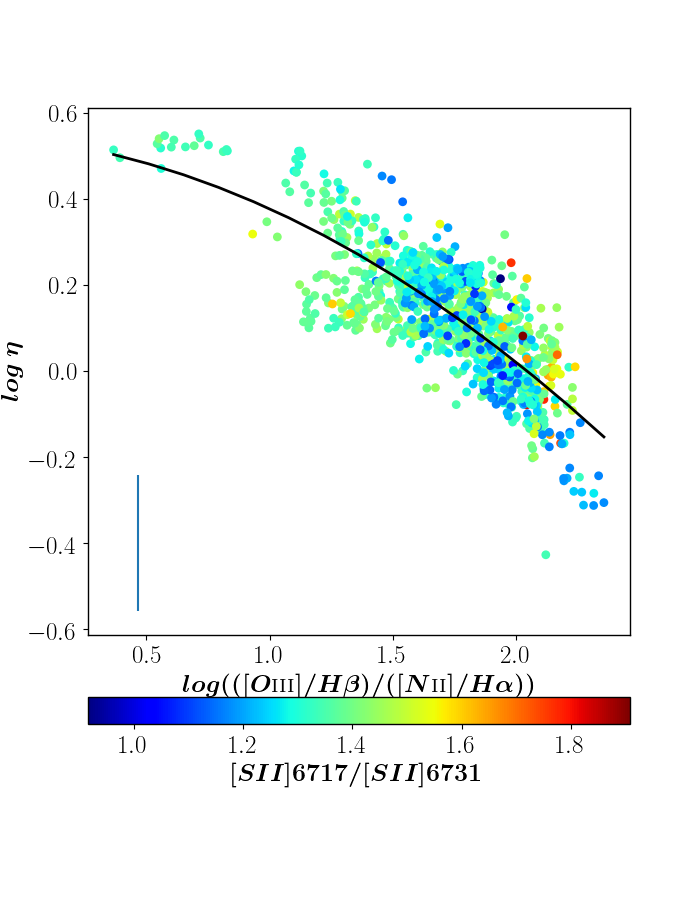}
	\includegraphics[width=0.33\textwidth, trim={0 2.5cm 0 0},clip]{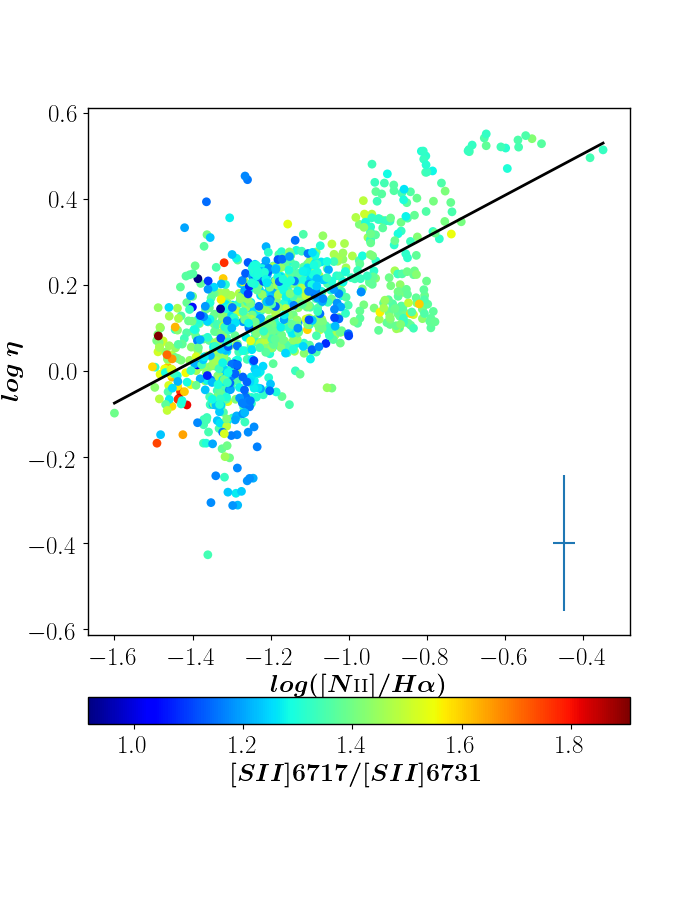}
	\includegraphics[width=0.33\textwidth, trim={0 2.5cm 0 0},clip]{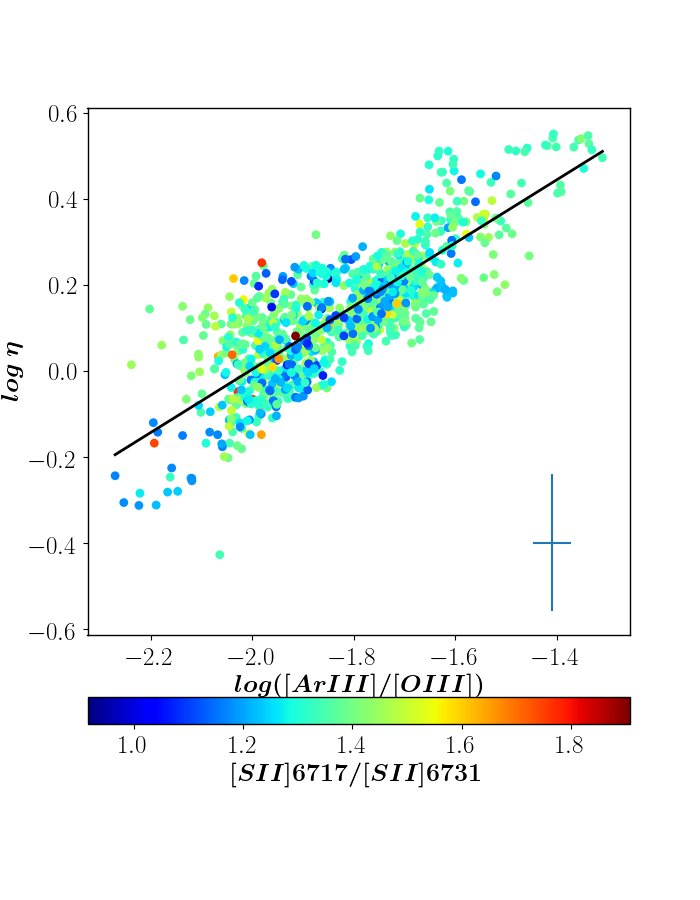}
	\caption{Relation between  softness parameter (log $\eta $) and emission line ratios sensitive to the abundance, O$_3$N$_2$ (left-hand panel), N$_2$ (middle panel) and Ar$_3$O$_3$ (right-hand panel), where data are colour-coded with respect to N$_e$-sensitive [S \textsc{ii}] doublet ratio. The median uncertainties on x-and y-axes are shown in the lower-left corner of the left-hand panel and in the lower-right corners of the other two panels. The median uncertainty on O$_3$N$_2$ is quite small. The solid black curve on each panel shows the best-fit curve to the spatially-resolved data points.}
	\label{fig:eta Z}
\end{figure*}

\subsection{Gas-phase Metallicity}
\label{sect:hardening metallicity}
\indent Figure \ref{fig:hardening metallicity} shows the O$_3$O$_2$-S$_3$S$_2$ plane for spatially-resolved data (on spaxel-by-spaxel basis) of all 67 galaxies in the sample where data points are colour-coded with respect to the abundance-sensitive emission line ratios, N$_2$, O$_3$N$_2$,  Ar$_3$O$_3$, S$_3$O$_3$, R$_{23}$ and S$_{23}$. On each panel, three diagonal dashed lines correspond to constant values of log $\eta \prime$ = $-$0.5, 0 and 0.5, where $\eta \prime$ is given by equation \ref{eq:eta prime}. Note that  log $\rm\eta\prime$=0 corresponds to spaxels with O$_3$O$_2$ = S$_3$S$_2$. A harder radiation field corresponds to lower values of log $\eta \prime$, and vice-versa. In the following, we discuss the variation in O$_3$O$_2$-S$_3$S$_2$ relation with each metallicity diagnostic mentioned above:

\begin{itemize}

\item \textbf{N$_2$} has been shown to increase with an increase in metallicity \citep[see e.g.][]{Pettini2004, Maiolino2008, Perez-Montero2009a, Marino2013, Curti2017, Maiolino2018} over a wide range of metallicities (7.6 $<$ 12 + log(O/H) $<$ 8.85) though it suffers from saturation of [N \textsc{ii}] at higher metallicities. In Figure \ref{fig:hardening metallicity} (upper-left-hand panel), N$_2$ shows a gradient on O$_3$O$_2$-S$_3$S$_2$ plane, where it decreases with a decrease in log $\eta \prime$. However, a constant value of log $\eta \prime$ does not correspond to a constant value of N$_2$, which indicates a secondary dependence on other unknown parameters. We find that at a constant S$_3$S$_2$, a lower value of N$_2$, i.e. a lower metallicity corresponds to a harder radiation field and vice-versa. However, at a constant O$_3$O$_2$, variation of N$_2$ with hardness of radiation field is less obvious. Both S$_3$S$_2$ and O$_3$O$_2$ trace ionization parameter (log $\mathcal{U}$), though S$_3$S$_2$ is shown to be a better diagnostic of log $\mathcal{U}$ than O$_3$O$_2$ \citep{Morisset2016}. 
	Hence, it appears that log $\eta \prime$ increases with an increase in metallicity (if traced by N$_2$) at constant ionization parameter (if traced by S$_3$S$_2$).

\indent \indent We also find that the highest values of O$_3$O$_2$ correspond to the lowest metallicities at a given value of S$_3$S$_2$. This result agrees with the findings of \citet{Kehrig2006} and \citet{Stasinska2015} for local blue compact dwarf galaxies with very high excitation. Moreover, the photoionization models of \citet[][Figure 1]{Kewley2002} also show an increase in metallicity with a decrease in O$_3$O$_2$ at a given ionization parameter (traced by S$_3$S$_2$).  We will further discuss log $\eta \prime$ and ionization parameter in Section \ref{sect:etaprime U}.

\item \textbf{O$_3$N$_2$} is known to decrease with an increase in metallicity \citep[see e.g.][]{Pettini2004, Maiolino2008, Marino2013, Curti2017, Maiolino2018}, and is deemed to be more useful than N$_2$ in the high metallicity regime, where [N \textsc{ii}] saturates but the strength of [O \textsc{iii}] continues to decrease with metallicity. In Figure 4 (lower-left-hand panel), O$_3$N$_2$ shows a similar gradient as seen in the case of N$_2$, indicating that metallicity and log $\eta \prime$ are correlated.

\item \textbf{Ar$_3$O$_3$} follows a monotonically increasing relation with the electron temperature and hence  with the metallicity \citep{Stasinska2006}. It is considered to be a more accurate metallicity diagnostic than N$_2$ at higher metallicity but needs a reliable reddening correction (unlike N$_2$). Moreover, it is unaffected by the presence of diffuse ionized gas (DIG) because of the use of high excitation lines and not the low excitation lines which may arise in both H \textsc{ii} regions and DIG. In Figure \ref{fig:hardening metallicity}, we study  O$_3$O$_2$-S$_3$S$_2$ plane where data points are colour-coded with respect to Ar$_3$O$_3$ ratio, where higher  Ar$_3$O$_3$ indicates higher metallicity. We find a similar gradient as N$_2$ and O$_3$N$_2$ though we note a weaker trend at lower values of O$_3$O$_2$  and S$_3$S$_2$, which might be simply because of fewer spaxels with enough S/N ($>$3) of [Ar \textsc{iii}] $\lambda$7135.

\item \textbf{S$_3$O$_3$} is posed as a good diagnostic of metallicity in both low and high metallicity regimes, which like Ar$_3$O$_3$, is unaffected by the presence of DIG \citep{Stasinska2006}. Hence we explored O$_3$O$_2$-S$_3$S$_2$ plane with a third variant as S$_3$O$_3$ line ratio in Figure \ref{fig:hardening metallicity} (lower-middle panel). We find a clear gradient in S$_3$O$_3$ across log $\eta \prime$ = -0.5--0.5, where S$_3$O$_3$ appears to be approximately constant at a given value of log $\eta \prime$ unlike our observation in previous plots. Note here that it might be simply because emission lines ([S~\textsc{iii}] and [O~\textsc{iii}]) involved in this metallicity diagnostic appear in the numerators of two axes on this plane, while the effect of the two emission lines in the denominator (i.e. [S \textsc{ii}] and [O \textsc{ii}]) nullify because of their very similar ionization potentials, and probably because oxygen and sulphur are produced in similar stars. We should therefore be cautious while using S$_3$O$_3$ as a metallicity indicator on the O$_3$O$_2$-S$_3$S$_2$ plane because the involved emission lines appear to be more sensitive to hardness of radiation fields.

\item \textbf{R$_{23}$} traces metallicity but there are two major caveats in its use: firstly, it is bimodal with metallicity, secondly it also depends on ionization parameter \citep[see e.g.,][]{Kewley2002}. This means that one needs to determine the metallicity regime (low or high) as well as ionization parameter to use R$_{23}$ for inferring a reliable value of metallicity. In Figure  \ref{fig:hardening metallicity} (upper-right-hand panel), we find that the variation of R$_{23}$ on the  O$_3$O$_2$-S$_3$S$_2$ plane is similar to other metallicity diagnostics though the trend is less obvious for lower values of O$_3$O$_2$ and S$_3$S$_2$. For example, at a constant value of S$_3$S$_2$ = -0.5, there is practically no trend in R$_{23}$. It is possible that the metallicites of these spaxels lie in the "knee", a region of confusion (i.e. 12 + log(O/H) $\sim$ 8.1-8.3) where R$_{23}$ peaks.

\item \textbf{S$_{23}$} has been used to trace metallicity \citep{Vilchez1996, Perez-Montero2005, Kehrig2006, Hagele2006}, and is similar to the R$_{23}$ parameter, however the knee (i.e. metallicity regime of confusion) appears at a higher metallicity \citep[12~+~log(O/H) $\sim$ 8.8,][]{Kewley2002} than R$_{23}$.   In Figure \ref{fig:hardening metallicity} (lower-right-hand panel), we do not observe a clear gradient  with respect to the S$_{23}$ emission line ratio. The absence of a clear gradient at higher metallicities may be because S$_{23}$ is double-valued with respect to metallicity  and is quite dependent on ionization parameter \citep{Kewley2002}. However, we also find that data points with extremely low-values of S$_{23}$ clearly show harder radiation field lying between the constant values of log $\eta \prime$ = $-$0.5 to 0. Those data-points (lying in the top-right corner of middle panel) are predominantly from the central region of a star-forming galaxy (Manga-8626-12704, see Figure \ref{fig:manga-8626-12704}) which shows prominent detection of the auroral line [O \textsc{iii}] $\lambda$4363. The detection of this weak emission line and the range of electron temperatures (Figure \ref{fig:manga-8626-12704}, lower-right-hand panel) show that the metallicities of these data points are low. Hence, in spite of the degeneracies related to the ionization parameter and double-valued nature of S$_{23}$, this plot is consistent with an inverse relation between metallicity and hardness of radiation field.

\end{itemize}

\indent In summary, we conclude that the gas-phase metallicity depends on the radiation hardness as traced by log $\rm\eta\prime$ (on O$_3$O$_2$-S$_3$S$_2$ plane), i.e. low-metallicity gas is associated with harder radiation field and vice-versa. This result agrees with those of \citet{Kehrig2006}, who found that H \textsc{ii} regions with lower gaseous metallicity present harder ionizing spectra. \citet{Kewley2013} pointed out several potential reasons which might cause the correlation between metallicity and radiation hardness. However, the diagnostics of metallicity and ionization parameter are correlated as well \citep{Dopita1986, Perez-Montero2014}, as such radiation hardness might be related to ionization parameter.

\subsection{Electron Temperature (T$_{e}$([O \textsc{iii}])) and Density (N$_e$)}
\label{sect:te}

\indent Figure \ref{fig:hardening Te} shows O$_3$O$_2$-S$_3$S$_2$ plane where data-points are colour-coded with respect to electron temperature T$_e$([O \textsc{iii}]). We estimated T$_e$([O \textsc{iii}]) on a spaxel-by-spaxel basis for all galaxies in the sample where auroral line [O \textsc{iii}] $\lambda$4363 was detected with S/N $>$ 3, by using the emission line ratio of ([O~\textsc{iii}] $\lambda\lambda$4959, 5007)/[O~\textsc{iii}]$\lambda$4363 with the prescriptions given in \citet{Perez-Montero2017}.  We restrict our analysis of  O$_3$O$_2$-S$_3$S$_2$ plane to only those spaxels with T$_{e}$([O \textsc{iii}]) lying in the range of 7000-25000 K, as the involved equations are only valid in the above-mentioned range\footnote{We also estimated T$_e$([O~\textsc{iii}]) by using emission line fluxes in \textsc{pyneb} \citep[v1.1.14,][]{Luridiana2013} and did not find any significant difference in the overall trend.}. We find that the majority of data points lies in the range of log $\eta \prime$ = $-$0.5 and 0. At first glance, it might appear that galaxies with [O \textsc{iii}] 4363 detection (and T$_e$([O \textsc{iii}]) estimates) exhibit harder radiation fields, though model-based analysis shows no such relation (Figure \ref{fig:models_A1}). The absence of data points with T$_e$ measurements on the lower-left corner in Figure \ref{fig:hardening Te} is likely due to an overall lower O$^{++}$/O.  


\indent \citet{Vilchez1988} show that ionic quotient ratio log $\rm\eta$ varies directly with the oxygen abundance. We explore this further by estimating log $\rm\eta$ using O$_3$O$_2$, S$_3$S$_2$ and T$_e$([O \textsc{iii}]) within equation \ref{eq:eta tes} and study its variation with respect to the abundance-sensitive emission line ratio O$_3$N$_2$, N$_2$ and Ar$_3$O$_3$ as shown in Figure \ref{fig:eta Z}. We find that softness parameter decreases with O$_3$N$_2$, but increases with N$_2$ and Ar$_3$O$_3$ implying that the hardness of radiation field varies proportionally with the metallicity traced by the three line ratios.  The result agrees with that of \citet{Vilchez1988} who shows an increase of softness parameter with the oxygen abundance. Since metal leads to cooling, we expect higher electron temperature for metal-poor gas.The result is in agreement with that in Section \ref{sect:hardening metallicity} where metallicity is found to have an inverse dependence on log $\eta \prime$  at a given log $\mathcal{U}$. 

\indent In Figure \ref{fig:eta Z}, we also fit the following polynomials between log $\rm\eta$ and abundance-sensitive line ratios O$_3$N$_2$, N$_2$ and Ar$_3$O$_3$ using an orthogonal distance regression and taking into account uncertainties on both axes:

 \begin{equation}
	\centering
	\rm log ~\eta =  (-0.096 \pm 0.014) O_3N_2^2 - (0.068 \pm 0.046) O_3N_2 + (0.540 \pm 0.035)
	\label{eq:eta-o3n2}
\end{equation}

\begin{equation}
    \centering
    \rm log ~\eta = (0.483 \pm 0.016) N_2 + (0.698 \pm 0.019)
    \label{eq:eta-n2}
\end{equation}

\begin{equation}
    \centering
    \rm log ~\eta = (0.734 \pm 0.015) Ar_3O_3 + (1.47 \pm 0.028)
    \label{eq:eta-ar3o3}
\end{equation}
 
\vspace{0.25cm}

\indent The above equations will allow future studies to estimate log $\rm\eta$ from O$_3$N$_2$, N$_2$ and Ar$_3$O$_3$ which are ratio of strong emission lines, thus extending the study of radiation hardness even in the systems where the temperature-sensitive weak auroral lines (e.g., [O \textsc{iii}] $\lambda$4363) are not detected. We do not claim that the line ratios O$_3$N$_2$, N$_2$ and Ar$_3$O$_3$ trace radiation hardness, however they can be used to estimate log $\rm\eta$ because of their sensitivity to metallicity and probably ionization parameter. So caution should be made while interpreting the log $\rm\eta$ as radiation hardness when estimated from O$_3$N$_2$,  N$_2$ or Ar$_3$O$_3$. In principle, similar relations can be found between log $\rm\eta$ and other abundance-sensitive line ratios shown in Figure \ref{fig:hardening metallicity}. However, we do not attempt to fit such a relation of $\rm\eta$ with the line ratios R$_{23}$ or S$_{23}$ since both of them are bi-modal in metallicity. We do not use S$_3$O$_3$ as we establish in Section \ref{sect:hardening metallicity} that the variation of $\rm\eta\prime$ with S$_3$O$_3$ might be a systematic effect of using the emission lines ratios of similar ionization potentials [O~\textsc{ii}] and [S~\textsc{ii}] in the definition of log $\rm\eta\prime$. 

\indent In Figure \ref{fig:eta Z}, data are colour-coded with respect to line ratio [S~\textsc{ii}]$\lambda$6717/[S \textsc{ii}]$\lambda$6731 which is sensitive to N$_e$, showing that there is no obvious relation between N$_e$ and radiation hardness. The result is consistent with the theoretical definition of log $\rm\eta$ (Equation \ref{eq:eta tes}) which do not show any first-order dependence on N$_e$. Furthermore, a majority of data exhibit the [S \textsc{ii}] doublet line ratio corresponding to the N$_e$ typical of H \textsc{ii} regions, i.e. there is not much variation of N$_e$ within our data set and hence no secondary effect is seen in log $\rm\eta$. Our results are in agreement with \citet{Hunt2010}, who find that N$_e$ is not correlated with radiation hardness, which they measure by [O \textsc{iv}]/[S \textsc{ii}] line ratio in their sample of dwarf galaxies. 

\indent 
Here, we do a qualitative analysis to infer whether T$\rm_{e}$ might be related to T$\rm_{eff}$, on the basis of previous studies which suggest that log $\eta \prime$ decreases  with the increase in T$\rm_{eff}$ \citep[e.g.][]{Kennicutt2000, Perez-Montero2009b, Perez-Montero2019}.  \citet{Kennicutt2000} also find that T$\rm_{eff}$\footnote{\cite{Kennicutt2000} uses the terminology T$_{\star}$ for T$\rm_{eff}$ when abundance is not within the calibrated range.} decreases with respect to the gas-phase metallicity in a given abundance range. Similarly, \citet{Hagele2006} argues that for a given stellar mass, stars of lower metallicity have higher effective temperature. Our work shows a low gas-phase metallicity or high T$\rm_e$ for low log $\rm\eta\prime$. Hence, this might imply  that the star-forming regions hosting hotter stars with harder ionizing radiation and lower stellar metallicity might have lower gas-phase metallicity which results in higher electron temperatures. 




\begin{figure}
	\centering
	\includegraphics[width=0.35\textwidth, trim={0 2.5cm 0 0},clip]{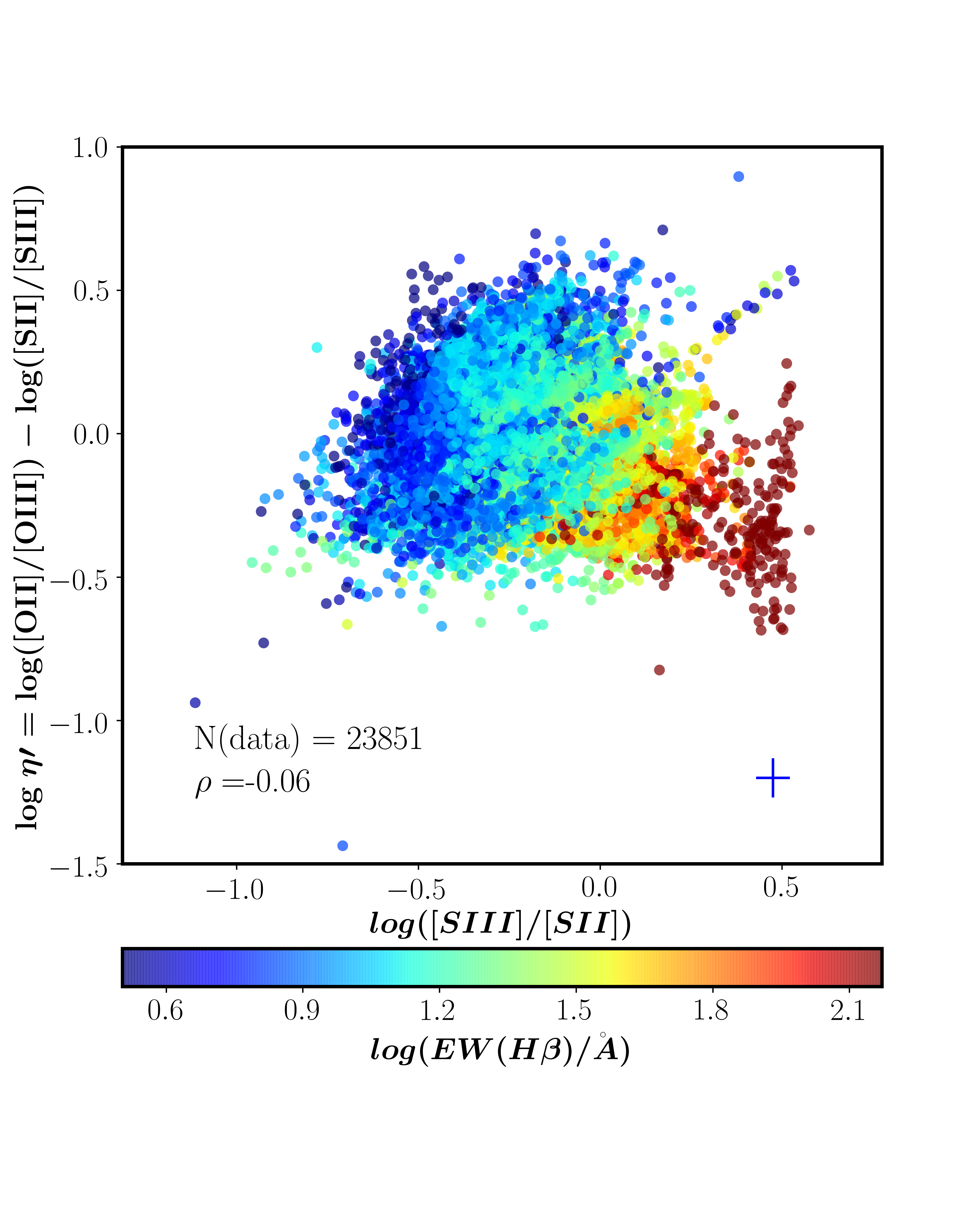}
	\includegraphics[width=0.35\textwidth, trim={0 2.5cm 0 0},clip]{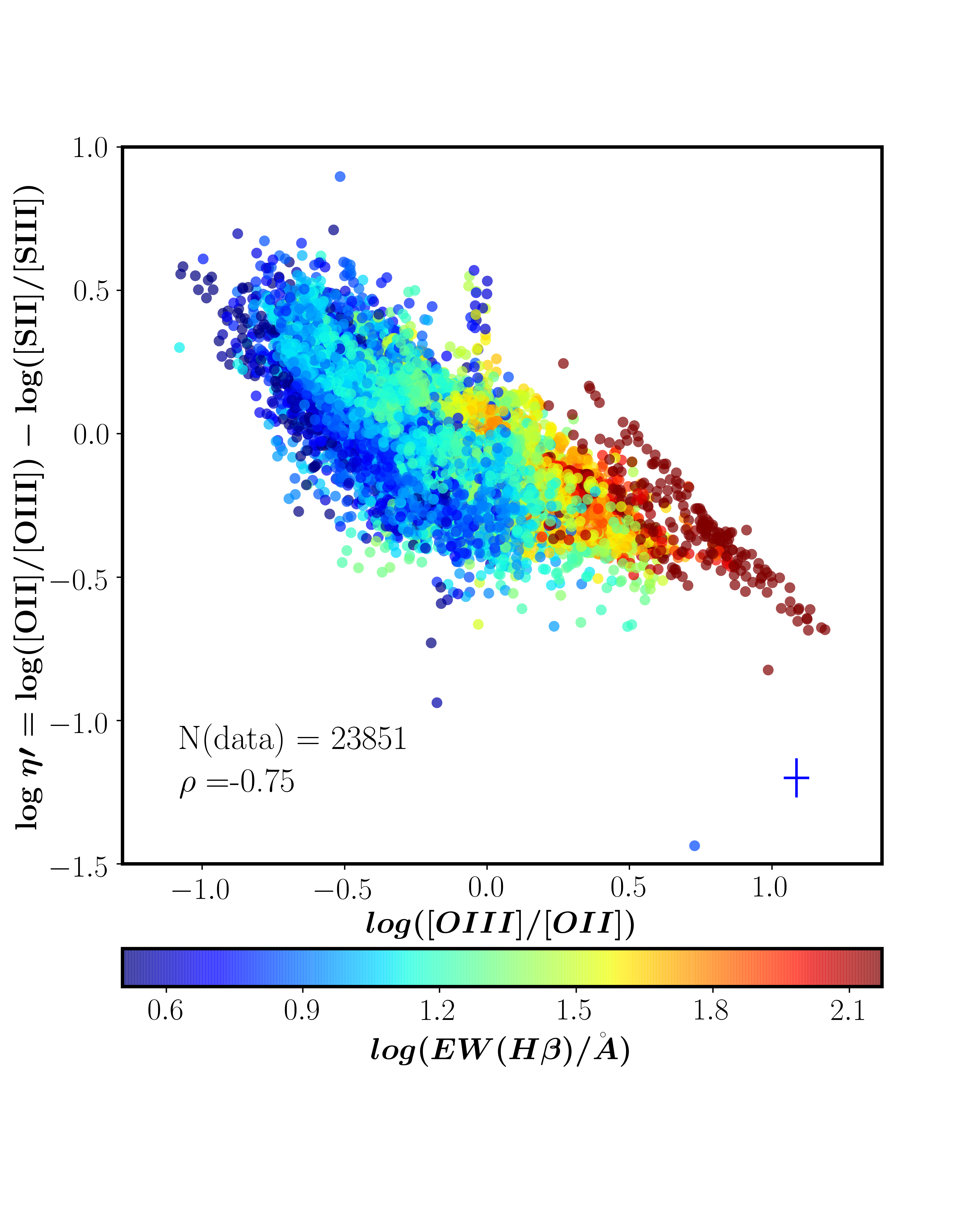}
	\caption{Relation between log $\rm\eta\prime$ and emission line ratios sensitive to the ionization parameter, S$_3$S$_2$ (upper panel) and O$_3$O$_2$ (lower panel), where data points are colour-coded with respect to EW(H$\beta$). The typical uncertainties on the variable on x- and y-axes are shown in the lower-right corner. `N(data)' represents the total number of plotted data points. $\rho$ denotes the Pearson correlation coefficient.}
	\label{fig:EW logU}
\end{figure}

\begin{figure}
	\centering
	\includegraphics[width=0.35\textwidth, trim={0 2.5cm 0 0},clip]{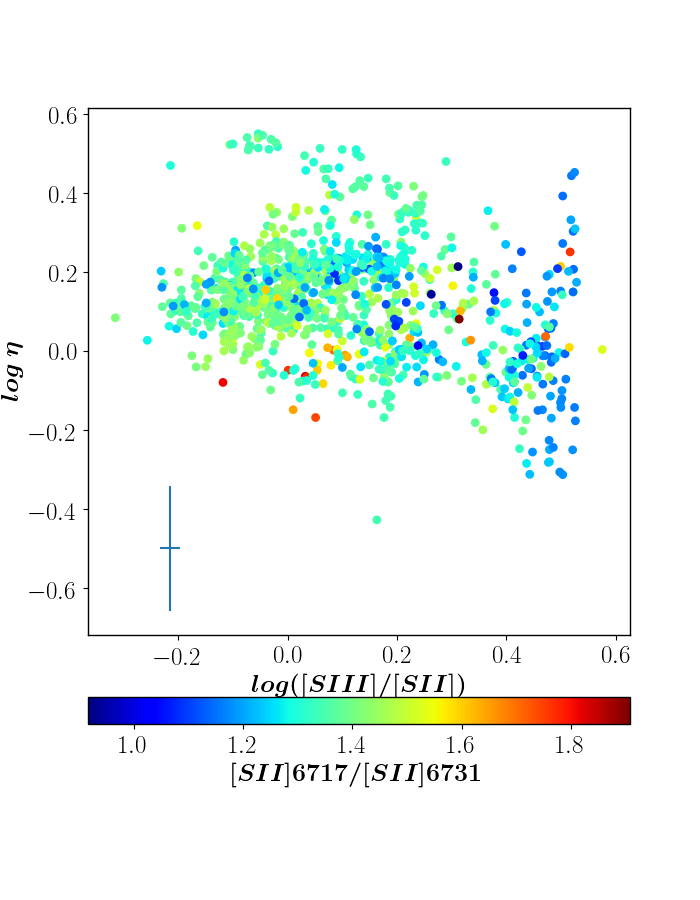}
	\includegraphics[width=0.35\textwidth, trim={0 2.5cm 0 0},clip]{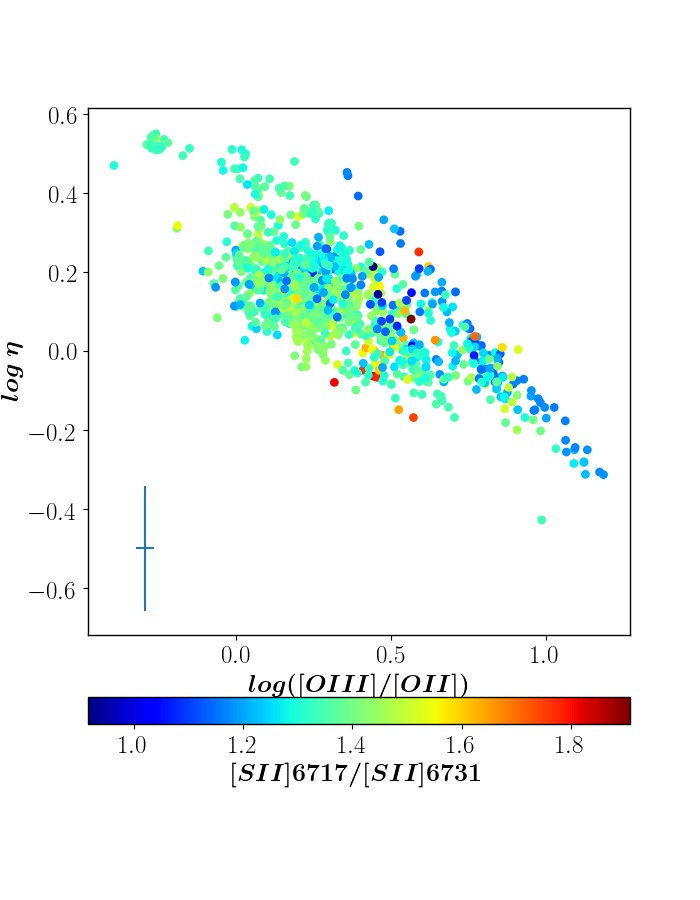}
	\caption{Relation between log $\rm\eta$ and emission line ratios sensitive to the ionization parameter, [S \textsc{iii}]/[S \textsc{ii}] (upper panel) and [O \textsc{iii}]/[O \textsc{ii}] (lower panel), where data are color-coded with respect to N$_e$-sensitive [S \textsc{ii}] doublet ratio. The typical uncertainties on the variable on x- and y-axes are shown in the lower-left corner. }
	\label{fig:eta logU}
\end{figure}

\begin{figure}
    \centering
    \includegraphics[width=0.35\textwidth]{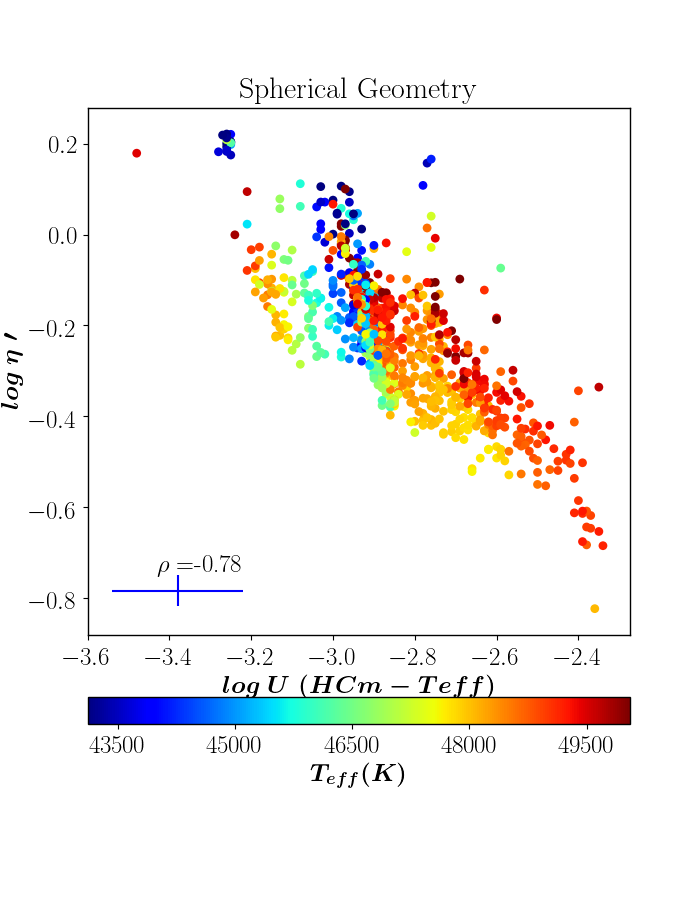}
    \caption{Relation between log $\rm\eta\prime$ and log $\mathcal{U}$ and colour-coded with respect to T$\rm_{eff}$ where log $\mathcal{U}$ and T$\rm_{eff}$ are determined from the code HCm-Teff for spherical geometry. $\rm \rho$ indicates the pearson correlation coefficient. The typical uncertainties on the variable on x- and y-axes are shown in the lower-left corner.}
    \label{fig:log U Teff}
\end{figure}

\begin{figure}
	\centering
	\includegraphics[width=0.33\textwidth, trim={0 2.5cm 0 0},clip]{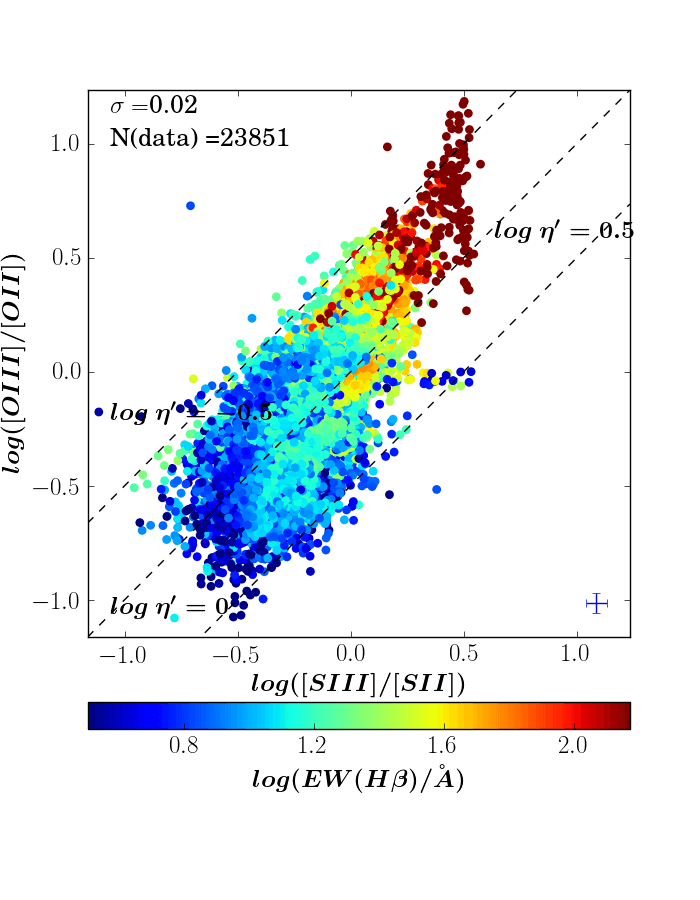}
	\includegraphics[width=0.33\textwidth, trim={0 2.5cm 0 0},clip]{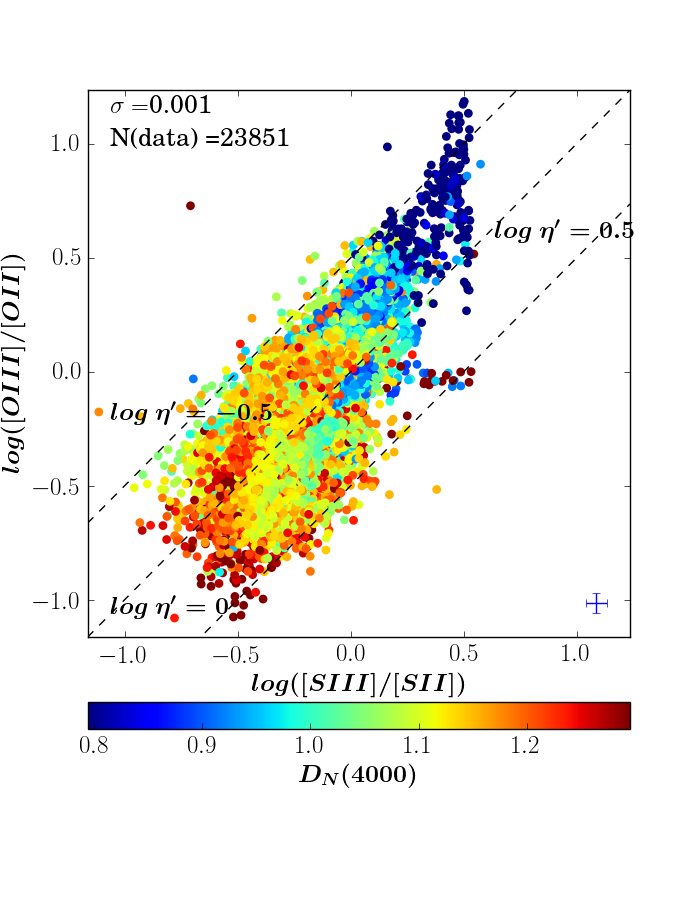}
	\caption{Spatially-resolved data of all 67 galaxies plotted on O$_3$O$_2$-S$_3$S$_2$ plane, where data points are colour-coded with respect to log (EW(H$\beta$)) (upper panel) and D$\rm{_N}$(4000) (lower panel). The diagonal dashed lines represent the constant values of log $\eta \prime$ = $-$0.5, 0, 0.5. The typical uncertainties on O$_3$O$_2$ and S$_3$S$_2$ are shown in the lower-right corner of each panel, while typical uncertainties on log (EW(H$\alpha$)) and D$\rm{_N}$(4000) are shown by $\sigma$ in the upper-left corner of upper and lower panels, respectively. `N(data)' represents the total number of plotted data points.}
	\label{fig:hardening ew}
\end{figure}

\subsection{Ionization parameter \& Equivalent Effective Temperature}
\label{sect:etaprime U}

\indent Figure \ref{fig:EW logU} shows the relation between log $\eta \prime$ and the emission line ratios sensitive to ionization parameter, S$_3$S$_2$ (upper panel) and O$_3$O$_2$ (lower panel), where data points are colour-coded with respect to EW(H$\beta$) and the Pearson correlation coefficient are mentioned in the bottom-left.  We find no correlation between log $\rm\eta\prime$ and S$_3$S$_2$ (upper panel, $\rho$ = -0.06) while log $\rm\eta\prime$ decreases with an increase in O$_3$O$_2$ (lower panel, $\rho$ = -0.75). S$_3$S$_2$ and O$_3$O$_2$ depend on log $\mathcal{U}$ and T$\rm_{eff}$ differently. While S$_3$S$_2$ depends more on log $\mathcal{U}$, O$_3$O$_2$ depends more on T$\rm_{eff}$ \citep{Perez-Montero2019}. Similarly, S$_3$S$_2$ has been shown to be a better diagnostic of ionization parameter than O$_3$O$_2$ via \textsc{cloudy} photoionization models \citep{Morisset2016}. As such, no correlation of S$_3$S$_2$ and log $\eta\prime$ might indicate that ionization parameter does not depend on radiation hardness as probed by log $\rm\eta\prime$. A similar behaviour is seen in Figure \ref{fig:eta logU} where we compute log $\rm\eta$ using equation \ref{eq:eta tes} and study its variation with respect to S$_3$S$_2$ (upper panel) and O$_3$O$_2$ (lower panel). However, we find in Section \ref{sect:hardening metallicity} that $\eta$ and $\eta\prime$ are correlated with strong line ratios such that N$_2$ and O$_3$N$_2$ which are not only sensitive to abundance but also to the ionization parameter. Hence, ionization parameter might be related to radiation hardness as well, as indicated by other works \citep[see e.g.][]{Perez-Montero2020}. We discuss this further in Section \ref{sect:comp pub models}. 

\indent We explore this further in Figure \ref{fig:log U Teff} which shows the relation between log $\rm\eta\prime$ and log $\mathcal{U}$ and color-coded with respect to the equivalent effective temperature T$\rm_{eff}$. Both log $\mathcal{U}$ and T$\rm_{eff}$ are computed from the python-based code \textsc{HCm-Teff} (v3.1)\footnote{https://www.iaa.csic.es/$\sim$epm/HII-CHI-mistry-Teff.html} \citep{Perez-Montero2019} using the direct-method metallicity and the emission line fluxes of [O \textsc{ii}] $\lambda$3727, [O \textsc{iii}]$\lambda$5007, [S~\textsc{ii}]$\lambda6717,6731$ and [S~\textsc{iii}] $\lambda$9069. We find a lower value of log $\mathcal{U}$ for a higher value of log $\eta \prime$ and vice-versa, though error bars are large. Hence, an interdependence between log $\mathcal{U}$ and log $\rm\eta\prime$ might be present. We discuss and explore this interdependence more via a comparison of data with models in Section \ref{sect:comp pub models}.  

\indent However, we also note that $\mathcal{U}$, by definition, is not related to the shape of radiation field but the production rate of the Lyman continuum photons, distance from the stars or stellar clusters and ionized or neutral hydrogen density. So, it is also possible that the emission line ratios S$_3$S$_2$ is rather tracing the ionization state of the gas which is determined by factors including radiation hardness, $\mathcal{U}$ and optical depth \citep{Ramambason2020}.

\indent In Figure \ref{fig:log U Teff}, we do not find any relation between log $\rm\eta\prime$ and equivalent effective temperature in contrast to the  previous studies which suggest that log $\eta \prime$ decreases with an increase in T$\rm_{eff}$ for galactic and extragalactic H \textsc{ii} regions \citep{Kennicutt2000} and for the disk-averaged radial profiles of external galaxies \citep{Perez-Montero2009b}. However, we can not rule out such a dependence because the typical uncertainties on T$\rm_{eff}$ estimated from \textsc{HCm-Teff} in our work is almost as large as the range of T$\rm_{eff}$. 

\indent We also studied the relation of log $\rm\eta$ with respect to equivalent effective temperature and ionization parameter, where the last two quantities were computed using HCm-Teff code as explained earlier. The trends are similar to that shown in Figure \ref{fig:log U Teff}. Note that the large errors on T$\rm_{eff}$ and lack of clear relation of log $\eta$ or log $\eta\prime$ with log $\mathcal{U}$ is likely due to the model grids used in \textsc{Hcm-Teff} (v3.1) based on single WM-Basic stars.  We show later in Section \ref{sect:comp pub models} that model grids corresponding to older age stellar populations satisfy MaNGA dataset which can be used with \textsc{HCm-Teff} to study the relations of radiation hardness with log $\mathcal{U}$ and T$\rm_{eff}$.

\subsection{Age of Stellar Populations}
\label{sect:hardening ew}

\indent Figure \ref{fig:hardening ew} shows O$_3$O$_2$-S$_3$S$_2$ plane where data-points are colour-coded with respect to equivalent width of H$\beta$ (EW(H$\beta$)) (upper panel) and the narrow index of 4000 \AA -break (D$\rm{_N}$(4000)) (lower panel), parameters sensitive to age of stellar populations.

\indent EW(H$\beta$) measures the ratio of young ionizing population traced by H$\beta$ emission line flux to the older non-ionizing population traced by the underlying continuum. As young and hot massive stars die, the supply of ionizing photons decreases which depletes the nebular content of any hydrogen recombination lines (including H$\beta$), while the level of continuum is determined by long-lived, cooler lower-mass stars. Hence, EW(H$\beta$) decreases as the age of stellar population producing ionizing photons increases. As such, EW(H$\beta$) is a good diagnostic for starburst age and is used in various IFS studies but its use depends on assumption on various other parameters such as metallicity, initial mass function, stellar mass loss rates and star formation history \citep{Leitherer1999, Levesque2013}. For instantaneous star-formation, EW(H$\beta$) is sensitive to stellar populations of ages up to 10 Myr (\textsc{straburst99}). In Figure \ref{fig:hardening ew} (upper panel), we find that EW(H$\beta$) spans a range of $\sim$ 1--100 \AA~ indicating that our sample include star-forming regions with a variety of young stellar populations, from relatively evolved to very recent starburst. However, our data do not show a clear correlation between EW(H$\beta$) and $\eta \prime$ which might be partly due to the H$\beta$ absorption. At high O$_3$O$_2$ and S$_3$S$_2$ we find the highest EW(H$\beta$), which corresponds to very young ($\sim$3--5 Myr) stellar population. Most of these data points show $\rm\eta\prime$ lower than zero, suggesting a harder radiation field compared to the average $\rm\eta\prime$ of lower EW points, and hence indicates that the lower age stellar population are characterized by harder radiation field. The result is consistent with that of  \citet{Kewley2001} which find that the ionizing spectrum becomes harder for lower age stellar population synthesis models of \textsc{Pegase v2.0} \citep{Fioc1997} and \textsc{starburst99} \citep{Leitherer1999}. In Figures \ref{fig:EW logU} and \ref{fig:hardening ew}, we find high O$_3$O$_2$ and high S$_3$S$_2$ for high EW(H$\beta$) overall, consistent with \citet{Campbell1986}, which find strong correlation between the O$_3$O$_2$ and EW(H$\beta$) in H \textsc{ii} galaxies indicating that high ionization parameter might be related to younger stellar populations. 


\indent D$\rm{_N}$(4000) represents the ratio of the narrow continuum bands \citep[3850--3950\AA~ and 4000--4100\AA,][]{Balogh1999}  around the break occurring at 4000 \AA~ which is caused by a strongly changing opacity of stellar atmospheres at this wavelength. In hot O and B stars, metal ions exist in highly ionized state which effectively decreases the opacity leading to weaker 4000 \AA-break than stars of other spectral types. As a result, old metal-rich galaxies have larger 4000 \AA-break than galaxies hosting younger stellar population. In Figure \ref{fig:hardening ew} (lower panel), we find that a majority of data points have D$\rm{_N}$(4000) lying in the range 0.8--1.3  (i.e. at 99\% confidence interval) indicating relatively young stellar populations, i.e. $\lesssim$ 500 Myr \citep[see e.g.][]{Noll2009, Winter2010}. We note here that unlike EW(H$\beta$), D$\rm{_N}$(4000) is independent of metallicity up to an age of 1 Gyr. Furthermore, unlike EW(H$\beta$), D$\rm{_N}$(4000) is sensitive to underlying older stellar population which might be co-spatial with the younger population traced by EW(H$\beta$). 
Despite such differences, both D$_N$(4000) and EW(H$\beta$) shows a broadly consistent variation on the O$_3$O$_2$-S$_3$S$_2$ plane, i.e. younger regions (with higher EW(H$\beta$) and low D$_N(4000)$) seem to have higher ionization parameter (i.e. higher O$_3$O$_2$ and higher S$_3$S$_2$) at a given value of $\rm\eta\prime$. 

\indent

\section{Discussion}
\label{sect:discussion}


\begin{figure}
	\centering
	\includegraphics[width=0.5\textwidth]{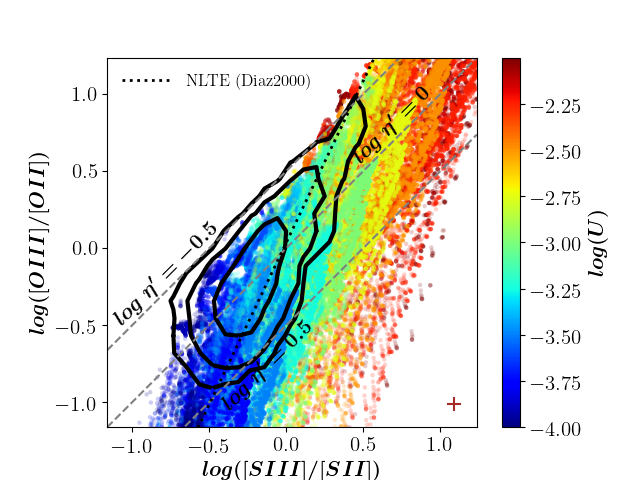}
	\caption{Photoionization models (colour points and black lines) and spatially-resolved data (grey contours) represented on the O$_3$O$_2$-S$_3$S$_2$ plane, where we have overlaid the three dashed lines corresponding to constant values of log $\eta \prime$ = -0.5, 0, 0.5. Inner to outer contours represent the 68\%, 95\% and 99.7\% of the datapoints shown in Figure~\ref{fig:hardening metallicity}, respectively. Both BOND and CALIFA models (see text for details) are colour-coded by the ionization parameter. The solid black line corresponds to equation \ref{eq:Diaz2000} and derived from the models of \citet{Diaz2000b}.}
	\label{fig:models}
\end{figure}

\begin{figure*}
	\centering
	\includegraphics[width=0.45\textwidth]{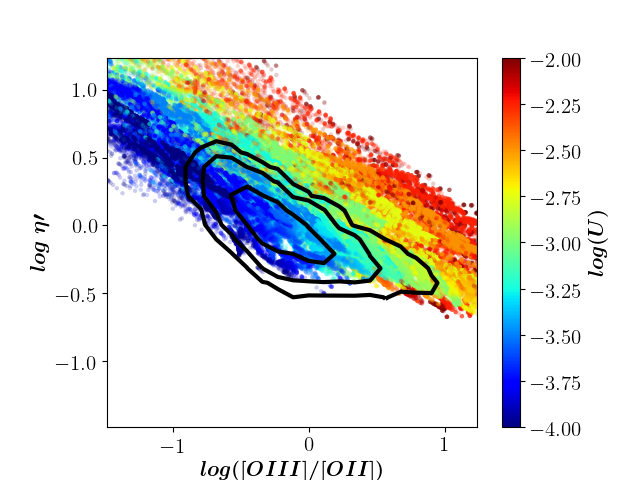}
	\includegraphics[width=0.45\textwidth]{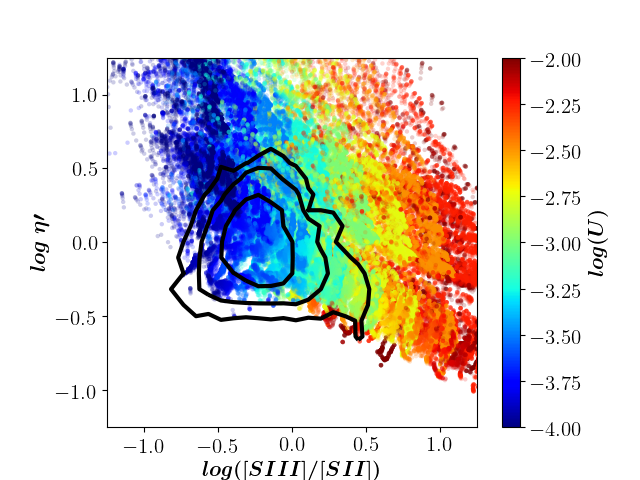}
	\caption{BOND and CALIFA photoionization models (colour points) and spatially-resolved data (black contours) showing the relation between log $\rm\eta\prime$ and O$_3$O$_2$ (Left-hand panel) and S$_3$S$_2$ (Right-hand panel). Inner to outer contours represent the 68\%, 95\% and 99.7\% of the datapoints shown in Figure~\ref{fig:EW logU}, respectively. Both BOND and CALIFA models (see text for details) are colour-coded by the ionization parameter.}
	\label{fig:models2}
\end{figure*}

\begin{figure*}
	\centering
	\includegraphics[width=0.45\textwidth]{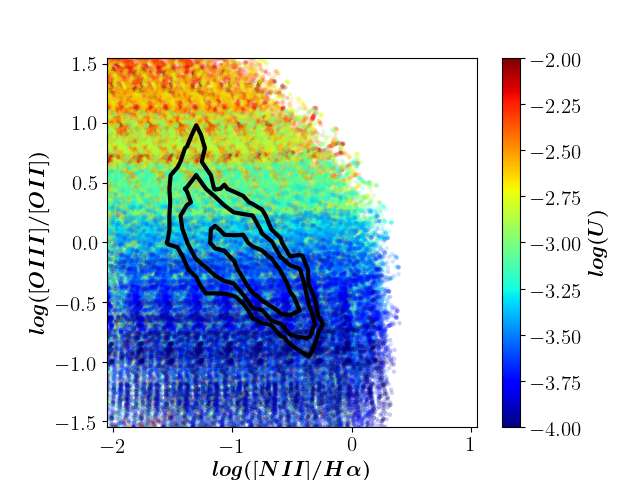}
	\includegraphics[width=0.45\textwidth]{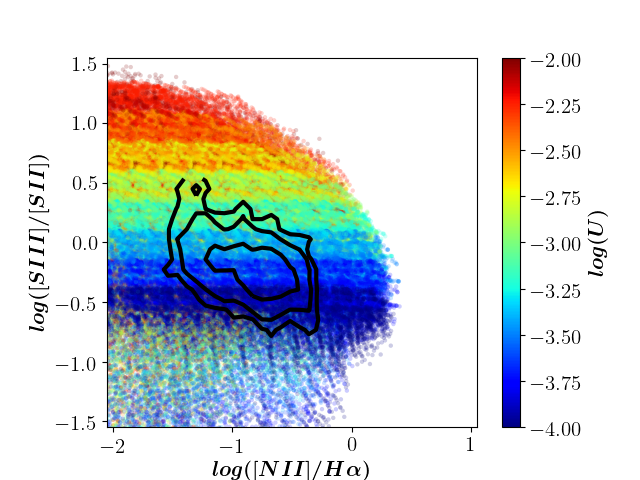}
	\caption{BOND and CALIFA photoionization models (colour points) and spatially-resolved data (black contours) showing the  O$_3$O$_2$ (Left-hand panel) and S$_3$S$_2$ (Right-hand panel) ratios as a function of N$_2$. Inner to outer contours represent the 68\%, 95\% and 99.7\% of the datapoints shown in Figure~\ref{fig:hardening metallicity}, respectively. Both BOND and CALIFA models (see text for details) are colour-coded by the ionization parameter.}
	\label{fig:models3}
\end{figure*}

\subsection{Comparison with photoionization models}
\label{sect:comp pub models}

In this section we  compare our observational results with  predictions from photoionization models. The goal of this exercise is to provide a qualitative interpretation of the observed sulphur line  ratios and discuss possible model limitations identified by previous works. In particular, some standard photoionization models appear to struggle in reproducing simultaneously  high and low ionization lines observed in star-forming galaxies using both long-slit and IFU data. For example,  \citet{Perez-Montero2009b} and \citet{Kehrig2006} specifically showed that such models fail to explain the hardness of ionizing radiation of most low-mass star-forming galaxies in the O3O2-S3S2 plane. More recently, the shortcomings of photoionization models to reproduce sulphur ratios of a more general population of star-forming galaxies using MaNGA data have been addressed by \citet{Mingozzi2020}, who suggest that this is due to limitations in the stellar atmosphere modelling and/or problems with sulphur line strengths due to inaccurate atomic data \citep[see also][and references therein]{Garnett1989,Kewley2019}. While the sample used for the comparison performed by \citet{Mingozzi2020} is larger than that used in this work, the explored range of line ratios is actually very similar. 
For their comparison they used four different models which include those of \citep{Dopita2013, Levesque2010, Byler2017, Perez-Montero2014}. The best agreement with data was found to be the grid of \textsc{Cloudy} photoionization models by \citet{Perez-Montero2014}, though these models did not cover a significant fraction of the data analysed by \citet{Mingozzi2020} and in particular the S$_3$S$_2$ ratio.


In order to explore these issues further, we have used the most recent \textsc{CLOUDY v17.02} photoionization models  available through the Mexican Million Models database \citep[3MdB,][]{Morisset2015}. One of the improvements of \textsc{Cloudy v17.02} compared to previous versions is the inclusion of updated atomic parameters for sulphur, which is relevant to our study. Here, we use two sets of model grids, namely BOND-2 and CALIFA-2 in the 3MdB-17 database\footnote{The full 3MdB database, including these models, is documented and can be accessed at \url{https://sites.google.com/site/mexicanmillionmodels/}} both of which adopt a spectral energy distribution obtained from the population synthesis code PopStar \citep{Molla2009} for a \citet{Chabrier2003} stellar IMF between 0.5 and 100 M$_{\odot}$. BOND-2 is an extension of the model grids developed for the code BOND \citep[Bayesian Oxygen and Nitrogen abundance Determinations, ][]{ValeAsari2016} for the starburst age going up to 6 Myr and is used to derive oxygen and nitrogen abundances simultaneously in giant H \textsc{ii} regions. On the other hand, CALIFA-2 is an extension of the model grids devised for the analysis of CALIFA galaxies \citep{Cid2013} and additionally uses the \textsc{starlight} spectral base of simple stellar populations of up to several Gyr \citep{Cid2014}.  The combination of BOND and CALIFA models span a wide range in gas-phase metallicity $\sim$\,0.01-1.5 solar, log $\mathcal{U}$ between $-4$ and $-1.5$, log(N/O) between $-$2 and 0.5 and a fixed electron density of 100\,cm$^{-3}$.

In Figure \ref{fig:models}, we show the full set of BOND and CALIFA models  predictions for the O$_3$O$_2$-S$_3$S$_2$ plane as a function of log $\mathcal{U}$. Overlaid,  we display contours representing the 68\%, 95\% and 99.7\% of the MaNGA spatially-resolved data shown in  Figure~\ref{fig:hardening metallicity}. Additionally, we include non-LTE model predictions from \citet{Diaz2000b}, whose derivation is shown in Appendix~\ref{A1}. We prefer here to show the full set of individual model predictions for each line ratio instead of the usual grid visualization in order to highlight differences and similarities with data and provide a qualitative interpretations of our results. Finally, we note that shock models, which may affect the line ratios in complicated ways, have not been considered. 

Conversely to previous works, the combination of model predictions from BOND and CALIFA do cover most data points in our sample. We note that this is mostly because of the inclusion of CALIFA models, as can be seen from the comparison shown in Figures~\ref{fig:models_A1}-\ref{fig:models_A4}. The use of BOND models alone produce a systematic mismatch with nearly half of the spaxels, which is removed when CALIFA models are adopted. Moreover, in Figure~\ref{fig:models} the mean relation obtained from \citet{Diaz2000b} models match the data relatively well, but a slight offset in the models towards higher S$_3$S$_2$ is apparent.

One of the main differences between the BOND and CALIFA models is the age of the ionizing stellar populations. In the former, this is restricted to OB stars in their first 6 Myr, while  the latter includes aged stellar populations of up to several Gyr which contribute to lower-ionization regions \citep[see e.g.][]{Morisset2016}. 
While such older age stellar population are generally found in the early type galaxies with only a few or negligible H \textsc{ii} regions, it is also possible that there is an underlying older stellar population along with the young stellar population within the galaxies under study. For example, a few galaxies in our sample are blue compact dwarfs which are known to host both young and old stellar populations \citep[see e.g.,][]{Aloisi2005, Amorin2009}. Another possibility is the presence of hot low-mass evolved stars \citep[HOLMES,][]{Flores-Fajardo2011} with hard radiation fields as shown in Figure \ref{fig:models_A4} where lower EW(H$\beta$) and lower log(Q0/Q1) indicative of older stellar population tend to have relatively lower log $\eta\prime$ indicative of harder radiation field. HOLMES  might be associated with the low surface brightness regions, mostly have low  log $\mathcal{U}$ and low log $\rm\eta\prime$ values, i.e. low O$_3$O$_2$ and low S$_3$S$_2$. As Figure~\ref{fig:models} shows, lower log $\mathcal{U}$ models match better with lower S$_3$S$_2$, whereas low  O$_3$O$_2$ does not necessarily indicate low ionization parameter, probably pointing to secondary dependencies \citep[e.g. metallicity;][]{Kewley2019}.


A good agreement between observations and models is also seen in Figure~\ref{fig:models2}, where we show the variation of log $\rm\eta\prime$ as a function of S$_3$S$_2$ (left-hand panel) and O$_3$O$_2$ (right-hand panel) and color-coded with respect to log $\mathcal{U}$. This figure shows that the BOND and CALIFA models predict nearly all possible values of ionization parameter for a fixed value of log $\rm\eta\prime$. This comparison also help us to understand the observational trends seen in Figure~\ref{fig:EW logU}. The MaNGA data with higher EWs can be identified with models having both larger log $\mathcal{U}$ and harder radiation field ($\rm\eta\prime$). Comparing the trends shown by S$_3$S$_2$ and O$_3$O$_2$, the former appears to be again a better diagnostic of the ionization parameter, less dependent of the hardness of the ionizing radiation. Moreover, Figure~\ref{fig:models3} shows that S$_3$S$_2$ has a negligible relation with the metallicity-sensitive N$_2$ ratio in the models, in contrast to the O$_3$O$_2$ ratio. Similar trends are seen in the data. Therefore,  S$_3$S$_2$ appears again as a more reliable tracer of log $\mathcal{U}$ than O$_3$O$_2$ as secondary dependencies with metallicity appear smaller.

The discrepancies found in previous works between data and models may come from a variety of factors. Models strongly rely on their simplified assumptions, such as the adopted geometry of the nebula (see Figure~\ref{fig:models}), simplified temperature and density structure, or the uncertain atomic data for sulphur \citep[e.g.][]{Kewley2019}. In the case of the line ratios under consideration, we find that models including the contribution to low-ionization gas by aged stellar populations, in particular HOLMES with hard radiation fields can significantly contribute to solve the discrepancies found by previous works relying on models in which the low and high ionization lines are powered only by very young stellar clusters  \citep[e.g.][]{Mingozzi2020}. 
Indeed, HOLMES have been proposed to explain line ratios sensitive to DIG in the past \citep[e.g.][]{Morisset2016}, as these stellar populations can significantly contribute to increase the strength of some low-ionization lines [S\textsc{ii}]  \citep{Sanders2017}. In Figure~\ref{fig:models} and Figure~\ref{fig:models2}, we see that data showing low S$_3$S$_2$ and moderate to low O$_3$O$_2$ (i.e. relatively hard radiation field but low ionization parameter) related to the low-excitation gas can be better described with CALIFA models than BOND models, which seems to match better data with higher log $\mathcal{U}$ and higher log $\rm\eta\prime$ values (Figure~\ref{A1} for a direct comparison). 

Although we selected our data points to have [S\textsc{iii}] detections and the contribution of DIG emission in our sample should not dominate in high EW data points probing high surface brightness regions (see examples of spatially resolved maps in Appendix~\ref{sect:individual}), the contribution of HOLMES is probably not negligible. This is especially true at low EW(H$\beta$) (characteristic of HOLMES), making the [S\textsc{ii}] emission higher and therefore decreasing the S$_3$S$_2$ ratios in Figure\ref{fig:models2} compared to models. Similarly, there is likely an effect of underlying older stellar population in these young star-forming galaxies. We principally performed this analysis on spaxels with [S \textsc{iii}] detection, some of which go beyond the maximum star-burst line in Figure \ref{fig:BPT all} and hence could be affected by DIG. To rule out any effects in sample selection, we investigated only those data points which lie below the maximum starburst line on [S \textsc{ii}]-BPT and found that there were still line ratios on O$_3$O$_2$-S$_3$S$_2$ plane which could be explained by the older age CALIFA model grids but not the BOND model grids.


On the other hand, we do not see any significant mismatch between data and the BOND models in the S$_3$-N$_2$ and S$_3$-S$_2$ diagnostics presented in Figure~\ref{fig:SIII-BPT all}, which are also reproduced by the CALIFA models (see Appendix~\ref{A1}, Figure~\ref{fig:models_A1}-\ref{fig:models_A2}). From these simple diagnostic, we can conclude that the S$_3$-S$_2$ plane (or the S$_3$S$_2$ ratio) is an excellent diagnostic for the ionization parameter which shows a little dependence with metallicity (Figure~\ref{fig:models3}), as also shown by previous models \citep[e.g.][]{Kewley2002, Kewley2019}.

Finally, we note that a recent approach to reconcile observed S$_3$S$_2$ and O$_3$O$_2$ ratios with models is presented by \citet{Ramambason2020}, who show that classic photoionization models, like the ones we use in Figure~\ref{fig:models2} and \ref{fig:models3}, can underpredict the low-excitation line emission of [S\textsc{ii}] arising from the outskirts of H \textsc{ii} regions. To alleviate this, \citet{Ramambason2020} propose a composite model combining high and low ionization parameters which is able to reconcile the S$_3$S$_2$ ratios of star-forming galaxies in SDSS using the same BONDS model grids adopted in our study. This is likely a much more reliable solution to model star-forming regions showing very high ionization, like the ones analysed by \citet{Ramambason2020}.

\begin{figure}
	\centering
	\includegraphics[width=0.45\textwidth]{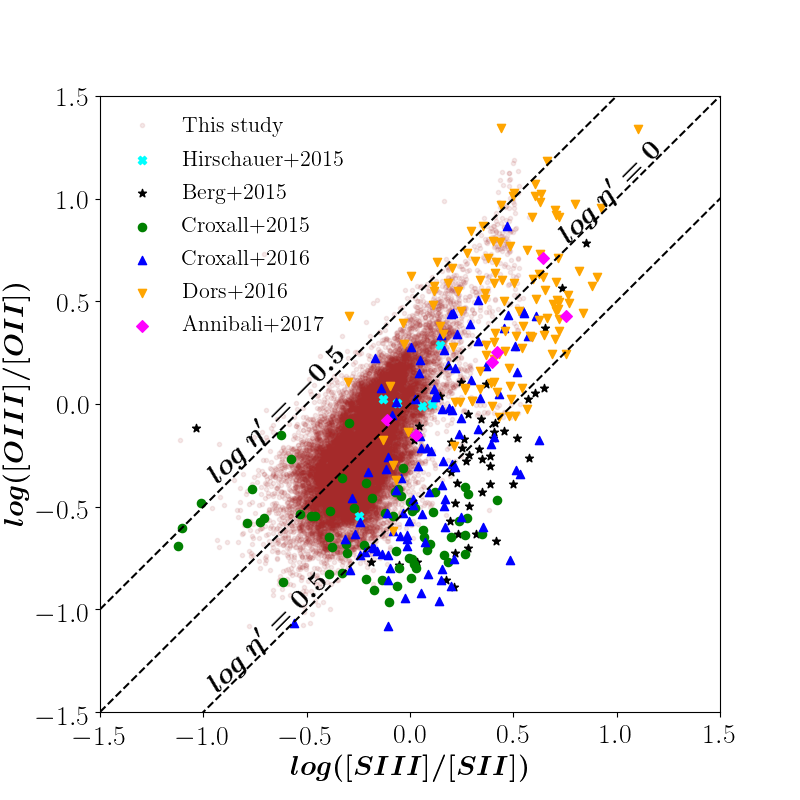}
	\caption{O$_3$O$_2$-S$_3$S$_2$ plane where we compare spaxel-by-spaxel MaNGA data from this study with published data of star-forming galaxies and H \textsc{ii} regions within star-forming galaxies. We have overlaid the three dashed lines corresponding to constant values of log $\rm\eta\prime$ = -0.5, 0, 0.5. }
	\label{fig:global local}
\end{figure}

\subsection{Comparison with previous works}
\label{sect:comp pre works}
\indent In Figure \ref{fig:global local}, we study the O$_3$O$_2$-S$_3$S$_2$ plane where we compare spatially-resolved data from this study (brown points) with published data from literature, including H \textsc{ii} regions along with global data of emission line galaxies. Table \ref{tab:compilation} shows data from literature along with references.

\indent We find that H \textsc{ii} regions from the nearby blue compact dwarf galaxy NGC4449 \citep{Annibali2017} are close to log $\eta \prime$ = 0. The sample of \citet{Dors2016} consists of H \textsc{ii} regions and star-forming galaxies and lie within the same range of log $\eta \prime$ as our spatially-resolved data, i.e. -0.5 $<$ log $\eta \prime$ $<$ 0.5. However, the majority of these data set have comparatively higher values of O$_3$O$_2$ and S$_3$S$_2$ than our spatially-resolved data set. Since these two line ratios are sensitive to ionization parameter, it is possible that a large part of the sample of \citet{Dors2016} consists of objects with comparatively higher values of ionization parameter. We also find that all galaxies from \citet{Hirschauer2015} except one lie close to log $\eta \prime$ = 0. We note that the galaxy from the sample of \citet{Hirschauer2015} with a relatively higher value of log $\eta \prime$ have a relatively lower value of [O \textsc{iii}] equivalent width and higher metallicity compared to the other galaxies in their sample.  

\indent In the case of spiral galaxies, NGC 0628 (black stars, \citealt{Berg2015}), NGC 5194 (green filled circles, \citealt{Croxall2015}) and NGC 5457 (blue triangles, \citealt{Croxall2016}), we find that a few H \textsc{ii} regions lie within -0.5 $<$ log $\rm\eta\prime$ $<$ 0.5 but a majority of H \textsc{ii} regions tend to have higher values of log $\rm\eta\prime$ than average of our spatially-resolved data. Spiral galaxies are known to present gradients in several physical properties. \citet{Perez-Montero2019} studied the radial profiles of gas-phase metallicity, log $\rm\eta\prime$, log $\mathcal{U}$ and T$\rm_{eff}$ for these three galaxies using the same data set used here. For NGC 5194, they find that the H \textsc{ii} regions in the central region ($<$ 4R$_e$) of this galaxy show a drop in metallicity and have log $\rm\eta\prime$ $<$0.5, while the H \textsc{ii} regions in the outskirts have higher log $\rm\eta\prime$ $>$ 0.5 and near-solar metallicity. In Figure \ref{fig:global local}, H \textsc{ii} regions in NGC 5194 (green points) with log $\rm\eta\prime$ $<$ 0.5 and sub-solar metallicity coincide with our sample. Hence, the offset of H \textsc{ii} regions in spiral galaxies with respect to the MaNGA spatially-resolved data on the O$_3$O$_2$-S$_3$S$_2$ plane in Figure \ref{fig:global local} indicate that our sample of  predominantly irregular galaxies (and hence relatively  lower gas-phase metallicities) have harder radiation fields on average compared to the higher-metallicity H ii regions of CHAOS spiral galaxies.  \citet{Perez-Montero2019} also mention that there might be some relation of radiation hardness with log $\mathcal{U}$ and T$\rm_{eff}$ for which they suggest doing more detailed analysis. Furthermore, another possibility of the mismatch between MaNGA spatially-resolved data and the overall location of H \textsc{ii} regions can be explained by the age of stellar populations. In Section \ref{sect:comp pub models}, we find that the older stellar population included as ionization source in CALIFA model grids are necessary to match the MaNGA spatially-resolved data, while \citet{Perez-Montero2019} show that the H \textsc{ii} regions within these spiral galaxies could be explained by the single star models.

\indent While it is hard to clearly find the relation between radiation hardness and fundamental nebular properties, the analyses here shown that the similarities and differences between MaNGA spatially-resolved data and other data set shown in Figure \ref{fig:global local} are likely related to gas-phase metallicities with secondary dependence of other properties such as ionization parameter and age and equivalent effective temperature of stellar populations and possibly DIG.

\begin{table}
	\caption{References for the literature data plotted in Figure \ref{fig:global local}.}
	\label{tab:compilation}
\begin{tabular}{ll}
	\toprule
	Reference & Object \\
	\midrule
	\citet{Berg2015} & NGC 0628 (H \textsc{ii} regions)\\
	\citet{Croxall2015} & NGC 5194 (H \textsc{ii} regions)\\
	\citet{Hirschauer2015} & Emission line galaxies\\
	\citet{Croxall2016} & NGC 5457(H \textsc{ii} regions)\\
	\citet{Dors2016} & H \textsc{ii} regions \& galaxies$^a$ \\
	\citet{Annibali2017} & NGC 4449 (H \textsc{ii} regions)\\ 
	\bottomrule
\end{tabular}

Notes:$^a$: Compiled from \citet{Kennicutt2003, Vermeij2002, Hagele2008, Bresolin2009, Vilchez2003, Hagele2006, Hagele2011, Izotov2006b, Guseva2011, Garnett1997, Vilchez1988, Skillman2013, Lopez-Hernandez2013, Zurita2012, Perez-Montero2003, Gonzalez-Delgado1995, Skillman1993, Russell1990, Hagele2012}.\\

\end{table}


\begin{figure}
	\centering
	\includegraphics[width=0.45\textwidth]{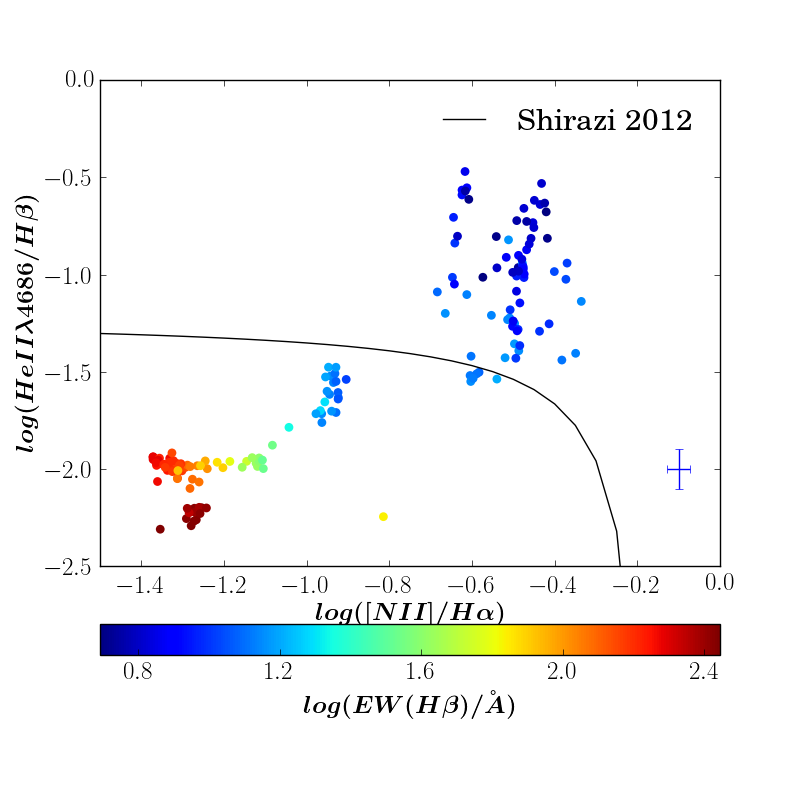}
	\caption{He \textsc{ii} $\lambda$4686/H$\beta$ versus [N \textsc{ii}]/H$\alpha$ on a spaxel-by-spaxel basis for all galaxies in the sample where He \textsc{ii} $\lambda$4686 was detected but no WR bump. These galaxies are marked with $\dagger$ in Table \ref{tab:properties}. The solid black curve represents the maximum starburst line taken from \citet{Shirazi2012}, and was devised to separate the star-forming galaxies from the AGNs or composite objects.} 
	\label{fig:He-BPT}
\end{figure}


\subsection{Radiation Hardness and Helium ionization}
\label{sect:etaprime He}

\indent The nebular Helium emission is often associated with hard radiation fields \citep[see e.g.,][]{Kehrig2015, Kehrig2018,Senchyna2020,Perez-Montero2020}. The He \textsc{ii} $\lambda$4686 line can only be produced by the sources of hard ionizing radiation because the ionization potential of He \textsc{ii} is very high (i.e. 54.4eV). Such hard radiation can be produced by AGNs \citep{Shirazi2012}, stellar sources, such as cool white dwarf stars \citep{Bergeron1997, Stasinska2008, Singh2013}, hot wolf-rayet (WR) stars \citep{Schaerer1996}, shocks \citep{Thuan2005}, massive X-ray binaries \citep{Garnett1991}, post-AGB stars, \citep{Binette1994, Papaderos2013}, rapidly rotating, metal-free massive stars and/or binary population of very metal-poor massive stars \citep{Kehrig2018}. Given the importance of Helium detection in the context of hardness, we  analyse seven galaxies separately here where He II 4686 was detected. These galaxies are marked with a $\dagger$ in Table \ref{tab:properties}, and relevant maps of this sample are shown in Section \ref{sect:He maps}. We have excluded two galaxies MaNGA-8250-3703 and MaNGA-8458-3702, in spite of  He \textsc{ii} $\lambda$4686 detection. The former is excluded because the He \textsc{ii} emission in this galaxy shows the Wolf-Rayet bump which is too broad to be considered as purely nebular. The latter is excluded because He \textsc{ii} and H$\beta$ are not co-spatial and as such prevents the analysis presented below.

\indent Figure \ref{fig:He-BPT} shows variation of He \textsc{ii} $\lambda$4686/H$\beta$ with [N~\textsc{ii}]/H$\alpha$ on a spaxel-by-spaxel basis in seven galaxies where He \textsc{ii} $\lambda$4686 are detected.  The solid black line corresponds to the maximum starburst line taken from \citet{Shirazi2012} and estimated from the \citet{Charlot2001} models based on population synthesis  codes of \citet{Bruzual2003} and \textsc{cloudy} photoionization models \citep{Ferland1996}. On the original diagram of \citet{Shirazi2012}, the region lying below the maximum starburst line correspond to star-forming galaxies while the region beyond this line correspond to He \textsc{ii}/H$\beta$ values which would  require the contribution of some non-thermal ionization mechanism, such as an AGN. In Figure \ref{fig:He-BPT}, we have colour-coded data points with respect to the age diagnostic, EW(H$\beta$) which shows a smooth gradient going from star-forming part of the diagram to the AGN/composite part. The data points lying beyond the demarcation line belong to three different galaxies (manga-8549-6104, manga-8553-3704 and manga-8613-12703), and these spaxels are located on the edges of the brightest region in the H$\alpha$ map, and have a relatively lower EW(H$\beta$). Since these spaxels do not coincide with the bright SF regions with high excitation but lie at their edges, it is possible that the enhanced He \textsc{ii}/H$\beta$ is due to the shocks \citep{Thuan2005, Shirazi2012}. However, an imperfect continuum subtraction might also be a problem as these spaxels with high He \textsc{ii}/H$\beta$ also have low EW(He $\textsc{ii}$) ($<$ 2\AA). 

\begin{figure}
	\centering
	\includegraphics[width=0.35\textwidth]{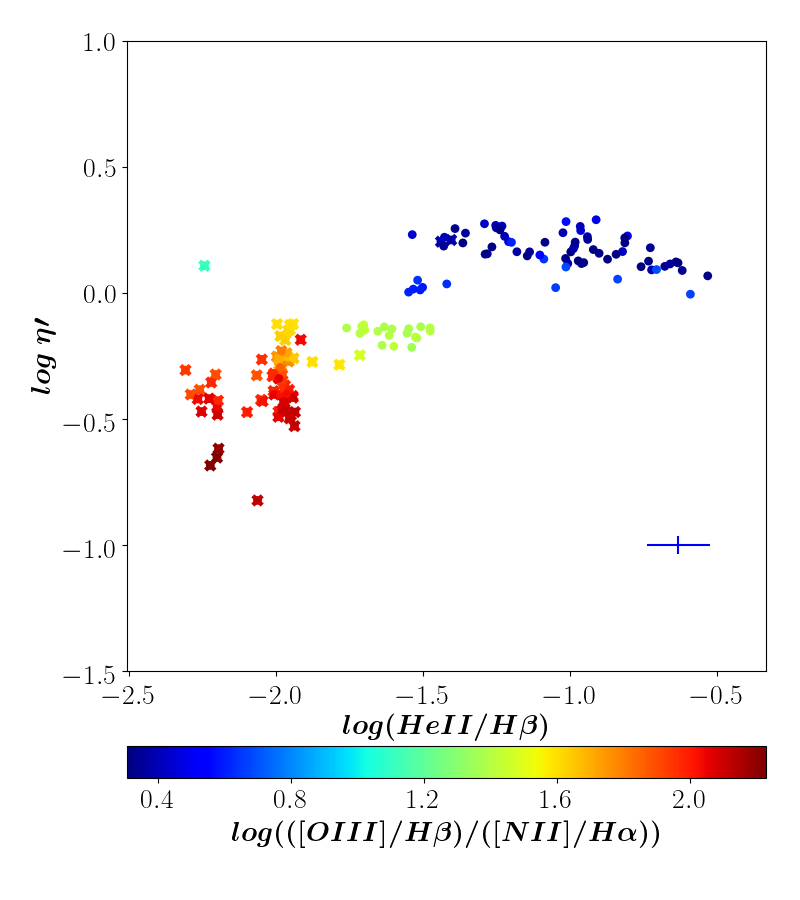}
	\caption{Relation between log $\eta \prime$ and He \textsc{ii} $\lambda$4686/H$\beta$, where data points are colour-coded with respect to abundance-sensitive line ratio O$_3$N$_2$. The typical uncertainties on the variable on x- and y-axes are shown in the lower-right corner. `N(data)' represents the total number of plotted data points. The data marked as stars have T$_e$([O \textsc{iii}]) lying between 7000--25000K.}
	\label{fig:etaprime He}
\end{figure}

\indent In Figure \ref{fig:etaprime He}, we study the relation between log $\eta \prime$ and He \textsc{ii} $\lambda$4686/H$\beta$\footnote{We could not do this experiment with other He lines because (i) He I 5876 overlapped with the Galactic sodium lines for one of the galaxies (ii) He I 6678 was not detected with S/N > 3.} on a spaxel-by-spaxel basis for those galaxies where He \textsc{ii} $\lambda$4686 is detected without any contamination by a WR broad stellar emission (see Table \ref{tab:properties}). log $\eta \prime$ increases with He \textsc{ii} $\lambda$4686/H$\beta$ up to log He \textsc{ii} $\lambda$4686/H$\beta$ $\sim$ -1.5, after which log $\eta \prime$ becomes constant, suggesting that lower values of  He \textsc{ii} $\lambda$4686/H$\beta$ correspond to harder radiation fields, contrary to the expectation that harder radiation field would result in nebular Helium lines. Further tests need to be done on a larger sample of galaxies where nebular helium lines are detected. The data marked as stars have [O \textsc{iii}] $\lambda$4363 detections and electron temperatures from the [O \textsc{iii}] 43636/[O \textsc{iii}] 5007 line ratio lie between 7000--25000K. The  data points in this Figure are colour-coded with respect to abundance-sensitive line ratio O$_3$N$_2$, thus showing a smooth increase in metallicity with increasing He \textsc{ii} $\lambda$4686/H$\beta$ and decreasing log $\eta \prime$. Though there is a strong dependence of O$_3$N$_2$ on the ionization parameter, we deem our choice of O$_3$N$_2$ to be the most appropriate here because the metallicity diagnostics such as R$_{23}$ and O$_3$S$_2$ \citep{Kumari2019a, Maiolino2019}, which have only a secondary dependence on log $\mathcal{U}$, are also bimodal. 

\indent In this study, we could not map all optical Helium lines (He I 5876, He I 6678, He I 7785), and He II 4686 was detected only in a few spaxels. Deeper S/N data would help us overcome challenges of this work, for which ground-based GMOS-IFU would be extremely useful. A possible avenue to investigate this work further is to study the UV Helium lines in low-metallicity local galaxies, for which COS and STIS instruments on HST would be extremely useful. With the launch of JWST, we will be able to apply the findings from the ground-based optical studies and space-based UV studies, to the high-z low-metallicity galaxies where harder ionization radiation fields are expected.   


\section{Summary}
\label{sect:summary}

\indent In this paper, we used the IFS MaNGA data of 67 nearby (0.02 $\lesssim$ z $\lesssim$ 0.06) star-forming galaxies to explore the relation between radiation hardness and various properties encoded in the emission lines emanating from the ionized gas within star-forming galaxies. In particular, we studied in detail softness parameter \citep[log $\rm\eta$, ][]{Vilchez1988} by investigating the observable quantity log $\rm\eta\prime$ which is the ratio of the two emission line ratios, O$_3$O$_2$ and S$_3$S$_2$ accessible by the long wavelength range of MaNGA covering the NIR sulphur lines such as [S \textsc{iii}] $\lambda\lambda$ 9069, 9532. In this analysis, we have considered various diagnostics sensitive to age, electron temperature, metallicity and ionization parameter in addition to exploring available models for explaining the observations. The main findings of this work are summarized below:

\begin{enumerate}
    \item  We find that log $\eta \prime$ is correlated to strong line metallicity diagnostics such as O$_3$N$_2$, N$_2$, Ar$_3$O$_3$, S$_3$O$_3$, R$_{23}$ and S$_{23}$. So, softness parameter, and consequently hardness of radiation fields can be  directly related to metallicity of ionized gas. We provide polynomial relations between log $\rm\eta$ and abundance-sensitive strong line ratios such as O$_3$N$_2$, N$_2$ and Ar$_3$O$_3$ which will allow us to study radiation hardness in galaxies where temperature-sensitive faint auroral lines are not detected thus preventing to estimate softness parameter from its definition. We caution here not to use these relations for high-metallicity spiral galaxies which are systematically offset with respect to the low-metallicity galaxies (7.12$<$12+log(O/H)$<$8.6) from which these relations are derived. 

\item We do not find any particular trend on the S$_3$S$_2$-O$_3$O$_2$ plot with respect to age diagnostics (EW(H$\beta$) and D$_N$(4000)), though there are signatures of harder radiation field for lower EW(H$\beta$). Similarly, we do not find direct evidence of a relation between radiation hardness and ionization parameter and equivalent effective temperature, but such possibilities can not be completely ruled out. No correlation is found between radiation hardness and electron density. 

\item We compare the spatially-resolved data with predictions from two publicly available \textsc{Cloudy (v.17)} photoionization models, which allow us to study the relation between the stellar age and radiation hardness. Photoionization models including both young and evolved stellar populations are able to predict the observed line ratios indicating that the hot and old low-mass stars such as HOLMES and underlying older stellar population in the star-forming galaxies might also be associated with hard radiation fields.

\item We compared the MaNGA data with published O$_3$O$_2$ and S$_3$S$_2$ emission line ratios for star-forming galaxies and H \textsc{ii} regions within star-forming galaxies, and compared them with MaNGA data set from this work. 
We find that higher metallicity H \textsc{ii} regions within CHAOS spiral galaxies have on average higher softness parameter than the relatively lower-metallicity MaNGA star-forming galaxies studied here.

\item Helium is generally associated with the harder radiation fields, hence we explored the relation between log $\rm\eta\prime$ and He \textsc{ii}/H$\beta$ in seven galaxies of our sample where He \textsc{ii}$\lambda$4686 is detected. We find that regions with softer ionizing radiation (i.e. higher $\rm\eta\prime$) with Helium detection tend to have higher He \textsc{ii}/H$\beta$ ratios, higher metallicity and lower EW(H$\beta$), and might be related to shocks.

\end{enumerate}

\indent Finally, the results of this study are useful in investigating the radiation hardness in high-z low-metallicity galaxies targeted by future ground and space-based telescopes such as James Webb Space Telescope (JWST) and European Extremely Large Telescope. This work is crucial in preparation for the upcoming high-z surveys exploiting the outstanding NIR facilities such as Near Infrared Spectrograph (NIRSPec) on JWST and Multi Object Optical and Near-infrared Spectrograph (MOONS) on Very Large Telescope (VLT). These future surveys will provide access to [S \textsc{iii}] lines for galaxies at intermediate and high redshifts, hence it is important to probe the usefulness of line ratios involving sulphur lines in local galaxies as potentially useful tracers of hardness, ionization, and metallicity. Moreover, the detailed spatially-resolved analysis of radiation hardness will further aid in understanding the extreme conditions in high redshift galaxies which host harder ionizing radiation \citep{Stark2015}.

\section*{Acknowledgements}

\indent We thank the referee for a thorough and constructive report, and for providing us with Equation \ref{eq:eta tes}. We also thank Claus Leitherer for a related discussion. NK acknowledges financial support from the Schlumberger foundation which facilitated her stay at the KICC during which a majority of work was carried out. RA acknowledges support from FONDECYT Regular Grant 1202007. RM acknowledges ERC Advanced Grant 695671 "QUENCH" and support by the Science and Technology Facilities Council (STFC). This project makes use of the MaNGA-Pipe3D dataproducts. We thank the IA-UNAM MaNGA team for creating this catalogue, and the ConaCyt-180125 project for supporting them \citep{Sanchez2016b}. This research made use of Marvin, a core Python package and web framework for MaNGA data, developed by Brian Cherinka, José Sánchez-Gallego, Brett Andrews, and Joel Brownstein \citep{Brian2018}; SAOImage DS9, developed by Smithsonian Astrophysical Observatory"; Astropy, a community-developed core Python package for Astronomy \citep{Astropy2013}.

\section*{Data availability}
\indent The data used in this work form part of the MaNGA DR14 Pipe3D value added catalog (Sánchez et al. 2016a,b;Sanchez et al. 2018) and are publicly available at \url{https://www.sdss.org/dr14/manga/manga-data/manga-pipe3d-value-added-catalog/}. 



\bibliographystyle{mnras}
\bibliography{bibliography} 

\section*{Supporting information}
\indent Supplementary data are available at MNRAS online.
\indent 

\noindent \textbf{Figures A1-A4}

\noindent \textbf{Figures B1-B3}

\noindent \textbf{Figures C1-C7}

\noindent \textbf{Figures D1-D8}


\appendix

\section{Photoionization models}
\label{A1}

\begin{figure*}
	\centering
\includegraphics[width=0.45\textwidth]{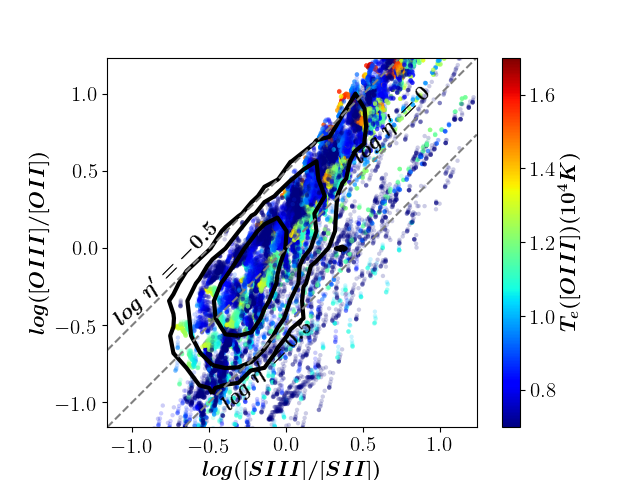}
\includegraphics[width=0.45\textwidth]{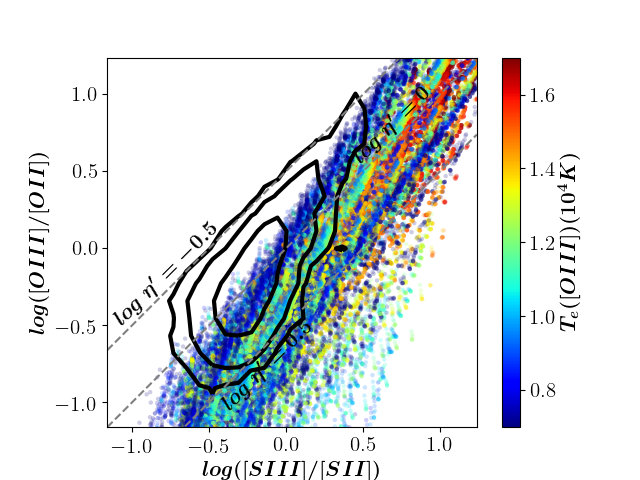}
	\caption{Photoionization models (colour points) and spatially-resolved data (black contours) represented on the O$_3$O$_2$-S$_3$S$_2$ plane, where we have overlaid the three dashed lines corresponding to constant values of log $\eta \prime$ = -0.5, 0, 0.5. Inner to outer contours represent the 68\%, 95\% and 99.7\% of the datapoints shown in Figure~\ref{fig:hardening metallicity}, respectively. CALIFA models (Left-hand panel) and BOND models (Right-hand panel) are colour-coded with respect to the electron temperature  (see Section~\ref{sect:te}).}
	\label{fig:models_A1}
\end{figure*}

\begin{figure*}
	\centering
\includegraphics[width=0.45\textwidth]{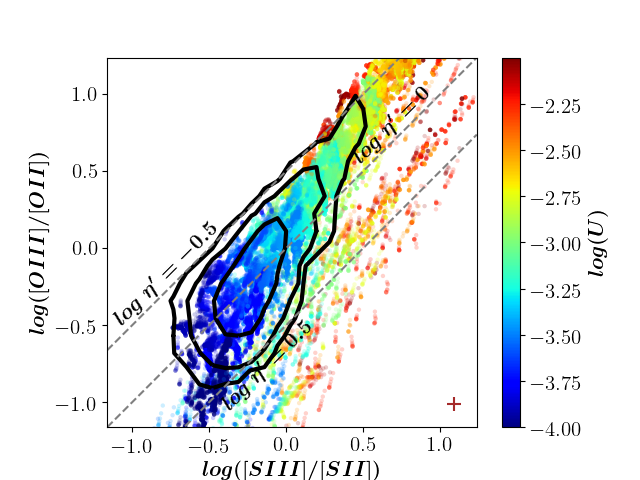}
\includegraphics[width=0.45\textwidth]{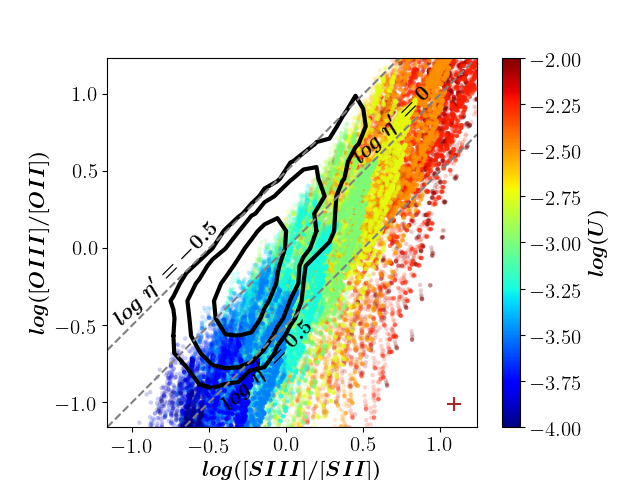}
	\caption{Photoionization models (colour points) and spatially-resolved data (gray contours) represented on the O$_3$O$_2$-S$_3$S$_2$ plane, where we have overlaid the three dashed lines corresponding to constant values of log $\eta \prime$ = -0.5, 0, 0.5. Inner to outer contours represent the 68\%, 95\% and 99.7\% of the datapoints shown in Figure~\ref{fig:hardening metallicity}, respectively. CALIFA models (Left-hand panel) and BOND models (Right-hand panel) are colour-coded by the ionization parameter  (see Section~\ref{sect:discussion}).}
	\label{fig:models_A2}
\end{figure*}

\begin{figure*}
	\centering

	\includegraphics[width=0.45\textwidth]{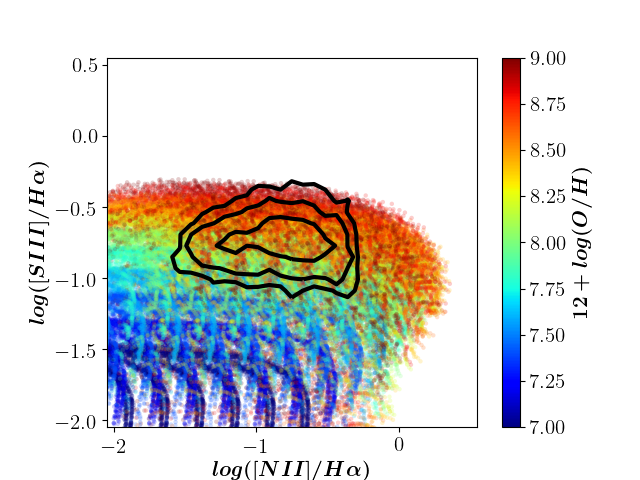} 
	\includegraphics[width=0.45\textwidth]{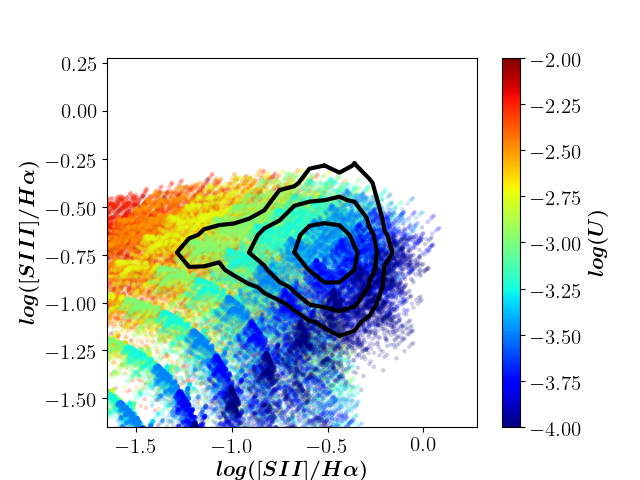}

	\caption{BOND photoionization models (colour points) and spatially-resolved data (black contours) represented on the S$_3$-N$_2$ (Left-hand panel) and S$_3$-S$_2$ plane (Right-hand panel), respectively. Inner to outer contours represent the 68\%, 95\% and 99.7\% of the data points shown in Figure~\ref{fig:SIII-BPT all} (lower-panel), respectively. BOND models are colour-coded by the gas-phase metallicity (Left-hand panel) and ionization parameter (Right-hand panel). See  Section~\ref{sect:discussion} for details on the models.}
	\label{fig:models_A3}
\end{figure*}

\begin{figure*}
\centering
\includegraphics[width=0.45\textwidth]{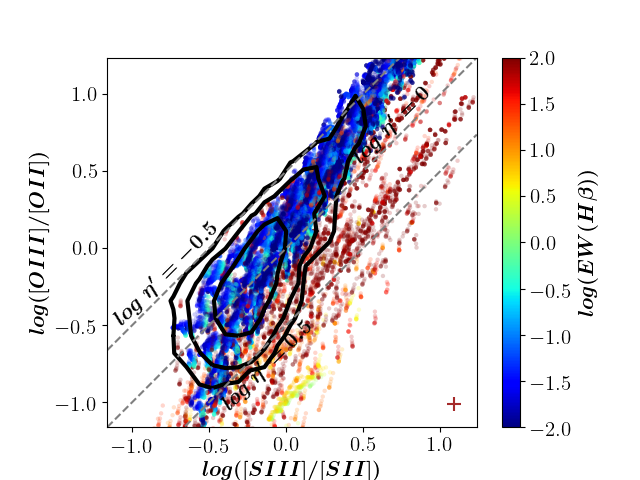}
\includegraphics[width=0.45\textwidth]{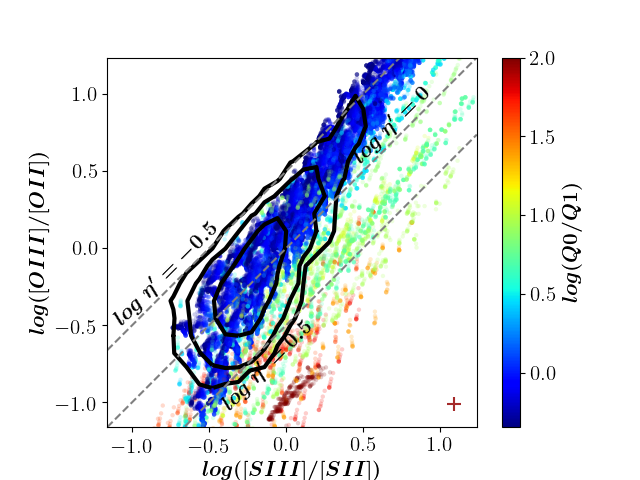}
\caption{CALIFA photoionization models represented on the O$_3$O$_2$-S$_3$S$_2$ plane and colour-coded with respect to the age sensitive parameters, log(EW(H$\beta$)) (left-hand panel) and log(Q0/Q1) (right-hand panel). We have overlaid spatially-resolved data (gray contours) and the three dashed lines corresponding to constant values of log $\eta \prime$ = -0.5, 0, 0.5. Inner to outer contours represent the 68\%, 95\% and 99.7\% of the data points shown in Figure~\ref{fig:hardening metallicity}, respectively.}
\label{fig:models_A4}
\end{figure*}

\indent Figures \ref{fig:models_A1}--\ref{fig:models_A4} show additional comparisons of photoionization models (BOND and CALIFA) and spatially-resolved data in different parameter space, including O$_3$O$_2$ versus S$_3$S$_2$ (Figure \ref{fig:hardening metallicity}), and the novel emission line diagnostic diagrams (Figure \ref{fig:SIII-BPT all}). These models are first presented in Figure \ref{fig:models}. In that figure, we also show model prediction (straight-line) from  \citet{Diaz2000b} and is described below.

\indent \citet{Diaz2000b} used a large grid of single-star \textsc{cloudy} photoionization models  \citep{Ferland1996}, with non-local thermodynamic equilibrium (LTE) single star stellar atmosphere models, closed geometry and constant particle density and derived the relation between log $\mathcal{U}$ and the emission line ratios sensitive to ionization parameter such as O$_3$O$_2$ and S$_3$S$_2$ (\citealp[see equations 7 and 8 in][]{Diaz2000b}). We combine the two equations to derive the following relation represented by the solid black line in Figure \ref{fig:models}, 

\begin{equation}
	\centering
	\rm O_3O_2 = 2.1 ~S_3S_2 + 0.0375
	\label{eq:Diaz2000}
\end{equation}

We note that \citet{Diaz2000b} discarded their relation of log $\mathcal{U}$-O$_3$O$_2$ as this formula underestimated the values of log $\mathcal{U}$ compared to other formulae for log $\mathrm{U}$ derived in their work.

\section{SDSS cutouts of all 67 galaxies in the sample}
\label{section:cutouts}
\begin{figure*}
	\centering
	\includegraphics[width=0.2\textwidth]{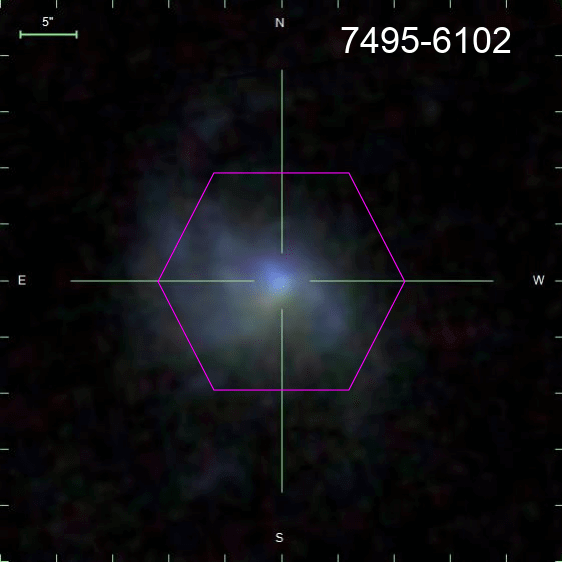}
	\includegraphics[width=0.2\textwidth]{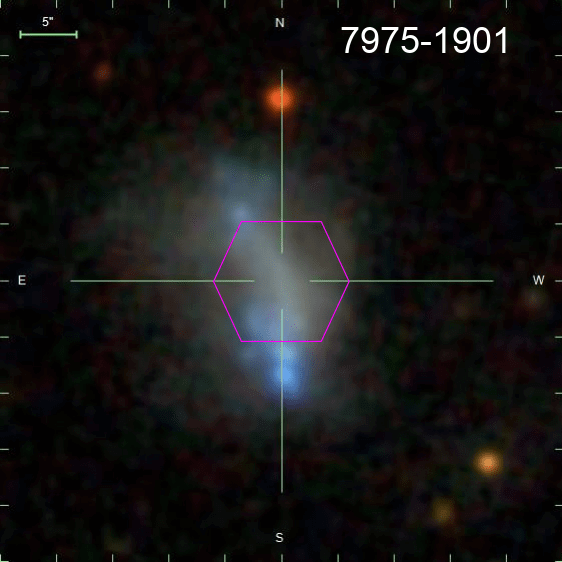}
	\includegraphics[width=0.2\textwidth]{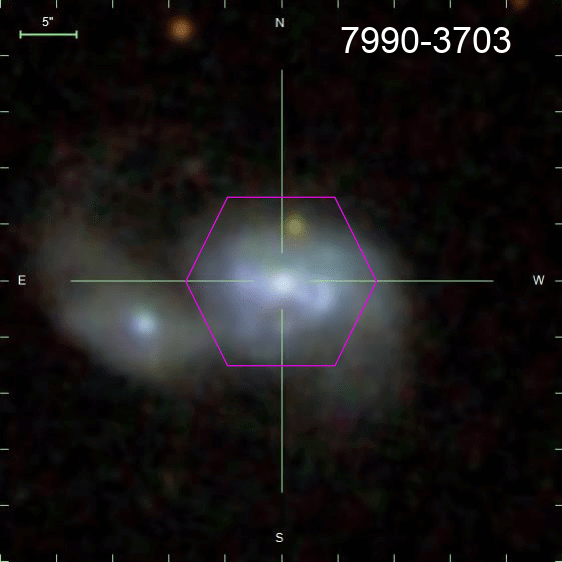}
	\includegraphics[width=0.2\textwidth]{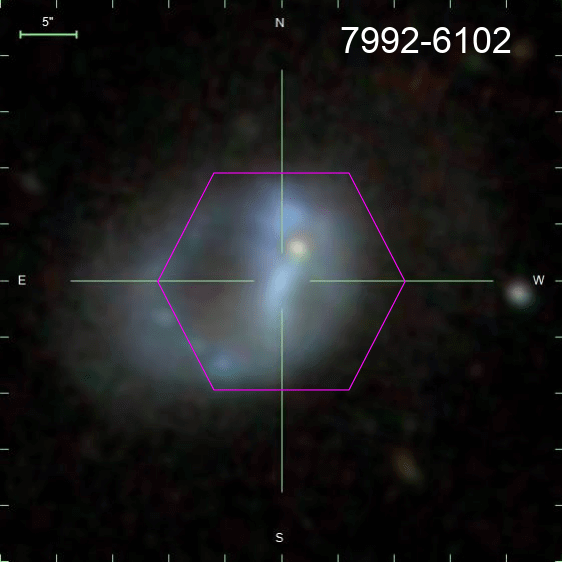}
	\centering
	\includegraphics[width=0.2\textwidth]{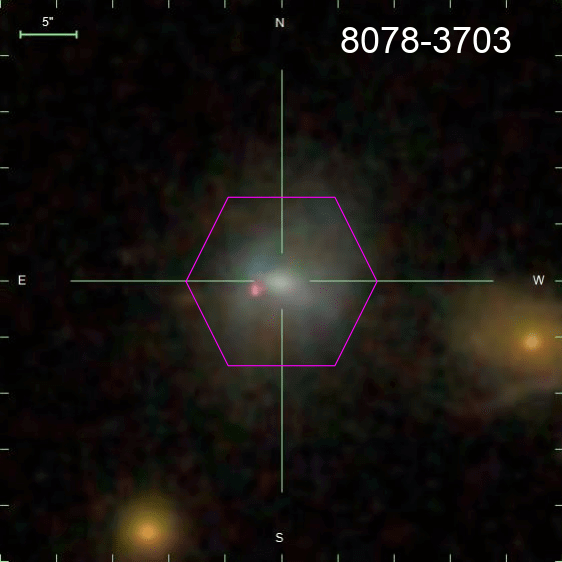}
	\includegraphics[width=0.2\textwidth]{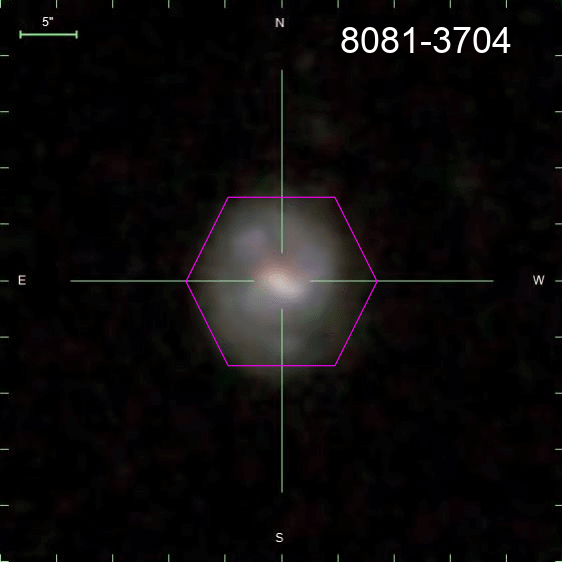}
	\includegraphics[width=0.2\textwidth]{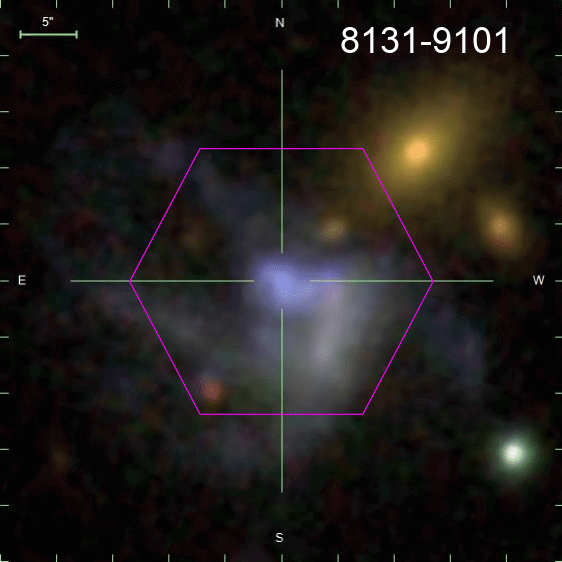}
	\includegraphics[width=0.2\textwidth]{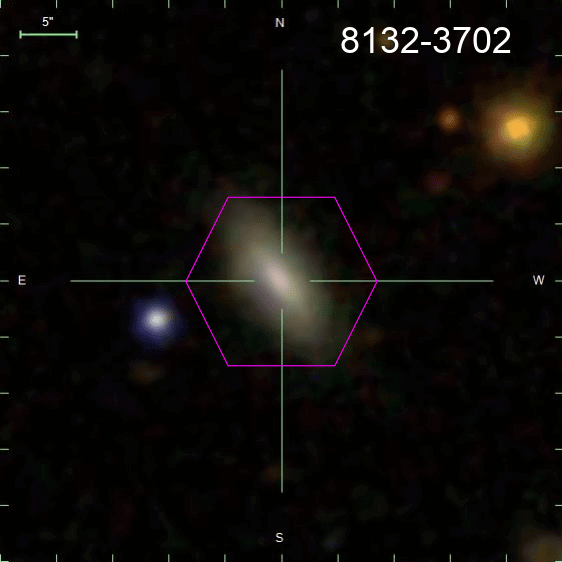}
	\centering
	\includegraphics[width=0.2\textwidth]{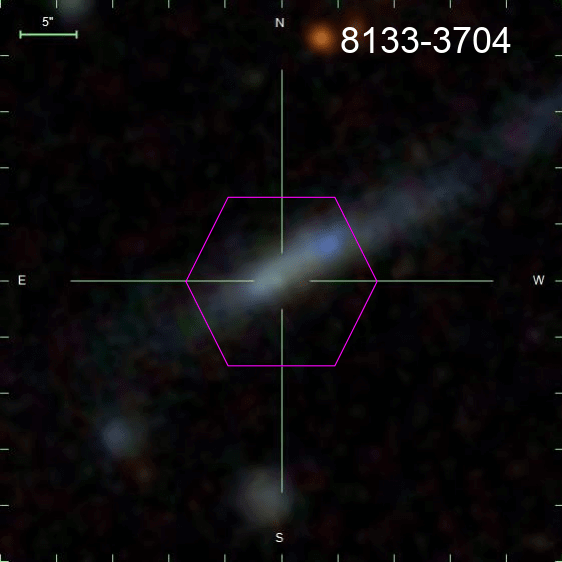}
	\includegraphics[width=0.2\textwidth]{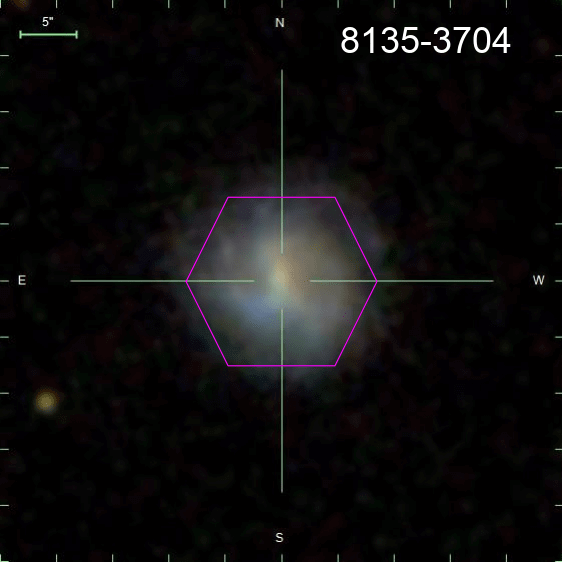}
	\includegraphics[width=0.2\textwidth]{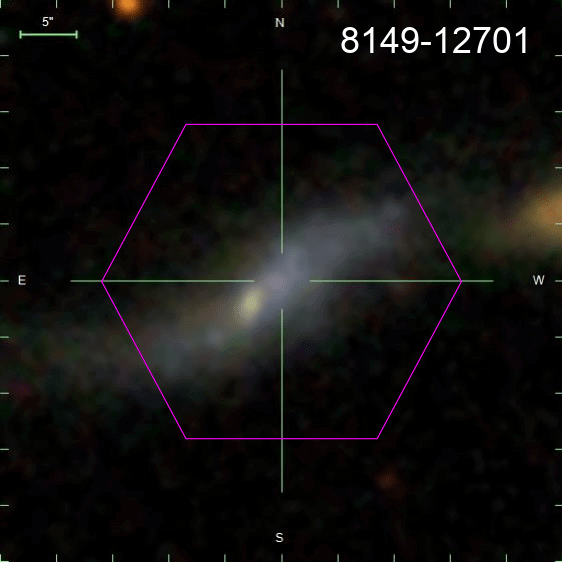}
	\includegraphics[width=0.2\textwidth]{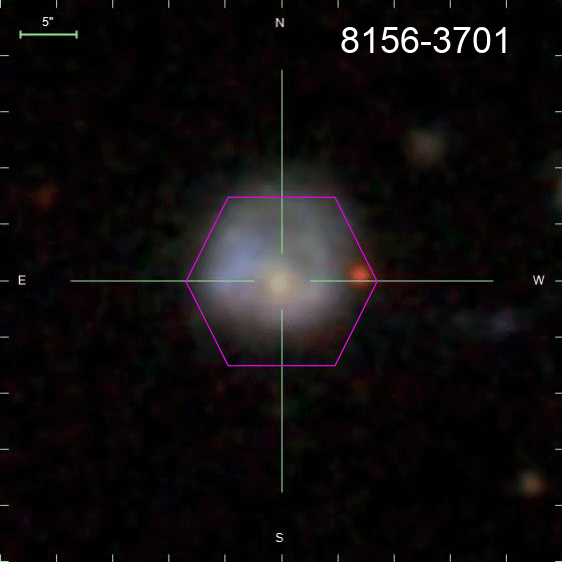}
	\centering
	\includegraphics[width=0.2\textwidth]{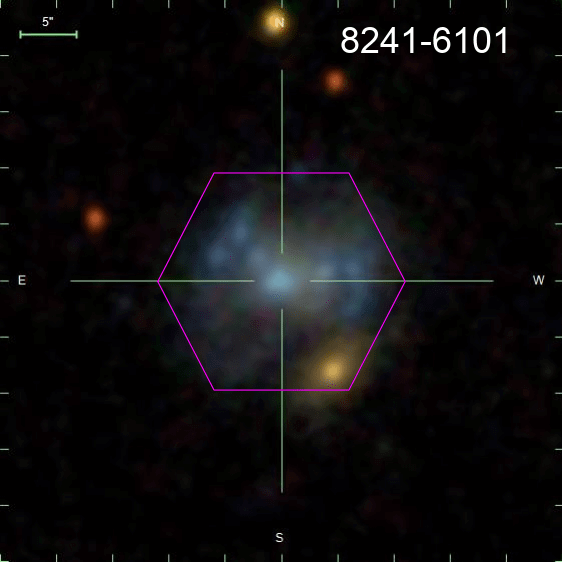}
	\includegraphics[width=0.2\textwidth]{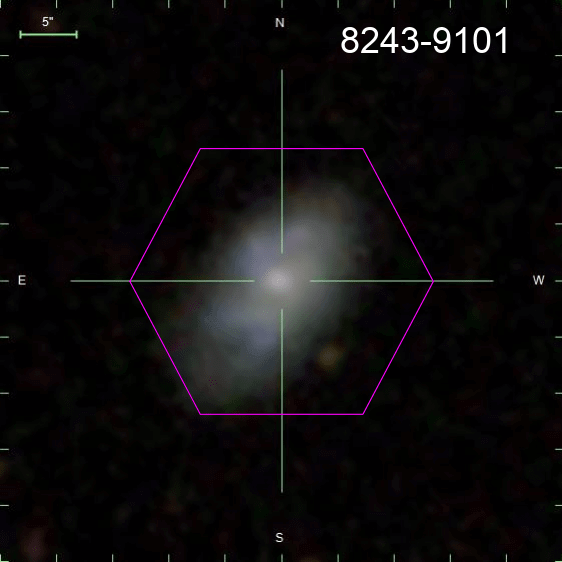}
	\includegraphics[width=0.2\textwidth]{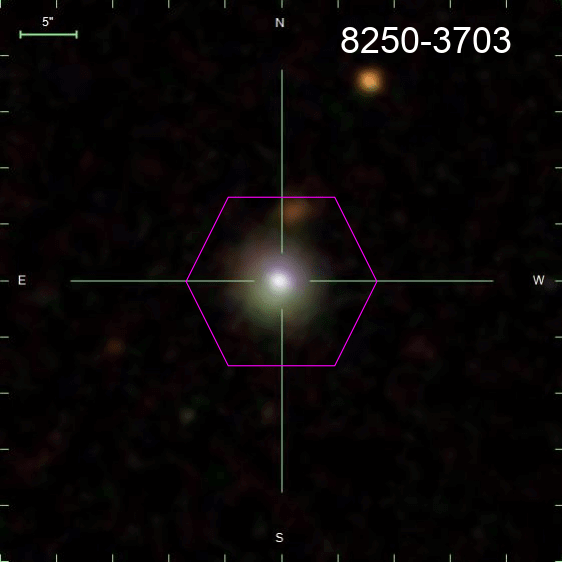}
	\includegraphics[width=0.2\textwidth]{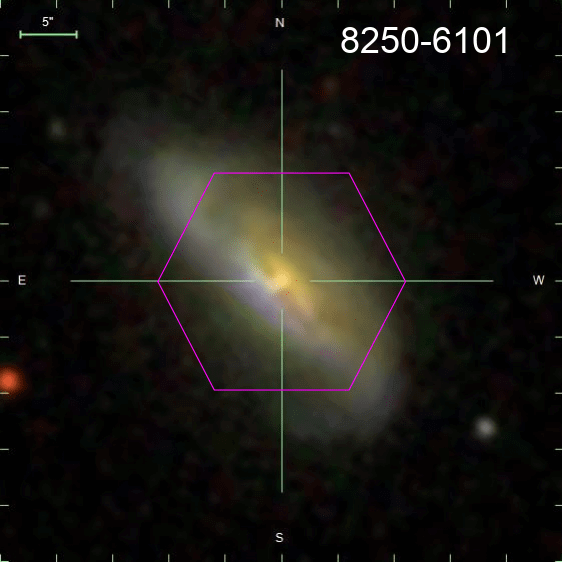}
	\centering
	\includegraphics[width=0.2\textwidth]{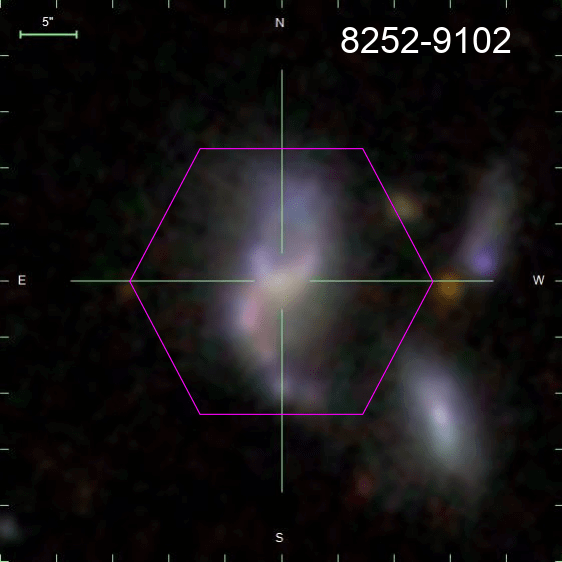}
	\includegraphics[width=0.2\textwidth]{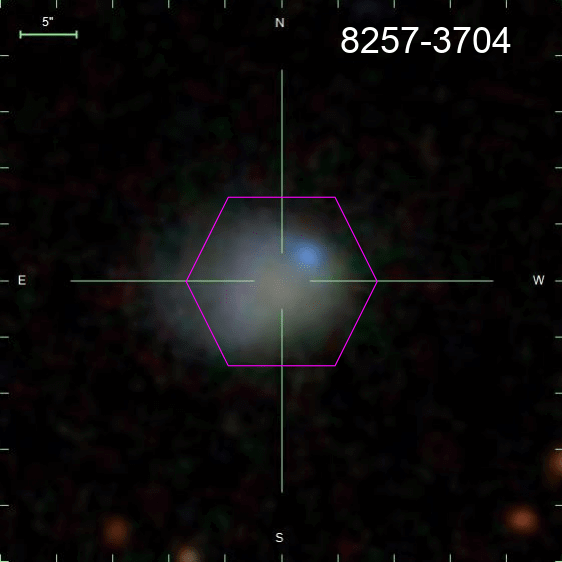}
	\includegraphics[width=0.2\textwidth]{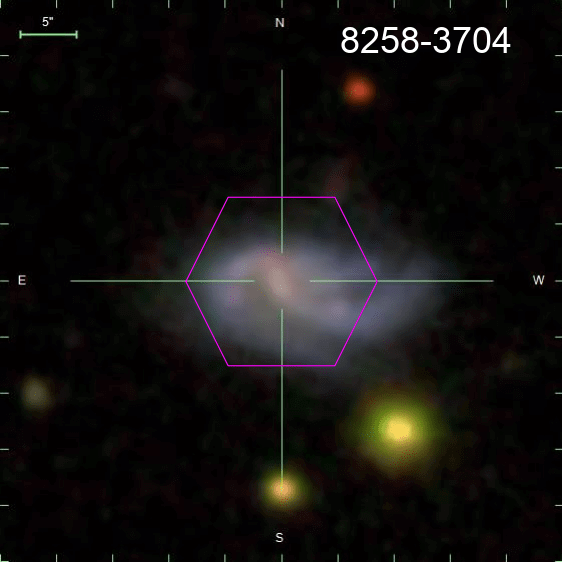}
	\includegraphics[width=0.2\textwidth]{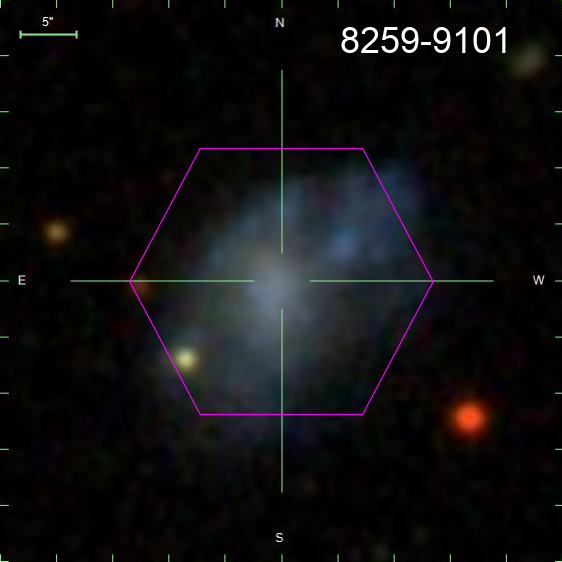}
	\centering
	\includegraphics[width=0.2\textwidth]{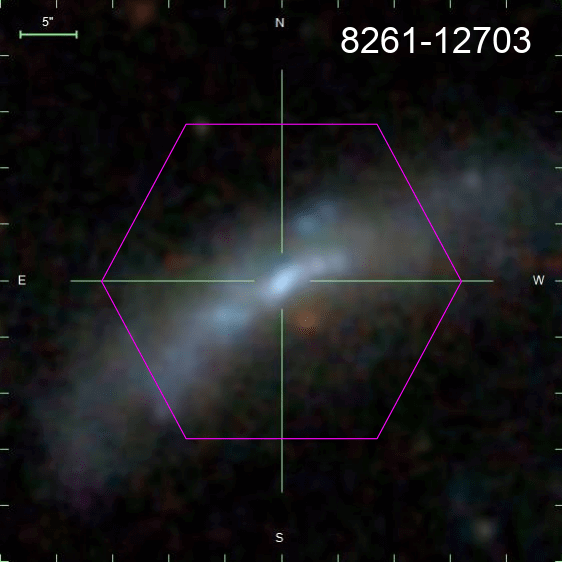}
	\includegraphics[width=0.2\textwidth]{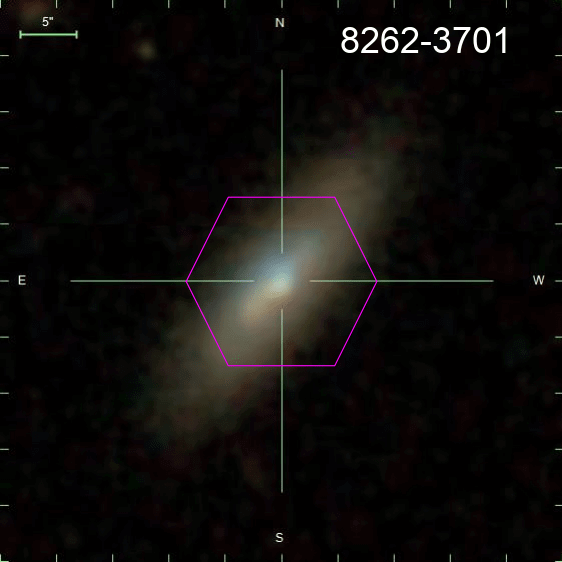}
	\includegraphics[width=0.2\textwidth]{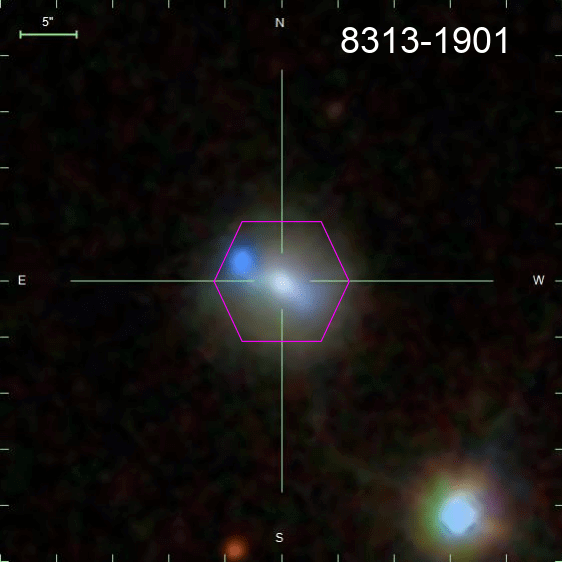}
	\includegraphics[width=0.2\textwidth]{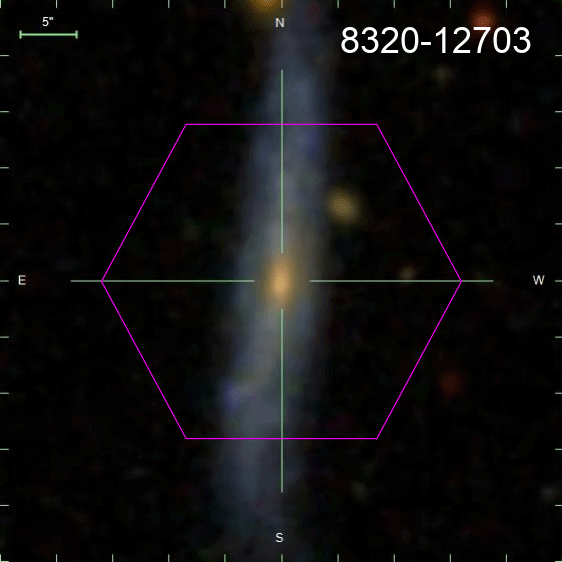}
\caption{SDSS cut-outs of our sample galaxies with their MaNGA ids mentioned in the top-right of each cut-out.}
\label{fig:cutout1}
\end{figure*}

\begin{figure*}
	\centering
	\includegraphics[width=0.2\textwidth]{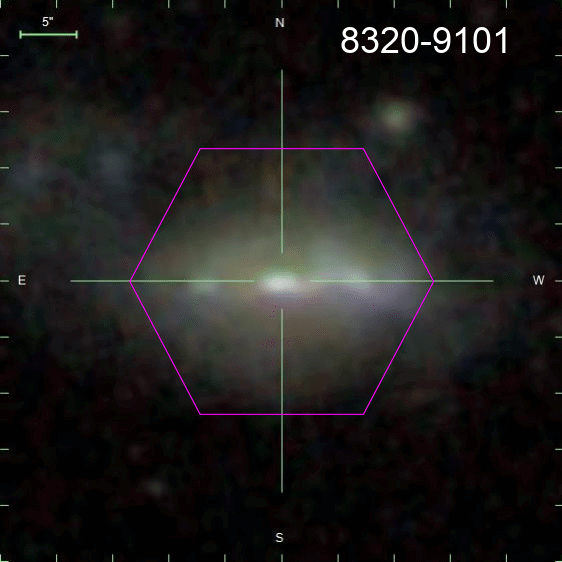}
	\includegraphics[width=0.2\textwidth]{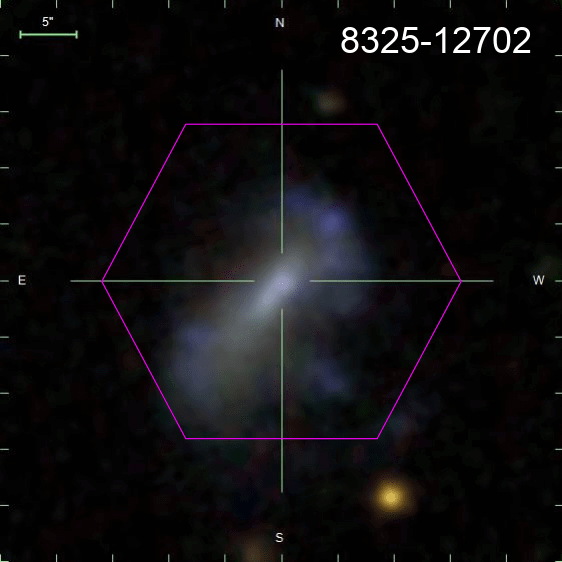}
	\includegraphics[width=0.2\textwidth]{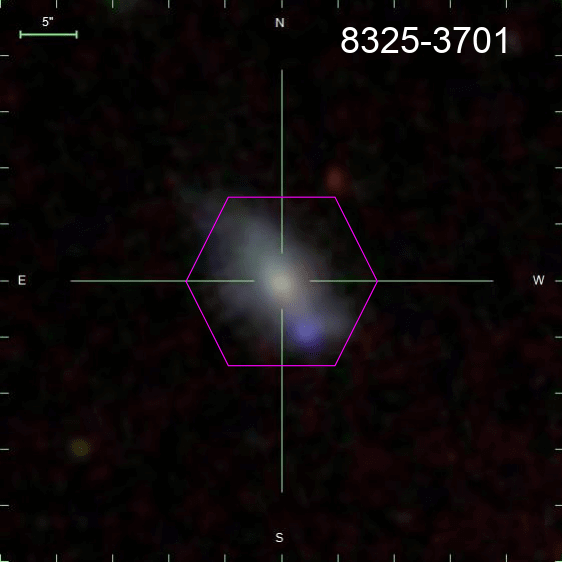}
	\includegraphics[width=0.2\textwidth]{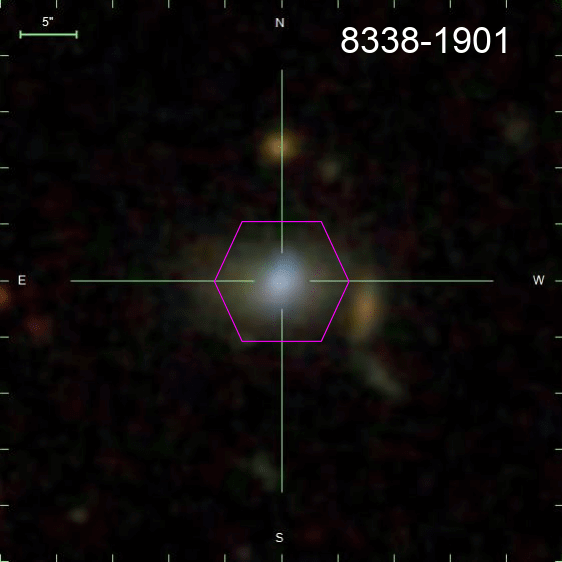}
	\centering
	\includegraphics[width=0.2\textwidth]{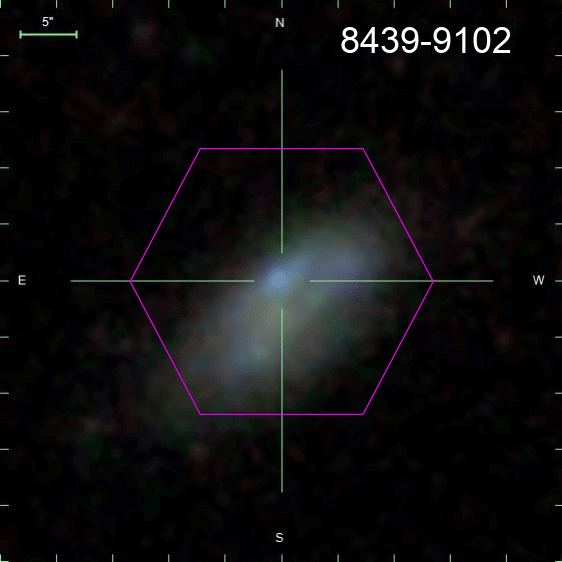}
	\includegraphics[width=0.2\textwidth]{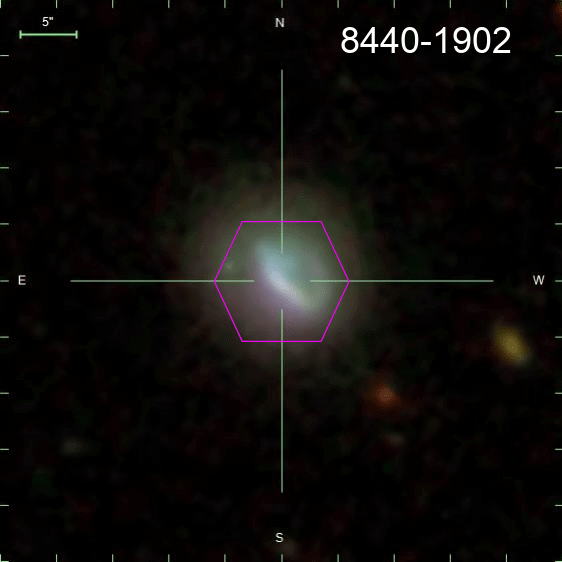}
	\includegraphics[width=0.2\textwidth]{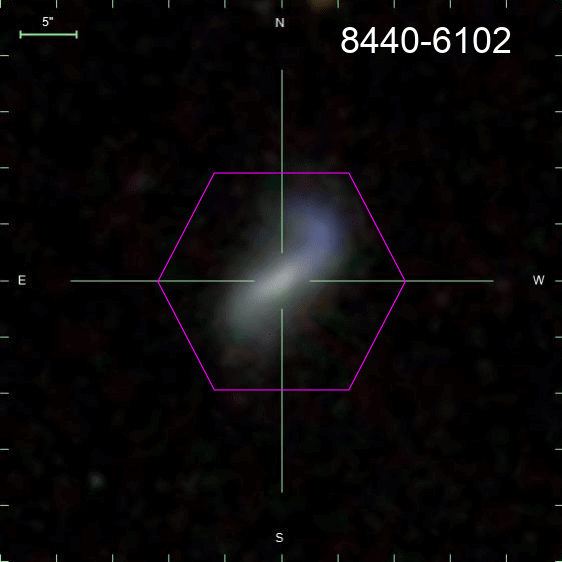}
	\includegraphics[width=0.2\textwidth]{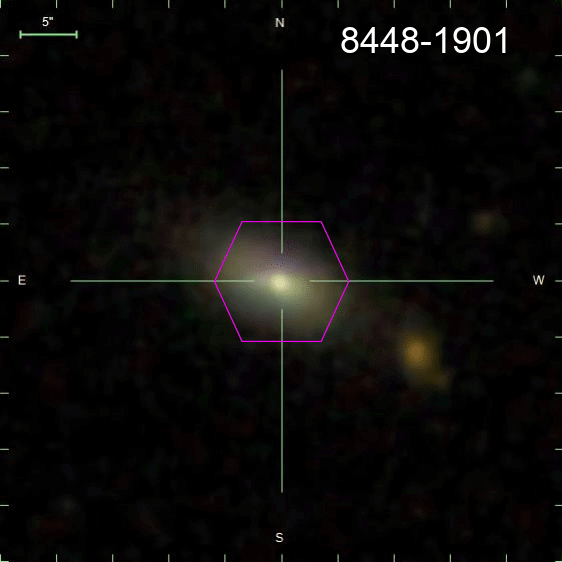}
	\centering
	\includegraphics[width=0.2\textwidth]{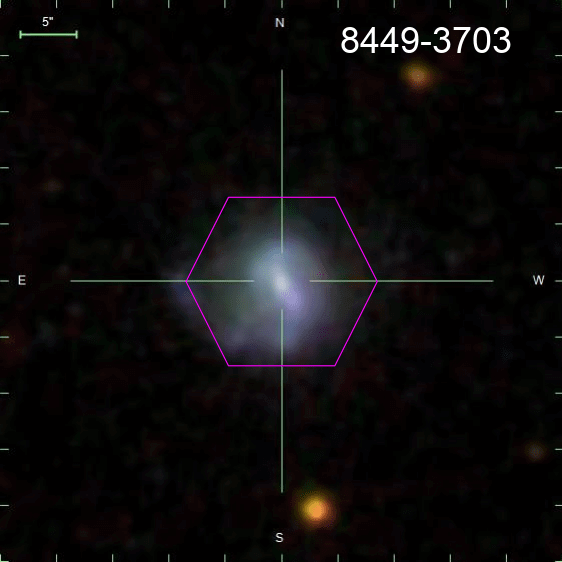}
	\includegraphics[width=0.2\textwidth]{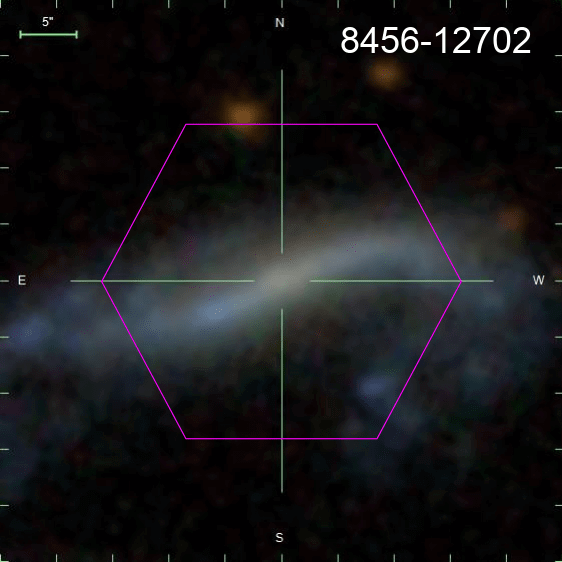}
	\includegraphics[width=0.2\textwidth]{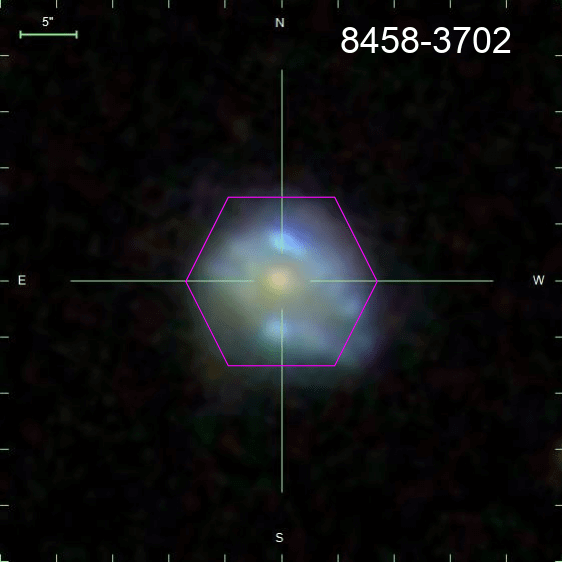}
	\includegraphics[width=0.2\textwidth]{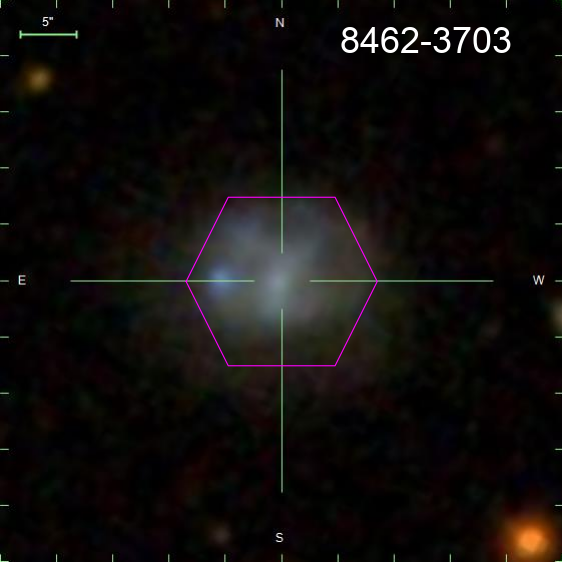}
	\centering
	\includegraphics[width=0.2\textwidth]{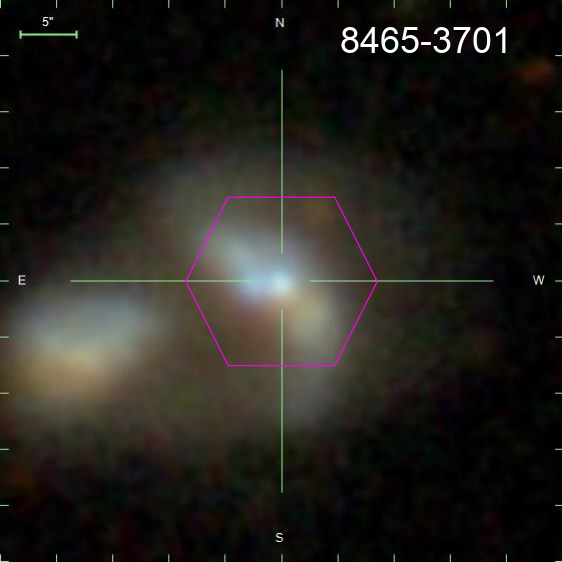}
	\includegraphics[width=0.2\textwidth]{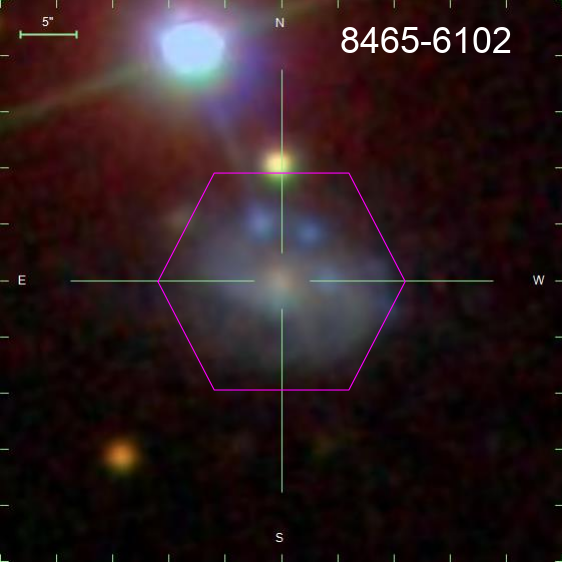}
	\includegraphics[width=0.2\textwidth]{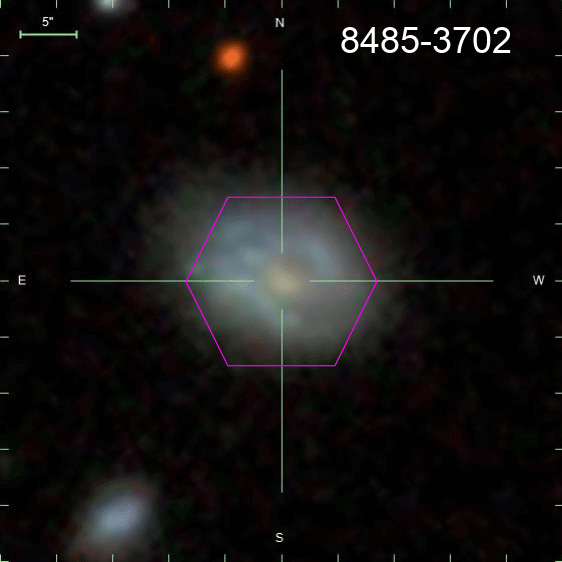}
	\includegraphics[width=0.2\textwidth]{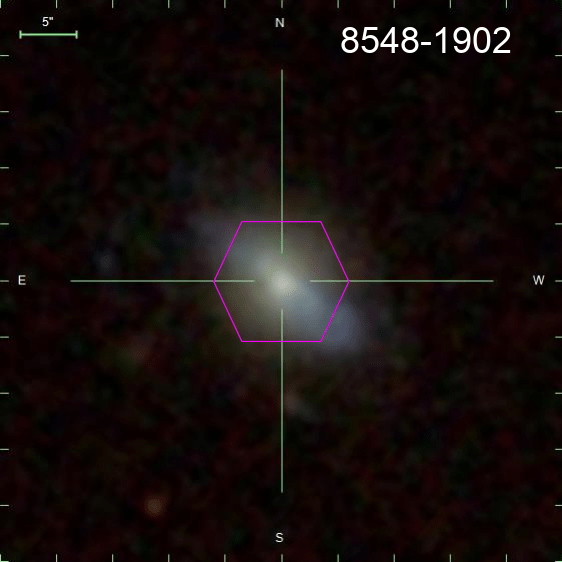}
	\centering
	\includegraphics[width=0.2\textwidth]{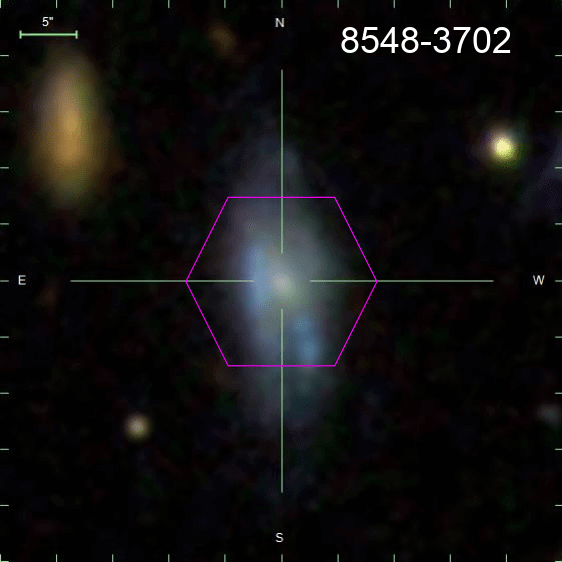}
	\includegraphics[width=0.2\textwidth]{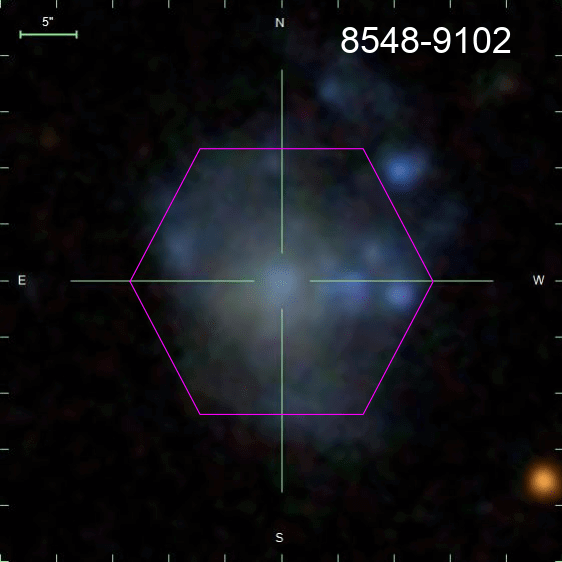}
	\includegraphics[width=0.2\textwidth]{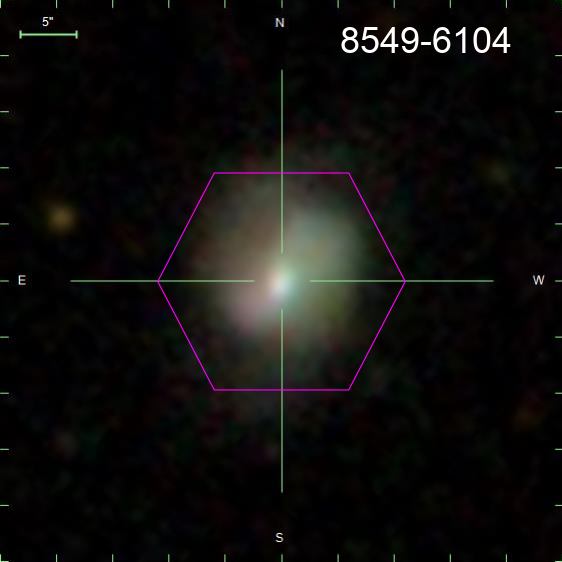}
	\includegraphics[width=0.2\textwidth]{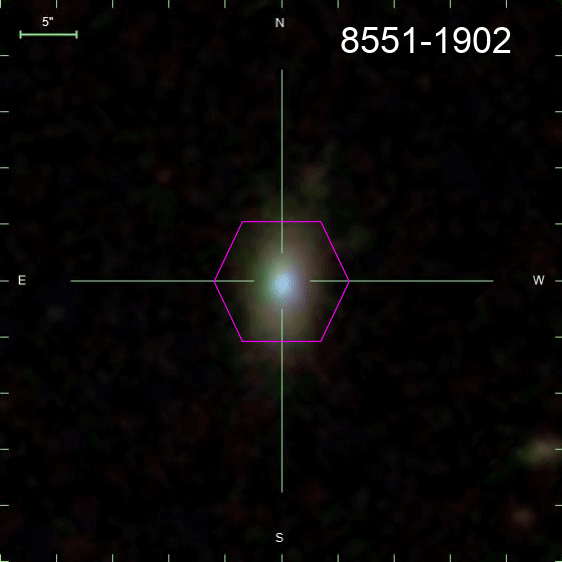}
	\centering
	\includegraphics[width=0.2\textwidth]{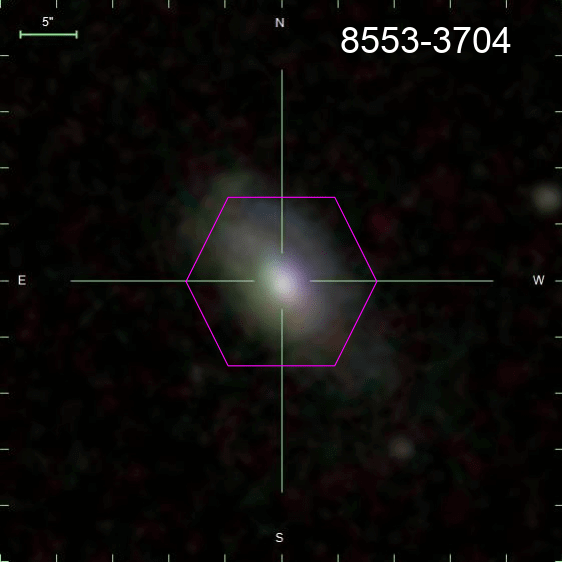}
	\includegraphics[width=0.2\textwidth]{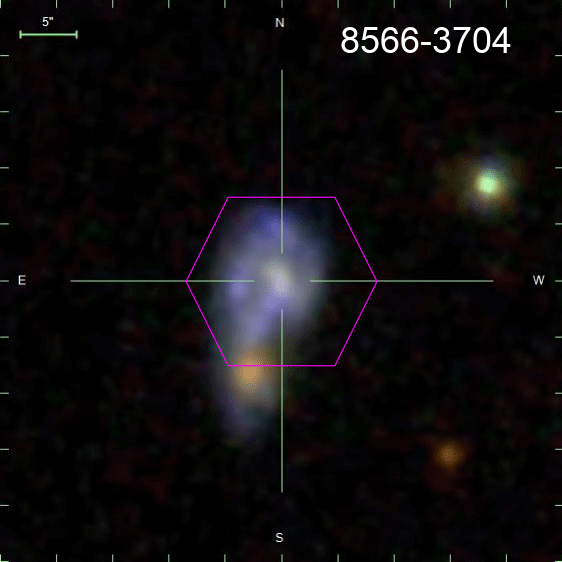}
	\includegraphics[width=0.2\textwidth]{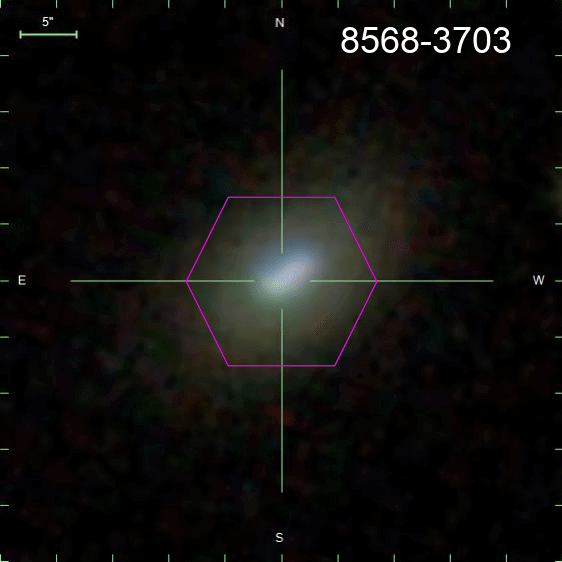}
	\includegraphics[width=0.2\textwidth]{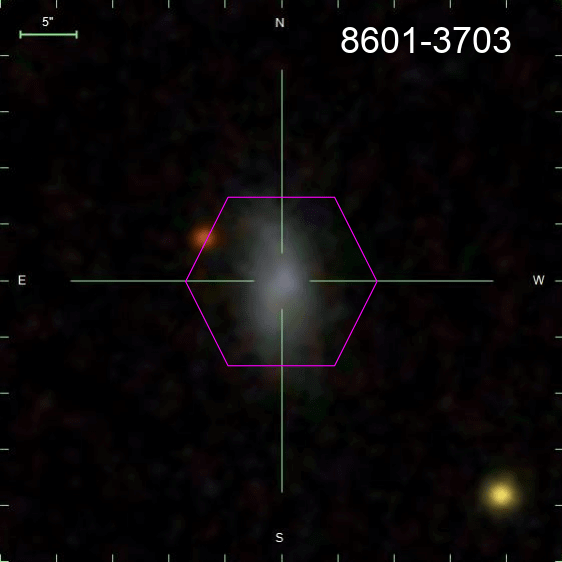}
\caption{SDSS cut-outs of our sample galaxies with their MaNGA ids mentioned in the top-right of each cut-out.}
\label{fig:cutout2}
\end{figure*}

\begin{figure*}
	\centering
	\includegraphics[width=0.2\textwidth]{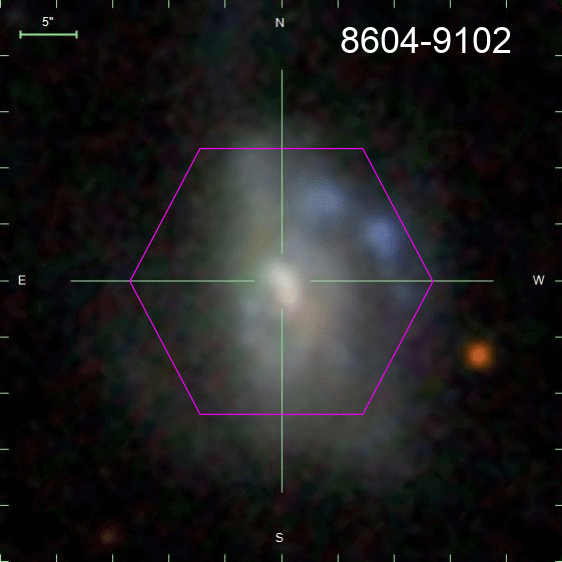}
	\includegraphics[width=0.2\textwidth]{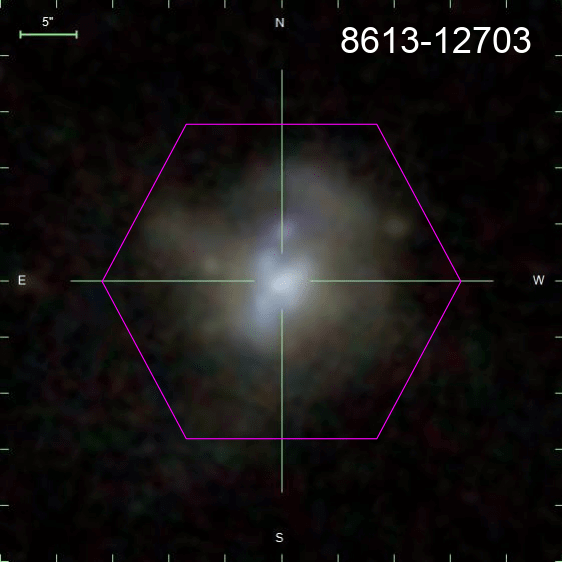}
	\includegraphics[width=0.2\textwidth]{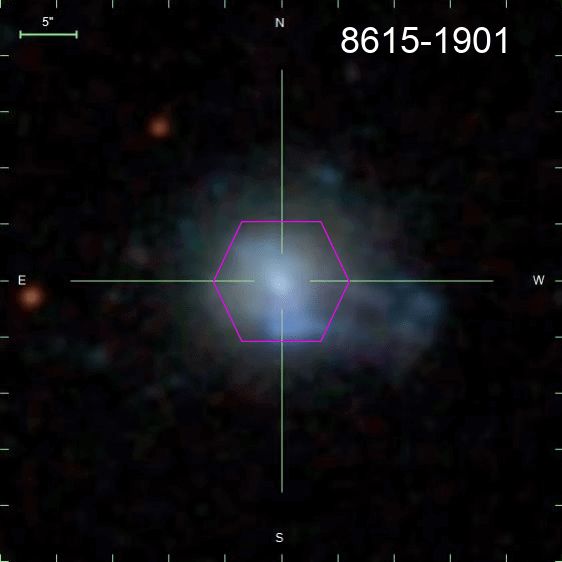}
	\includegraphics[width=0.2\textwidth]{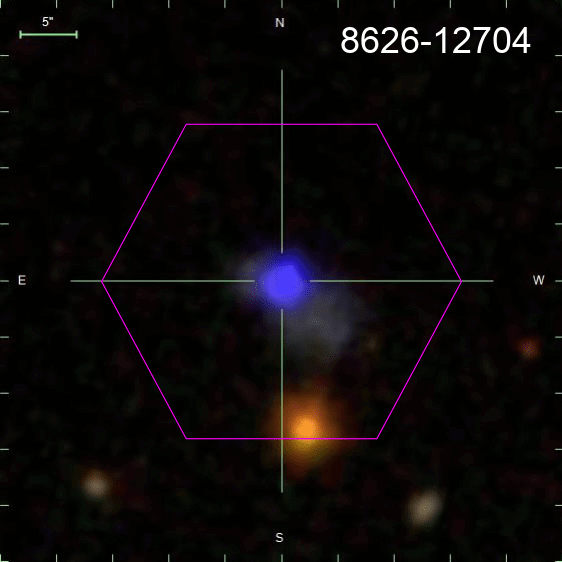}
	\centering
	\includegraphics[width=0.2\textwidth]{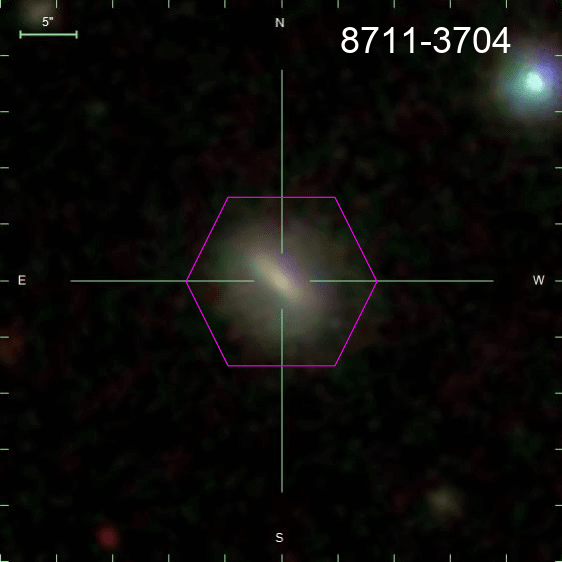}
	\includegraphics[width=0.2\textwidth]{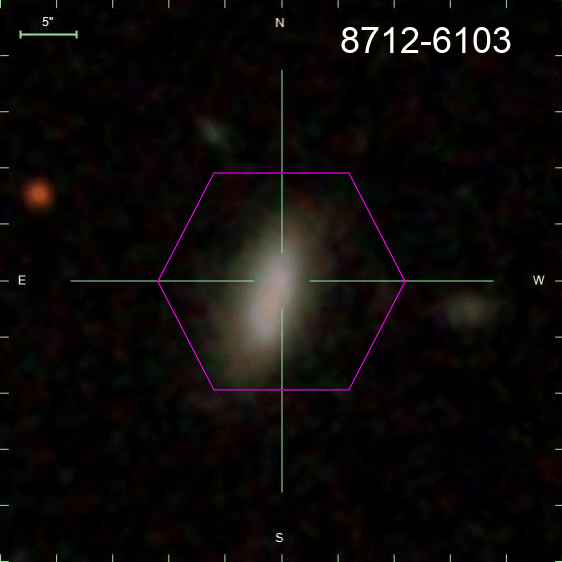}
	\includegraphics[width=0.2\textwidth]{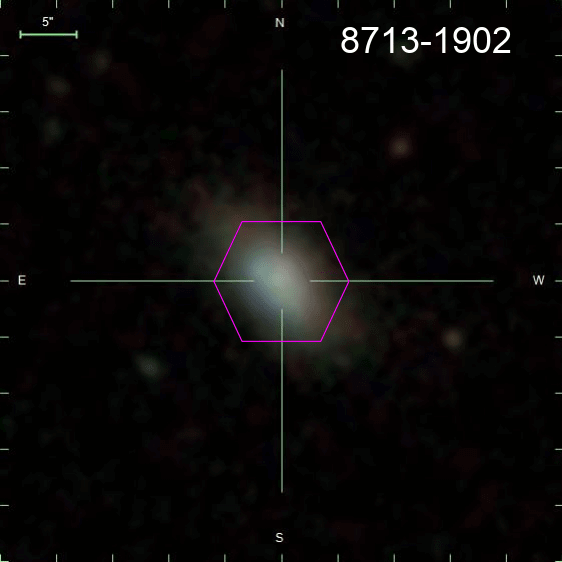}
	\includegraphics[width=0.2\textwidth]{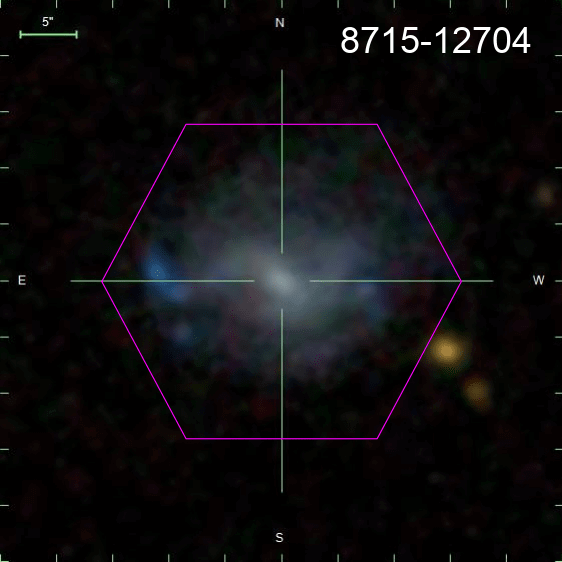}
	\centering
	\includegraphics[width=0.2\textwidth]{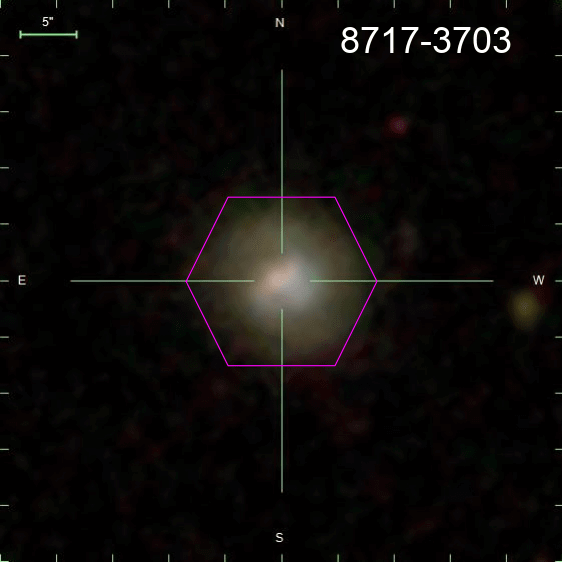}
	\includegraphics[width=0.2\textwidth]{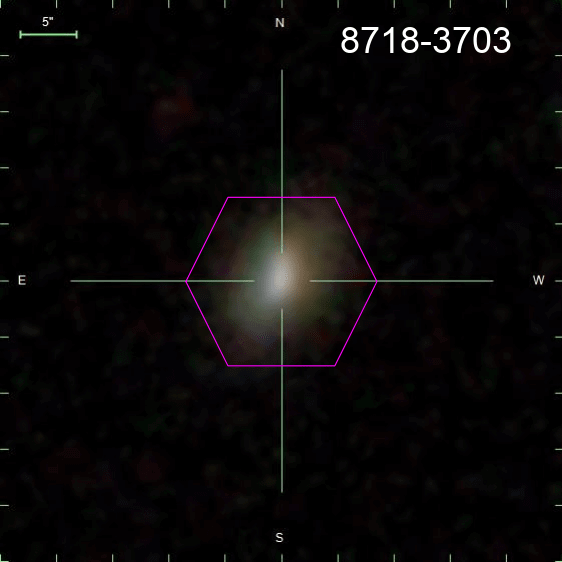}
	\includegraphics[width=0.2\textwidth]{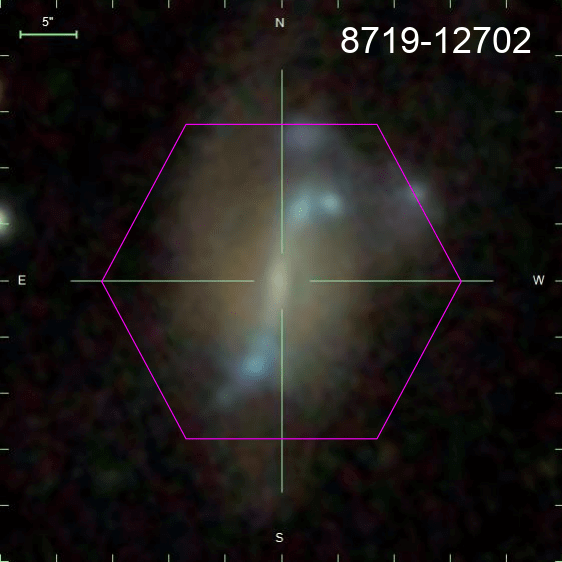}
	\includegraphics[width=0.2\textwidth]{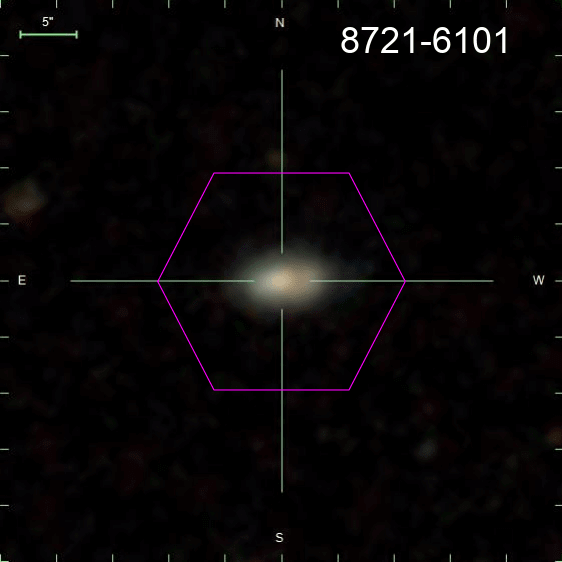}
	\centering
	\includegraphics[width=0.2\textwidth]{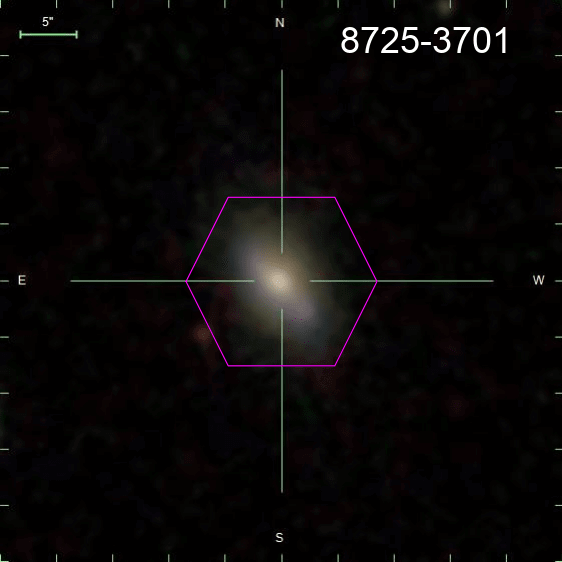}
	\includegraphics[width=0.2\textwidth]{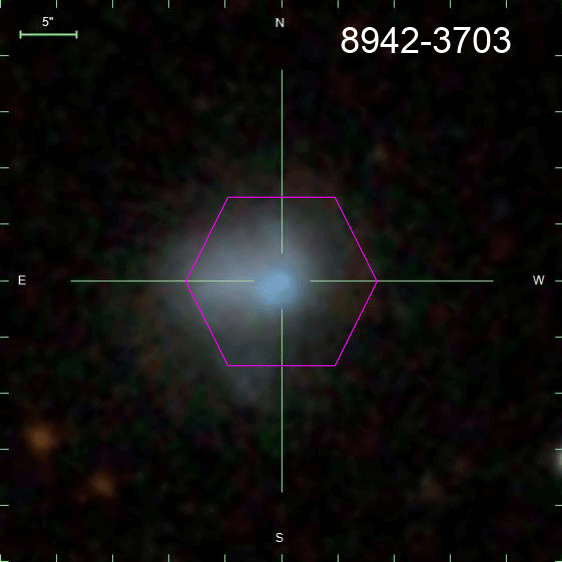}
	\includegraphics[width=0.2\textwidth]{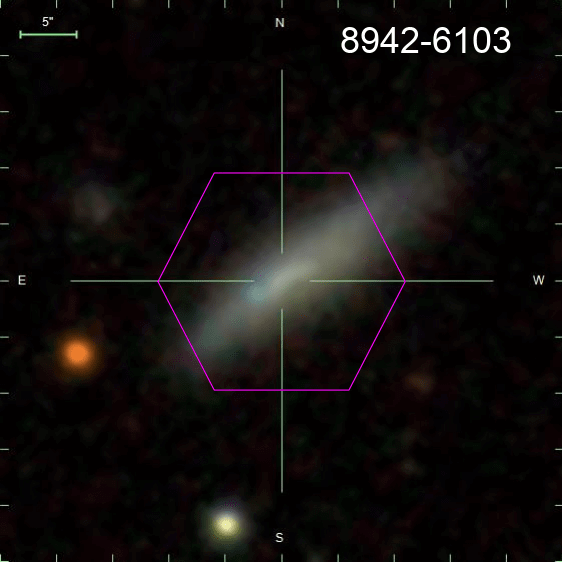}
	\includegraphics[width=0.2\textwidth]{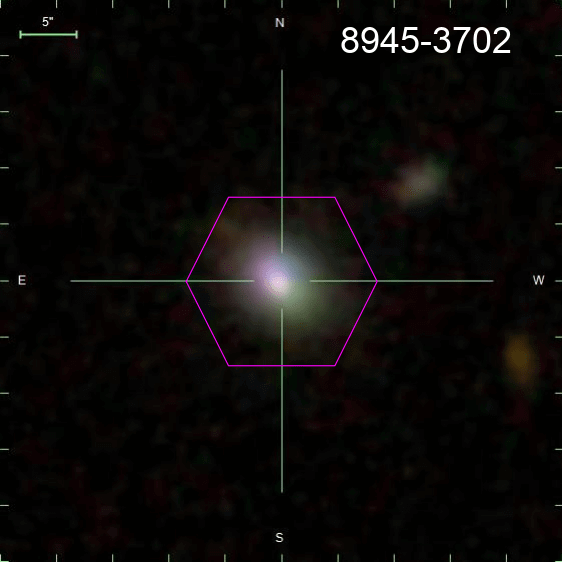}
	\centering
	\includegraphics[width=0.2\textwidth]{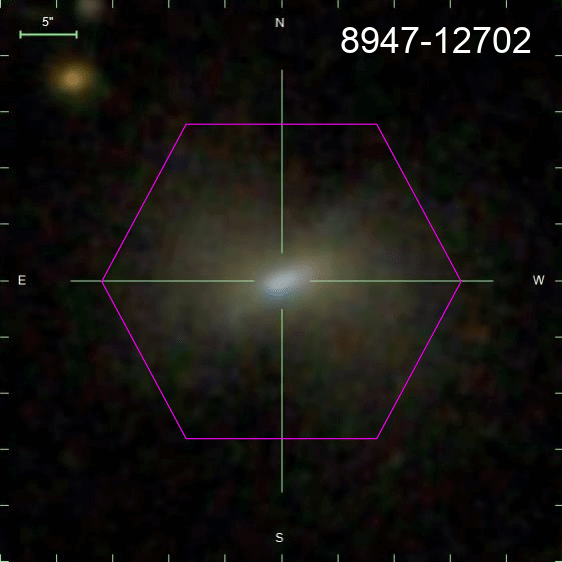}
	\includegraphics[width=0.2\textwidth]{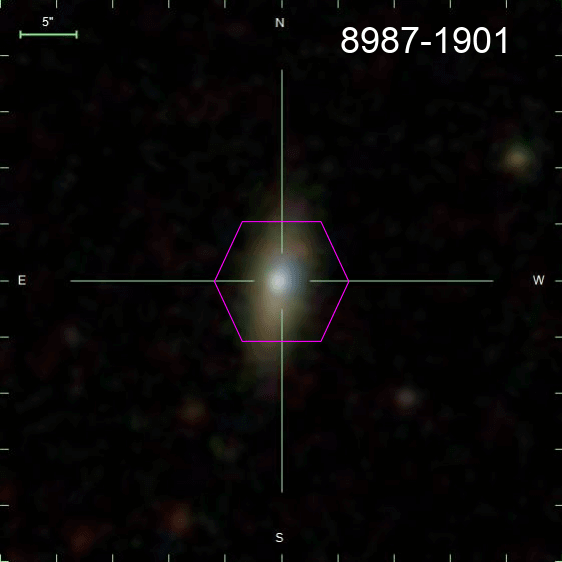}
	\includegraphics[width=0.2\textwidth]{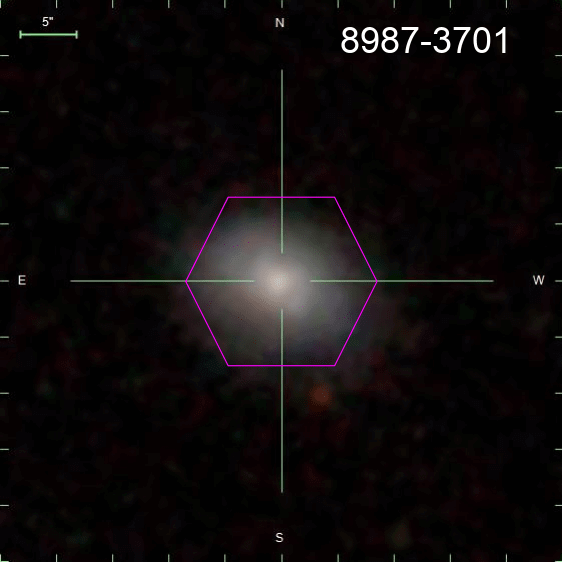}
	\caption{SDSS cut-outs of our sample galaxies with their MaNGA ids mentioned in the top-right of each cut-out.}
	\label{fig:cutout3}
\end{figure*}

Figures \ref{fig:cutout1}-\ref{fig:cutout3} show the SDSS cutouts of all 67 galaxies in our sample where the magenta hexagon indicates the FOV of MaNGA and the horizontal green line indicates an angular scale of 5 arcsec. The four directions, north (N), south (S), east (E) and west (W) are shown as the vertical and horizontal green lines in the centre of each image. The MaNGA id is mentioned at the upper-right corner of each cutout. 


\section{Galaxies with Helium detection}
\label{sect:He maps}
\begin{figure*}
    \centering
    \includegraphics[width=0.185\textwidth, trim={4.0cm 0 4.0cm 0}, clip]{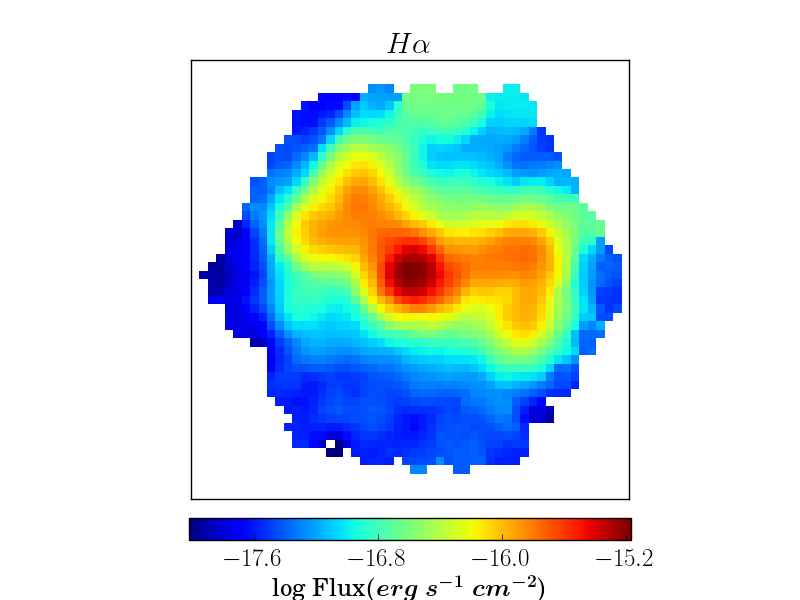}
	\includegraphics[width=0.185\textwidth, trim={4.0cm 0 4.0cm 0}, clip]{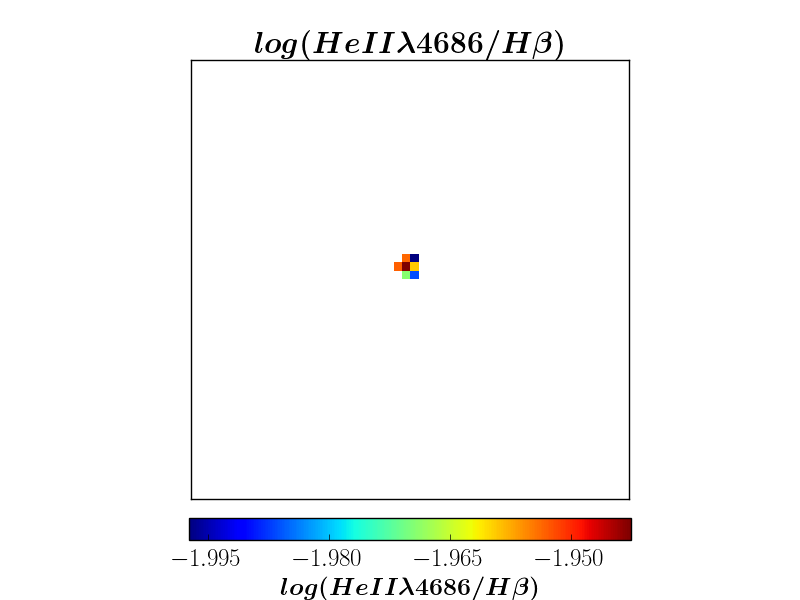}
	\includegraphics[width=0.185\textwidth, trim={2.15cm 0 2.15cm 0}, clip]{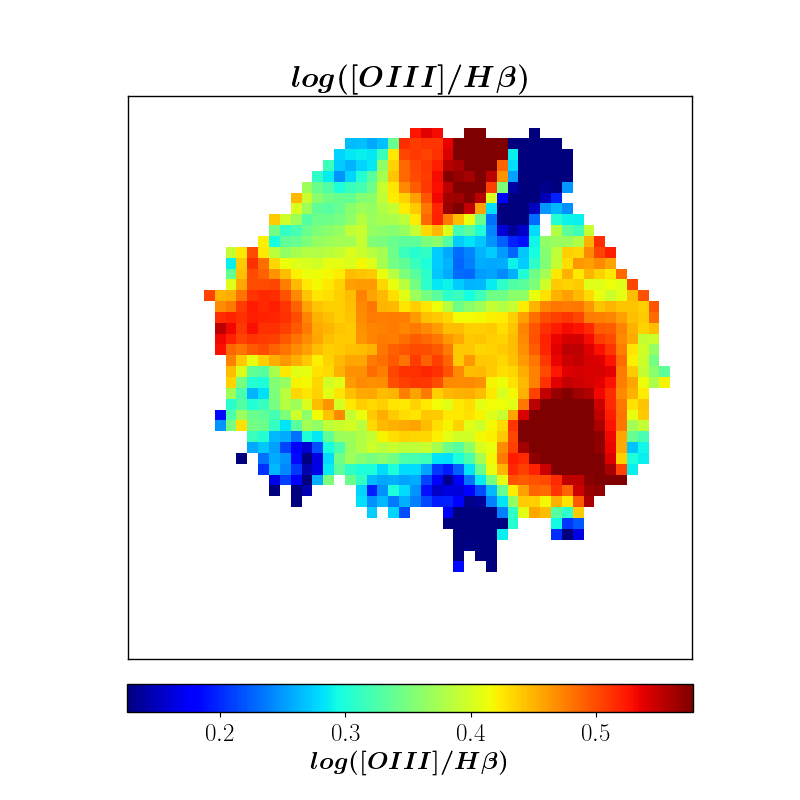}
	\includegraphics[width=0.185\textwidth, trim={4.0cm 0 4.0cm 0}, clip]{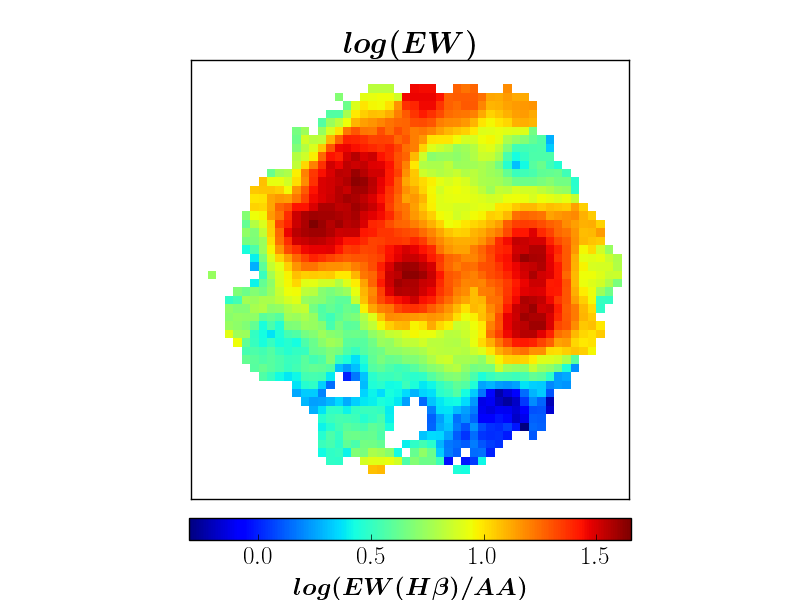}
	\includegraphics[width=0.185\textwidth, trim={4.0cm 0 4.0cm 0}, clip]{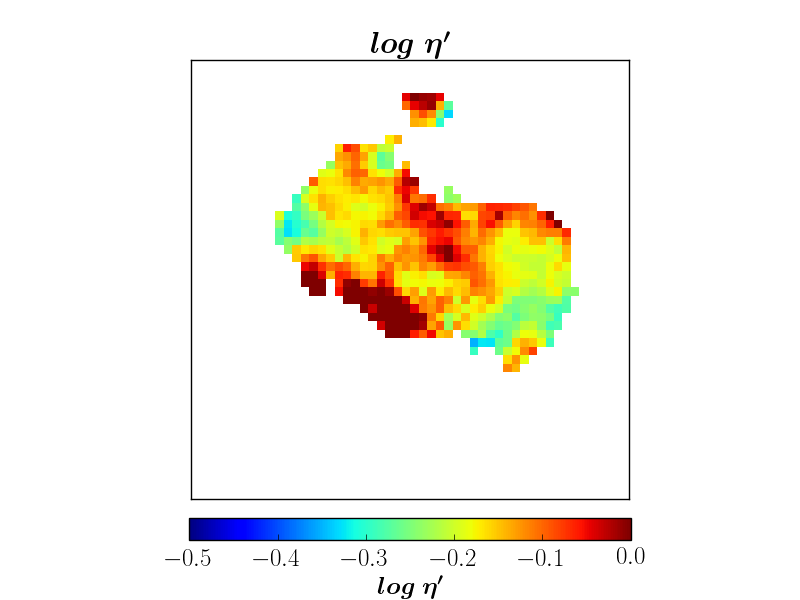}
    \caption{manga-8241-6101}
    \label{fig:manga-8241-6101}
    
    \includegraphics[width=0.185\textwidth, trim={4.0cm 0 4.0cm 0}, clip]{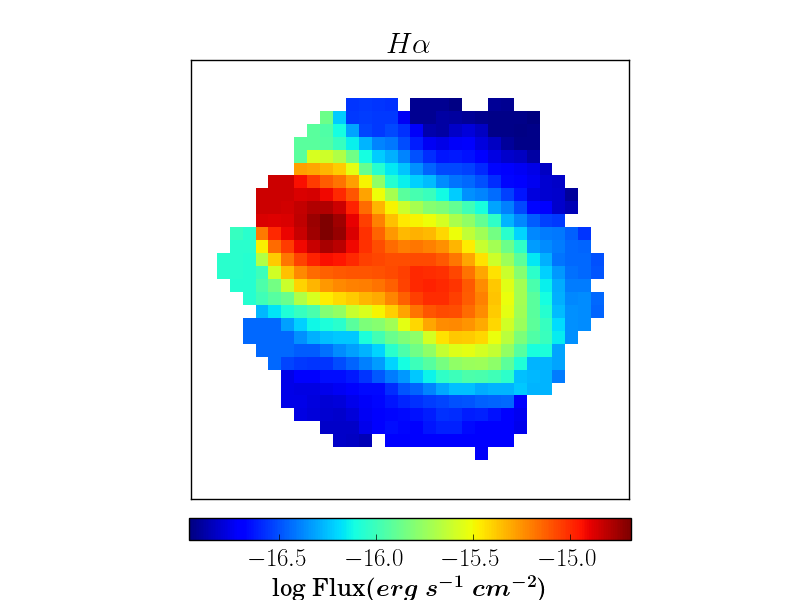}
	\includegraphics[width=0.185\textwidth, trim={4.0cm 0 4.0cm 0},clip]{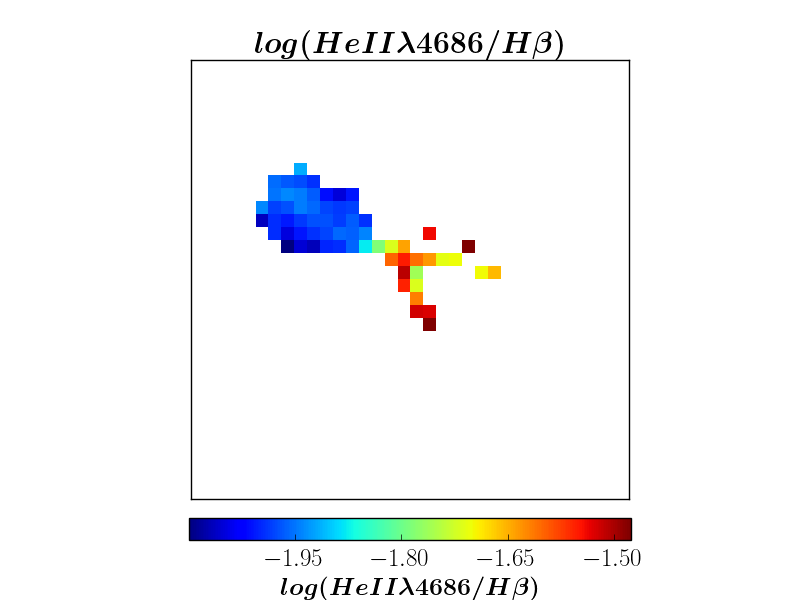}
	\includegraphics[width=0.185\textwidth, trim={2.15cm 0 2.15cm 0},clip]{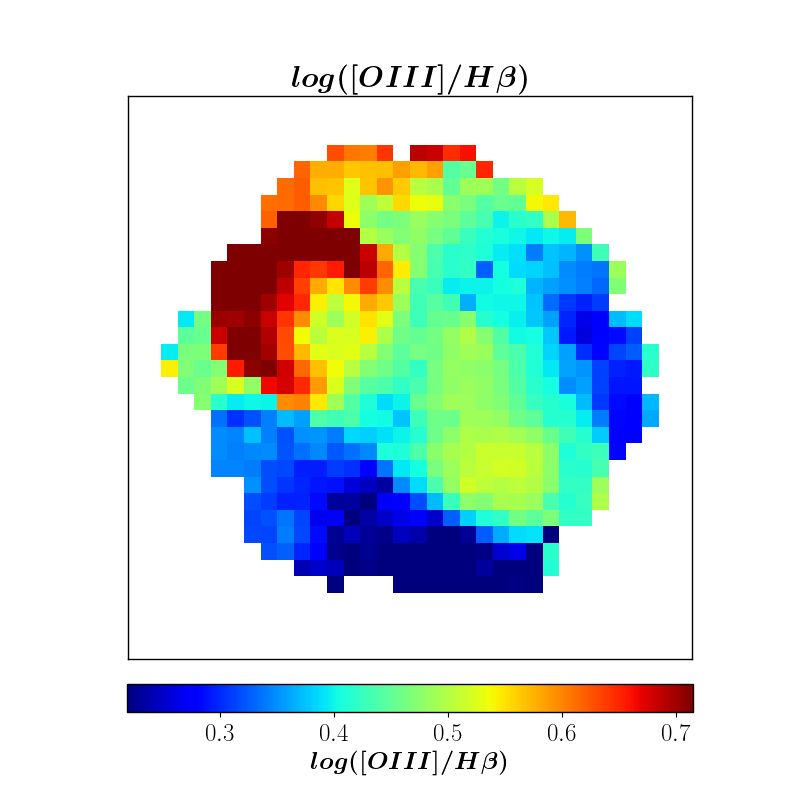}
	\includegraphics[width=0.185\textwidth, trim={4.0cm 0 4.0cm 0}, clip]{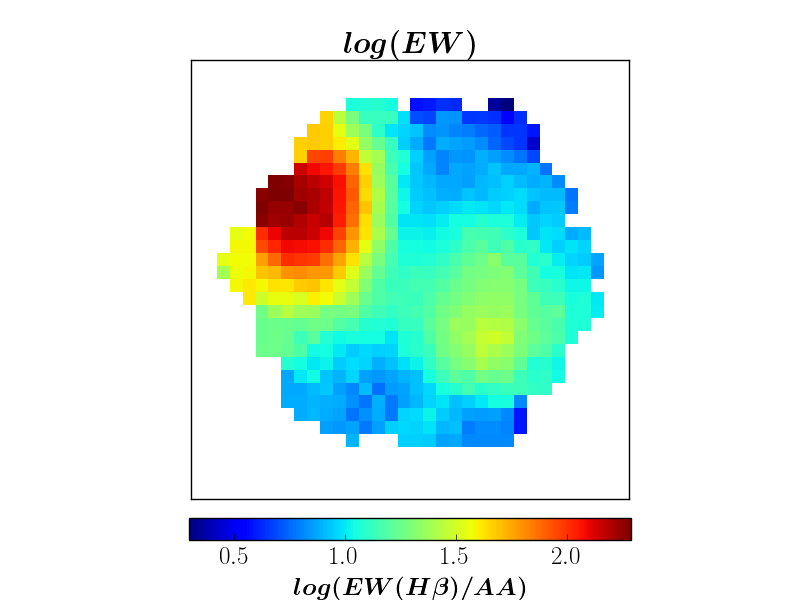}
	\includegraphics[width=0.185\textwidth, trim={4.0cm 0 4.0cm 0},clip]{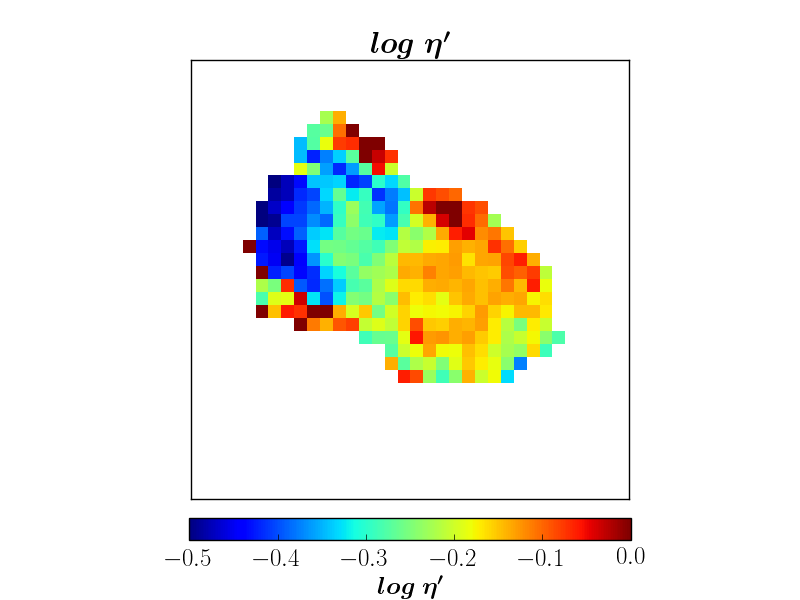}
    \caption{manga-8313-1901}
    \end{figure*}

Figures \ref{fig:manga-8241-6101}-\ref{fig:manga-8718-3703} present the relavant maps of those galaxies in our sample where helium is detected. These maps include those of H$\alpha$ line flux, He \textsc{ii} $\lambda$4686/H$\beta$, [O \textsc{iii}]$\lambda$5007/H$\beta$, equivalent width of H$\beta$ and softness parameter (all quantities in logarithmic). 
  \begin{figure*}
    \centering  
    
    \includegraphics[width=0.185\textwidth, trim={4.0cm 0 4.0cm 0}, clip]{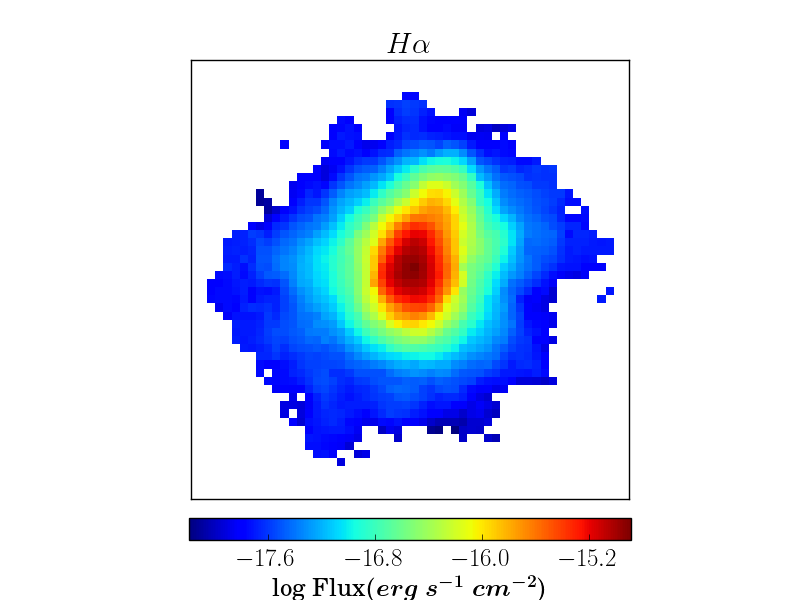}
    \includegraphics[width=0.185\textwidth, trim={4.0cm 0 4.0cm 0},clip]{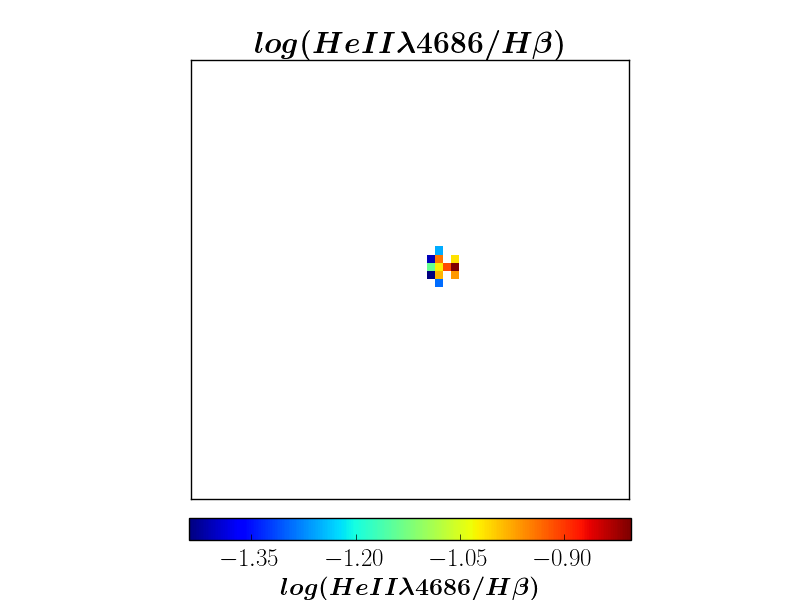}
    \includegraphics[width=0.185\textwidth, trim={2.15cm 0 2.15cm 0},clip]{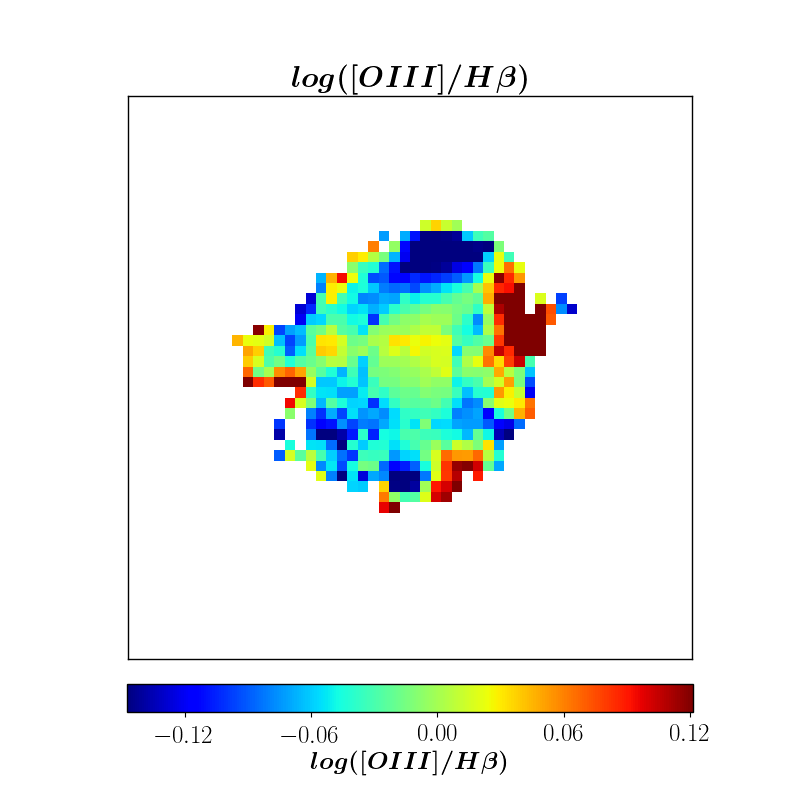}
    \includegraphics[width=0.185\textwidth, trim={4.0cm 0 4.0cm 0}, clip]{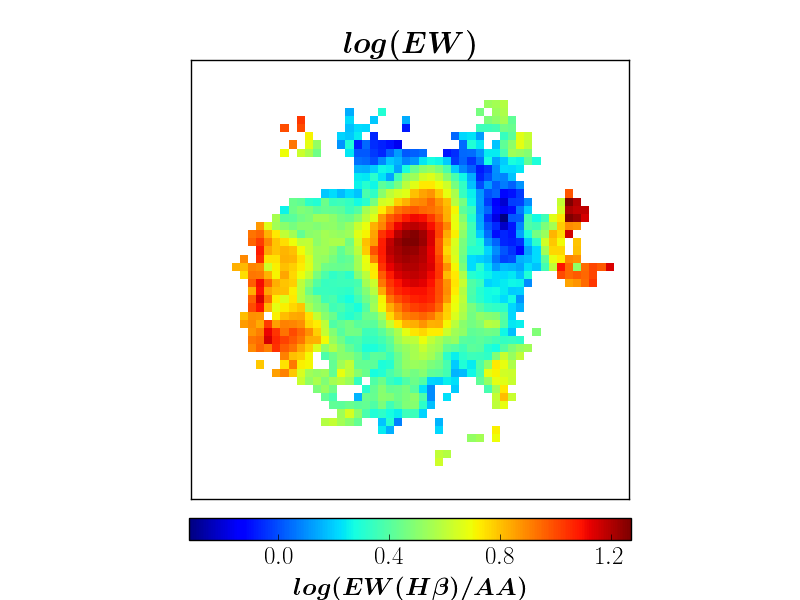}
    \includegraphics[width=0.185\textwidth, trim={2.15cm 0 2.15cm 0},clip]{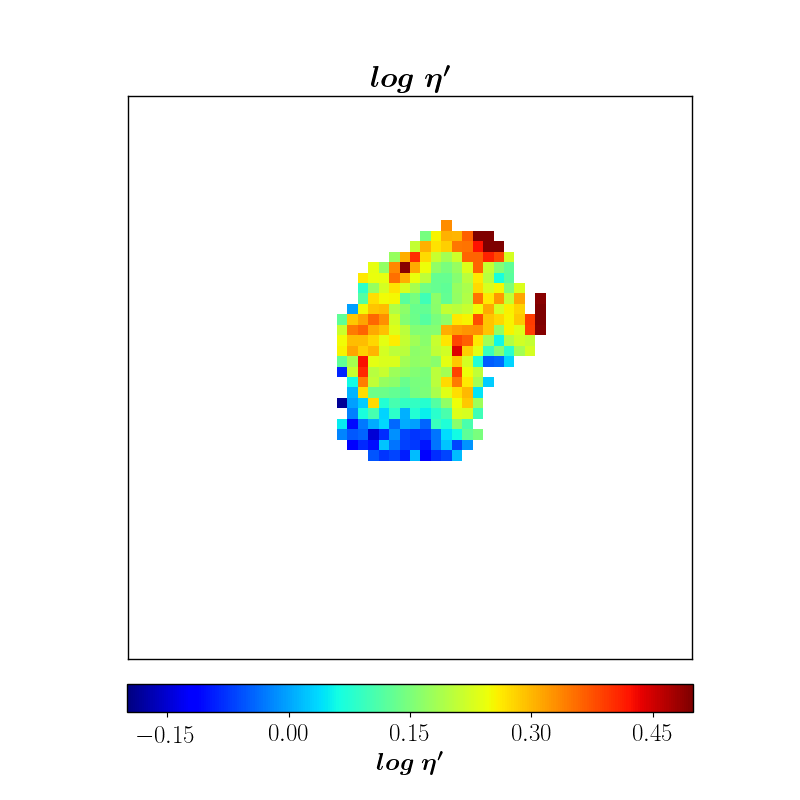}
    \caption{manga-8549-6104}
    
    \includegraphics[width=0.185\textwidth, trim={4.0cm 0 4.0cm 0}, clip]{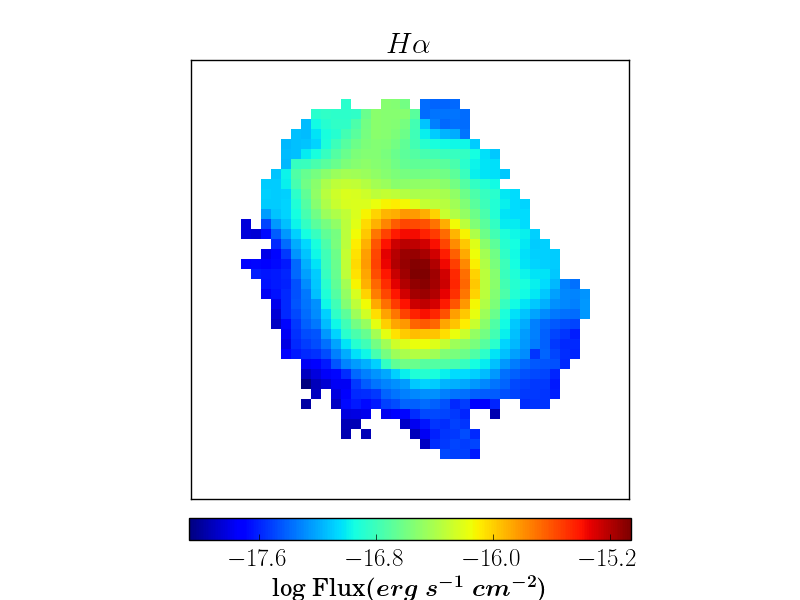}
    \includegraphics[width=0.185\textwidth, trim={4.0cm 0 4.0cm 0},clip]{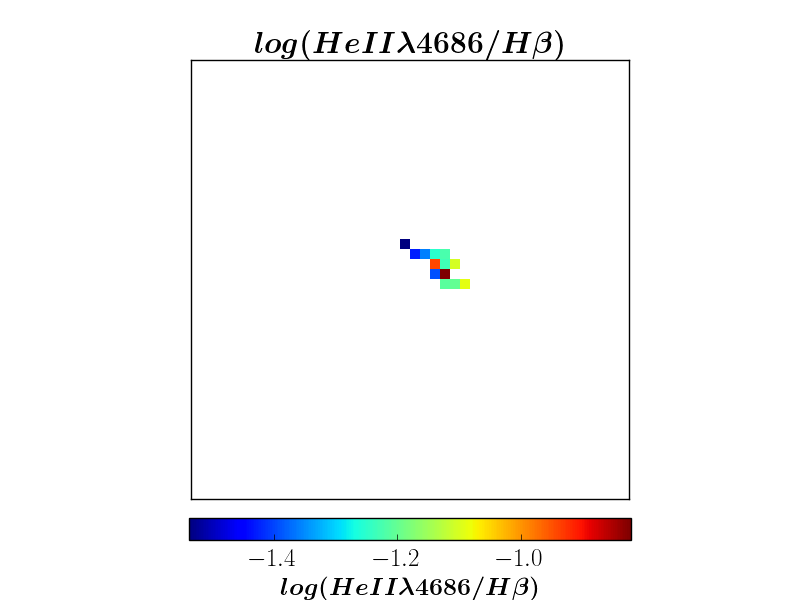}
     \includegraphics[width=0.185\textwidth, trim={2.15cm 0 2.15cm 0},clip]{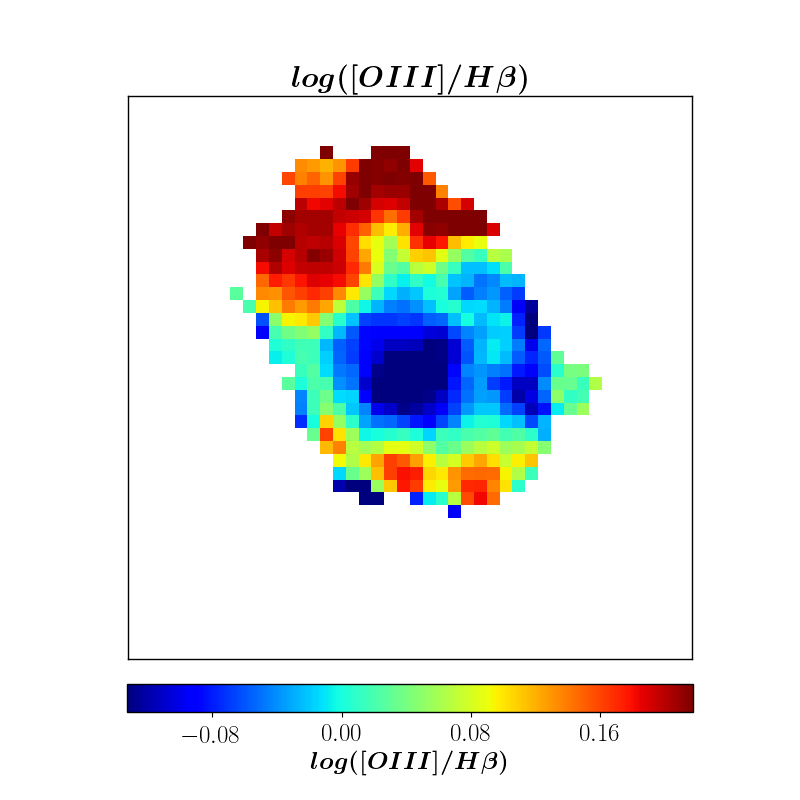}
    \includegraphics[width=0.185\textwidth, trim={4.0cm 0 4.0cm 0}, clip]{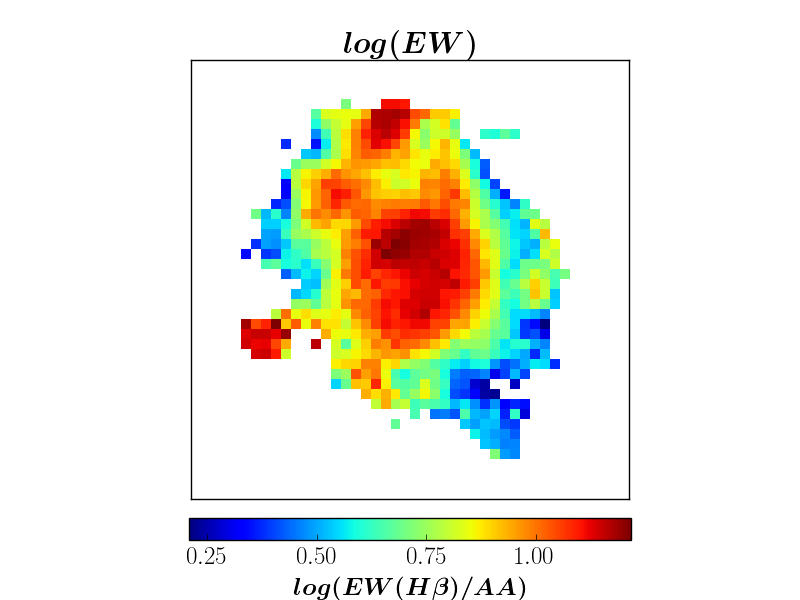}
    \includegraphics[width=0.185\textwidth, trim={2.15cm 0 2.15cm 0},clip]{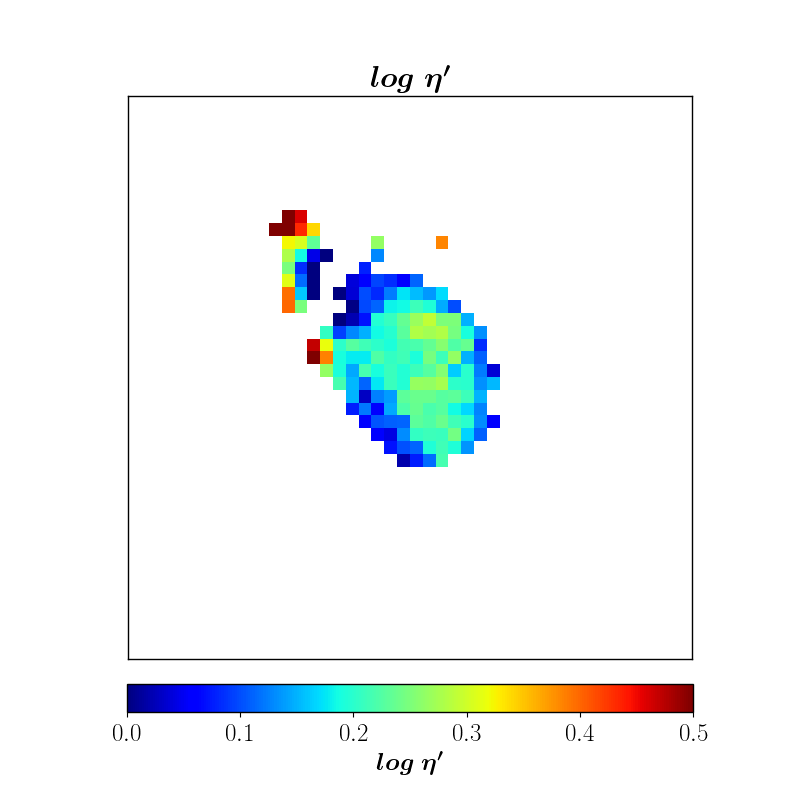}
    \caption{manga-8553-3704}
    
    \end{figure*}
    
   \begin{figure*}
    \centering 
    
    \includegraphics[width=0.185\textwidth, trim={4.0cm 0 4.0cm 0}, clip]{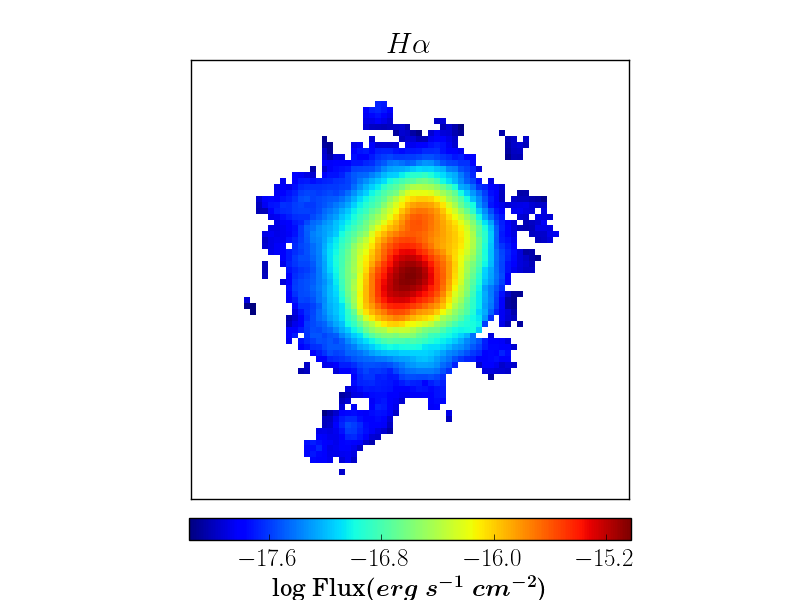}
	\includegraphics[width=0.185\textwidth, trim={4.0cm 0 4.0cm 0}, clip]{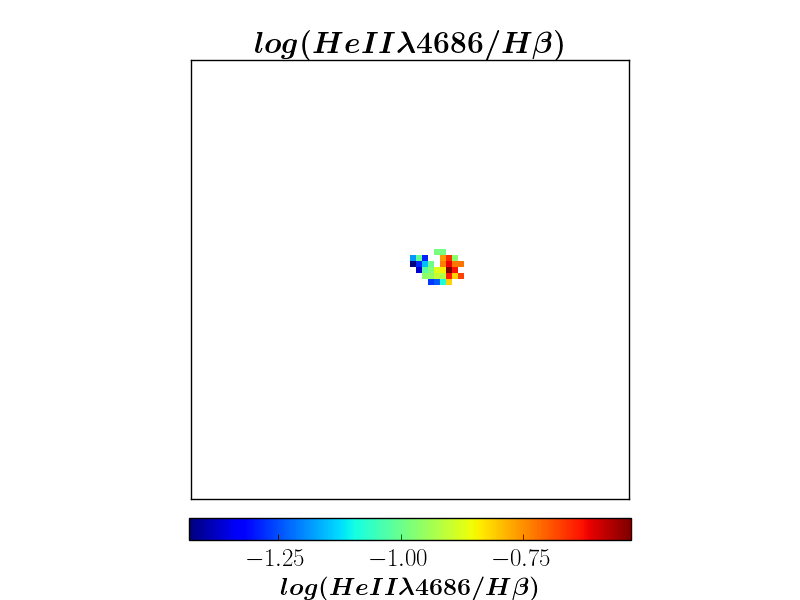}
	\includegraphics[width=0.185\textwidth, trim={2.15cm 0 2.15cm 0}, clip]{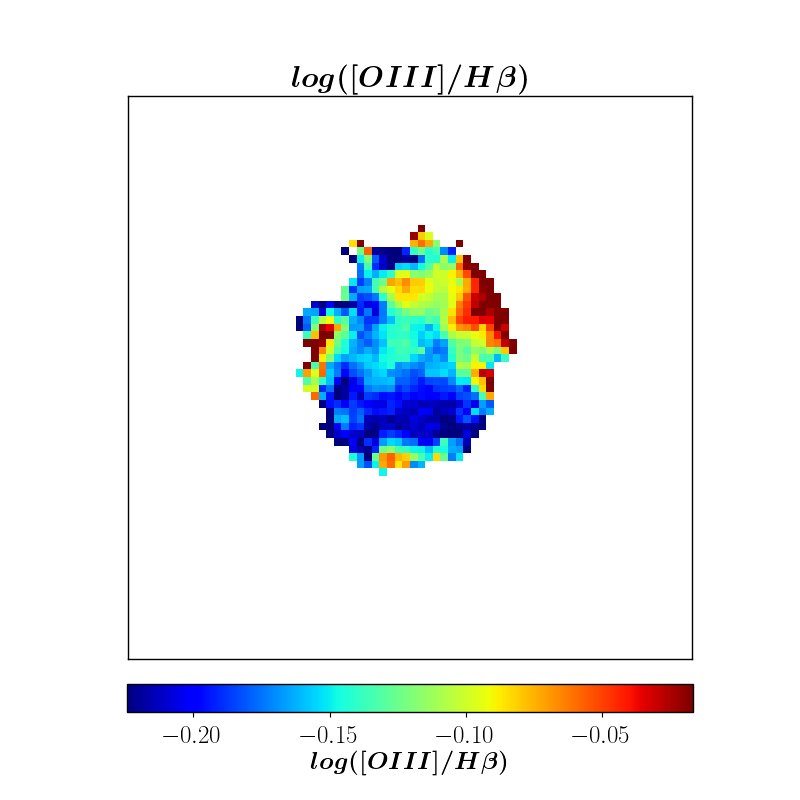}
	\includegraphics[width=0.185\textwidth, trim={4.0cm 0 4.0cm 0}, clip]{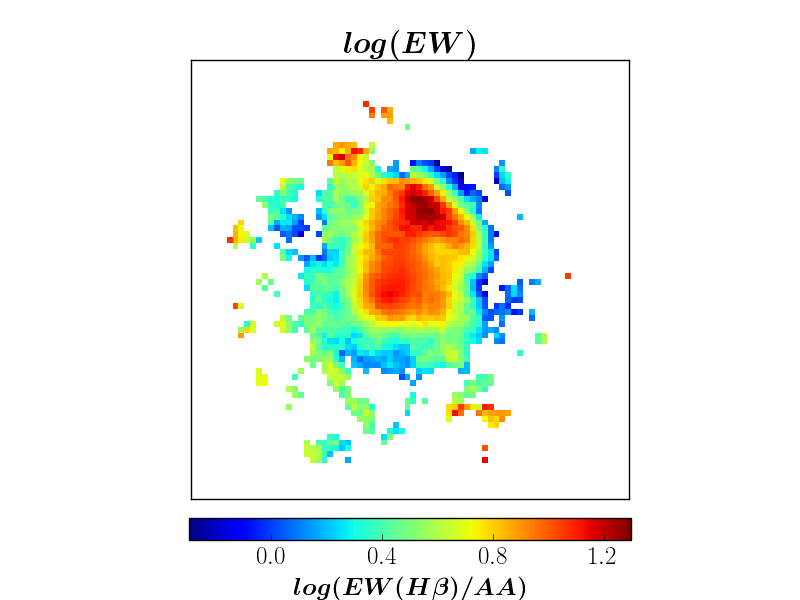}
	\includegraphics[width=0.185\textwidth, trim={2.15cm 0 2.15cm 0}, clip]{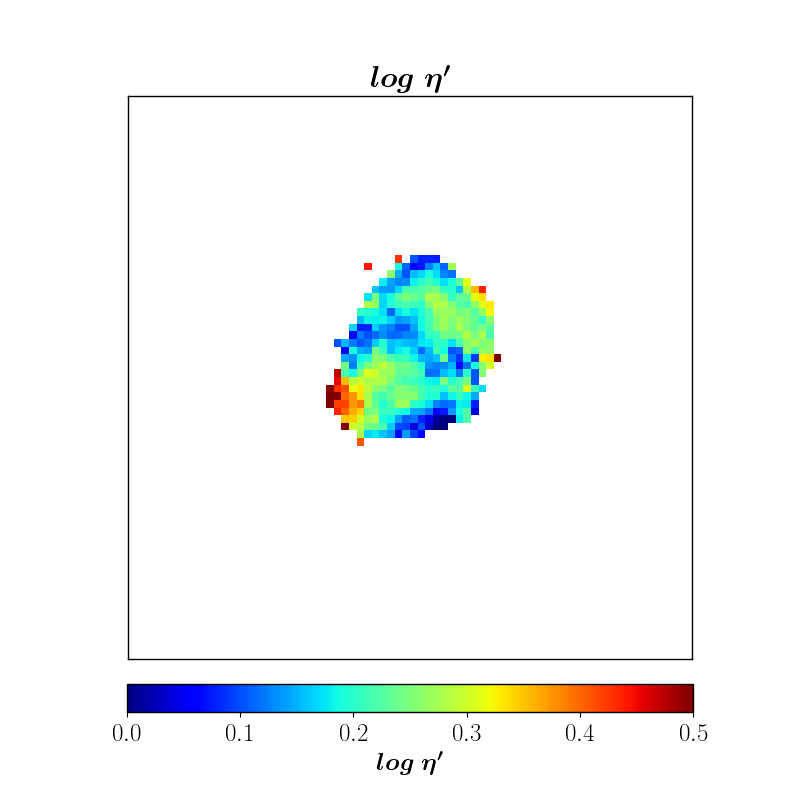}
    \caption{manga-8613-12703}
    
    \includegraphics[width=0.185\textwidth, trim={4.0cm 0 4.0cm 0}, clip]{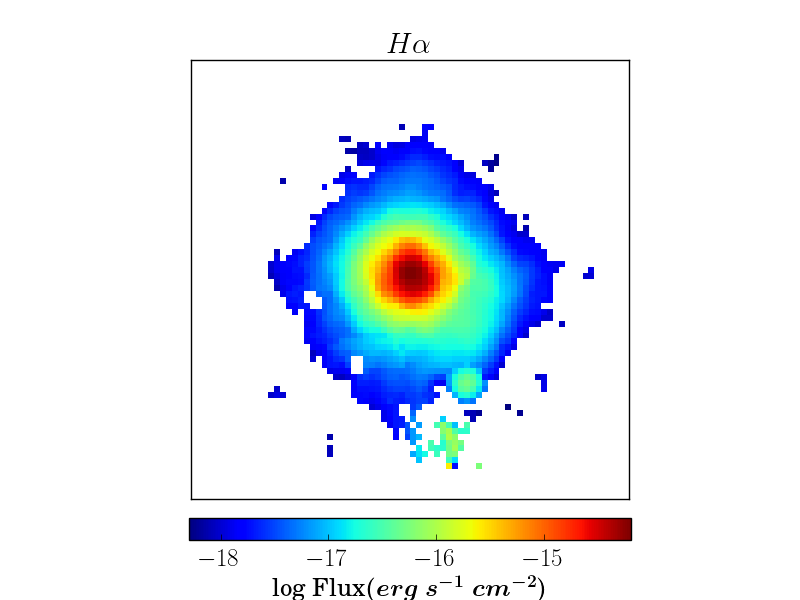}
	\includegraphics[width=0.185\textwidth, trim={4.0cm 0 4.0cm 0}, clip]{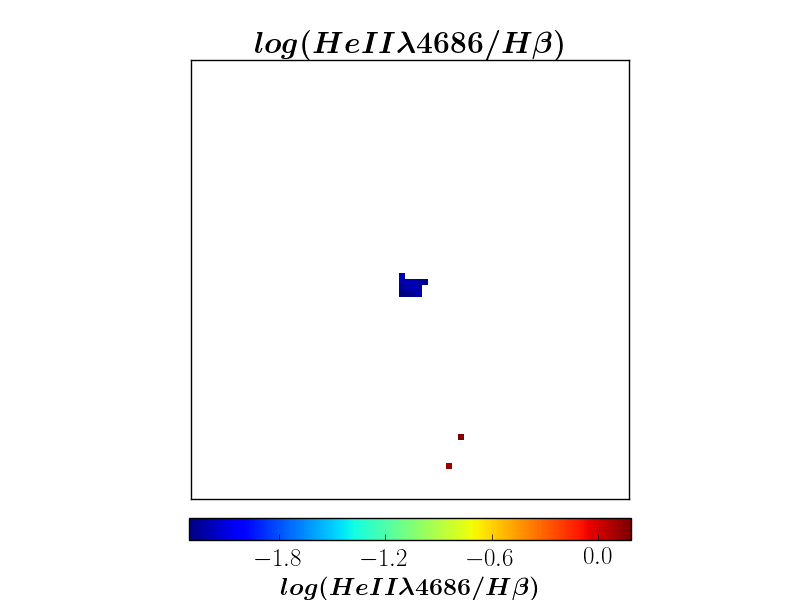}
	\includegraphics[width=0.185\textwidth, trim={2.15cm 0 2.15cm 0}, clip]{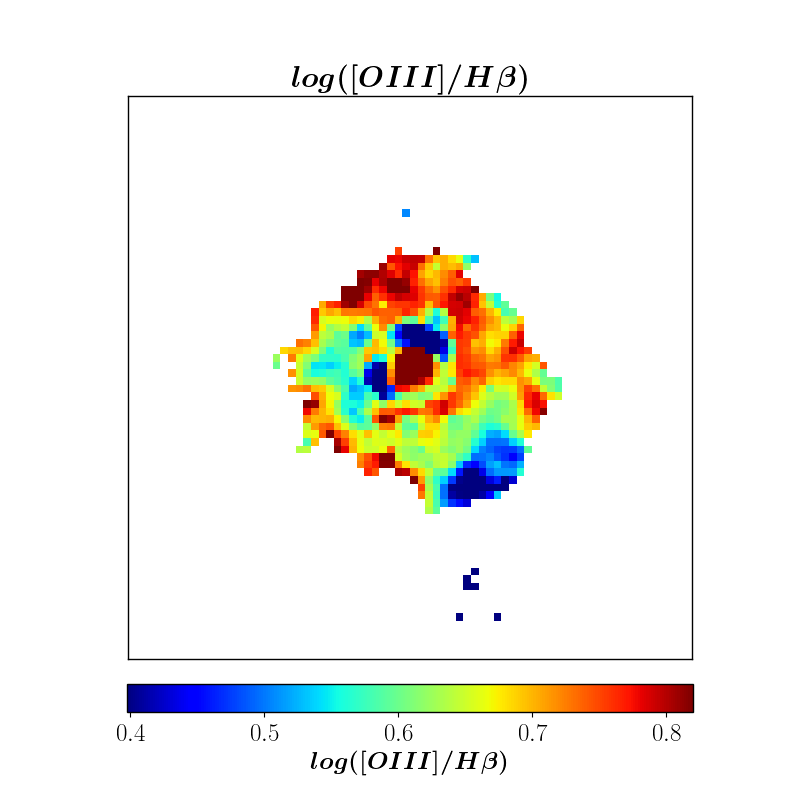}
	\includegraphics[width=0.185\textwidth, trim={2.15cm 0 2.15cm 0}, clip]{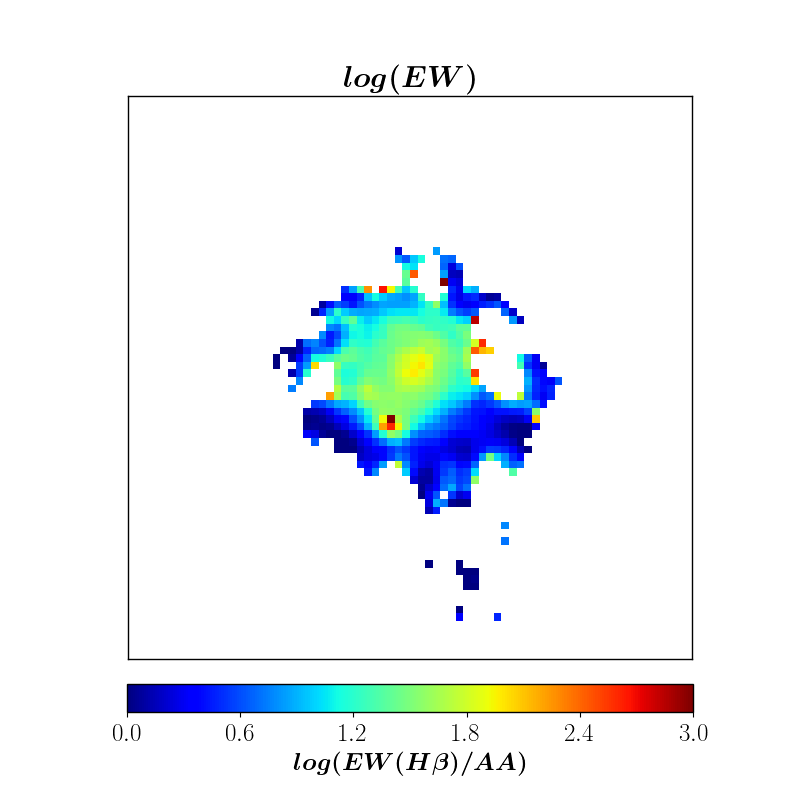}
	\includegraphics[width=0.185\textwidth, trim={4.0cm 0 4.0cm 0}, clip]{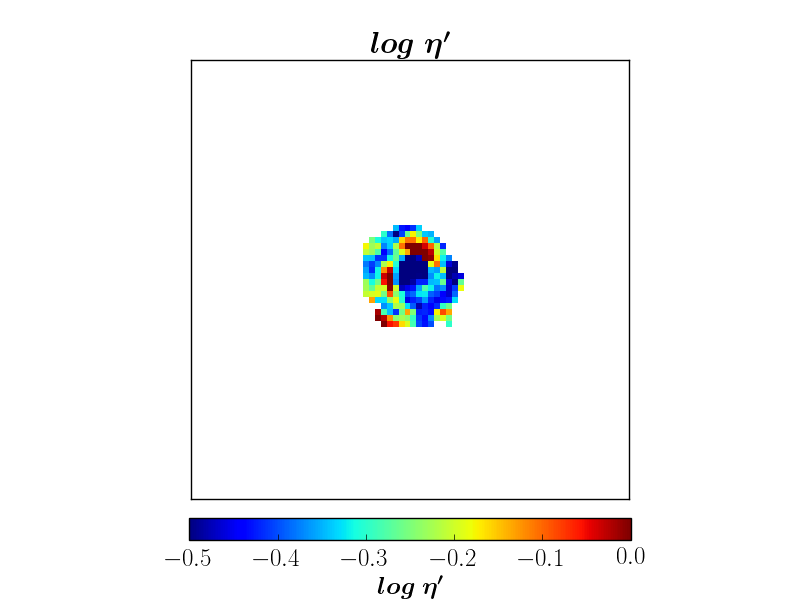}
    \caption{manga-8626-12704}
     \end{figure*}
  \begin{figure*}
    \centering  
     \includegraphics[width=0.185\textwidth, trim={4.0cm 0 4.0cm 0}, clip]{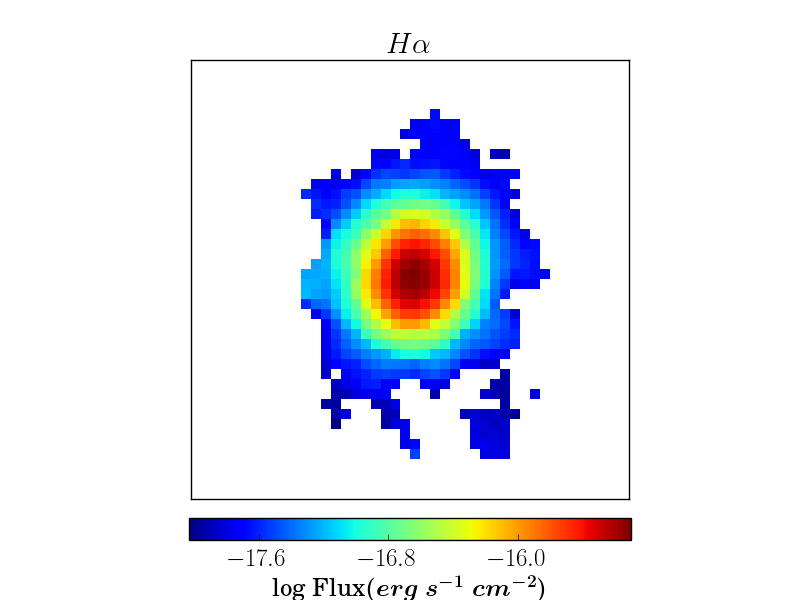}
	\includegraphics[width=0.185\textwidth, trim={4.0cm 0 4.0cm 0},clip]{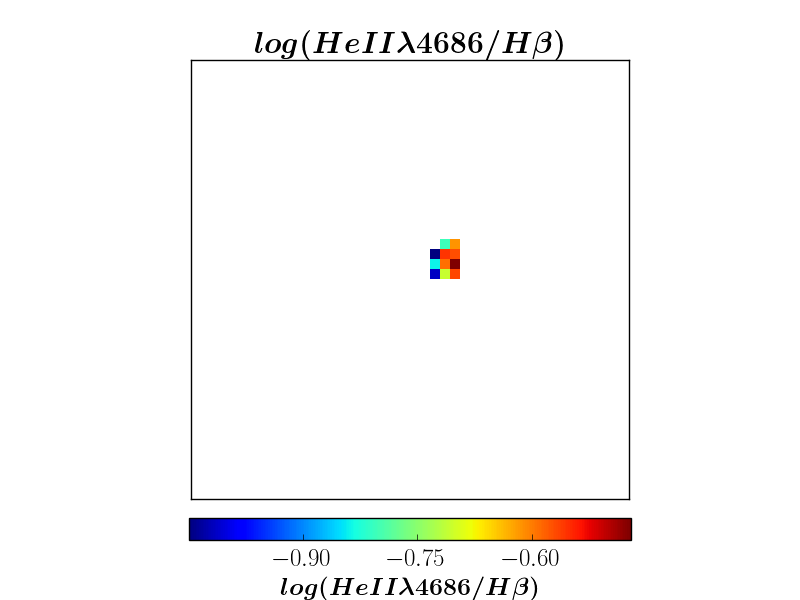}
	\includegraphics[width=0.185\textwidth, trim={2.15cm 0 2.15cm 0},clip]{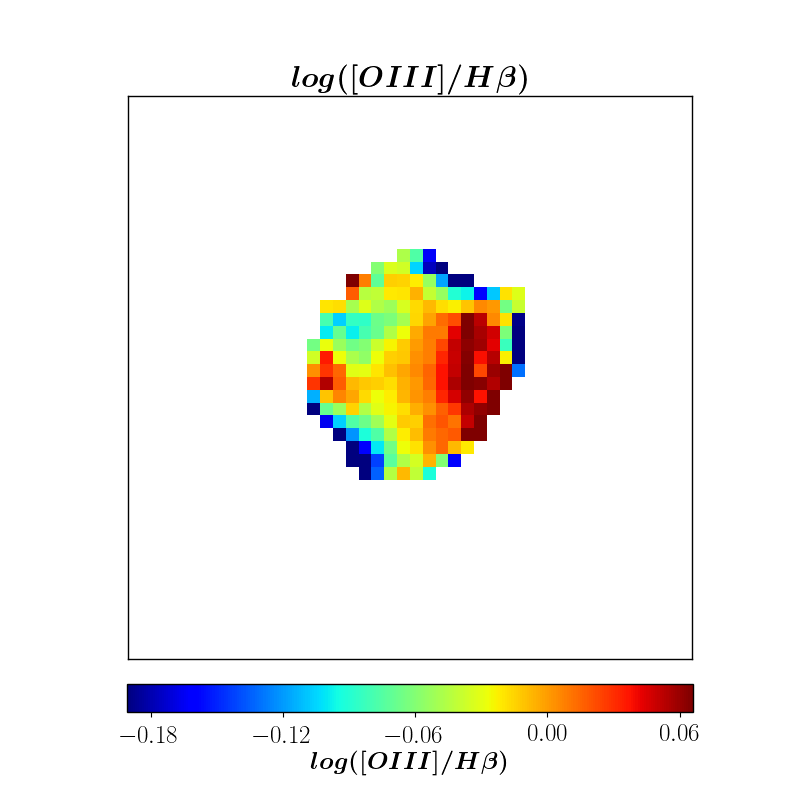}
	\includegraphics[width=0.185\textwidth, trim={4.0cm 0 4.0cm 0}, clip]{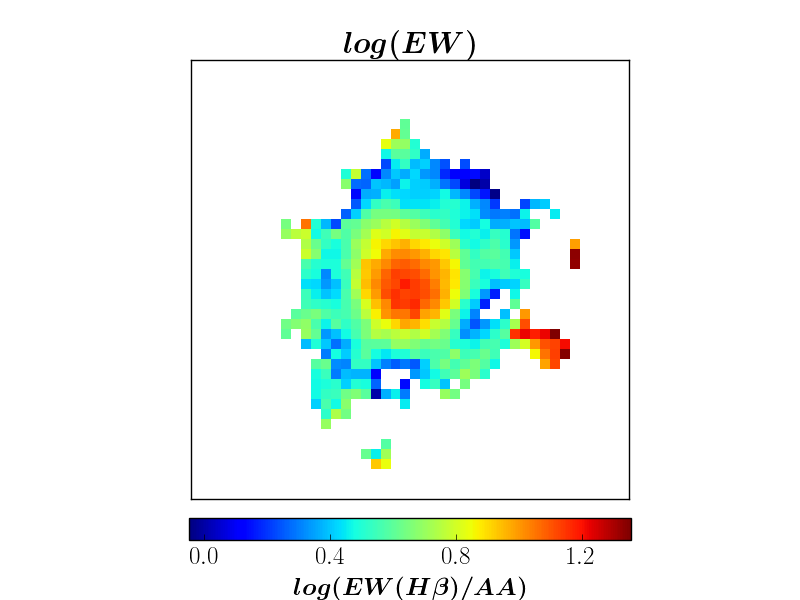}
	\includegraphics[width=0.185\textwidth, trim={2.15cm 0 2.15cm 0},clip]{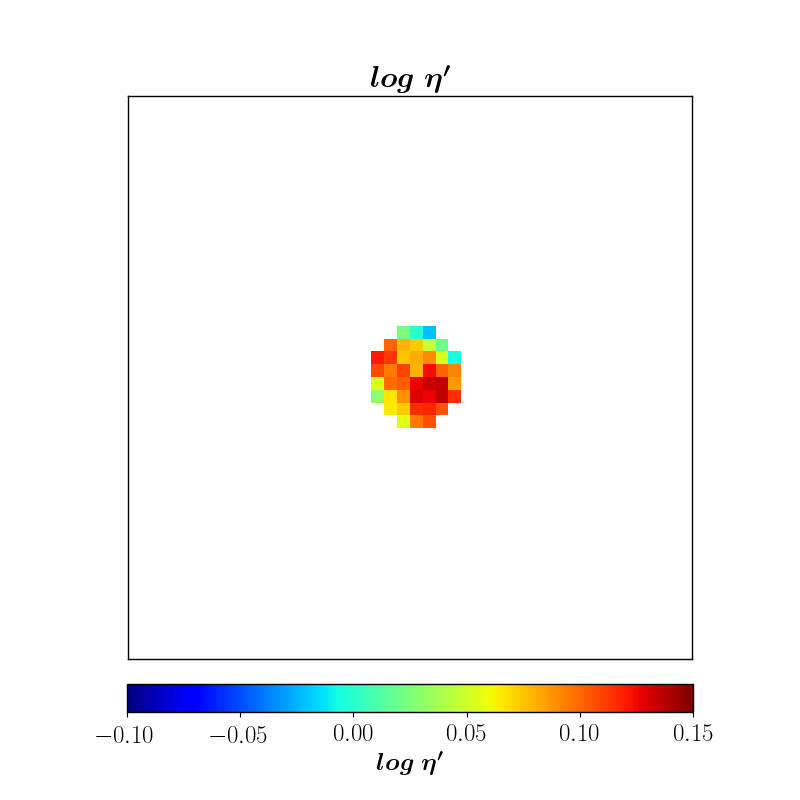}
    \caption{manga-8718-3703}
    \label{fig:manga-8718-3703}
\end{figure*}

\section{Galaxies with [O \textsc{iii}] $\lambda$4363 detection}
\label{sect:individual}
\begin{figure*}
	\centering
	\includegraphics[width=0.185\textwidth, trim={4.0cm 0 4.0cm 0}, clip]{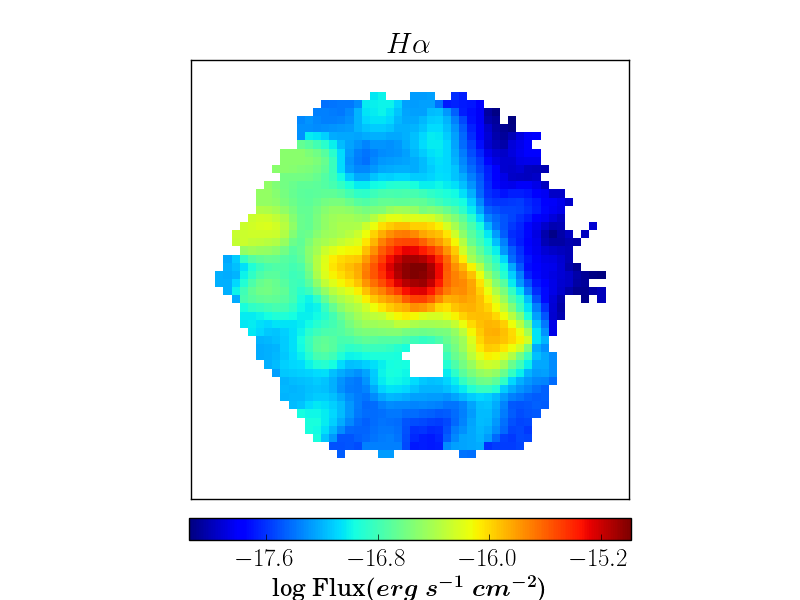}
	\includegraphics[width=0.185\textwidth, trim={4.0cm 0 4.0cm 0}, clip]{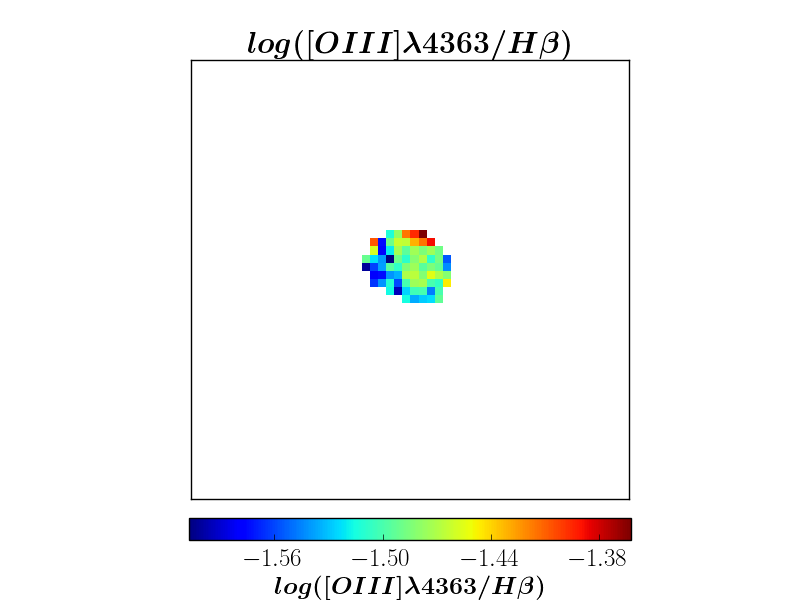}
	\includegraphics[width=0.185\textwidth, trim={4.0cm 0 4.0cm 0}, clip]{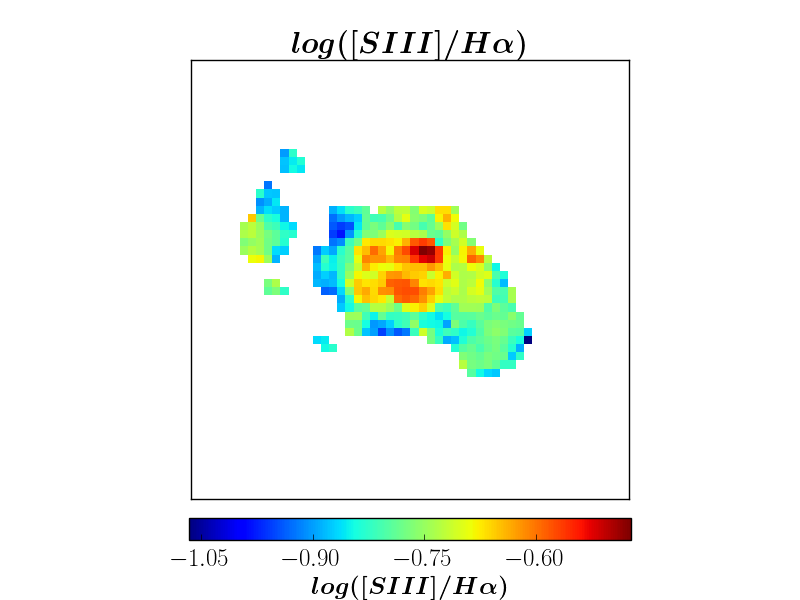}
	\includegraphics[width=0.185\textwidth, trim={4.0cm 0 4.0cm 0}, clip]{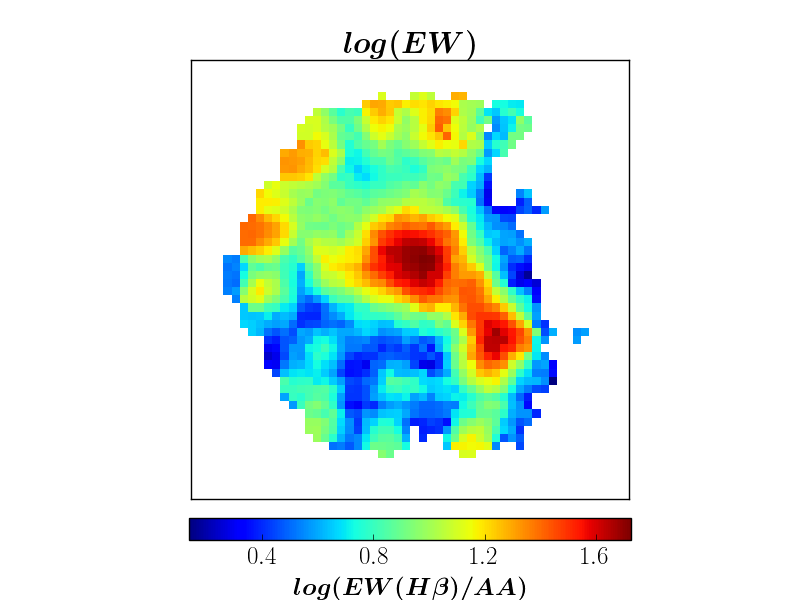}
	\includegraphics[width=0.185\textwidth, trim={4.0cm 0 4.0cm 0}, clip]{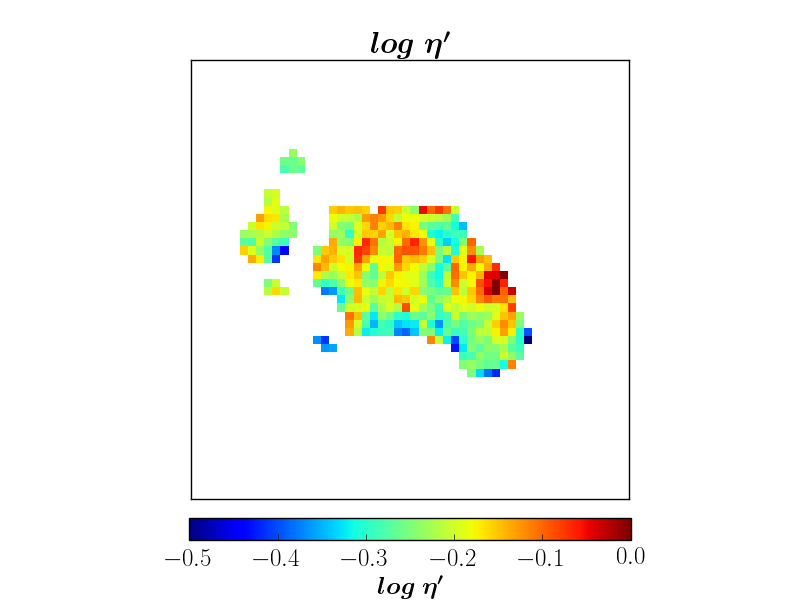}

	\includegraphics[width=0.21\textwidth, trim={0 2.0cm 1.5cm 1.0cm}, clip]{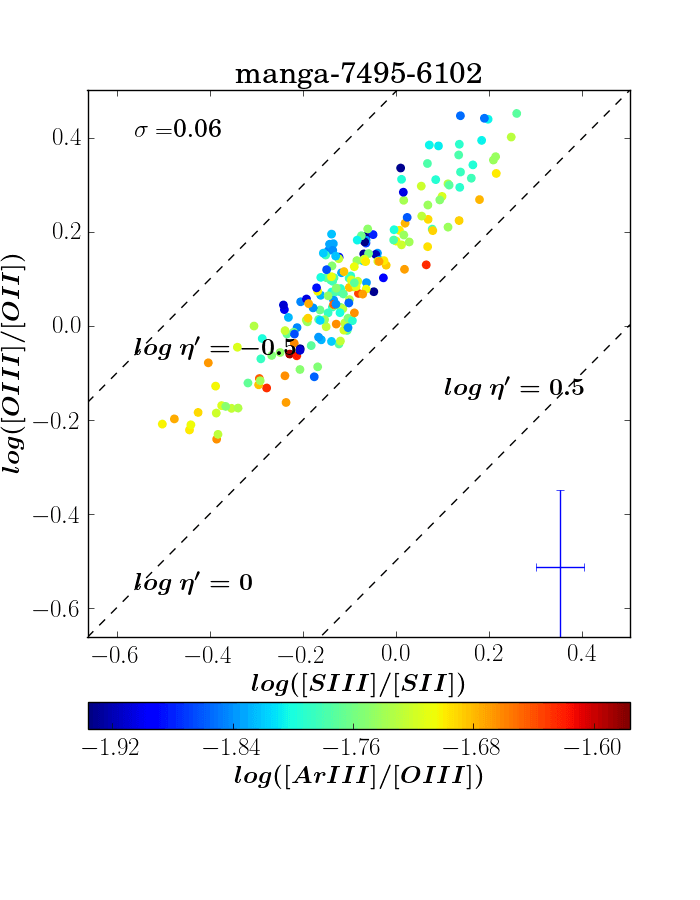}
	\includegraphics[width=0.185\textwidth, trim={2.1cm 2.0cm 1.5cm 1.0cm}, clip]{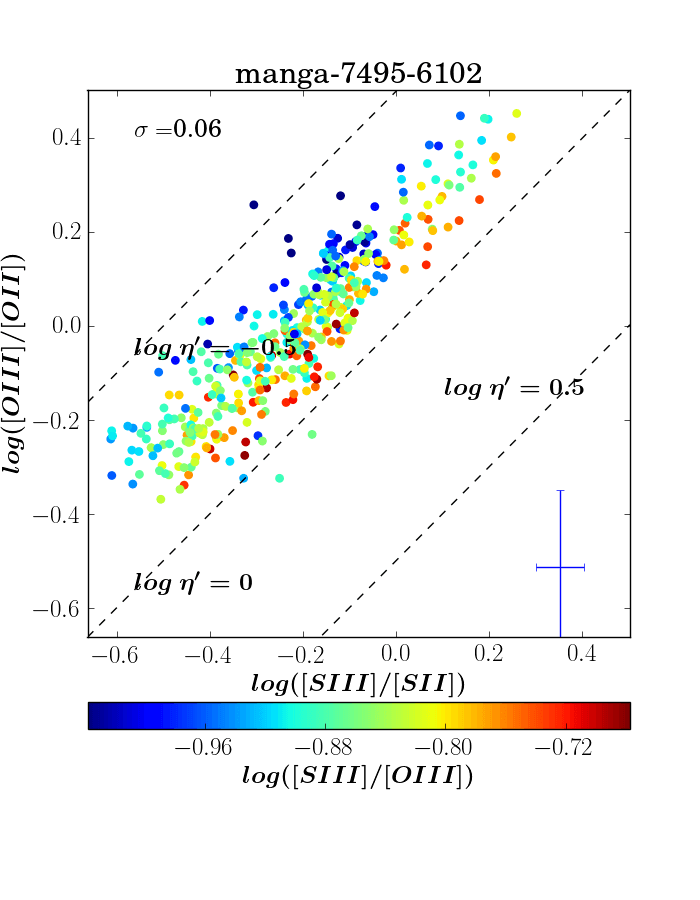}
	\includegraphics[width=0.185\textwidth, trim={2.1cm 2.0cm 1.5cm 1.0cm}, clip]{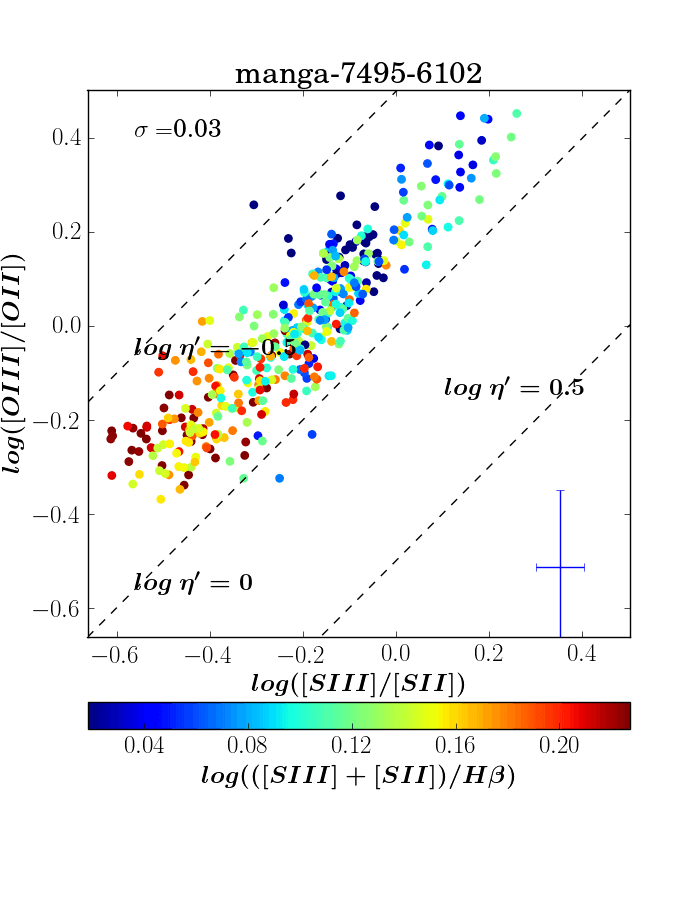}
	\includegraphics[width=0.185\textwidth, trim={2.1cm 2.0cm 1.5cm 1.0cm}, clip]{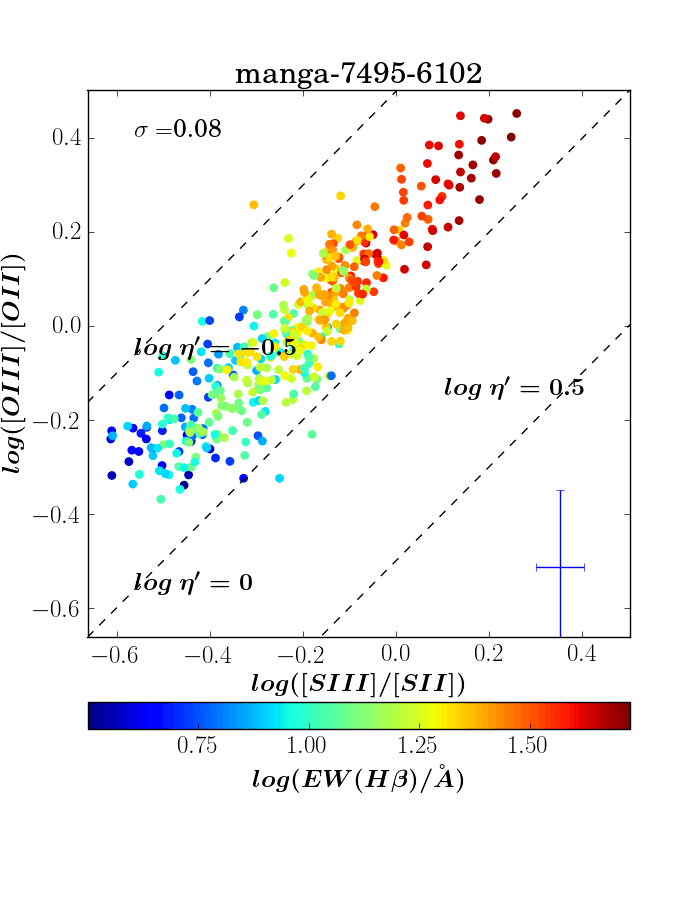}
	\includegraphics[width=0.185\textwidth, trim={2.1cm 2.0cm 1.5cm 1.0cm}, clip]{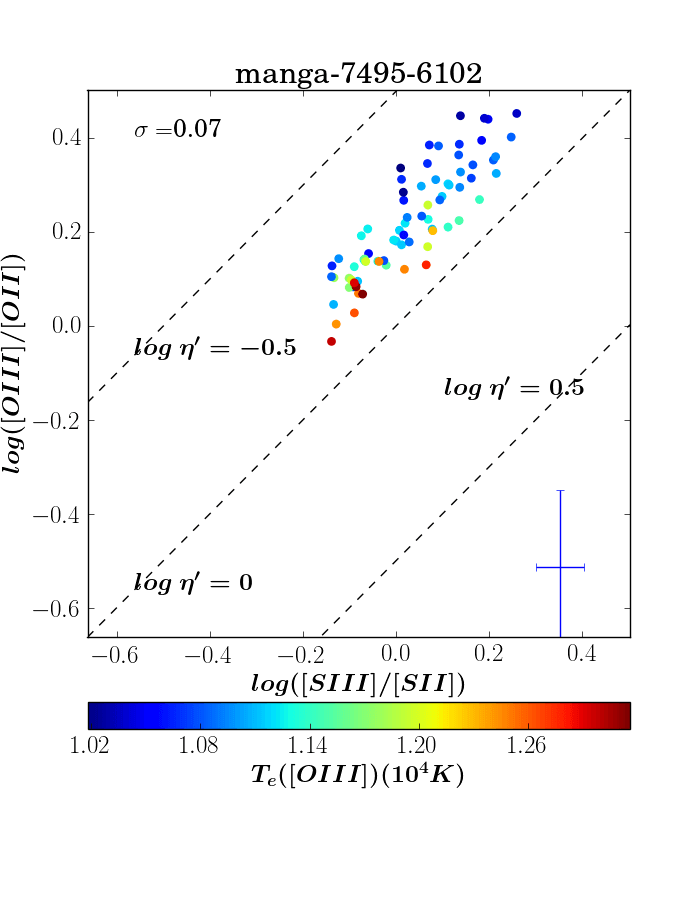}
	\caption{Manga-7495-6102.  Upper panel: Maps of H$\alpha$ emission line (left), [O \textsc{iii}] $\lambda$4363/H$\beta$ line ratio (2nd panel from left), [S \textsc{iii}] $\lambda$9532/H$\alpha$ (3rd panel from left), equivalent width of H$\alpha$ (4th panel from left) and softness parameter (right-hand panel). Lower panel: O$_3$O$_2$ versus S$_3$S$_2$ colour-coded with respect to Ar$_3$/O$_3$ (left), S$_3$/O$_3$ (2nd panel from left), S$_{23}$ (3rd panel from left), EW(H$\beta$) (4th panel from left)  and T$_e$([O \textsc{iii}]) (right-hand panel). On each panel, the three dashed black lines correspond to log $\rm\eta\prime$ = $-$0.5, 0 and 0.5. The median uncertainties on O$_3$O$_2$ and S$_3$S$_2$ are shown in lower-right corner on each panel, and median uncertainties ($\sigma$) on Ar$_3$/O$_3$, S$_3$/O$_3$, S$_{23}$, EW(H$\beta$) and T$_e$([O \textsc{iii}]) are shown on upper-left corner of each panel. Note that the number of plotted points are different in each panel.}
	\label{fig:manga-7495-6102}
\end{figure*}

\begin{figure*}
	\centering
	\includegraphics[width=0.185\textwidth, trim={4.0cm 0 4.0cm 0}, clip]{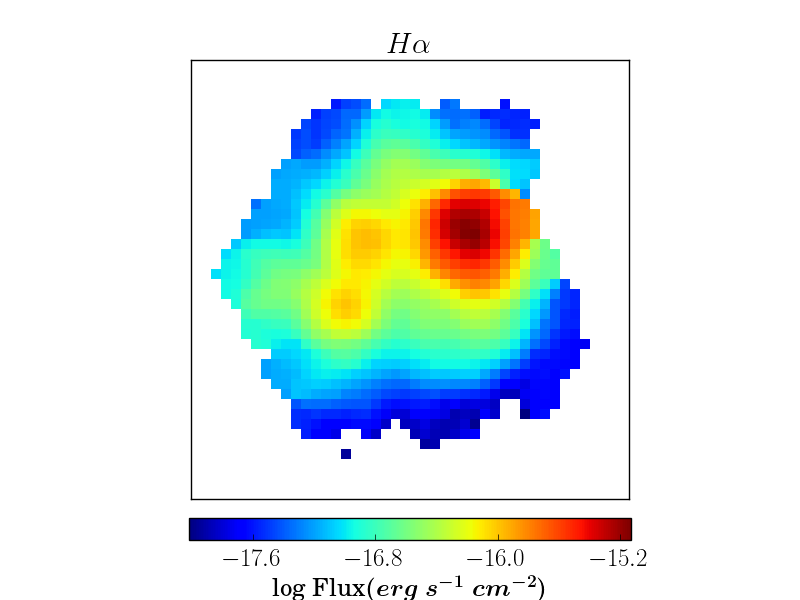}
	\includegraphics[width=0.185\textwidth, trim={4.0cm 0 4.0cm 0}, clip]{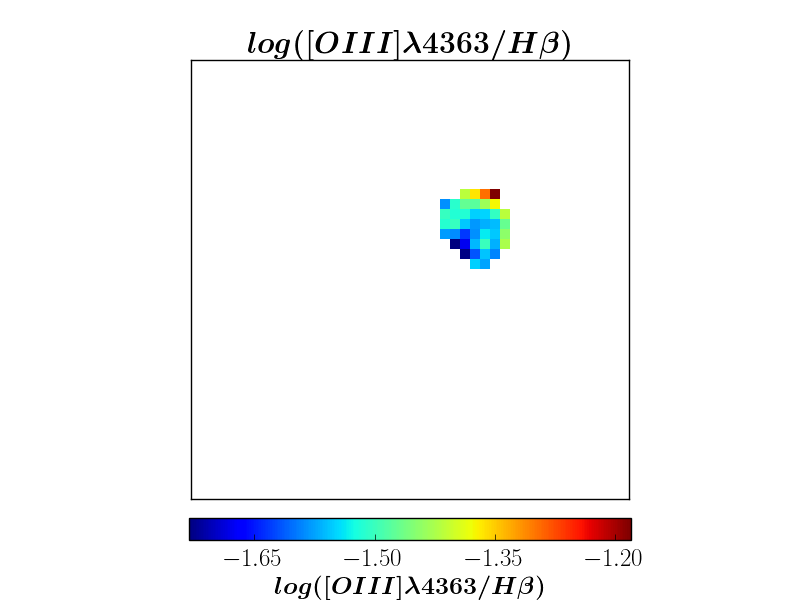}
	\includegraphics[width=0.185\textwidth, trim={4.0cm 0 4.0cm 0}, clip]{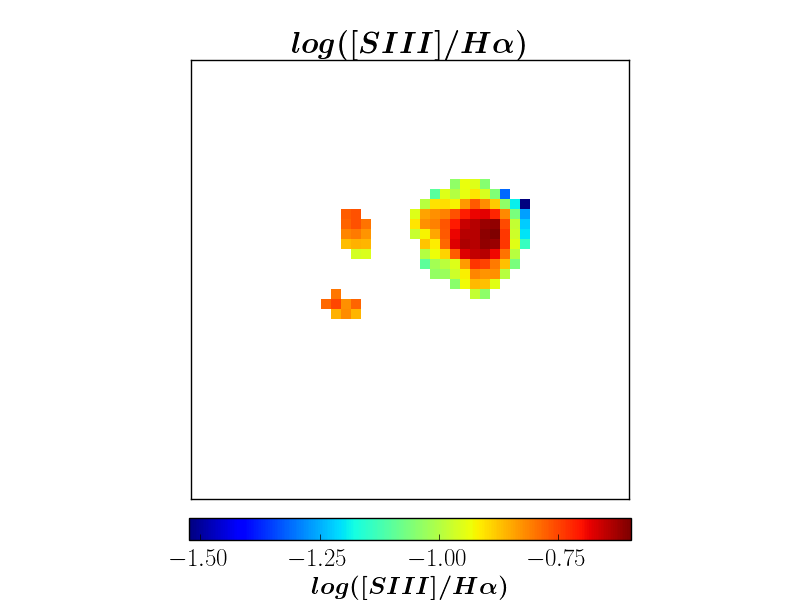}
	\includegraphics[width=0.185\textwidth, trim={4.0cm 0 4.0cm 0}, clip]{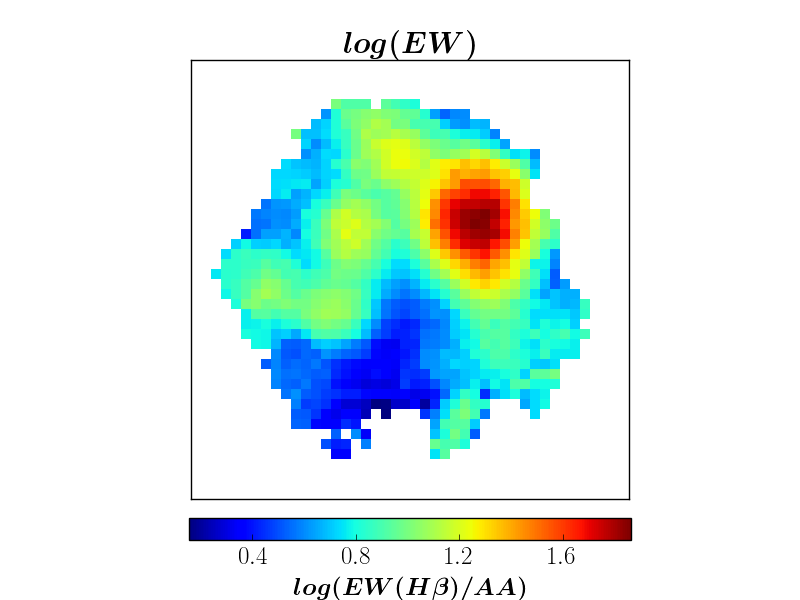}
	\includegraphics[width=0.185\textwidth, trim={4.0cm 0 4.0cm 0}, clip]{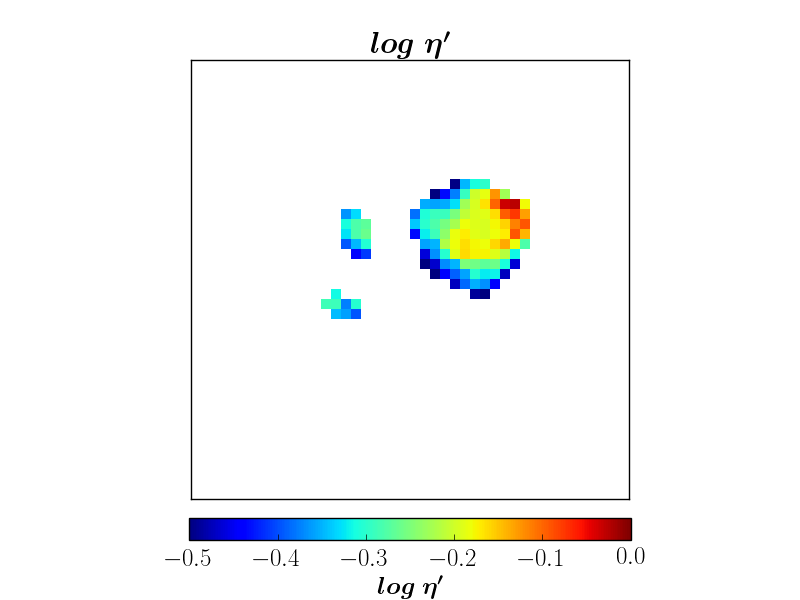}

	\includegraphics[width=0.21\textwidth, trim={0 2.0cm 1.5cm 1.0cm}, clip]{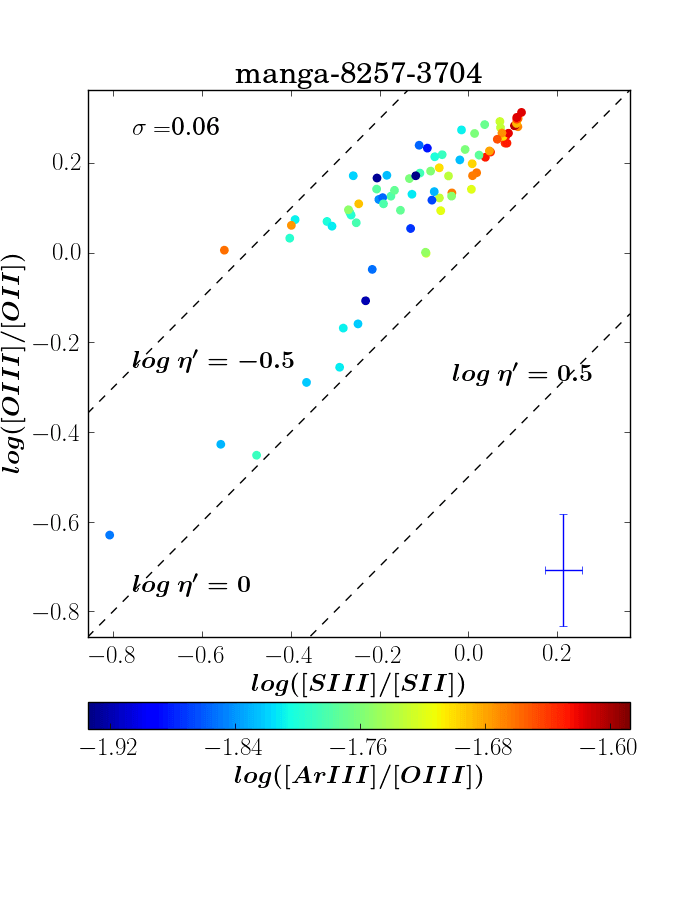}
	\includegraphics[width=0.185\textwidth, trim={2.1cm 2.0cm 1.5cm 1.0cm}, clip]{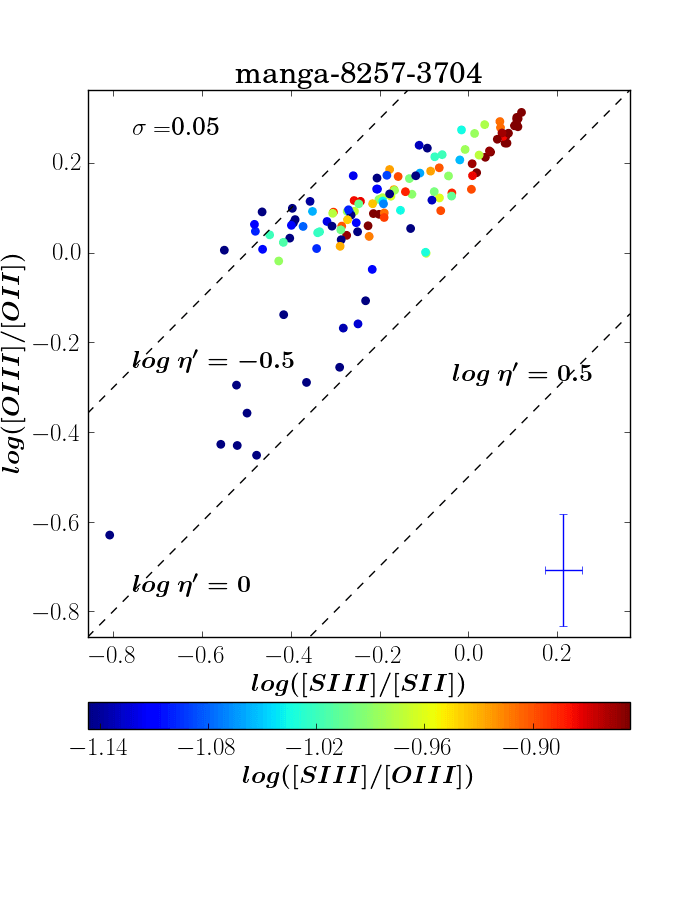}
	\includegraphics[width=0.185\textwidth, trim={2.1cm 2.0cm 1.5cm 1.0cm}, clip]{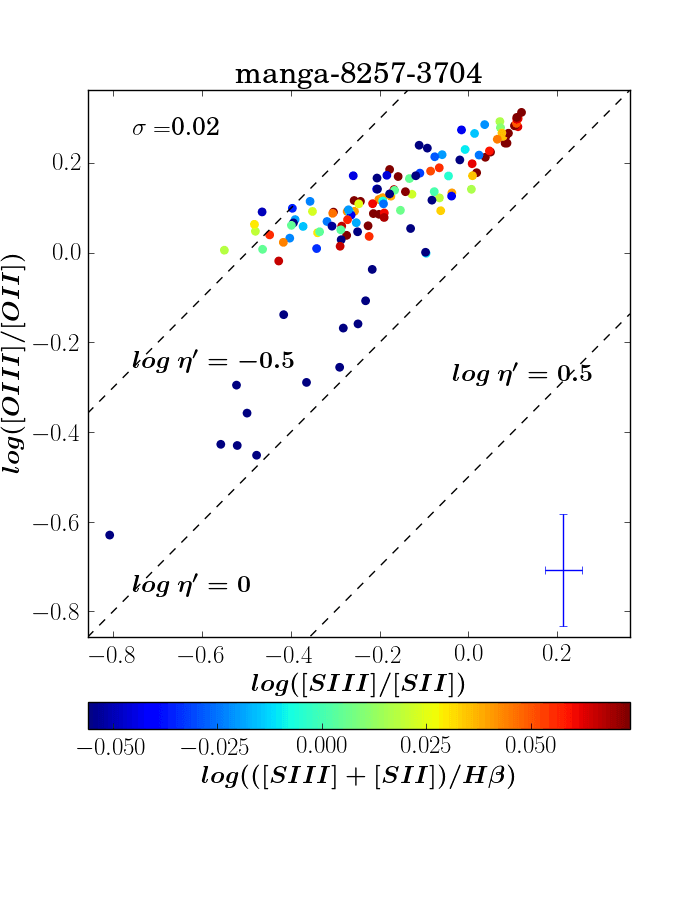}
	\includegraphics[width=0.185\textwidth, trim={2.1cm 2.0cm 1.5cm 1.0cm}, clip]{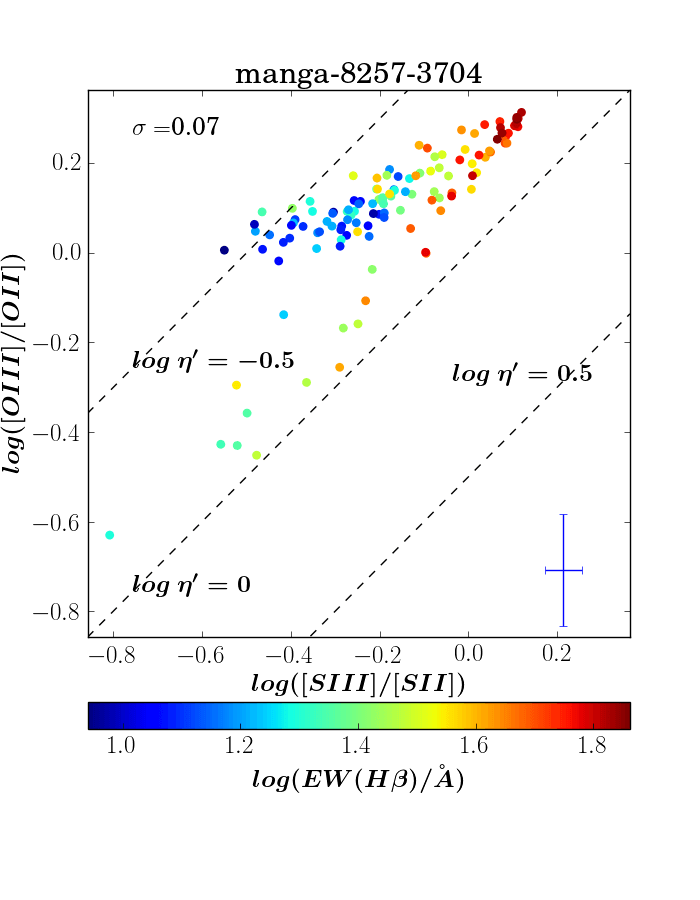}
	\includegraphics[width=0.185\textwidth, trim={2.1cm 2.0cm 1.5cm 1.0cm}, clip]{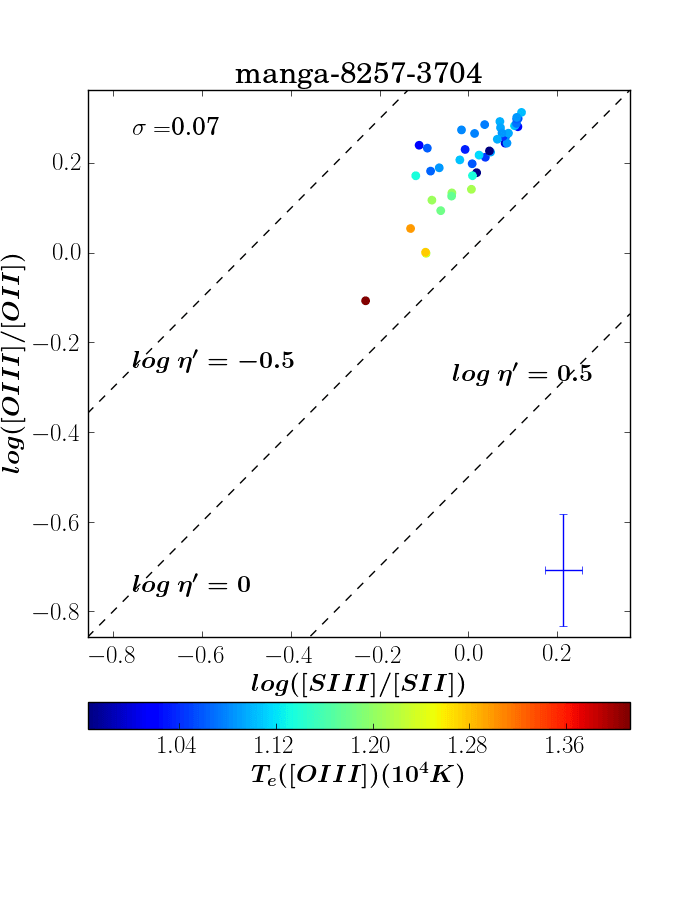}
	\caption{Manga-8257-3704 See caption of Figure \ref{fig:manga-7495-6102} for details.}
	\label{fig:manga-8257-3704}
\end{figure*}

\begin{figure*}
	\centering
	\includegraphics[width=0.185\textwidth, trim={4.0cm 0 4.0cm 0}, clip]{flux_Ha_manga-8313-1901.png}
	\includegraphics[width=0.185\textwidth, trim={4.0cm 0 4.0cm 0}, clip]{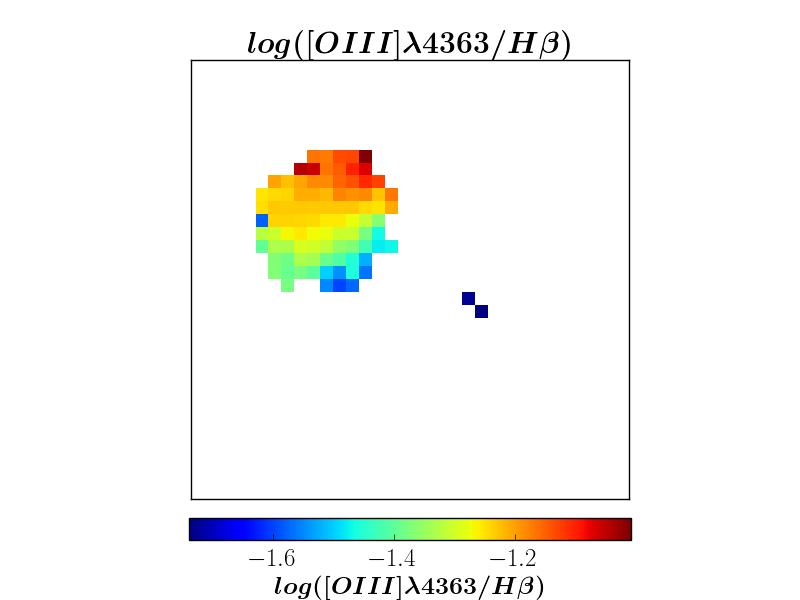}
	\includegraphics[width=0.185\textwidth, trim={4.0cm 0 4.0cm 0}, clip]{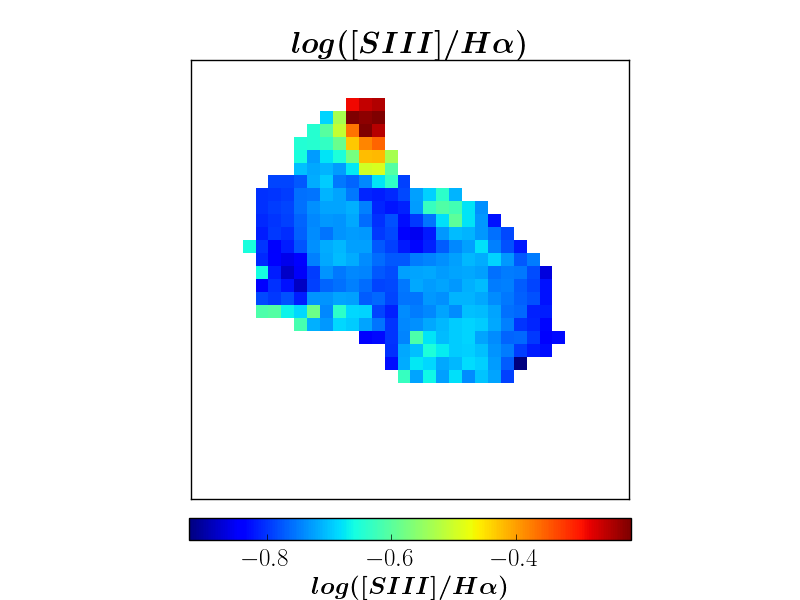}
	\includegraphics[width=0.185\textwidth, trim={4.0cm 0 4.0cm 0}, clip]{log_ew_hb_manga-8313-1901.png}
	\includegraphics[width=0.185\textwidth, trim={4.0cm 0 4.0cm 0}, clip]{eta_p_manga-8313-1901.png}

	\includegraphics[width=0.21\textwidth, trim={0 2.0cm 1.5cm 1.0cm}, clip]{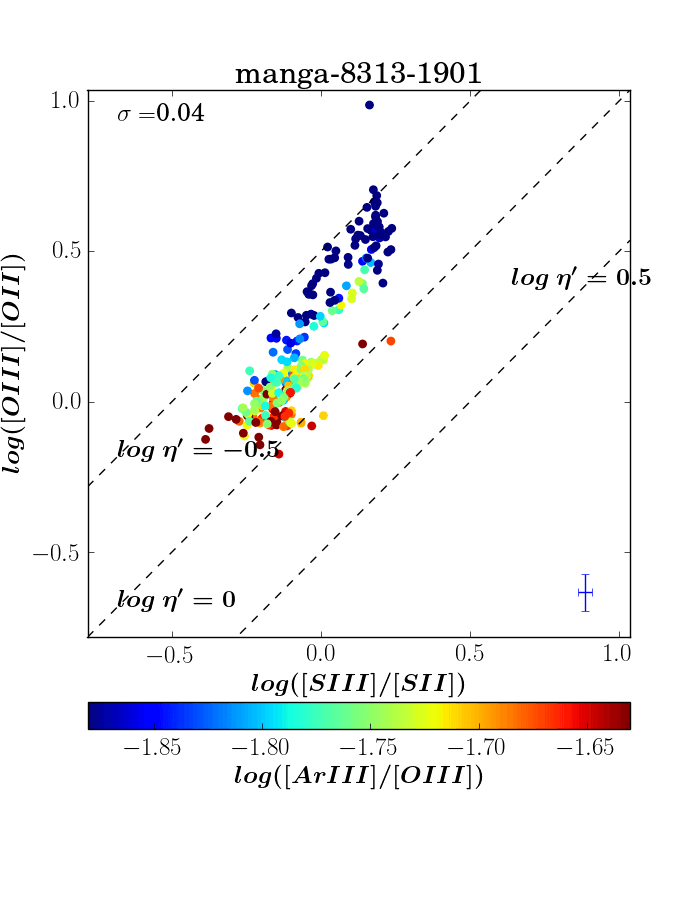}
	\includegraphics[width=0.185\textwidth, trim={2.1cm 2.0cm 1.5cm 1.0cm}, clip]{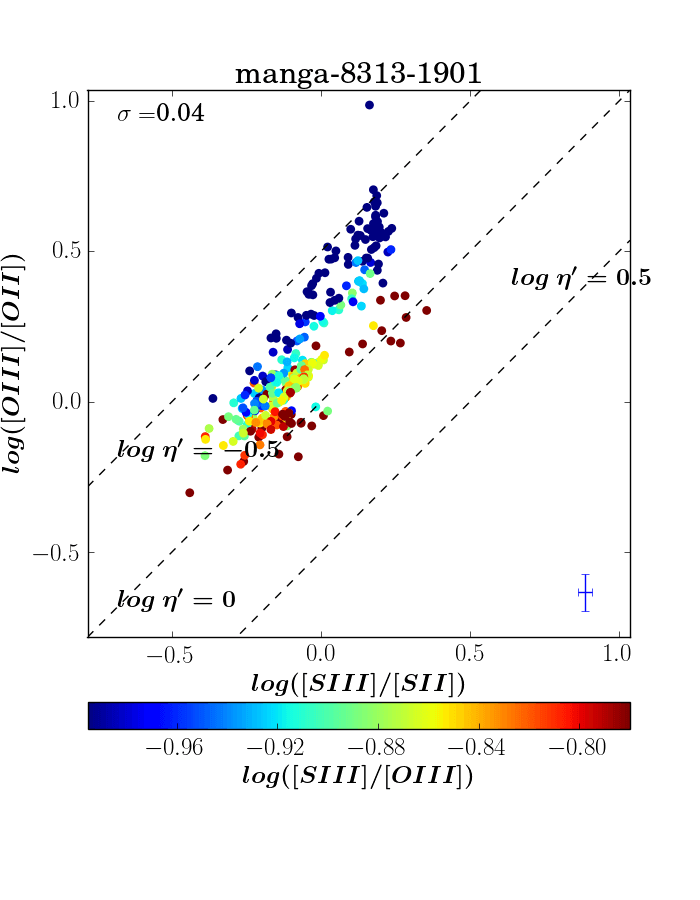}
	\includegraphics[width=0.185\textwidth, trim={2.1cm 2.0cm 1.5cm 1.0cm}, clip]{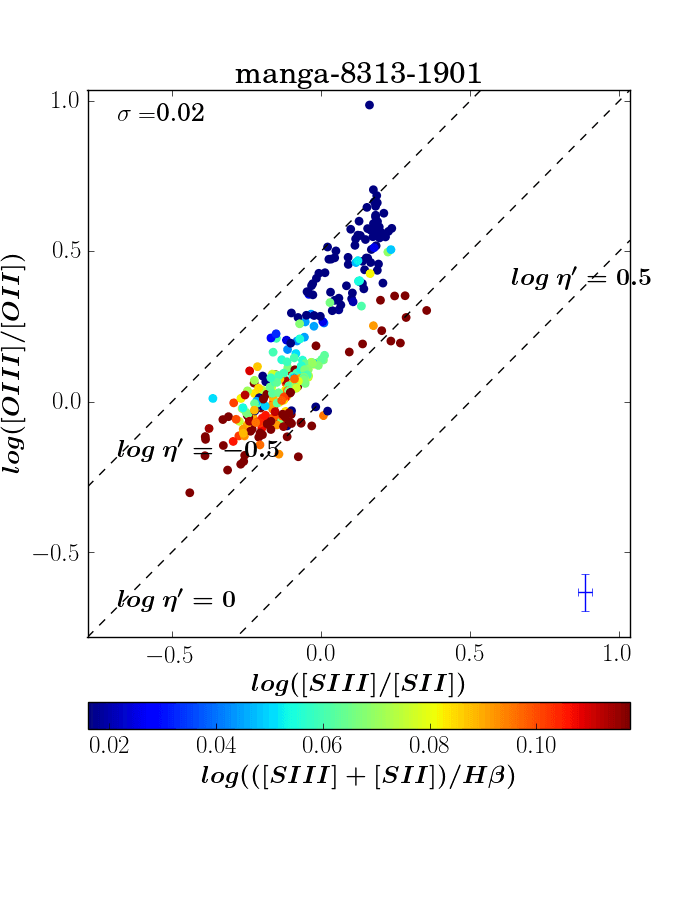}
	\includegraphics[width=0.185\textwidth, trim={2.1cm 2.0cm 1.5cm 1.0cm}, clip]{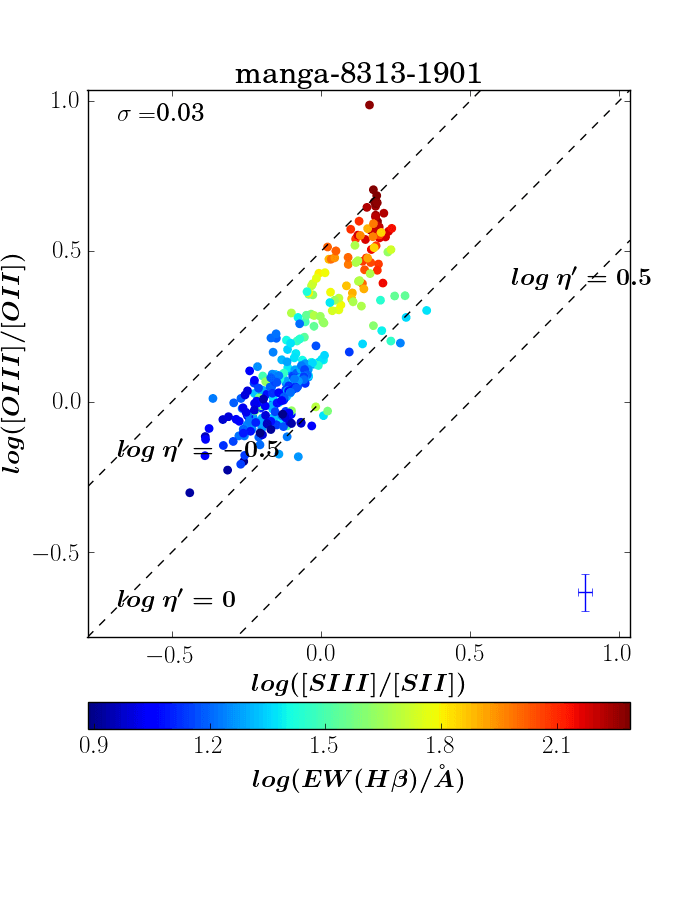}
	\includegraphics[width=0.185\textwidth, trim={2.1cm 2.0cm 1.5cm 1.0cm}, clip]{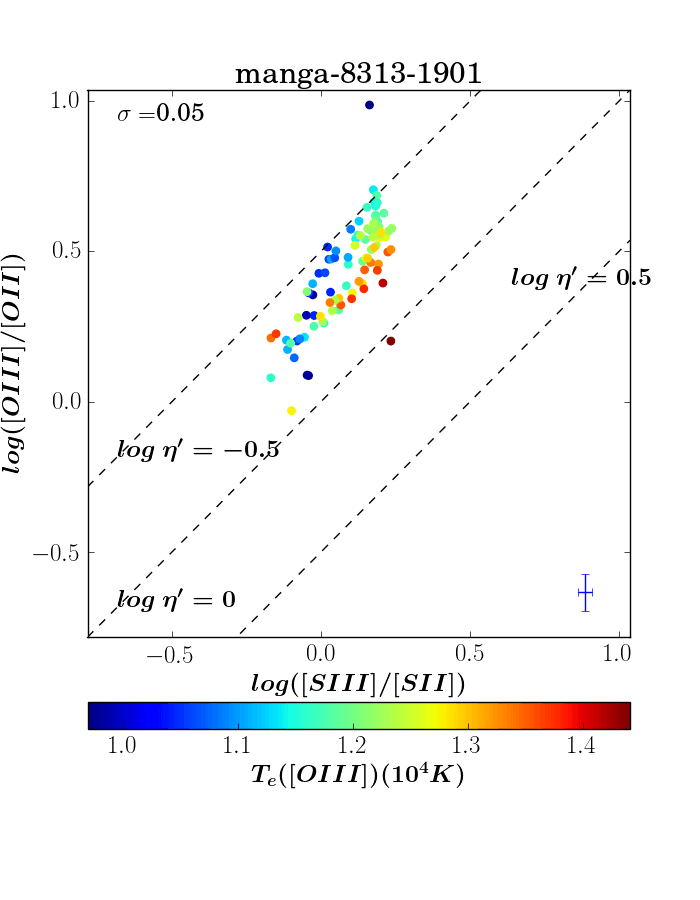}
	\caption{Manga-8313-1901 See caption of Figure \ref{fig:manga-7495-6102} for details.}
	\label{fig:manga-8313-1901}
\end{figure*}

\begin{figure*}
	\centering
	\includegraphics[width=0.185\textwidth, trim={4.0cm 0 4.0cm 0}, clip]{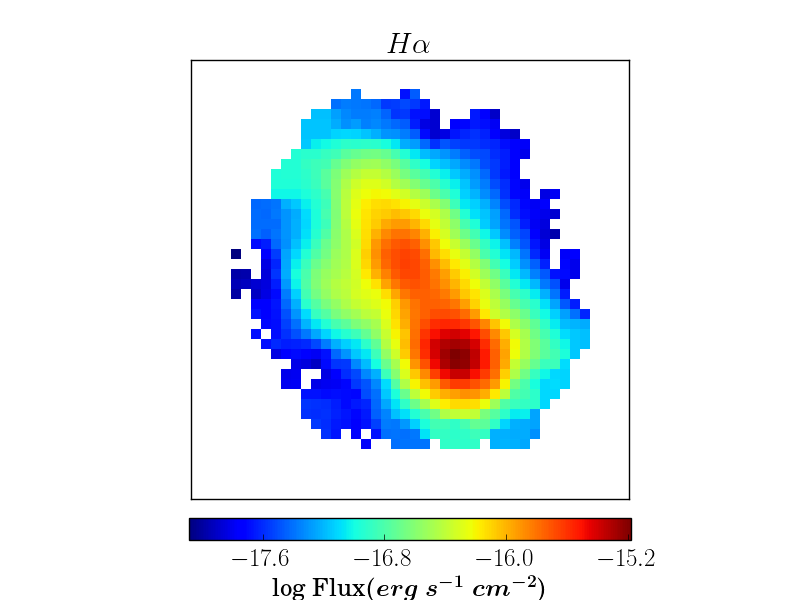}
	\includegraphics[width=0.185\textwidth, trim={4.0cm 0 4.0cm 0}, clip]{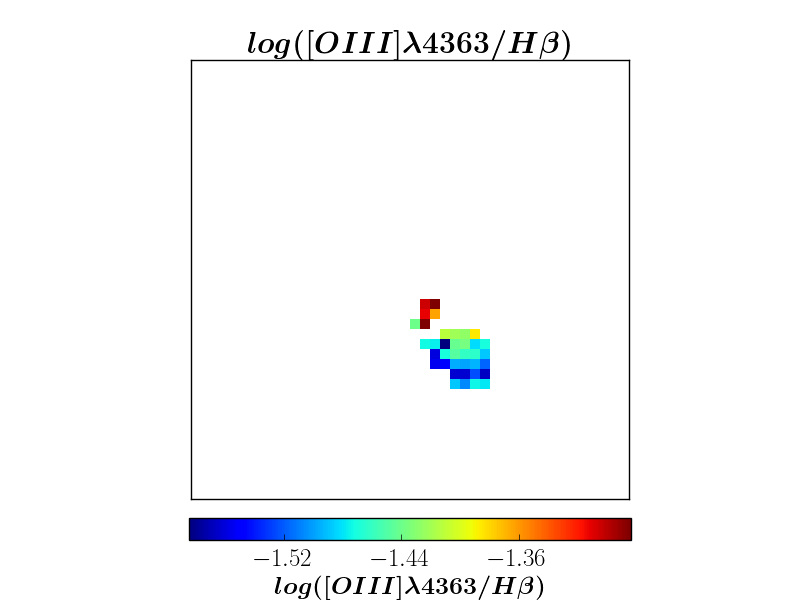}
	\includegraphics[width=0.185\textwidth, trim={4.0cm 0 4.0cm 0}, clip]{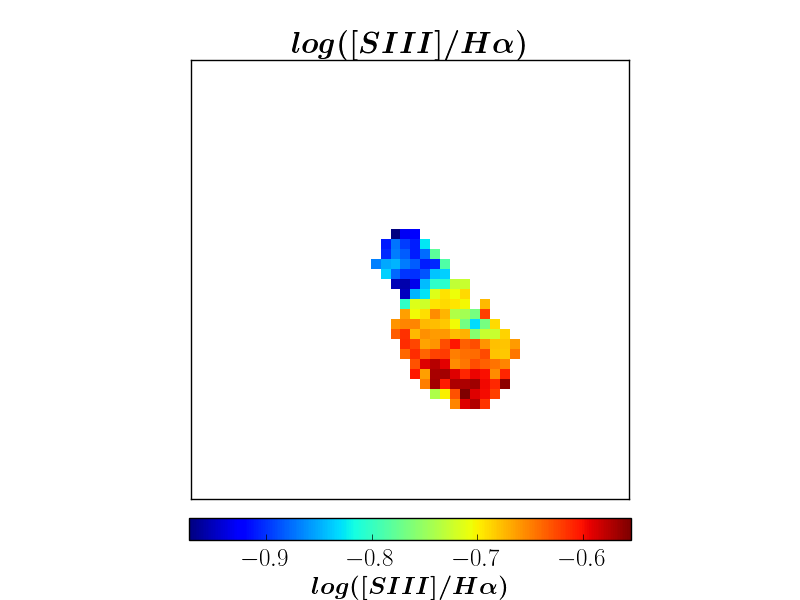}
	\includegraphics[width=0.185\textwidth, trim={4.0cm 0 4.0cm 0}, clip]{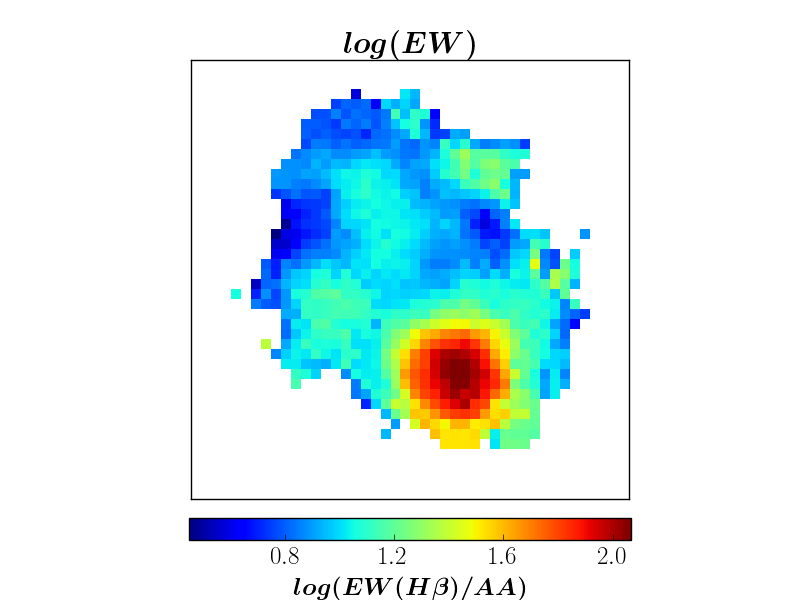}
	\includegraphics[width=0.185\textwidth, trim={4.0cm 0 4.0cm 0}, clip]{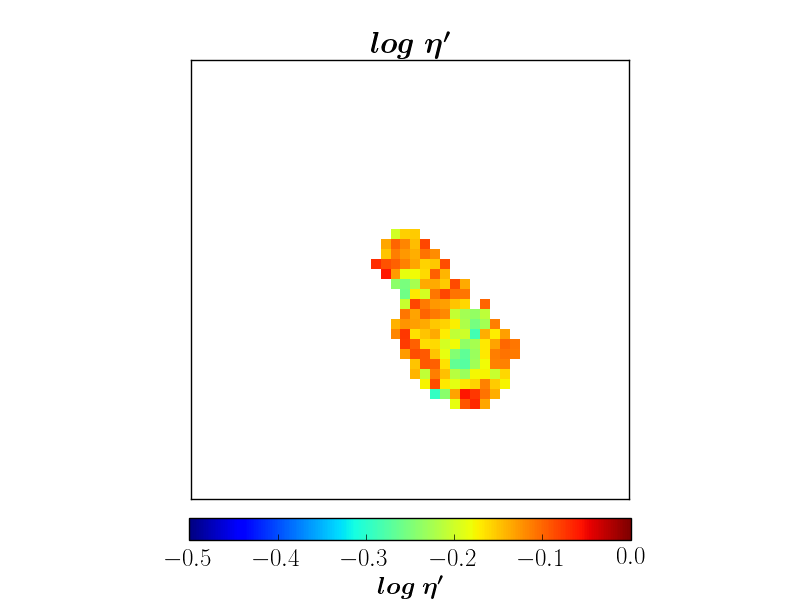}

	\includegraphics[width=0.21\textwidth, trim={0 2.0cm 1.5cm 1.0cm}, clip]{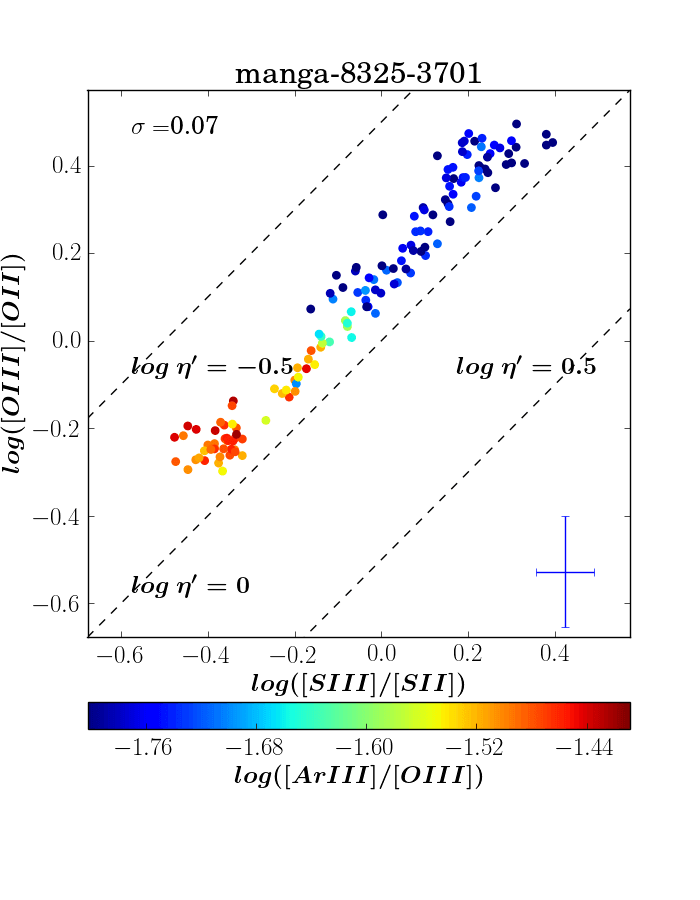}
	\includegraphics[width=0.185\textwidth, trim={2.1cm 2.0cm 1.5cm 1.0cm}, clip]{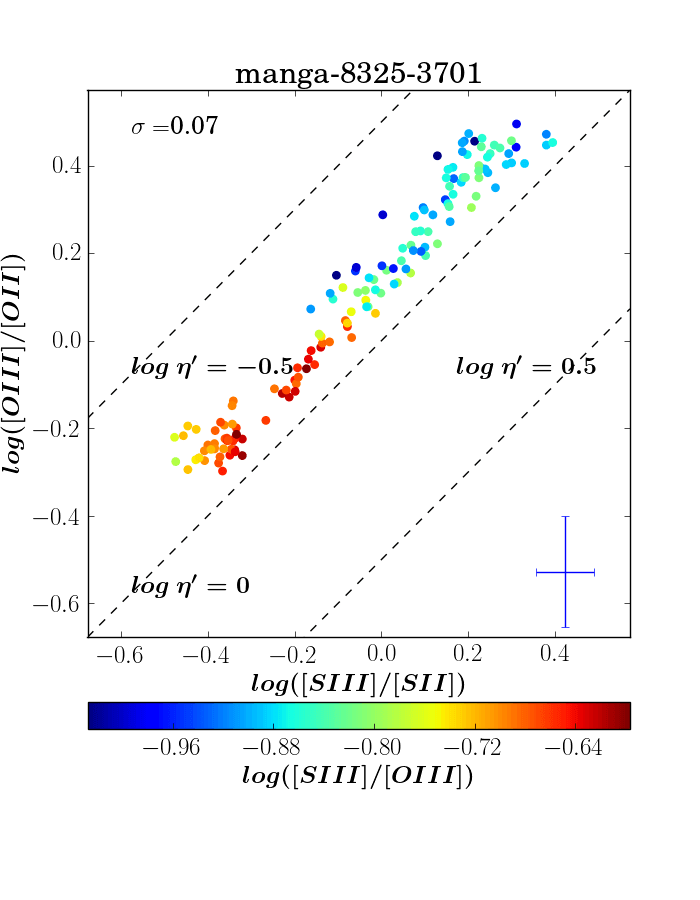}
	\includegraphics[width=0.185\textwidth, trim={2.1cm 2.0cm 1.5cm 1.0cm}, clip]{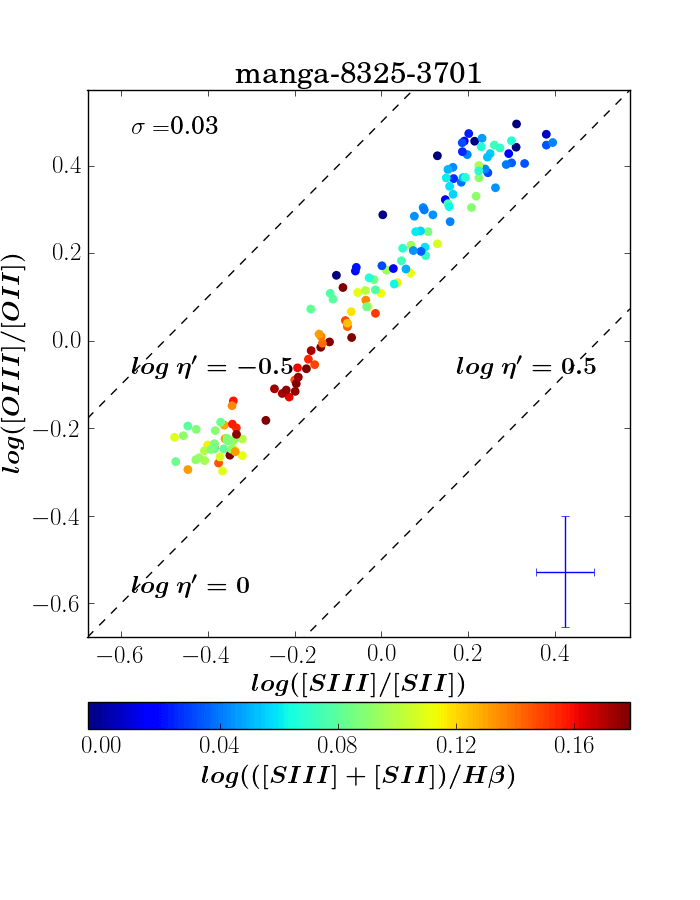}
	\includegraphics[width=0.185\textwidth, trim={2.1cm 2.0cm 1.5cm 1.0cm}, clip]{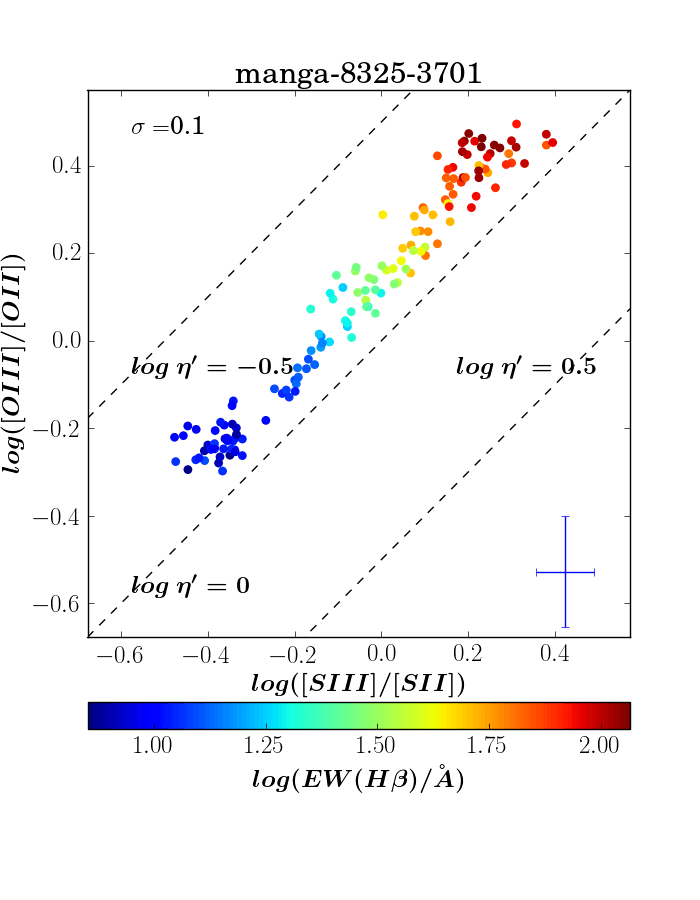}
	\includegraphics[width=0.185\textwidth, trim={2.1cm 2.0cm 1.5cm 1.0cm}, clip]{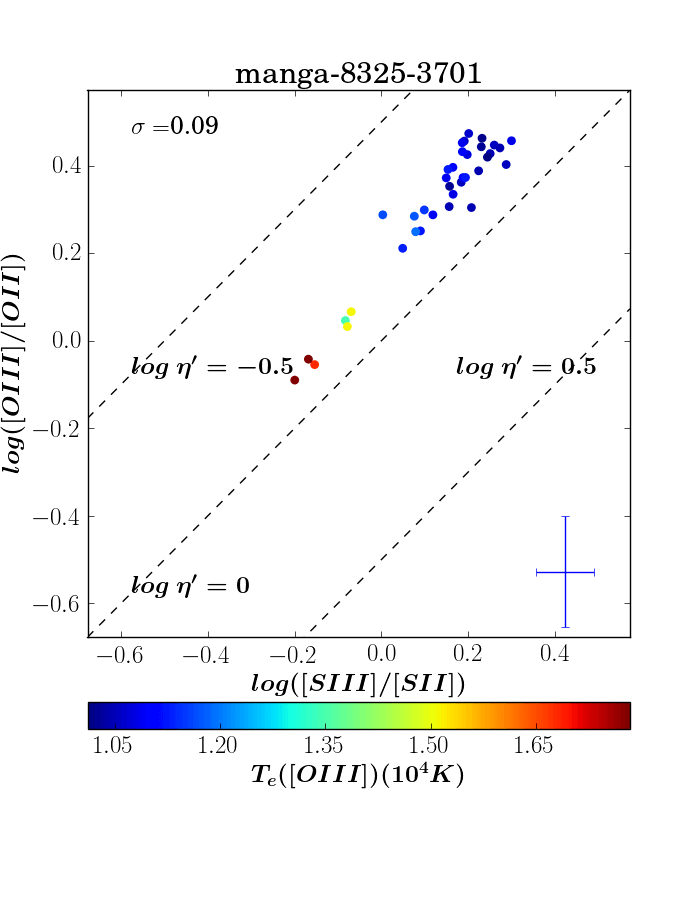}
	\caption{Manga-8325-3701 See caption of Figure \ref{fig:manga-7495-6102} for details.}
	\label{fig:manga-8325-3701}
\end{figure*}

\begin{figure*}
	\centering
	\includegraphics[width=0.185\textwidth, trim={4.0cm 0 4.0cm 0}, clip]{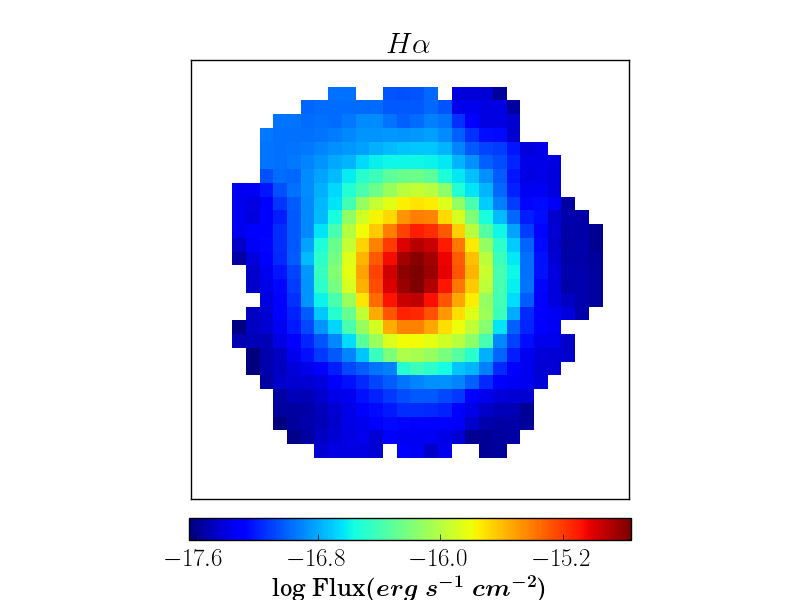}
	\includegraphics[width=0.185\textwidth, trim={4.0cm 0 4.0cm 0}, clip]{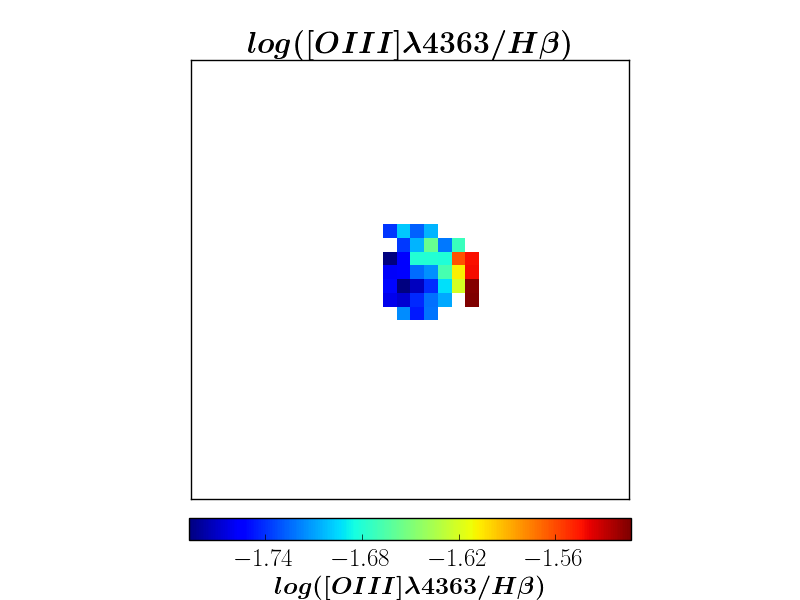}
	\includegraphics[width=0.185\textwidth, trim={4.0cm 0 4.0cm 0}, clip]{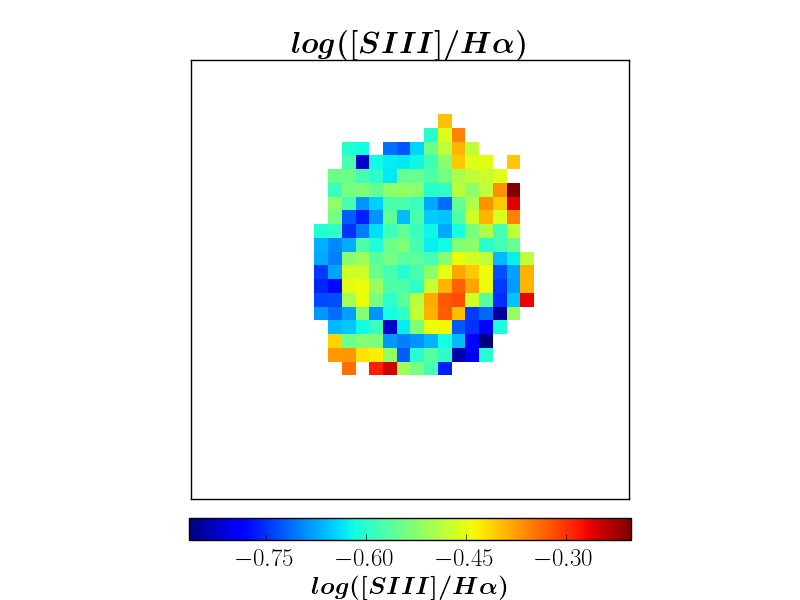}
	\includegraphics[width=0.185\textwidth, trim={4.0cm 0 4.0cm 0}, clip]{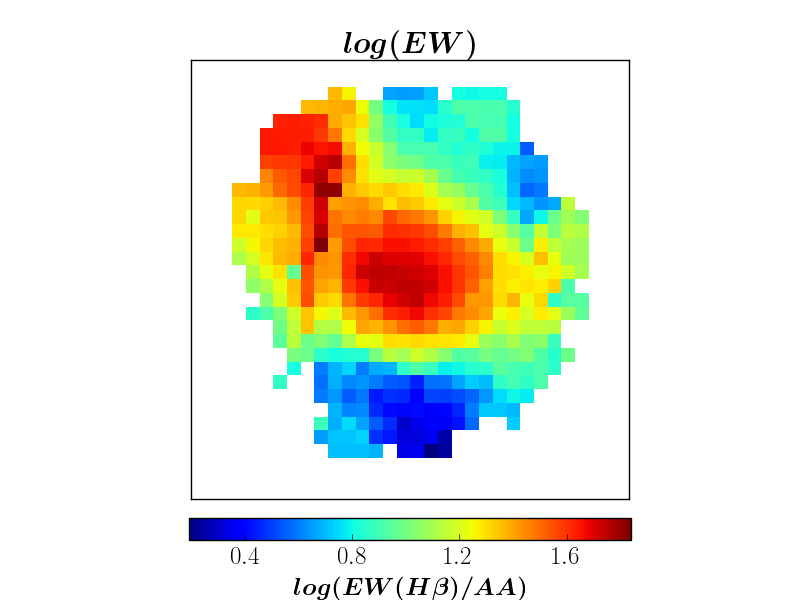}
	\includegraphics[width=0.185\textwidth, trim={4.0cm 0 4.0cm 0}, clip]{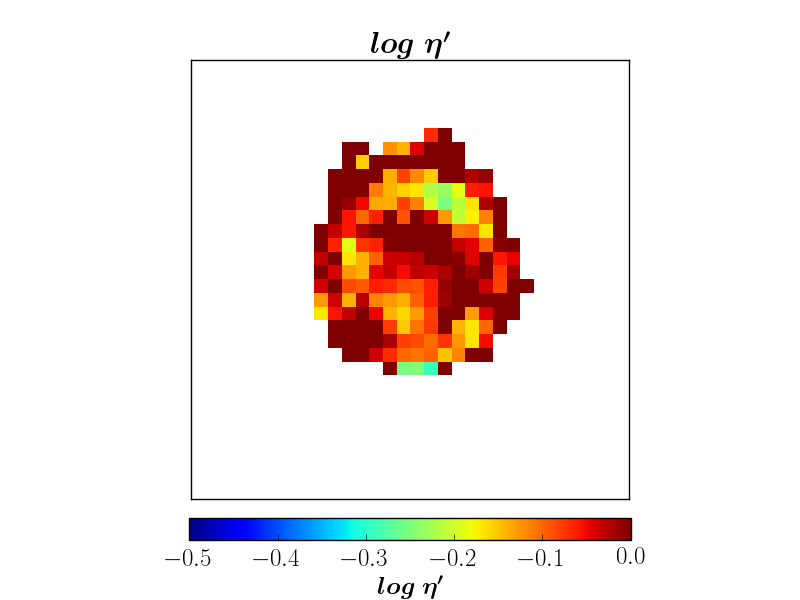}

	\includegraphics[width=0.21\textwidth, trim={0 2.0cm 1.5cm 1.0cm}, clip]{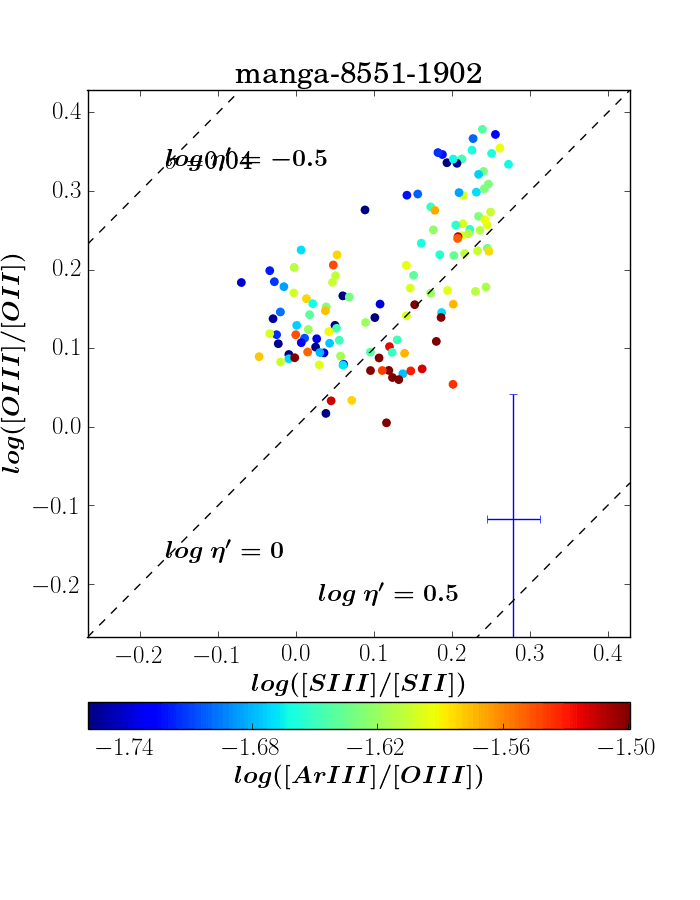}
	\includegraphics[width=0.185\textwidth, trim={2.1cm 2.0cm 1.5cm 1.0cm}, clip]{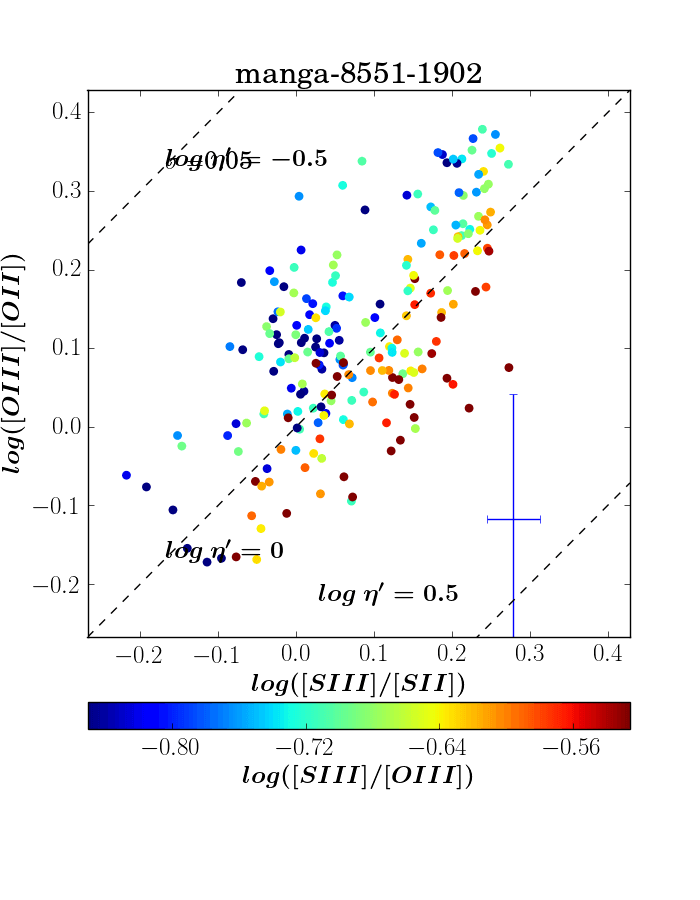}
	\includegraphics[width=0.185\textwidth, trim={2.1cm 2.0cm 1.5cm 1.0cm}, clip]{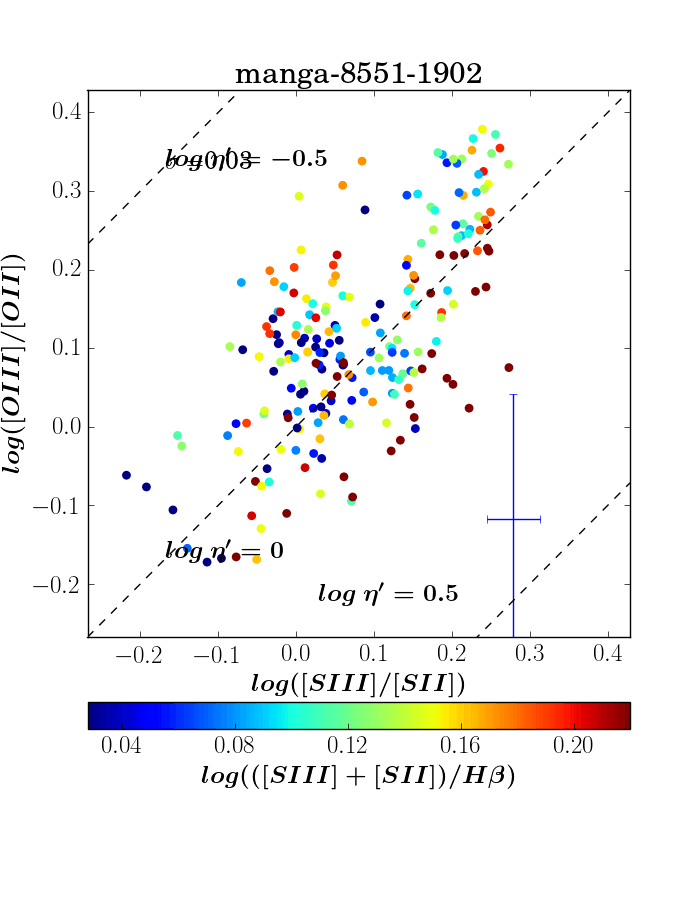}
	\includegraphics[width=0.185\textwidth, trim={2.1cm 2.0cm 1.5cm 1.0cm}, clip]{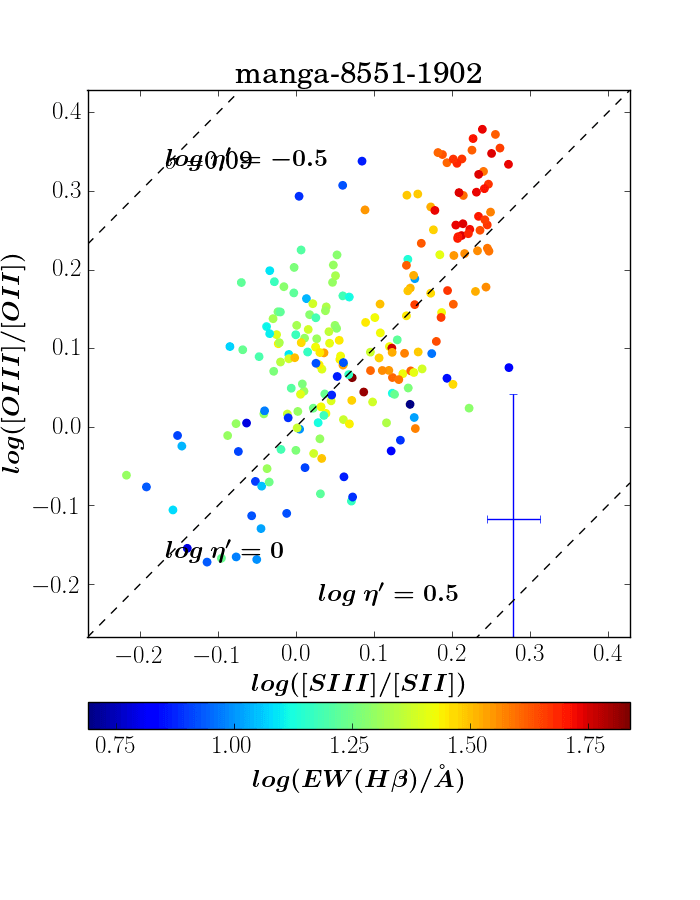}
	\includegraphics[width=0.185\textwidth, trim={2.1cm 2.0cm 1.5cm 1.0cm}, clip]{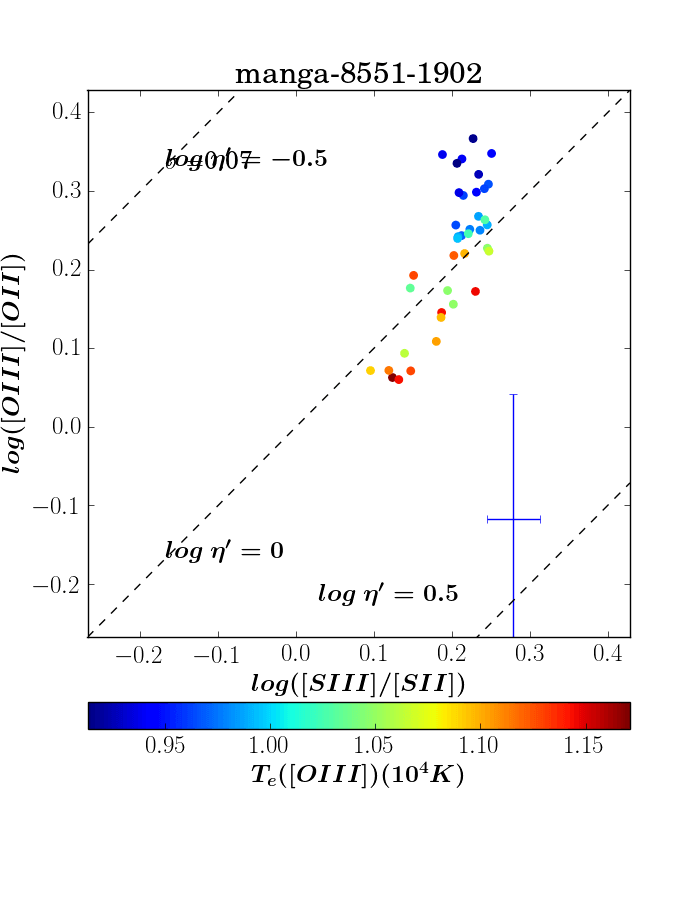}
	\caption{Manga-8551-1902 See caption of Figure \ref{fig:manga-7495-6102} for details.}
	\label{fig:manga-8551-1902}
\end{figure*}

\begin{figure*}
	\centering
	\includegraphics[width=0.185\textwidth, trim={4.0cm 0 4.0cm 0}, clip]{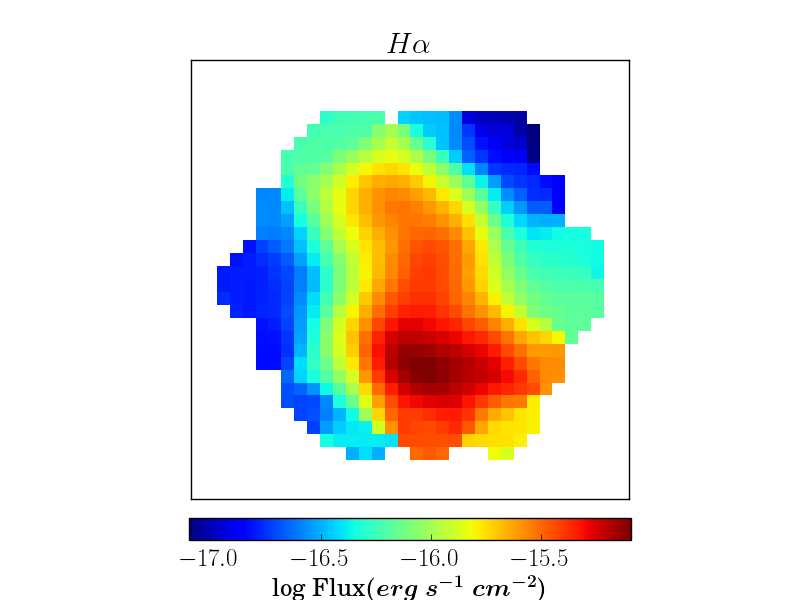}
	\includegraphics[width=0.185\textwidth, trim={4.0cm 0 4.0cm 0}, clip]{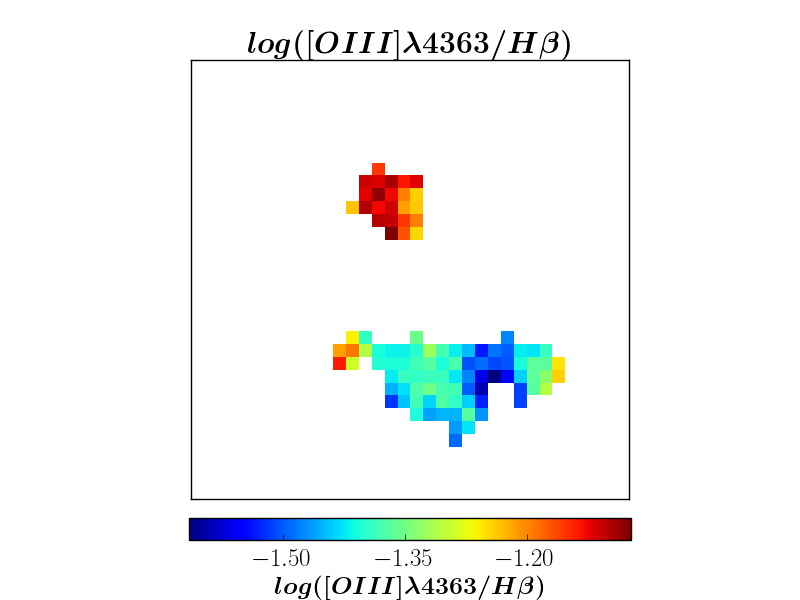}
	\includegraphics[width=0.185\textwidth, trim={4.0cm 0 4.0cm 0}, clip]{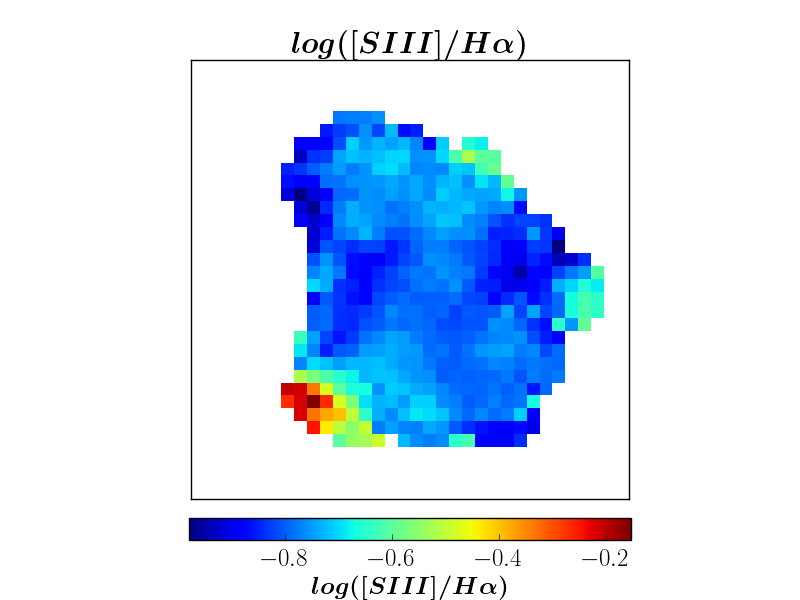}
	\includegraphics[width=0.185\textwidth, trim={4.0cm 0 4.0cm 0}, clip]{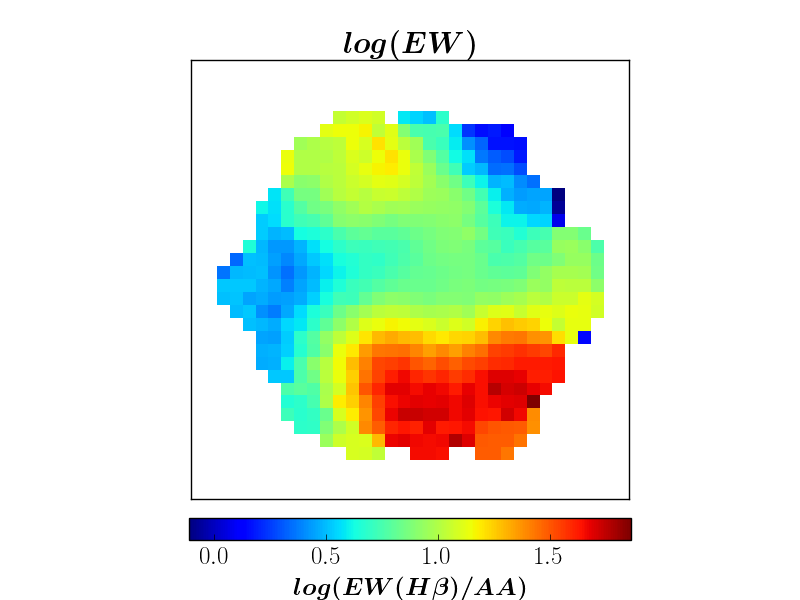}
	\includegraphics[width=0.185\textwidth, trim={4.0cm 0 4.0cm 0}, clip]{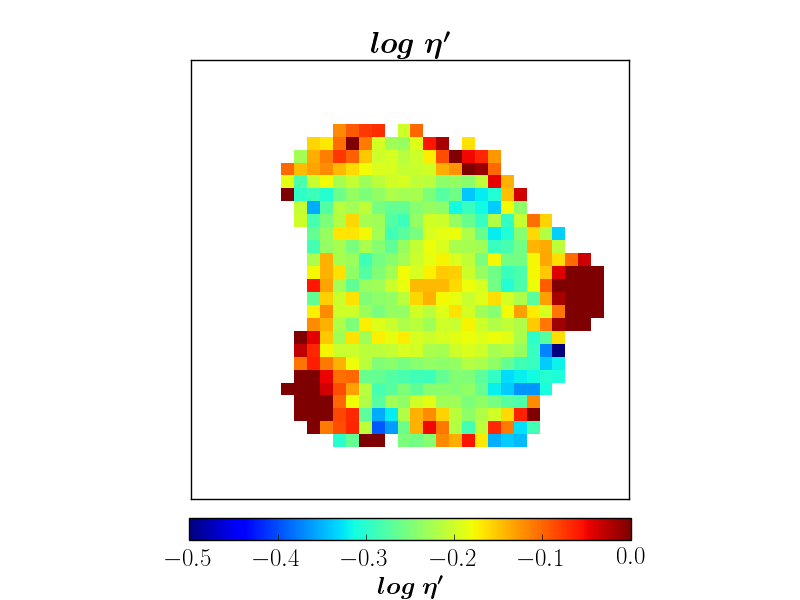}

	\includegraphics[width=0.21\textwidth, trim={0 2.0cm 1.5cm 1.0cm}, clip]{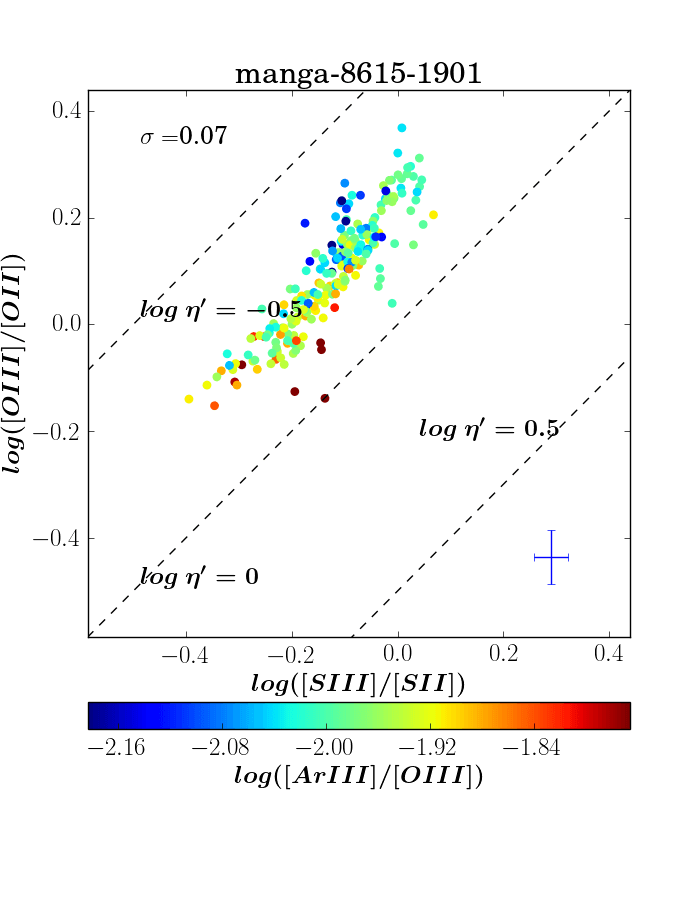}
	\includegraphics[width=0.185\textwidth, trim={2.1cm 2.0cm 1.5cm 1.0cm}, clip]{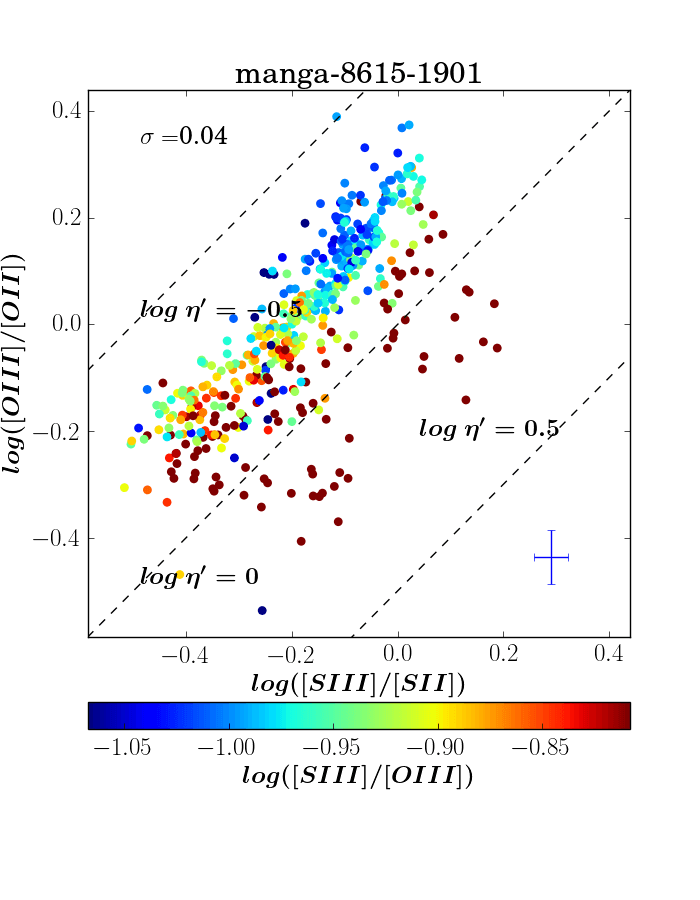}
	\includegraphics[width=0.185\textwidth, trim={2.1cm 2.0cm 1.5cm 1.0cm}, clip]{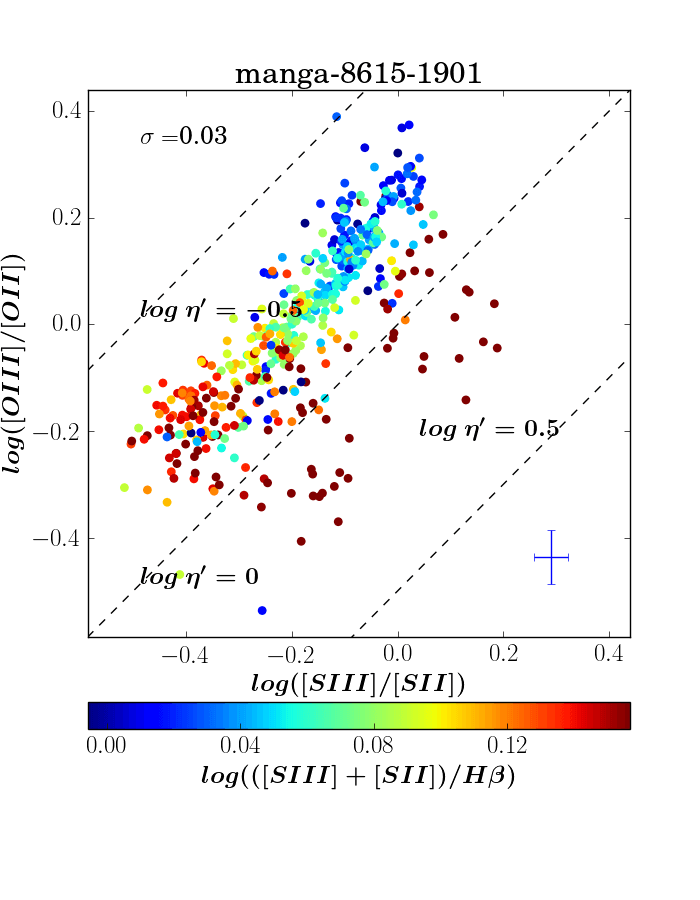}
	\includegraphics[width=0.185\textwidth, trim={2.1cm 2.0cm 1.5cm 1.0cm}, clip]{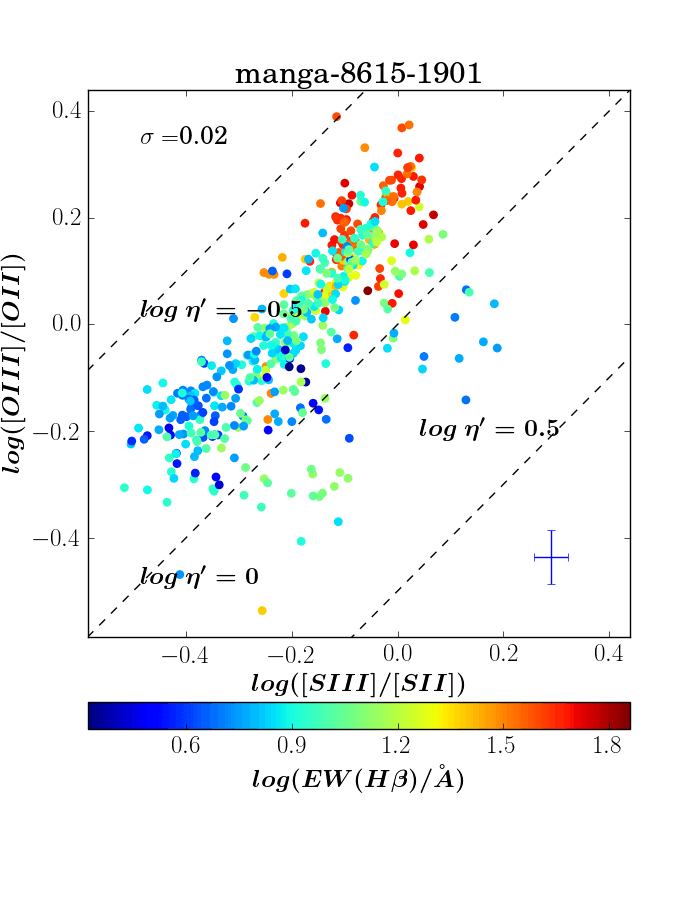}
	\includegraphics[width=0.185\textwidth, trim={2.1cm 2.0cm 1.5cm 1.0cm}, clip]{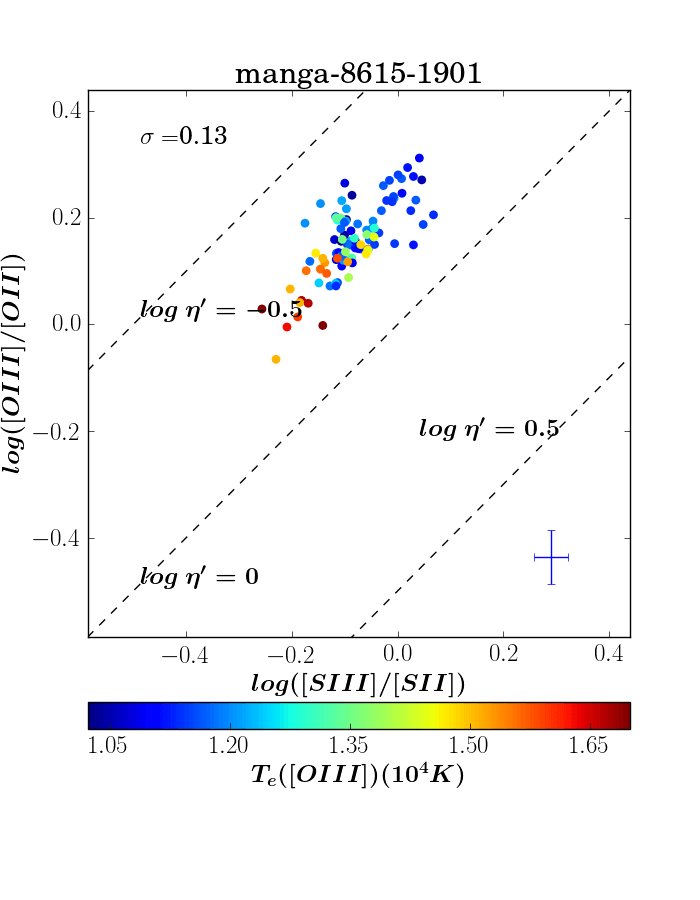}
	\caption{Manga-8615-1901 See caption of Figure \ref{fig:manga-7495-6102} for details.}
	\label{fig:manga-8615-1901}
\end{figure*}

\begin{figure*}
	\centering
	\includegraphics[width=0.185\textwidth, trim={4.0cm 0 4.0cm 0}, clip]{flux_Ha_manga-8626-12704.png}
	\includegraphics[width=0.185\textwidth, trim={4.0cm 0 4.0cm 0}, clip]{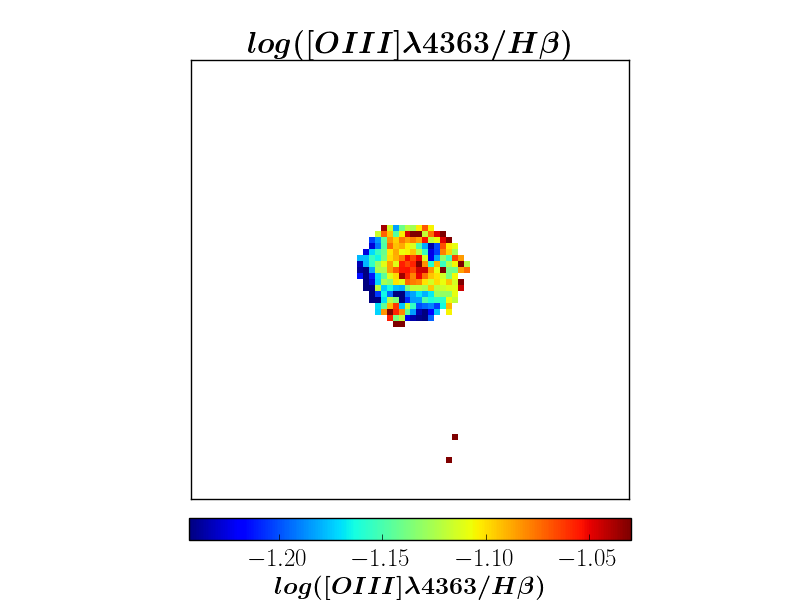}
	\includegraphics[width=0.185\textwidth, trim={4.0cm 0 4.0cm 0}, clip]{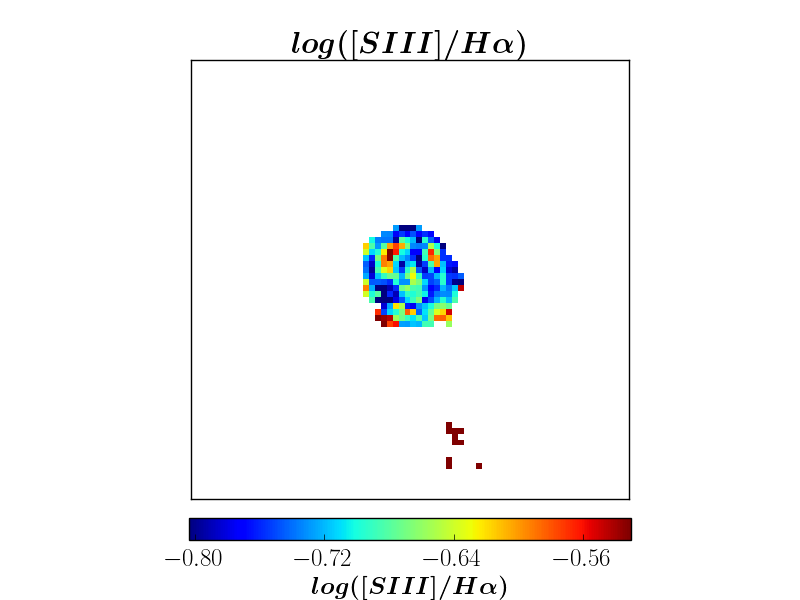}
	\includegraphics[width=0.185\textwidth, trim={4.0cm 0 4.0cm 0}, clip]{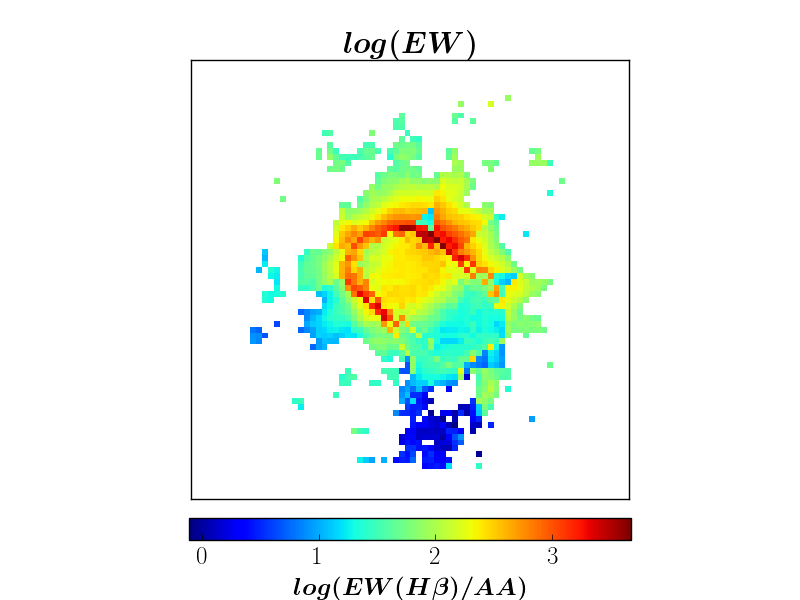}
	\includegraphics[width=0.185\textwidth, trim={4.0cm 0 4.0cm 0}, clip]{eta_p_manga-8626-12704.png}

	\includegraphics[width=0.21\textwidth, trim={0 2.0cm 1.5cm 1.0cm}, clip]{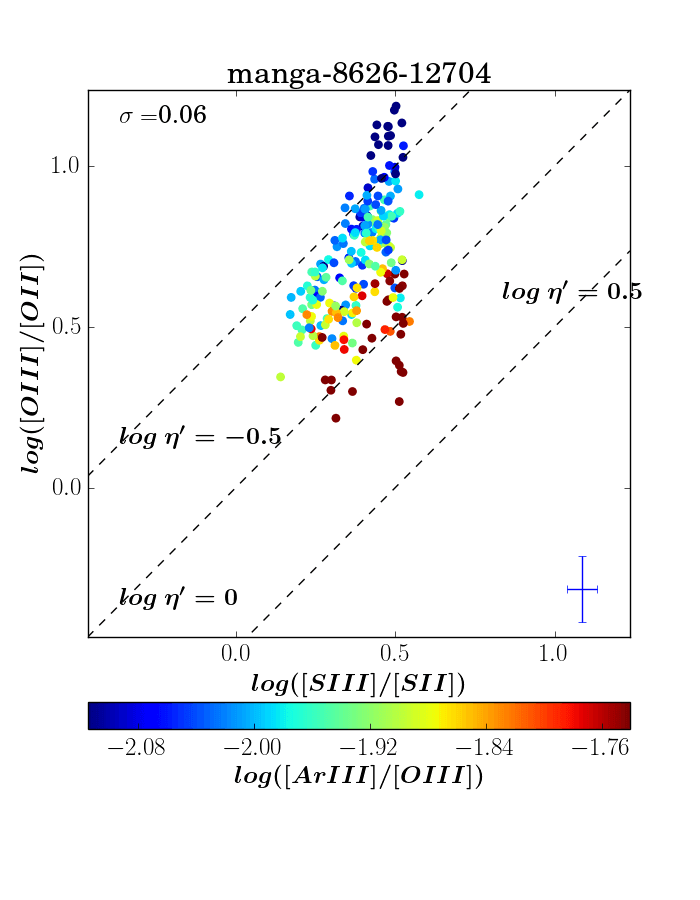}
	\includegraphics[width=0.185\textwidth, trim={2.1cm 2.0cm 1.5cm 1.0cm}, clip]{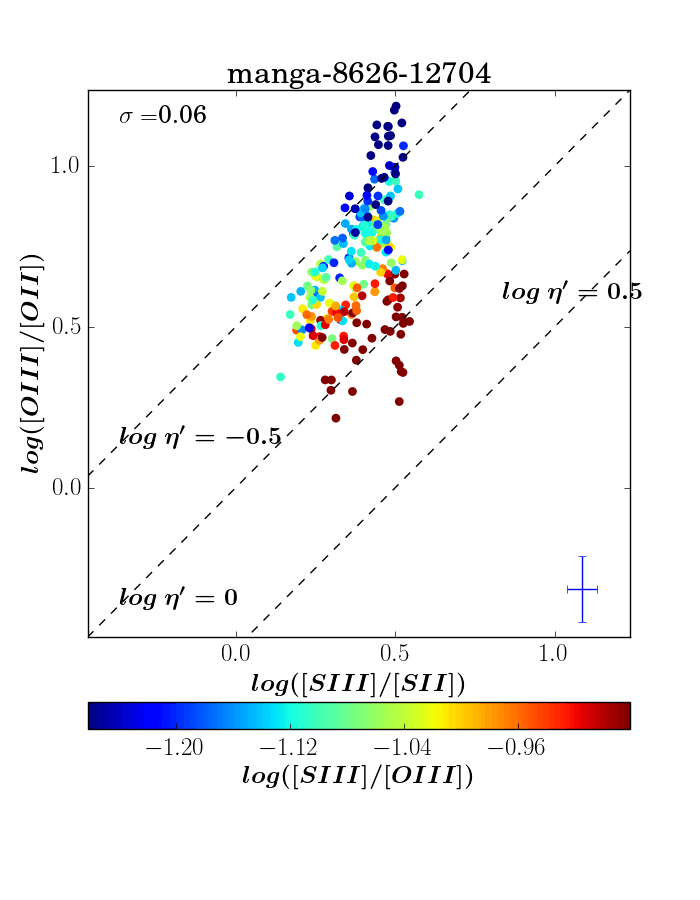}
	\includegraphics[width=0.185\textwidth, trim={2.1cm 2.0cm 1.5cm 1.0cm}, clip]{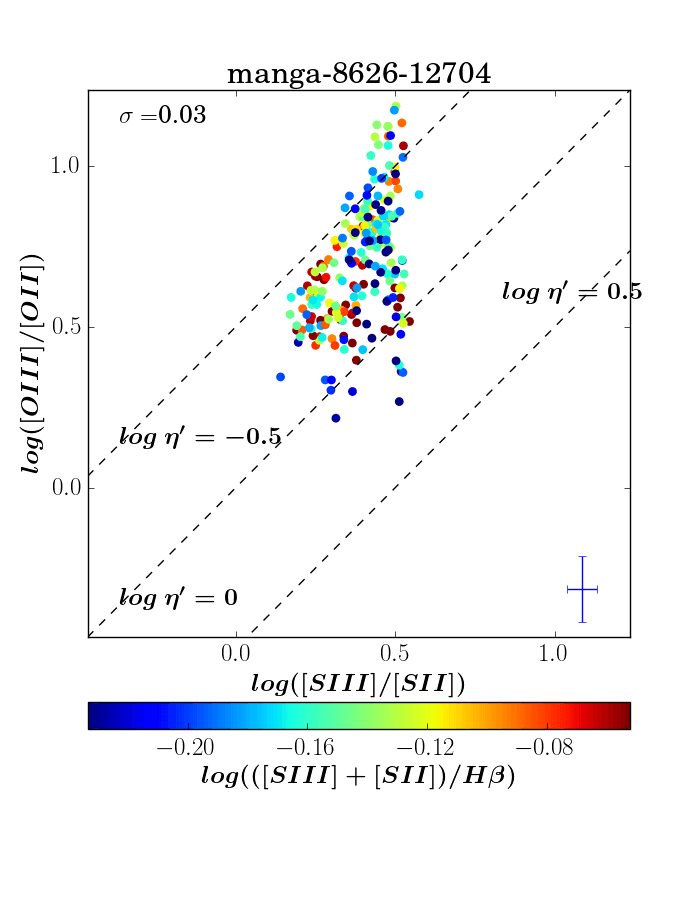}
	\includegraphics[width=0.185\textwidth, trim={2.1cm 2.0cm 1.5cm 1.0cm}, clip]{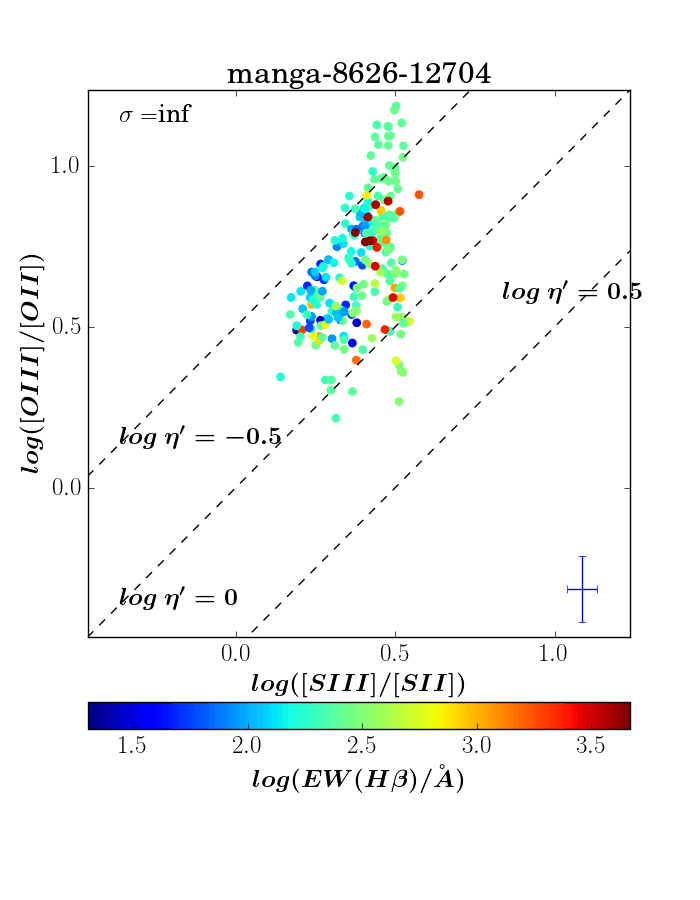}
	\includegraphics[width=0.185\textwidth, trim={2.1cm 2.0cm 1.5cm 1.0cm}, clip]{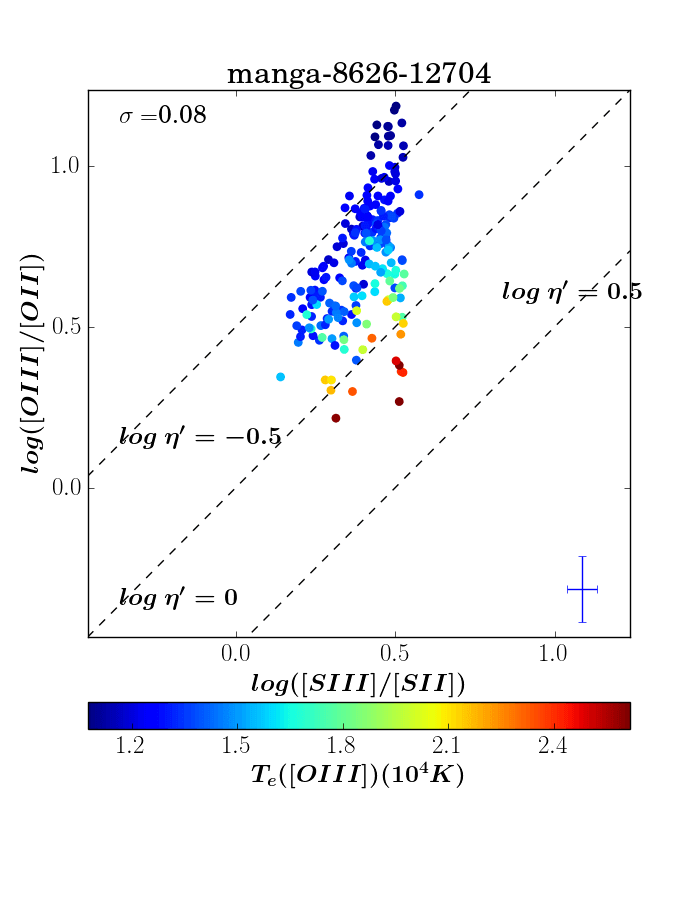}
	\caption{Manga-8626-12704 See caption of Figure \ref{fig:manga-7495-6102} for details}
	\label{fig:manga-8626-12704}
\end{figure*}

\begin{figure*}
	\centering
	\includegraphics[width=0.185\textwidth, trim={4.0cm 0 4.0cm 0}, clip]{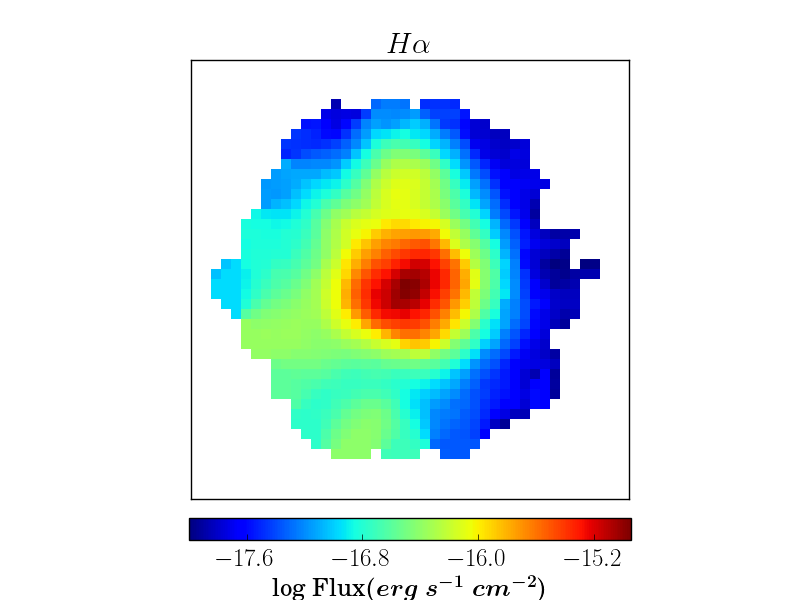}
	\includegraphics[width=0.185\textwidth, trim={4.0cm 0 4.0cm 0}, clip]{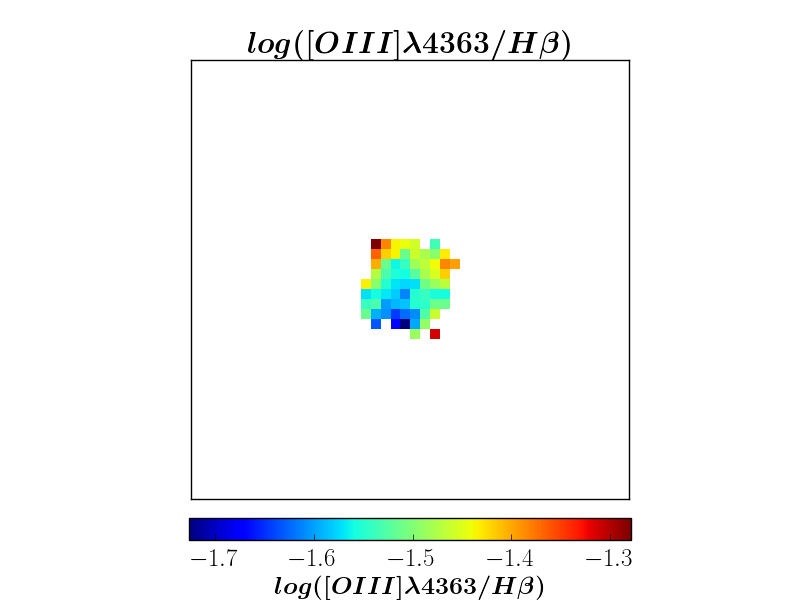}
	\includegraphics[width=0.185\textwidth, trim={4.0cm 0 4.0cm 0}, clip]{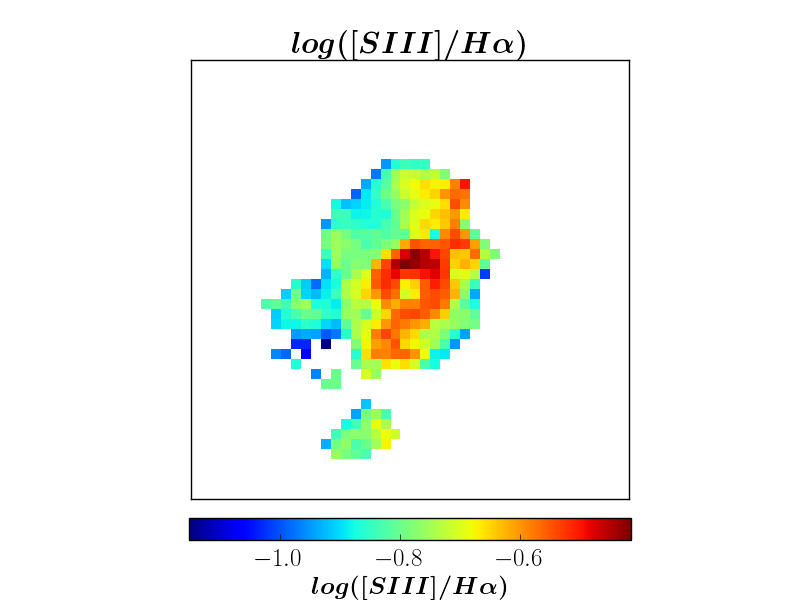}
	\includegraphics[width=0.185\textwidth, trim={4.0cm 0 4.0cm 0}, clip]{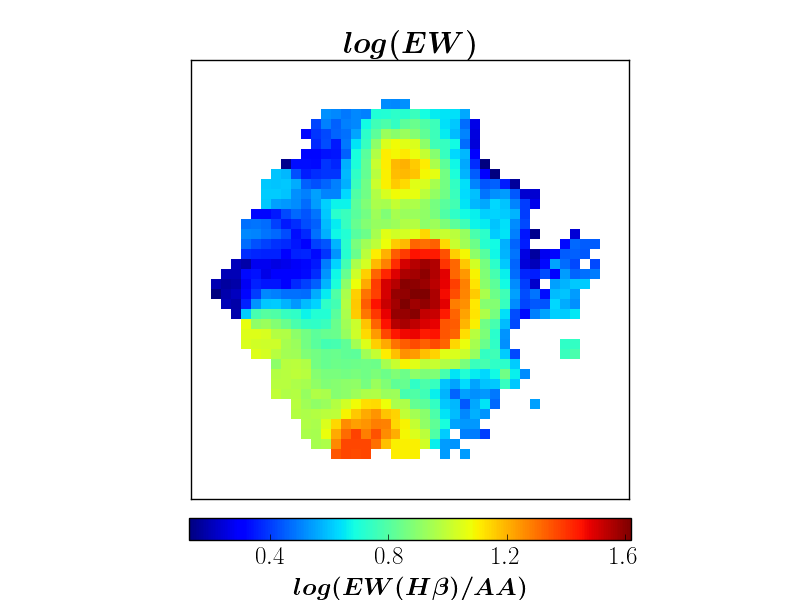}
	\includegraphics[width=0.185\textwidth, trim={4.0cm 0 4.0cm 0}, clip]{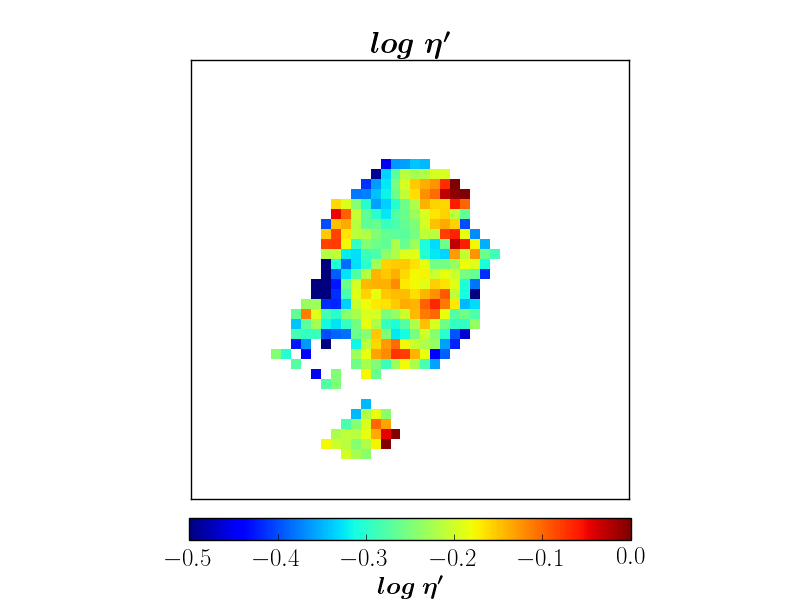}

	\includegraphics[width=0.21\textwidth,  trim={0 2.0cm 1.5cm 1.0cm}, clip]{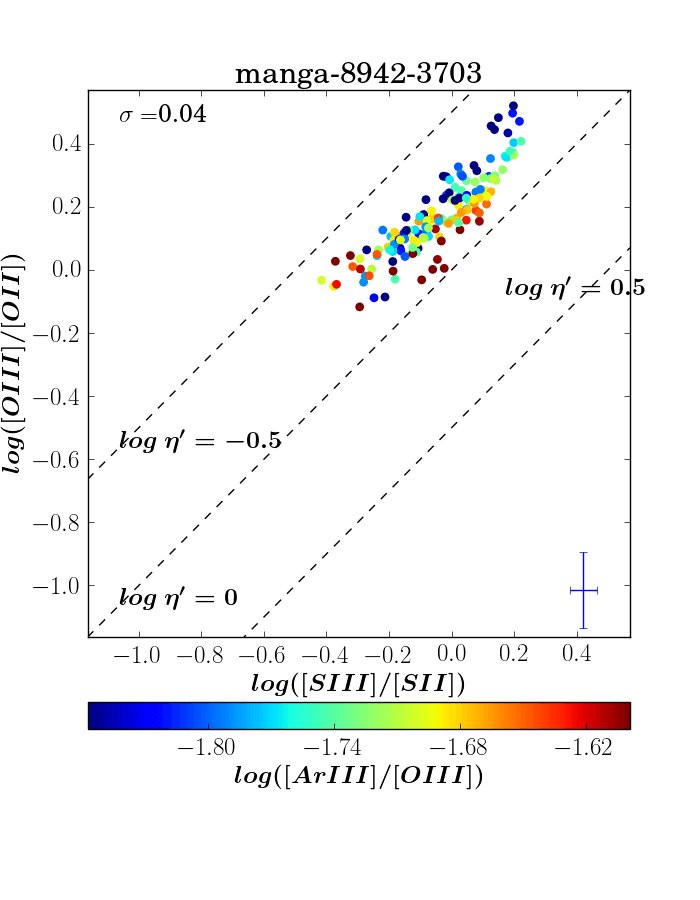}
	\includegraphics[width=0.185\textwidth, trim={2.1cm 2.0cm 1.5cm 1.0cm}, clip]{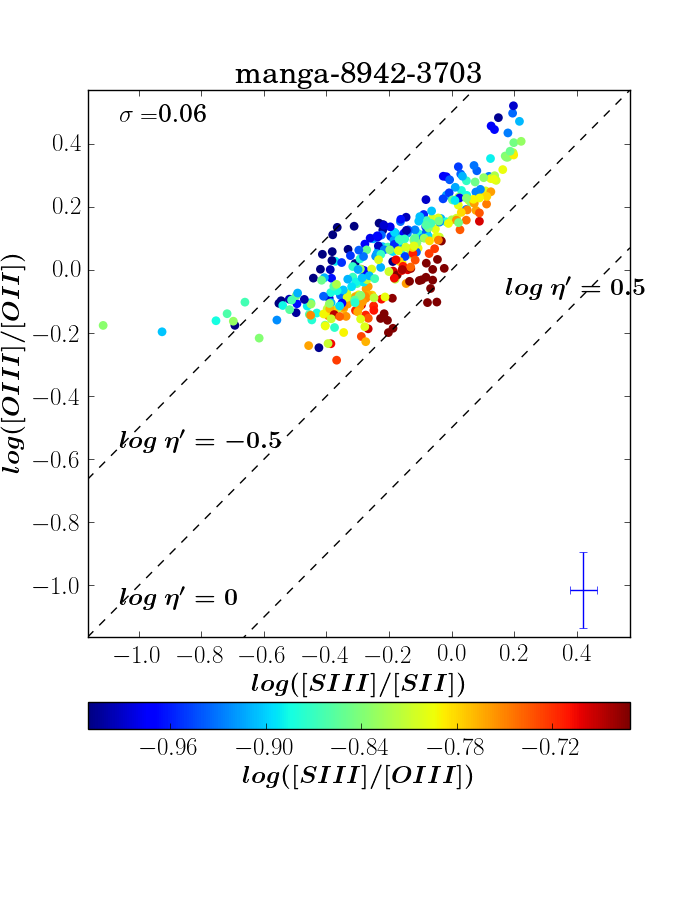}
	\includegraphics[width=0.185\textwidth, trim={2.1cm 2.0cm 1.5cm 1.0cm}, clip]{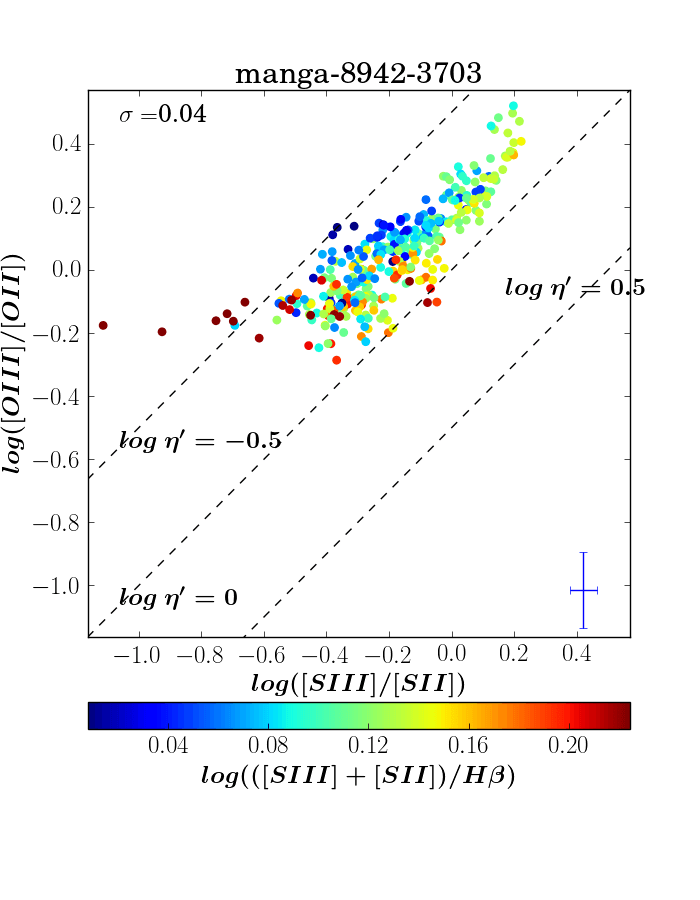}
	\includegraphics[width=0.185\textwidth, trim={2.1cm 2.0cm 1.5cm 1.0cm}, clip]{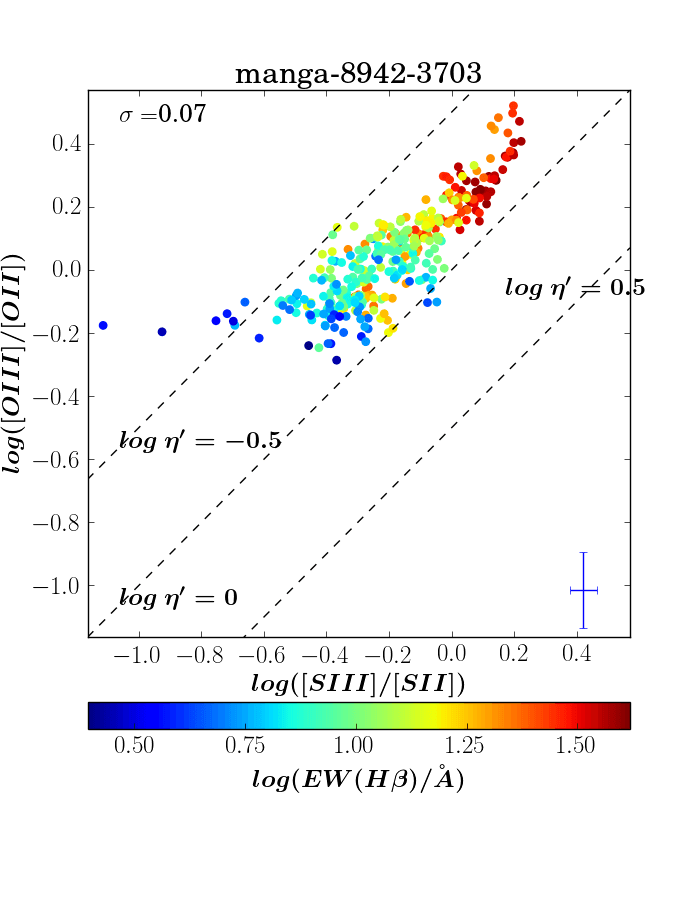}
	\includegraphics[width=0.185\textwidth, trim={2.1cm 2.0cm 1.5cm 1.0cm}, clip]{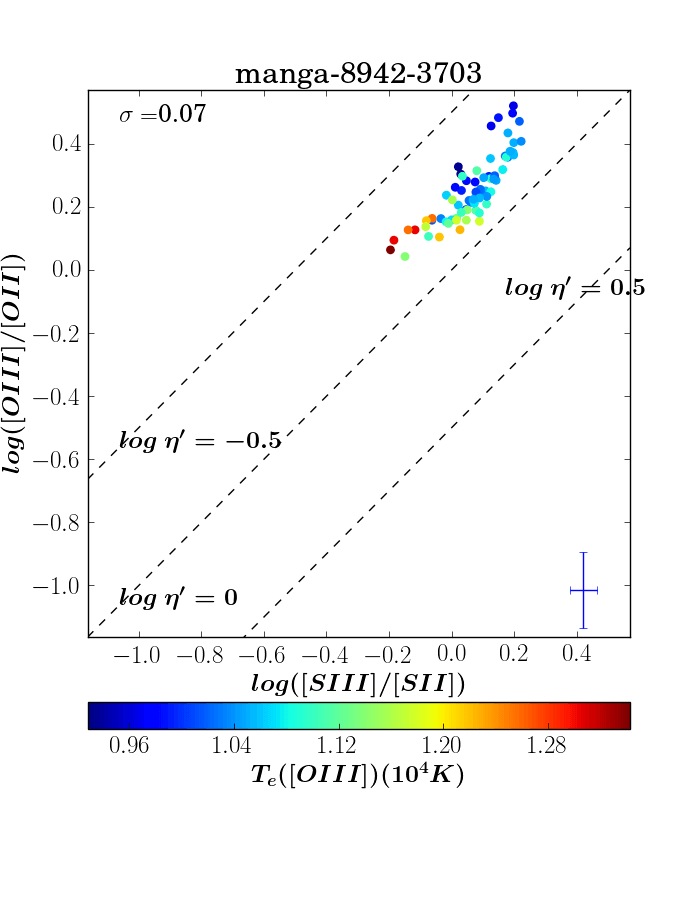}
	\caption{Manga-8942-3703 See caption of Figure \ref{fig:manga-7495-6102} for details.}
	\label{fig:manga-8942-3703}
\end{figure*}

\indent Figures \ref{fig:manga-7495-6102}--\ref{fig:manga-8942-3703} show the spatially-resolved maps of several properties of galaxies with extended [O \textsc{iii}] $\lambda$4363 detection. We have specifically chosen to show these data set because [O \textsc{iii}] $\lambda$4363 detection allow us to determine T$_e$ and hence log $\rm\eta$ which is the proxy for radiation hardness. In addition, these figures also show the O$_3$O$_2$--S$_3$S$_2$ plane for each galaxies colour-coded with respect to parameters sensitive to abundance, T$_e$, and stellar age.


\bsp	
\label{lastpage}
\end{document}